\documentclass[acmsmall]{acmart}

\setcopyright{rightsretained}
\acmDOI{10.1145/3643777}
\acmYear{2024}
\copyrightyear{2024}
\acmSubmissionID{fse24main-p1391-p}
\acmJournal{PACMSE}
\acmVolume{1}
\acmNumber{FSE}
\acmArticle{51}
\acmMonth{7}
\received{2023-09-29}
\received[accepted]{2024-01-23}

\usepackage{xspace}
\usepackage{graphicx}
\usepackage{xcolor}
\usepackage[normalem]{ulem} %
\usepackage{amsmath}
\usepackage{enumitem}
\usepackage{semantic} %
\usepackage{siunitx}
\usepackage{scalefnt}
\usepackage{multicol}

\usepackage{hyperref}
\usepackage{breakurl}

\usepackage{booktabs}
\usepackage{multirow}
\usepackage{makecell}
\usepackage{ragged2e}

\usepackage{caption}
\usepackage{subcaption}
\usepackage{wrapfig}
\usepackage{adjustbox}

\usepackage{pifont}
\usepackage{upquote} %
\usepackage{listings}

\usepackage{algorithm}
\usepackage[noend]{algpseudocode} %

\usepackage{tikz}
\usepackage{forest} %
\usetikzlibrary{arrows.meta}
\usetikzlibrary{backgrounds}
\usetikzlibrary{calc}
\usetikzlibrary{decorations.pathreplacing}
\usetikzlibrary{fit}
\usetikzlibrary{positioning}
\usetikzlibrary{shapes.geometric}

\usepackage{flushend} %
\usepackage{datetime}
\usepackage{fmtcount} %

\newcommand{\XSpace}[1]{}
\newcommand{\XComment}[1]{}
\newcommand{\Fix}[1]{\textcolor{red}{#1}}
\newcommand{\mg}[1]{\textcolor{purple}{mg: #1}}
\newcommand{\DefMacro}[2]{\expandafter\newcommand\csname rmk-#1\endcsname{#2}}
\newcommand{\UseMacro}[1]{\csname rmk-#1\endcsname}

\newcommand{\MyPara}[1]{\vspace{2pt}\noindent\textbf{#1}.}
\newcommand{\MyParaOnly}[1]{\noindent\textbf{#1}}
\newcommand{\reducedstrut}{\vrule width 0pt height .9\ht\strutbox depth .9\dp\strutbox\relax}
\newcommand{\InputWithSpace}[1]{\bgroup\def\arraystretch{1.0}\input{#1}\egroup}
\newcommand{\Code}[1]{{\ifmmode{\mathtt{#1}}\else$\mathtt{#1}$\fi}}
\newcommand{\CodeIn}[1]{\texttt{\small #1}}
\newcommand{\CodeInAlgo}[1]{\texttt{\scriptsize #1}}
\newcommand{\ColorBack}[1]{%
\begingroup
\setlength{\fboxsep}{0pt}%
\colorbox{purple!20}{\reducedstrut#1\/}%
\endgroup
}
\newcommand{\BigNum}[1]{\num[group-separator = {,}, group-four-digits]{#1}} %
\newcommand{\Num}[1]{#1}    %
\newcommand{\Hrule}[1][gray]{%
\par\addvspace{1pt}%
\begingroup\color{#1}%
\hrule
\endgroup
\addvspace{1pt}%
}

\newcommand{\specialcell}[2][c]{%
\begin{tabular}[#1]{@{}c@{}}#2\end{tabular}}
\newcolumntype{R}[1]{>{\RaggedLeft\arraybackslash}p{#1}}
\newcolumntype{L}[1]{>{\RaggedRight\arraybackslash}p{#1}}

\newcommand{\ltrue}{\top} %
\newcommand{\lfalse}{\bot} %

\newcommand{\limply}{\rightarrow} %
\newcommand{\liff}{\leftrightarrow} %
\newcommand{\lcofactor}{\downarrow} %

\newcommand{\llimply}{\Rightarrow} %
\newcommand{\lliff}{\Leftrightarrow} %
\newcommand{\lsat}{\vDash} %
\newcommand{\lnsat}{\nvDash} %

\newcommand{\lqall}[1]{\forall #1.~} %
\newcommand{\lqexist}[1]{\exists #1.~} %
\newcommand{\lqallOnly}[1]{\forall #1 } %
\newcommand{\lqexistOnly}[1]{\exists #1 } %

\newcommand{\lAnd}[1]{\bigwedge_{\substack{#1}}} %
\newcommand{\lOr}[1]{\bigvee_{\substack{#1}}} %

\newcommand{\set}[1]{\{#1\}} %
\newcommand{\Set}[1]{\left\{#1\right\}} %
\newcommand{\tuple}[1]{\langle#1\rangle} %
\newcommand{\Tuple}[1]{\left\langle#1\right\rangle} %

\newcommand{\sunion}{\cup} %
\newcommand{\sinter}{\cap} %

\definecolor{gray}{RGB}{211,211,211}
\newcommand{\jbasicstyle}{\small\sffamily} %
\newcommand{\textcode}[1]{{#1}}
\newcommand{\jnumberstyle}{\scriptsize}
\newcommand{\Hilight}{\makebox[0pt][l]{\color{gray}\rule[-3pt]{0.80\linewidth}{9pt}}}
\newcommand{\lstbg}[3][0pt]{{\fboxsep#1\colorbox{#2}{\strut #3}}}

\let\origthelstnumber\thelstnumber
\makeatletter
\newcommand*\Suppressnumber{%
\lst@AddToHook{OnNewLine}{%
\let\thelstnumber\relax%
\advance\c@lstnumber-\@ne\relax%
}%
}
\newcommand*\Reactivatenumber[1]{%
\setcounter{lstnumber}{\numexpr#1-1\relax}
\lst@AddToHook{OnNewLine}{%
\let\thelstnumber\origthelstnumber%
\refstepcounter{lstnumber}
}%
}
\makeatother

\makeatletter
\lst@Key{countblanklines}{true}[t]%
{\lstKV@SetIf{#1}\lst@ifcountblanklines}

\lst@AddToHook{OnEmptyLine}{%
\lst@ifnumberblanklines\else%
\lst@ifcountblanklines\else%
\advance\c@lstnumber-\@ne\relax%
\fi%
\fi}
\makeatother

\lstdefinelanguage{bash}
{
language=sh,
numbers=none,
basicstyle=\footnotesize\ttfamily,
breaklines=true,
columns=fullflexible,
showstringspaces=false,
keywordstyle={},
escapeinside={(*@}{@*)}
}

\lstdefinelanguage{bug}
{
language=java-pretty,
numbers=left,
basicstyle=\scriptsize\ttfamily,
numberstyle=\scriptsize,
numberblanklines=false,
countblanklines=true,
numbersep=0.5em,
xleftmargin=16pt,
}

\lstdefinelanguage{example}
{
language=jattack,
numbers=left,
basicstyle=\footnotesize\ttfamily,
numberstyle=\footnotesize,
numberblanklines=false,
countblanklines=true,
numbersep=0.5em,
xleftmargin=16pt,
}

\lstdefinelanguage{example-half-width}
{
language=jattack,
numbers=left,
basicstyle=\scriptsize\ttfamily,
numberstyle=\scriptsize,
numberblanklines=false,
countblanklines=true,
numbersep=0.5em,
xleftmargin=16pt,
}

\lstdefinelanguage{jattack-display}
{
language=jattack,
numbers=none,
basicstyle=\footnotesize\ttfamily,
numberstyle=\footnotesize,
xleftmargin=0pt
}

\lstdefinelanguage{jattack}
{
language=java,
numbers=left,
basicstyle=\scriptsize\ttfamily,
numberstyle=\scriptsize,
breaklines=true,
columns=fullflexible,
xleftmargin=16pt,
showstringspaces=false,
emph={intId,intVal,doubleId,doubleVal,charVal,arithmetic,relation,logic,refId,cast,charArrAcc,@Entry,@Argument,@Arguments}, emphstyle=\color{blue},
emph={[2]eval}, emphstyle={[2]\color{purple}\bfseries},
escapeinside={(*@}{@*)}
}

\lstdefinelanguage{syntax}
{
numbers=none,
basicstyle=\footnotesize\ttfamily,
breaklines=true,
columns=fullflexible,
xleftmargin=16pt,
showstringspaces=false,
}

\lstdefinelanguage{semantics}
{
morekeywords={if,else,where}, keywordstyle=\bfseries,
numbers=none,
basicstyle=\footnotesize\ttfamily,
breaklines=true,
columns=fullflexible,
xleftmargin=16pt,
showstringspaces=false,
}

\lstdefinelanguage{api-list}
{
language=java,
numbers=none,
basicstyle=\footnotesize\ttfamily,
breaklines=true,
columns=fullflexible,
showstringspaces=false,
emph={boolVal,intVal,boolId,intId,arithmetic,relation,logic,alt,refId,exprStmt,ifStmt,whileStmt,tryStmt,assign}, emphstyle=\color{blue},
}

\lstdefinelanguage{java-pretty}
{
language=java,
numbers=left,
basicstyle=\footnotesize\ttfamily,
numberstyle=\footnotesize,
breaklines=true,
columns=fullflexible,
xleftmargin=16pt,
showstringspaces=false,
}

\lstdefinelanguage{java-display}
{
language=java-pretty,
numbers=none
}

\makeatletter
\let\OldStatex\Statex
\renewcommand{\Statex}[1][3]{%
\setlength\@tempdima{\algorithmicindent}%
\OldStatex\hskip\dimexpr#1\@tempdima\relax}
\makeatother

\algnewcommand\OR{\textbf{or} }

\algnewcommand\AND{\textbf{and} }

\algnewcommand\True{\textbf{true}}

\algnewcommand\False{\textbf{false}}

\algnewcommand\Null{\textbf{null}}

\algnewcommand\Break{\textbf{break}}%

\algdef{SE}[DOWHILE]{Do}{EndDoWhile}{\algorithmicdo}[1]{\algorithmicwhile\ #1}

\algnewcommand\algorithmicswitch{\textbf{switch}}
\algnewcommand\algorithmiccase{\textbf{case}}
\algdef{SE}[SWITCH]{Switch}{EndSwitch}[1]{\algorithmicswitch\ #1\ \algorithmicdo}{\algorithmicend\ \algorithmicswitch}%
\algdef{SE}[CASE]{Case}{EndCase}[1]{\algorithmiccase\ #1:}{\algorithmicend\ \algorithmiccase}%
\makeatletter
\ifthenelse{\equal{\ALG@noend}{t}}%
{%
\algtext*{EndSwitch}%
\algtext*{EndCase}%
}{}%
\makeatother

\algdef{SL}[INPUT]{Input}{0}{\textbf{Input: }}
\algdef{SL}[OUTPUT]{Output}{0}{\textbf{Output: }}

\makeatletter
\algrenewcommand\alglinenumber[1]{\footnotesize #1:} %
\algrenewcommand\ALG@beginalgorithmic{\footnotesize}
\algrenewcommand\algorithmiccomment[2][\footnotesize]{{#1\hfill\(\triangleright\) #2}} %
\makeatother

\algrenewcommand\algorithmicindent{1em}%

\newcommand{\rememberlines}{\xdef\rememberedlines{\number\value{ALG@line}}}
\newcommand{\resumenumbering}{\setcounter{ALG@line}{\rememberedlines}}

\newcommand{\Title}{Java \JIT Testing with Template Extraction}
\newcommand{\Tool}{\textsc{LeJit}\xspace}

\newcommand{\JOpFuzzer}{JOpFuzzer\xspace}
\newcommand{\JavaFuzzer}{Java* Fuzzer\xspace}
\newcommand{\JITfuzz}{JITfuzz\xspace}
\newcommand{\JAttack}{\textsc{JAttack}\xspace}
\newcommand{\Randoop}{Randoop\xspace}
\newcommand{\Csmith}{Csmith\xspace}
\newcommand{\JavaTailor}{JavaTailor\xspace}

\newcommand{\ToolTestBased}{$\mathrm{\Tool}_t$\xspace}
\newcommand{\ToolPoolBased}{$\mathrm{\Tool}_p$\xspace}
\newcommand{\ToolOnlyRandoop}{$\mathrm{\Tool}_{\mathrm{NoTmpl}}$\xspace}
\newcommand{\ToolWoRandoop}{$\mathrm{\Tool}_{\mathrm{NoPool}}$\xspace}

\newcommand{\HotSpot}{HotSpot\xspace}
\newcommand{\OpenJNine}{OpenJ9\xspace}
\newcommand{\GraalVM}{GraalVM\xspace}
\newcommand{\JIT}{JIT\xspace}
\newcommand{\COne}{C1\xspace}
\newcommand{\CTwo}{C2\xspace}
\newcommand{\JVMs}{JVMs\xspace}
\newcommand{\JVM}{JVM\xspace}
\newcommand{\API}{API\xspace}

\newcommand{\argumentmethod}{argument method\xspace}
\newcommand{\argumentmethods}{argument methods\xspace}
\newcommand{\Argumentmethod}{Argument method\xspace}
\newcommand{\Argumentmethods}{Argument methods\xspace}
\newcommand{\entrymethod}{entry method\xspace}
\newcommand{\entrymethods}{entry methods\xspace}
\newcommand{\Entrymethod}{Entry method\xspace}
\newcommand{\Entrymethods}{Entry methods\xspace}
\newcommand{\generatedprogram}{generated program\xspace}
\newcommand{\generatedprograms}{generated programs\xspace}
\newcommand{\Generatedprogram}{Generated program\xspace}
\newcommand{\Generatedprograms}{Generated programs\xspace}
\newcommand{\GeneratedPrograms}{Generated Programs\xspace}
\newcommand{\hole}{hole\xspace}
\newcommand{\holes}{holes\xspace}
\newcommand{\Hole}{Hole\xspace}
\newcommand{\Holes}{Holes\xspace}
\newcommand{\template}{template\xspace}
\newcommand{\templates}{templates\xspace}
\newcommand{\Template}{Template\xspace}
\newcommand{\Templates}{Templates\xspace}
\newcommand{\Phasecollection}{collection\xspace}
\newcommand{\PhaseCollection}{Collection\xspace}
\newcommand{\Phaseextraction}{extraction\xspace}
\newcommand{\PhaseExtraction}{Extraction\xspace}
\newcommand{\Phasegeneration}{generation\xspace}
\newcommand{\PhaseGeneration}{Generation\xspace}
\newcommand{\Phasetesting}{testing\xspace}
\newcommand{\PhaseTesting}{Testing\xspace}
\newcommand{\Phasepruning}{pruning\xspace}
\newcommand{\PhasePruning}{Pruning\xspace}
\newcommand{\ApproachtestBased}{test-based\xspace}
\newcommand{\ApproachTestBased}{Test-based\xspace}
\newcommand{\ApproachpoolBased}{pool-based\xspace}
\newcommand{\ApproachPoolBased}{Pool-based\xspace}

\newcommand{\JITconfig}{JIT configuration\xspace}
\newcommand{\JITconfigs}{JIT configurations\xspace}
\newcommand{\JITConfig}{JIT Configuration\xspace}
\newcommand{\JITConfigs}{JIT Configurations\xspace}

\newcommand{\BugOne}{Arithmetic mis-compilation\xspace}
\newcommand{\BugTwo}{Incorrect elimination of range checks\xspace}
\newcommand{\BugThree}{Erroneous loop condition evaluation\xspace}
\newcommand{\BugFour}{Standard library mis-compilation\xspace}

\newcommand{\aHole}[1]{\ensuremath{[\![}#1\ensuremath{]\!]}}
\newcommand{\aConfig}[1]{\ensuremath{\langle} #1 \ensuremath{\rangle}}

\let\oldding\ding%
\renewcommand{\ding}[2][1]{\scalebox{#1}{\oldding{#2}}}%
\newcommand{\cicrlenumber}[1]{\ding[0.9]{\number\numexpr181 + #1}}

\newcommand{\annotation}[2]{\cicrlenumber{#1}\label{#2}}
\newcommand{\highlightApi}[1]{{\color{blue}#1}}
\newcommand{\highlightEval}[1]{{\color{purple}\textbf{#1}}}
\colorlet{colorhole1}{red!20}
\colorlet{colorhole2}{green!20}
\colorlet{colorhole3}{yellow!20}
\colorlet{colorhole4}{teal!20}
\colorlet{colorhole5}{brown!20}

\newcommand{\aExp}{\ensuremath{\overline{e}}\xspace}
\newcommand{\aTemplate}{\ensuremath{\overline{t}}\xspace}
\newcommand{\aType}{\ensuremath{\tau}\xspace}
\newcommand{\aEntry}{\ensuremath{M}\xspace}
\newcommand{\aClass}{\ensuremath{C}\xspace}

\DefMacro{table-filtering-row0-run_id}{results-20231115}
\DefMacro{table-filtering-row0-failures}{-1}
\DefMacro{table-filtering-row0-false_positives-in_run}{-1}
\DefMacro{table-filtering-row0-false_positives-post_run}{1}
\DefMacro{table-filtering-row0-false_positives-manual}{0}
\DefMacro{table-filtering-row0-bugs-reported}{0}
\DefMacro{table-filtering-row0-bugs-duplicate}{0}
\DefMacro{table-filtering-row1-run_id}{results-20231108}
\DefMacro{table-filtering-row1-failures}{1,716}
\DefMacro{table-filtering-row1-false_positives-in_run}{1,577}
\DefMacro{table-filtering-row1-false_positives-post_run}{138}
\DefMacro{table-filtering-row1-false_positives-manual}{1}
\DefMacro{table-filtering-row1-bugs-reported}{0}
\DefMacro{table-filtering-row1-bugs-duplicate}{0}
\DefMacro{table-filtering-row2-run_id}{results-20231018}
\DefMacro{table-filtering-row2-failures}{759}
\DefMacro{table-filtering-row2-false_positives-in_run}{704}
\DefMacro{table-filtering-row2-false_positives-post_run}{18}
\DefMacro{table-filtering-row2-false_positives-manual}{20}
\DefMacro{table-filtering-row2-bugs-reported}{0}
\DefMacro{table-filtering-row2-bugs-duplicate}{17}
\DefMacro{table-filtering-row3-run_id}{results-20231013}
\DefMacro{table-filtering-row3-failures}{950}
\DefMacro{table-filtering-row3-false_positives-in_run}{889}
\DefMacro{table-filtering-row3-false_positives-post_run}{17}
\DefMacro{table-filtering-row3-false_positives-manual}{33}
\DefMacro{table-filtering-row3-bugs-reported}{0}
\DefMacro{table-filtering-row3-bugs-duplicate}{11}
\DefMacro{table-filtering-row4-run_id}{results-20231009}
\DefMacro{table-filtering-row4-failures}{4,022}
\DefMacro{table-filtering-row4-false_positives-in_run}{3,832}
\DefMacro{table-filtering-row4-false_positives-post_run}{146}
\DefMacro{table-filtering-row4-false_positives-manual}{14}
\DefMacro{table-filtering-row4-bugs-reported}{0}
\DefMacro{table-filtering-row4-bugs-duplicate}{30}
\DefMacro{table-filtering-row5-run_id}{results-20231002}
\DefMacro{table-filtering-row5-failures}{1,300}
\DefMacro{table-filtering-row5-false_positives-in_run}{1,261}
\DefMacro{table-filtering-row5-false_positives-post_run}{34}
\DefMacro{table-filtering-row5-false_positives-manual}{1}
\DefMacro{table-filtering-row5-bugs-reported}{0}
\DefMacro{table-filtering-row5-bugs-duplicate}{4}
\DefMacro{table-filtering-row6-run_id}{results-20230930}
\DefMacro{table-filtering-row6-failures}{2,396}
\DefMacro{table-filtering-row6-false_positives-in_run}{2,267}
\DefMacro{table-filtering-row6-false_positives-post_run}{86}
\DefMacro{table-filtering-row6-false_positives-manual}{25}
\DefMacro{table-filtering-row6-bugs-reported}{0}
\DefMacro{table-filtering-row6-bugs-duplicate}{18}
\DefMacro{table-filtering-row7-run_id}{results-20230928 2}
\DefMacro{table-filtering-row7-failures}{1,023}
\DefMacro{table-filtering-row7-false_positives-in_run}{951}
\DefMacro{table-filtering-row7-false_positives-post_run}{0}
\DefMacro{table-filtering-row7-false_positives-manual}{6}
\DefMacro{table-filtering-row7-bugs-reported}{1}
\DefMacro{table-filtering-row7-bugs-duplicate}{65}
\DefMacro{table-filtering-row8-run_id}{results-20230928}
\DefMacro{table-filtering-row8-failures}{1,545}
\DefMacro{table-filtering-row8-false_positives-in_run}{1,516}
\DefMacro{table-filtering-row8-false_positives-post_run}{0}
\DefMacro{table-filtering-row8-false_positives-manual}{10}
\DefMacro{table-filtering-row8-bugs-reported}{0}
\DefMacro{table-filtering-row8-bugs-duplicate}{19}
\DefMacro{table-filtering-row9-run_id}{results-20230924}
\DefMacro{table-filtering-row9-failures}{1,980}
\DefMacro{table-filtering-row9-false_positives-in_run}{1,874}
\DefMacro{table-filtering-row9-false_positives-post_run}{0}
\DefMacro{table-filtering-row9-false_positives-manual}{82}
\DefMacro{table-filtering-row9-bugs-reported}{20}
\DefMacro{table-filtering-row9-bugs-duplicate}{4}
\DefMacro{table-filtering-row10-run_id}{results-20230910}
\DefMacro{table-filtering-row10-failures}{4,020}
\DefMacro{table-filtering-row10-false_positives-in_run}{3,951}
\DefMacro{table-filtering-row10-false_positives-post_run}{0}
\DefMacro{table-filtering-row10-false_positives-manual}{27}
\DefMacro{table-filtering-row10-bugs-reported}{0}
\DefMacro{table-filtering-row10-bugs-duplicate}{42}
\DefMacro{table-filtering-row11-run_id}{results-20230904}
\DefMacro{table-filtering-row11-failures}{983}
\DefMacro{table-filtering-row11-false_positives-in_run}{962}
\DefMacro{table-filtering-row11-false_positives-post_run}{0}
\DefMacro{table-filtering-row11-false_positives-manual}{13}
\DefMacro{table-filtering-row11-bugs-reported}{0}
\DefMacro{table-filtering-row11-bugs-duplicate}{8}
\DefMacro{table-filtering-row12-run_id}{results-20230505}
\DefMacro{table-filtering-row12-failures}{1,076}
\DefMacro{table-filtering-row12-false_positives-in_run}{923}
\DefMacro{table-filtering-row12-false_positives-post_run}{0}
\DefMacro{table-filtering-row12-false_positives-manual}{153}
\DefMacro{table-filtering-row12-bugs-reported}{0}
\DefMacro{table-filtering-row12-bugs-duplicate}{0}
\DefMacro{table-filtering-row13-run_id}{results-20230423}
\DefMacro{table-filtering-row13-failures}{2,447}
\DefMacro{table-filtering-row13-false_positives-in_run}{2,294}
\DefMacro{table-filtering-row13-false_positives-post_run}{0}
\DefMacro{table-filtering-row13-false_positives-manual}{144}
\DefMacro{table-filtering-row13-bugs-reported}{0}
\DefMacro{table-filtering-row13-bugs-duplicate}{9}
\DefMacro{table-filtering-row14-run_id}{results-20230421}
\DefMacro{table-filtering-row14-failures}{886}
\DefMacro{table-filtering-row14-false_positives-in_run}{840}
\DefMacro{table-filtering-row14-false_positives-post_run}{0}
\DefMacro{table-filtering-row14-false_positives-manual}{30}
\DefMacro{table-filtering-row14-bugs-reported}{1}
\DefMacro{table-filtering-row14-bugs-duplicate}{15}
\DefMacro{table-filtering-row15-run_id}{results-20230419}
\DefMacro{table-filtering-row15-failures}{668}
\DefMacro{table-filtering-row15-false_positives-in_run}{638}
\DefMacro{table-filtering-row15-false_positives-post_run}{0}
\DefMacro{table-filtering-row15-false_positives-manual}{29}
\DefMacro{table-filtering-row15-bugs-reported}{1}
\DefMacro{table-filtering-row15-bugs-duplicate}{0}
\DefMacro{table-filtering-row16-run_id}{results-20230418}
\DefMacro{table-filtering-row16-failures}{417}
\DefMacro{table-filtering-row16-false_positives-in_run}{400}
\DefMacro{table-filtering-row16-false_positives-post_run}{0}
\DefMacro{table-filtering-row16-false_positives-manual}{16}
\DefMacro{table-filtering-row16-bugs-reported}{0}
\DefMacro{table-filtering-row16-bugs-duplicate}{1}
\DefMacro{table-filtering-row17-run_id}{results-20230417}
\DefMacro{table-filtering-row17-failures}{303}
\DefMacro{table-filtering-row17-false_positives-in_run}{277}
\DefMacro{table-filtering-row17-false_positives-post_run}{0}
\DefMacro{table-filtering-row17-false_positives-manual}{26}
\DefMacro{table-filtering-row17-bugs-reported}{0}
\DefMacro{table-filtering-row17-bugs-duplicate}{0}
\DefMacro{table-filtering-row18-run_id}{results-20230412 + results-20230415}
\DefMacro{table-filtering-row18-failures}{2,067}
\DefMacro{table-filtering-row18-false_positives-in_run}{1,939}
\DefMacro{table-filtering-row18-false_positives-post_run}{0}
\DefMacro{table-filtering-row18-false_positives-manual}{82}
\DefMacro{table-filtering-row18-bugs-reported}{1}
\DefMacro{table-filtering-row18-bugs-duplicate}{45}
\DefMacro{table-filtering-row19-run_id}{results-20230409 + results-20230411}
\DefMacro{table-filtering-row19-failures}{1,286}
\DefMacro{table-filtering-row19-false_positives-in_run}{1,260}
\DefMacro{table-filtering-row19-false_positives-post_run}{0}
\DefMacro{table-filtering-row19-false_positives-manual}{14}
\DefMacro{table-filtering-row19-bugs-reported}{3}
\DefMacro{table-filtering-row19-bugs-duplicate}{9}
\DefMacro{table-filtering-row20-run_id}{$\sum$}
\DefMacro{table-filtering-row20-failures}{29,844}
\DefMacro{table-filtering-row20-false_positives-in_run}{28,355}
\DefMacro{table-filtering-row20-false_positives-post_run}{440}
\DefMacro{table-filtering-row20-false_positives-manual}{726}
\DefMacro{table-filtering-row20-bugs-reported}{27}
\DefMacro{table-filtering-row20-bugs-duplicate}{297}

\DefMacro{table-filtering}{Table of filtering. "Failures": number of diff and crashes before any filtering. "In-Run": false positives filtered out by main run. "Post-Run": FPs filtered out by the (new) post-run script. "Manual": FPs filtered out by manual inspection. "New Report": Number of classes of a new bug type which has not been previously reported. "Duplicate": Number of classes of a bug type which has been previously reported. \label{tab:filtering}}
\DefMacro{table-filtering-col-run_id}{Run ID}
\DefMacro{table-filtering-col-failures}{Failures}
\DefMacro{table-filtering-col-false_positives}{False Positives}
\DefMacro{table-filtering-col-real_bugs}{Real Bugs}
\DefMacro{table-filtering-col-false_positives-in_run}{In-Run}
\DefMacro{table-filtering-col-false_positives-post_run}{Post-Run}
\DefMacro{table-filtering-col-false_positives-manual}{Manual}
\DefMacro{table-filtering-col-real_bugs-reported}{New Report}
\DefMacro{table-filtering-col-real_bugs-duplicate}{Duplicate}
\DefMacro{table-filtering-col-real_bugs-to_rename_2}{RB 2}

\DefMacro{table-toc_numbers_of_programs_and_bugs-row0-table}{Table~\ref{tab:numbers-of-programs-and-bugs-20230409_20230411}}
\DefMacro{table-toc_numbers_of_programs_and_bugs-row0-desc}{\UseMacro{results-20230409_20230411-desc}}
\DefMacro{table-toc_numbers_of_programs_and_bugs-row1-table}{Table~\ref{tab:numbers-of-programs-and-bugs-20230423}}
\DefMacro{table-toc_numbers_of_programs_and_bugs-row1-desc}{\UseMacro{results-20230423-desc}}
\DefMacro{table-toc_numbers_of_programs_and_bugs-row2-table}{Table~\ref{tab:numbers-of-programs-and-bugs-20230904}}
\DefMacro{table-toc_numbers_of_programs_and_bugs-row2-desc}{\UseMacro{results-20230904-desc}}
\DefMacro{table-toc_numbers_of_programs_and_bugs-row3-table}{Table~\ref{tab:numbers-of-programs-and-bugs-20230412_20230415}}
\DefMacro{table-toc_numbers_of_programs_and_bugs-row3-desc}{\UseMacro{results-20230412_20230415-desc}}
\DefMacro{table-toc_numbers_of_programs_and_bugs-row4-table}{Table~\ref{tab:numbers-of-programs-and-bugs-20230910}}
\DefMacro{table-toc_numbers_of_programs_and_bugs-row4-desc}{\UseMacro{results-20230910-desc}}
\DefMacro{table-toc_numbers_of_programs_and_bugs-row5-table}{Table~\ref{tab:numbers-of-programs-and-bugs-20231009}}
\DefMacro{table-toc_numbers_of_programs_and_bugs-row5-desc}{\UseMacro{results-20231009-desc}}
\DefMacro{table-toc_numbers_of_programs_and_bugs-row6-table}{Table~\ref{tab:numbers-of-programs-and-bugs-20230417}}
\DefMacro{table-toc_numbers_of_programs_and_bugs-row6-desc}{\UseMacro{results-20230417-desc}}
\DefMacro{table-toc_numbers_of_programs_and_bugs-row7-table}{Table~\ref{tab:numbers-of-programs-and-bugs-20230418}}
\DefMacro{table-toc_numbers_of_programs_and_bugs-row7-desc}{\UseMacro{results-20230418-desc}}
\DefMacro{table-toc_numbers_of_programs_and_bugs-row8-table}{Table~\ref{tab:numbers-of-programs-and-bugs-20230419}}
\DefMacro{table-toc_numbers_of_programs_and_bugs-row8-desc}{\UseMacro{results-20230419-desc}}
\DefMacro{table-toc_numbers_of_programs_and_bugs-row9-table}{Table~\ref{tab:numbers-of-programs-and-bugs-20230421}}
\DefMacro{table-toc_numbers_of_programs_and_bugs-row9-desc}{\UseMacro{results-20230421-desc}}
\DefMacro{table-toc_numbers_of_programs_and_bugs-row10-table}{Table~\ref{tab:numbers-of-programs-and-bugs-20231013}}
\DefMacro{table-toc_numbers_of_programs_and_bugs-row10-desc}{\UseMacro{results-20231013-desc}}
\DefMacro{table-toc_numbers_of_programs_and_bugs-row11-table}{Table~\ref{tab:numbers-of-programs-and-bugs-20231018}}
\DefMacro{table-toc_numbers_of_programs_and_bugs-row11-desc}{\UseMacro{results-20231018-desc}}
\DefMacro{table-toc_numbers_of_programs_and_bugs-row12-table}{Table~\ref{tab:numbers-of-programs-and-bugs-20230924}}
\DefMacro{table-toc_numbers_of_programs_and_bugs-row12-desc}{\UseMacro{results-20230924-desc}}
\DefMacro{table-toc_numbers_of_programs_and_bugs-row13-table}{Table~\ref{tab:numbers-of-programs-and-bugs-20230928}}
\DefMacro{table-toc_numbers_of_programs_and_bugs-row13-desc}{\UseMacro{results-20230928-desc}}
\DefMacro{table-toc_numbers_of_programs_and_bugs-row14-table}{Table~\ref{tab:numbers-of-programs-and-bugs-20231002}}
\DefMacro{table-toc_numbers_of_programs_and_bugs-row14-desc}{\UseMacro{results-20231002-desc}}
\DefMacro{table-toc_numbers_of_programs_and_bugs-row15-table}{Table~\ref{tab:numbers-of-programs-and-bugs-20230928_2}}
\DefMacro{table-toc_numbers_of_programs_and_bugs-row15-desc}{\UseMacro{results-20230928_2-desc}}
\DefMacro{table-toc_numbers_of_programs_and_bugs-row16-table}{Table~\ref{tab:numbers-of-programs-and-bugs-20230930}}
\DefMacro{table-toc_numbers_of_programs_and_bugs-row16-desc}{\UseMacro{results-20230930-desc}}
\DefMacro{table-toc_numbers_of_programs_and_bugs-row17-table}{Table~\ref{tab:numbers-of-programs-and-bugs-20230505}}
\DefMacro{table-toc_numbers_of_programs_and_bugs-row17-desc}{\UseMacro{results-20230505-desc}}
\DefMacro{table-toc_numbers_of_programs_and_bugs-row18-table}{Table~\ref{tab:numbers-of-programs-and-bugs-20231108}}
\DefMacro{table-toc_numbers_of_programs_and_bugs-row18-desc}{\UseMacro{results-20231108-desc}}
\DefMacro{table-toc_numbers_of_programs_and_bugs-row19-table}{Table~\ref{tab:numbers-of-programs-and-bugs-20231115}}
\DefMacro{table-toc_numbers_of_programs_and_bugs-row19-desc}{\UseMacro{results-20231115-desc}}

\DefMacro{table-toc-numbers_of_programs_and_bugs-caption}{Table of Contents. \label{tab:toc-numbers-of-programs-and-bugs}}
\DefMacro{table-toc_numbers_of_programs_and_bugs-col-table}{Table}
\DefMacro{table-toc_numbers_of_programs_and_bugs-col-desc}{Note}
\DefMacro{results-20230904-desc}{\ToolTestBased Run \#3}
\DefMacro{results-20230423-desc}{\ToolTestBased Run \#2}
\DefMacro{results-20230409_20230411-desc}{\ToolTestBased Run \#1}

\DefMacro{results-20231009-desc}{\ToolPoolBased Run \#3}
\DefMacro{results-20230910-desc}{\ToolPoolBased Run \#2}
\DefMacro{results-20230412_20230415-desc}{\ToolPoolBased Run \#1}

\DefMacro{results-20230419-desc}{\ToolOnlyRandoop Run \#3}
\DefMacro{results-20230418-desc}{\ToolOnlyRandoop Run \#2}
\DefMacro{results-20230417-desc}{\ToolOnlyRandoop Run \#1}

\DefMacro{results-20231018-desc}{\ToolWoRandoop Run \#3}
\DefMacro{results-20231013-desc}{\ToolWoRandoop Run \#2}
\DefMacro{results-20230421-desc}{\ToolWoRandoop Run \#1}

\DefMacro{results-20231002-desc}{Arithmetic \& Shift \Holes}
\DefMacro{results-20230930-desc}{All \Holes}
\DefMacro{results-20230928_2-desc}{Relation \& Logic \Holes}
\DefMacro{results-20230928-desc}{Val \Holes}
\DefMacro{results-20230924-desc}{Id \Holes}

\DefMacro{results-20231115-desc}{\Tool Matching \JITfuzz Run \#3}
\DefMacro{results-20231108-desc}{\Tool Matching \JITfuzz Run \#2}
\DefMacro{results-20230505-desc}{\Tool Matching \JITfuzz Run \#1}

\DefMacro{table-numbers_of_programs_and_bugs-20230904-row0-project}{lang}
\DefMacro{table-numbers_of_programs_and_bugs-20230904-row0-num_unit_tests}{1,869}
\DefMacro{table-numbers_of_programs_and_bugs-20230904-row0-num_unique_entry_methods}{1,123}
\DefMacro{table-numbers_of_programs_and_bugs-20230904-row0-num_unique_classes}{96}
\DefMacro{table-numbers_of_programs_and_bugs-20230904-row0-convert-fail}{93}
\DefMacro{table-numbers_of_programs_and_bugs-20230904-row0-convert-timeout}{0}
\DefMacro{table-numbers_of_programs_and_bugs-20230904-row0-convert-not_compile}{0}
\DefMacro{table-numbers_of_programs_and_bugs-20230904-row0-convert-no_original}{0}
\DefMacro{table-numbers_of_programs_and_bugs-20230904-row0-gen-fail}{48}
\DefMacro{table-numbers_of_programs_and_bugs-20230904-row0-gen-timeout}{84}
\DefMacro{table-numbers_of_programs_and_bugs-20230904-row0-gen-no_hole}{38}
\DefMacro{table-numbers_of_programs_and_bugs-20230904-row0-gen-no_gen}{347}
\DefMacro{table-numbers_of_programs_and_bugs-20230904-row0-num_working_templates}{1,391}
\DefMacro{table-numbers_of_programs_and_bugs-20230904-row0-num_unique_entry_methods_out_of_working_templates}{830}
\DefMacro{table-numbers_of_programs_and_bugs-20230904-row0-num_unique_classes_out_of_working_templates}{76}
\DefMacro{table-numbers_of_programs_and_bugs-20230904-row0-exec-total}{11,393}
\DefMacro{table-numbers_of_programs_and_bugs-20230904-row0-exec-not_compile}{0}
\DefMacro{table-numbers_of_programs_and_bugs-20230904-row0-exec-timeout}{99}
\DefMacro{table-numbers_of_programs_and_bugs-20230904-row0-exec-ok}{10,885}
\DefMacro{table-numbers_of_programs_and_bugs-20230904-row0-exec-skipped_diff}{387}
\DefMacro{table-numbers_of_programs_and_bugs-20230904-row0-exec-skipped_crash}{7}
\DefMacro{table-numbers_of_programs_and_bugs-20230904-row0-exec-final_diff}{6}
\DefMacro{table-numbers_of_programs_and_bugs-20230904-row0-exec-final_crash}{9}
\DefMacro{table-numbers_of_programs_and_bugs-20230904-row1-project}{math}
\DefMacro{table-numbers_of_programs_and_bugs-20230904-row1-num_unit_tests}{1,072}
\DefMacro{table-numbers_of_programs_and_bugs-20230904-row1-num_unique_entry_methods}{814}
\DefMacro{table-numbers_of_programs_and_bugs-20230904-row1-num_unique_classes}{210}
\DefMacro{table-numbers_of_programs_and_bugs-20230904-row1-convert-fail}{14}
\DefMacro{table-numbers_of_programs_and_bugs-20230904-row1-convert-timeout}{0}
\DefMacro{table-numbers_of_programs_and_bugs-20230904-row1-convert-not_compile}{21}
\DefMacro{table-numbers_of_programs_and_bugs-20230904-row1-convert-no_original}{0}
\DefMacro{table-numbers_of_programs_and_bugs-20230904-row1-gen-fail}{124}
\DefMacro{table-numbers_of_programs_and_bugs-20230904-row1-gen-timeout}{21}
\DefMacro{table-numbers_of_programs_and_bugs-20230904-row1-gen-no_hole}{34}
\DefMacro{table-numbers_of_programs_and_bugs-20230904-row1-gen-no_gen}{221}
\DefMacro{table-numbers_of_programs_and_bugs-20230904-row1-num_working_templates}{782}
\DefMacro{table-numbers_of_programs_and_bugs-20230904-row1-num_unique_entry_methods_out_of_working_templates}{596}
\DefMacro{table-numbers_of_programs_and_bugs-20230904-row1-num_unique_classes_out_of_working_templates}{173}
\DefMacro{table-numbers_of_programs_and_bugs-20230904-row1-exec-total}{6,695}
\DefMacro{table-numbers_of_programs_and_bugs-20230904-row1-exec-not_compile}{0}
\DefMacro{table-numbers_of_programs_and_bugs-20230904-row1-exec-timeout}{57}
\DefMacro{table-numbers_of_programs_and_bugs-20230904-row1-exec-ok}{6,586}
\DefMacro{table-numbers_of_programs_and_bugs-20230904-row1-exec-skipped_diff}{47}
\DefMacro{table-numbers_of_programs_and_bugs-20230904-row1-exec-skipped_crash}{2}
\DefMacro{table-numbers_of_programs_and_bugs-20230904-row1-exec-final_diff}{0}
\DefMacro{table-numbers_of_programs_and_bugs-20230904-row1-exec-final_crash}{3}
\DefMacro{table-numbers_of_programs_and_bugs-20230904-row2-project}{text}
\DefMacro{table-numbers_of_programs_and_bugs-20230904-row2-num_unit_tests}{1,676}
\DefMacro{table-numbers_of_programs_and_bugs-20230904-row2-num_unique_entry_methods}{504}
\DefMacro{table-numbers_of_programs_and_bugs-20230904-row2-num_unique_classes}{39}
\DefMacro{table-numbers_of_programs_and_bugs-20230904-row2-convert-fail}{25}
\DefMacro{table-numbers_of_programs_and_bugs-20230904-row2-convert-timeout}{0}
\DefMacro{table-numbers_of_programs_and_bugs-20230904-row2-convert-not_compile}{0}
\DefMacro{table-numbers_of_programs_and_bugs-20230904-row2-convert-no_original}{0}
\DefMacro{table-numbers_of_programs_and_bugs-20230904-row2-gen-fail}{14}
\DefMacro{table-numbers_of_programs_and_bugs-20230904-row2-gen-timeout}{36}
\DefMacro{table-numbers_of_programs_and_bugs-20230904-row2-gen-no_hole}{54}
\DefMacro{table-numbers_of_programs_and_bugs-20230904-row2-gen-no_gen}{322}
\DefMacro{table-numbers_of_programs_and_bugs-20230904-row2-num_working_templates}{1,275}
\DefMacro{table-numbers_of_programs_and_bugs-20230904-row2-num_unique_entry_methods_out_of_working_templates}{433}
\DefMacro{table-numbers_of_programs_and_bugs-20230904-row2-num_unique_classes_out_of_working_templates}{32}
\DefMacro{table-numbers_of_programs_and_bugs-20230904-row2-exec-total}{11,819}
\DefMacro{table-numbers_of_programs_and_bugs-20230904-row2-exec-not_compile}{0}
\DefMacro{table-numbers_of_programs_and_bugs-20230904-row2-exec-timeout}{87}
\DefMacro{table-numbers_of_programs_and_bugs-20230904-row2-exec-ok}{11,575}
\DefMacro{table-numbers_of_programs_and_bugs-20230904-row2-exec-skipped_diff}{156}
\DefMacro{table-numbers_of_programs_and_bugs-20230904-row2-exec-skipped_crash}{0}
\DefMacro{table-numbers_of_programs_and_bugs-20230904-row2-exec-final_diff}{0}
\DefMacro{table-numbers_of_programs_and_bugs-20230904-row2-exec-final_crash}{1}
\DefMacro{table-numbers_of_programs_and_bugs-20230904-row3-project}{numbers}
\DefMacro{table-numbers_of_programs_and_bugs-20230904-row3-num_unit_tests}{3,138}
\DefMacro{table-numbers_of_programs_and_bugs-20230904-row3-num_unique_entry_methods}{44}
\DefMacro{table-numbers_of_programs_and_bugs-20230904-row3-num_unique_classes}{4}
\DefMacro{table-numbers_of_programs_and_bugs-20230904-row3-convert-fail}{121}
\DefMacro{table-numbers_of_programs_and_bugs-20230904-row3-convert-timeout}{0}
\DefMacro{table-numbers_of_programs_and_bugs-20230904-row3-convert-not_compile}{0}
\DefMacro{table-numbers_of_programs_and_bugs-20230904-row3-convert-no_original}{0}
\DefMacro{table-numbers_of_programs_and_bugs-20230904-row3-gen-fail}{0}
\DefMacro{table-numbers_of_programs_and_bugs-20230904-row3-gen-timeout}{81}
\DefMacro{table-numbers_of_programs_and_bugs-20230904-row3-gen-no_hole}{0}
\DefMacro{table-numbers_of_programs_and_bugs-20230904-row3-gen-no_gen}{1}
\DefMacro{table-numbers_of_programs_and_bugs-20230904-row3-num_working_templates}{3,016}
\DefMacro{table-numbers_of_programs_and_bugs-20230904-row3-num_unique_entry_methods_out_of_working_templates}{41}
\DefMacro{table-numbers_of_programs_and_bugs-20230904-row3-num_unique_classes_out_of_working_templates}{3}
\DefMacro{table-numbers_of_programs_and_bugs-20230904-row3-exec-total}{29,495}
\DefMacro{table-numbers_of_programs_and_bugs-20230904-row3-exec-not_compile}{0}
\DefMacro{table-numbers_of_programs_and_bugs-20230904-row3-exec-timeout}{3}
\DefMacro{table-numbers_of_programs_and_bugs-20230904-row3-exec-ok}{29,492}
\DefMacro{table-numbers_of_programs_and_bugs-20230904-row3-exec-skipped_diff}{0}
\DefMacro{table-numbers_of_programs_and_bugs-20230904-row3-exec-skipped_crash}{0}
\DefMacro{table-numbers_of_programs_and_bugs-20230904-row3-exec-final_diff}{0}
\DefMacro{table-numbers_of_programs_and_bugs-20230904-row3-exec-final_crash}{0}
\DefMacro{table-numbers_of_programs_and_bugs-20230904-row4-project}{vectorz}
\DefMacro{table-numbers_of_programs_and_bugs-20230904-row4-num_unit_tests}{1,144}
\DefMacro{table-numbers_of_programs_and_bugs-20230904-row4-num_unique_entry_methods}{920}
\DefMacro{table-numbers_of_programs_and_bugs-20230904-row4-num_unique_classes}{157}
\DefMacro{table-numbers_of_programs_and_bugs-20230904-row4-convert-fail}{0}
\DefMacro{table-numbers_of_programs_and_bugs-20230904-row4-convert-timeout}{0}
\DefMacro{table-numbers_of_programs_and_bugs-20230904-row4-convert-not_compile}{0}
\DefMacro{table-numbers_of_programs_and_bugs-20230904-row4-convert-no_original}{0}
\DefMacro{table-numbers_of_programs_and_bugs-20230904-row4-gen-fail}{60}
\DefMacro{table-numbers_of_programs_and_bugs-20230904-row4-gen-timeout}{45}
\DefMacro{table-numbers_of_programs_and_bugs-20230904-row4-gen-no_hole}{108}
\DefMacro{table-numbers_of_programs_and_bugs-20230904-row4-gen-no_gen}{278}
\DefMacro{table-numbers_of_programs_and_bugs-20230904-row4-num_working_templates}{758}
\DefMacro{table-numbers_of_programs_and_bugs-20230904-row4-num_unique_entry_methods_out_of_working_templates}{634}
\DefMacro{table-numbers_of_programs_and_bugs-20230904-row4-num_unique_classes_out_of_working_templates}{145}
\DefMacro{table-numbers_of_programs_and_bugs-20230904-row4-exec-total}{5,617}
\DefMacro{table-numbers_of_programs_and_bugs-20230904-row4-exec-not_compile}{0}
\DefMacro{table-numbers_of_programs_and_bugs-20230904-row4-exec-timeout}{176}
\DefMacro{table-numbers_of_programs_and_bugs-20230904-row4-exec-ok}{5,222}
\DefMacro{table-numbers_of_programs_and_bugs-20230904-row4-exec-skipped_diff}{217}
\DefMacro{table-numbers_of_programs_and_bugs-20230904-row4-exec-skipped_crash}{0}
\DefMacro{table-numbers_of_programs_and_bugs-20230904-row4-exec-final_diff}{2}
\DefMacro{table-numbers_of_programs_and_bugs-20230904-row4-exec-final_crash}{0}
\DefMacro{table-numbers_of_programs_and_bugs-20230904-row5-project}{statistics}
\DefMacro{table-numbers_of_programs_and_bugs-20230904-row5-num_unit_tests}{1,868}
\DefMacro{table-numbers_of_programs_and_bugs-20230904-row5-num_unique_entry_methods}{374}
\DefMacro{table-numbers_of_programs_and_bugs-20230904-row5-num_unique_classes}{27}
\DefMacro{table-numbers_of_programs_and_bugs-20230904-row5-convert-fail}{165}
\DefMacro{table-numbers_of_programs_and_bugs-20230904-row5-convert-timeout}{0}
\DefMacro{table-numbers_of_programs_and_bugs-20230904-row5-convert-not_compile}{0}
\DefMacro{table-numbers_of_programs_and_bugs-20230904-row5-convert-no_original}{0}
\DefMacro{table-numbers_of_programs_and_bugs-20230904-row5-gen-fail}{132}
\DefMacro{table-numbers_of_programs_and_bugs-20230904-row5-gen-timeout}{26}
\DefMacro{table-numbers_of_programs_and_bugs-20230904-row5-gen-no_hole}{0}
\DefMacro{table-numbers_of_programs_and_bugs-20230904-row5-gen-no_gen}{599}
\DefMacro{table-numbers_of_programs_and_bugs-20230904-row5-num_working_templates}{1,104}
\DefMacro{table-numbers_of_programs_and_bugs-20230904-row5-num_unique_entry_methods_out_of_working_templates}{252}
\DefMacro{table-numbers_of_programs_and_bugs-20230904-row5-num_unique_classes_out_of_working_templates}{24}
\DefMacro{table-numbers_of_programs_and_bugs-20230904-row5-exec-total}{8,864}
\DefMacro{table-numbers_of_programs_and_bugs-20230904-row5-exec-not_compile}{0}
\DefMacro{table-numbers_of_programs_and_bugs-20230904-row5-exec-timeout}{16}
\DefMacro{table-numbers_of_programs_and_bugs-20230904-row5-exec-ok}{8,737}
\DefMacro{table-numbers_of_programs_and_bugs-20230904-row5-exec-skipped_diff}{111}
\DefMacro{table-numbers_of_programs_and_bugs-20230904-row5-exec-skipped_crash}{0}
\DefMacro{table-numbers_of_programs_and_bugs-20230904-row5-exec-final_diff}{0}
\DefMacro{table-numbers_of_programs_and_bugs-20230904-row5-exec-final_crash}{0}
\DefMacro{table-numbers_of_programs_and_bugs-20230904-row6-project}{codec}
\DefMacro{table-numbers_of_programs_and_bugs-20230904-row6-num_unit_tests}{1,332}
\DefMacro{table-numbers_of_programs_and_bugs-20230904-row6-num_unique_entry_methods}{351}
\DefMacro{table-numbers_of_programs_and_bugs-20230904-row6-num_unique_classes}{49}
\DefMacro{table-numbers_of_programs_and_bugs-20230904-row6-convert-fail}{15}
\DefMacro{table-numbers_of_programs_and_bugs-20230904-row6-convert-timeout}{0}
\DefMacro{table-numbers_of_programs_and_bugs-20230904-row6-convert-not_compile}{8}
\DefMacro{table-numbers_of_programs_and_bugs-20230904-row6-convert-no_original}{0}
\DefMacro{table-numbers_of_programs_and_bugs-20230904-row6-gen-fail}{98}
\DefMacro{table-numbers_of_programs_and_bugs-20230904-row6-gen-timeout}{14}
\DefMacro{table-numbers_of_programs_and_bugs-20230904-row6-gen-no_hole}{0}
\DefMacro{table-numbers_of_programs_and_bugs-20230904-row6-gen-no_gen}{193}
\DefMacro{table-numbers_of_programs_and_bugs-20230904-row6-num_working_templates}{1,116}
\DefMacro{table-numbers_of_programs_and_bugs-20230904-row6-num_unique_entry_methods_out_of_working_templates}{295}
\DefMacro{table-numbers_of_programs_and_bugs-20230904-row6-num_unique_classes_out_of_working_templates}{37}
\DefMacro{table-numbers_of_programs_and_bugs-20230904-row6-exec-total}{5,829}
\DefMacro{table-numbers_of_programs_and_bugs-20230904-row6-exec-not_compile}{0}
\DefMacro{table-numbers_of_programs_and_bugs-20230904-row6-exec-timeout}{211}
\DefMacro{table-numbers_of_programs_and_bugs-20230904-row6-exec-ok}{5,583}
\DefMacro{table-numbers_of_programs_and_bugs-20230904-row6-exec-skipped_diff}{23}
\DefMacro{table-numbers_of_programs_and_bugs-20230904-row6-exec-skipped_crash}{12}
\DefMacro{table-numbers_of_programs_and_bugs-20230904-row6-exec-final_diff}{0}
\DefMacro{table-numbers_of_programs_and_bugs-20230904-row6-exec-final_crash}{0}
\DefMacro{table-numbers_of_programs_and_bugs-20230904-row7-project}{$\sum$}
\DefMacro{table-numbers_of_programs_and_bugs-20230904-row7-num_unit_tests}{12,099}
\DefMacro{table-numbers_of_programs_and_bugs-20230904-row7-num_unique_entry_methods}{4,130}
\DefMacro{table-numbers_of_programs_and_bugs-20230904-row7-num_unique_classes}{582}
\DefMacro{table-numbers_of_programs_and_bugs-20230904-row7-convert-fail}{433}
\DefMacro{table-numbers_of_programs_and_bugs-20230904-row7-convert-timeout}{0}
\DefMacro{table-numbers_of_programs_and_bugs-20230904-row7-convert-not_compile}{29}
\DefMacro{table-numbers_of_programs_and_bugs-20230904-row7-convert-no_original}{0}
\DefMacro{table-numbers_of_programs_and_bugs-20230904-row7-gen-fail}{476}
\DefMacro{table-numbers_of_programs_and_bugs-20230904-row7-gen-timeout}{307}
\DefMacro{table-numbers_of_programs_and_bugs-20230904-row7-gen-no_hole}{234}
\DefMacro{table-numbers_of_programs_and_bugs-20230904-row7-gen-no_gen}{1,961}
\DefMacro{table-numbers_of_programs_and_bugs-20230904-row7-num_working_templates}{9,442}
\DefMacro{table-numbers_of_programs_and_bugs-20230904-row7-num_unique_entry_methods_out_of_working_templates}{3,081}
\DefMacro{table-numbers_of_programs_and_bugs-20230904-row7-num_unique_classes_out_of_working_templates}{490}
\DefMacro{table-numbers_of_programs_and_bugs-20230904-row7-exec-total}{79,712}
\DefMacro{table-numbers_of_programs_and_bugs-20230904-row7-exec-not_compile}{0}
\DefMacro{table-numbers_of_programs_and_bugs-20230904-row7-exec-timeout}{649}
\DefMacro{table-numbers_of_programs_and_bugs-20230904-row7-exec-ok}{78,080}
\DefMacro{table-numbers_of_programs_and_bugs-20230904-row7-exec-skipped_diff}{941}
\DefMacro{table-numbers_of_programs_and_bugs-20230904-row7-exec-skipped_crash}{21}
\DefMacro{table-numbers_of_programs_and_bugs-20230904-row7-exec-final_diff}{8}
\DefMacro{table-numbers_of_programs_and_bugs-20230904-row7-exec-final_crash}{13}

\DefMacro{table-numbers_of_programs_and_bugs-20230423-row0-project}{lang}
\DefMacro{table-numbers_of_programs_and_bugs-20230423-row0-num_unit_tests}{1,894}
\DefMacro{table-numbers_of_programs_and_bugs-20230423-row0-num_unique_entry_methods}{1,091}
\DefMacro{table-numbers_of_programs_and_bugs-20230423-row0-num_unique_classes}{95}
\DefMacro{table-numbers_of_programs_and_bugs-20230423-row0-convert-fail}{106}
\DefMacro{table-numbers_of_programs_and_bugs-20230423-row0-convert-timeout}{0}
\DefMacro{table-numbers_of_programs_and_bugs-20230423-row0-convert-not_compile}{2}
\DefMacro{table-numbers_of_programs_and_bugs-20230423-row0-convert-no_original}{0}
\DefMacro{table-numbers_of_programs_and_bugs-20230423-row0-gen-fail}{49}
\DefMacro{table-numbers_of_programs_and_bugs-20230423-row0-gen-timeout}{61}
\DefMacro{table-numbers_of_programs_and_bugs-20230423-row0-gen-no_hole}{44}
\DefMacro{table-numbers_of_programs_and_bugs-20230423-row0-gen-no_gen}{289}
\DefMacro{table-numbers_of_programs_and_bugs-20230423-row0-num_working_templates}{1,453}
\DefMacro{table-numbers_of_programs_and_bugs-20230423-row0-num_unique_entry_methods_out_of_working_templates}{810}
\DefMacro{table-numbers_of_programs_and_bugs-20230423-row0-num_unique_classes_out_of_working_templates}{76}
\DefMacro{table-numbers_of_programs_and_bugs-20230423-row0-exec-total}{12,457}
\DefMacro{table-numbers_of_programs_and_bugs-20230423-row0-exec-not_compile}{0}
\DefMacro{table-numbers_of_programs_and_bugs-20230423-row0-exec-timeout}{131}
\DefMacro{table-numbers_of_programs_and_bugs-20230423-row0-exec-ok}{12,071}
\DefMacro{table-numbers_of_programs_and_bugs-20230423-row0-exec-skipped_diff}{218}
\DefMacro{table-numbers_of_programs_and_bugs-20230423-row0-exec-skipped_crash}{16}
\DefMacro{table-numbers_of_programs_and_bugs-20230423-row0-exec-final_diff}{11}
\DefMacro{table-numbers_of_programs_and_bugs-20230423-row0-exec-final_crash}{10}
\DefMacro{table-numbers_of_programs_and_bugs-20230423-row1-project}{math}
\DefMacro{table-numbers_of_programs_and_bugs-20230423-row1-num_unit_tests}{1,061}
\DefMacro{table-numbers_of_programs_and_bugs-20230423-row1-num_unique_entry_methods}{797}
\DefMacro{table-numbers_of_programs_and_bugs-20230423-row1-num_unique_classes}{210}
\DefMacro{table-numbers_of_programs_and_bugs-20230423-row1-convert-fail}{16}
\DefMacro{table-numbers_of_programs_and_bugs-20230423-row1-convert-timeout}{0}
\DefMacro{table-numbers_of_programs_and_bugs-20230423-row1-convert-not_compile}{11}
\DefMacro{table-numbers_of_programs_and_bugs-20230423-row1-convert-no_original}{0}
\DefMacro{table-numbers_of_programs_and_bugs-20230423-row1-gen-fail}{167}
\DefMacro{table-numbers_of_programs_and_bugs-20230423-row1-gen-timeout}{22}
\DefMacro{table-numbers_of_programs_and_bugs-20230423-row1-gen-no_hole}{40}
\DefMacro{table-numbers_of_programs_and_bugs-20230423-row1-gen-no_gen}{272}
\DefMacro{table-numbers_of_programs_and_bugs-20230423-row1-num_working_templates}{722}
\DefMacro{table-numbers_of_programs_and_bugs-20230423-row1-num_unique_entry_methods_out_of_working_templates}{553}
\DefMacro{table-numbers_of_programs_and_bugs-20230423-row1-num_unique_classes_out_of_working_templates}{170}
\DefMacro{table-numbers_of_programs_and_bugs-20230423-row1-exec-total}{5,918}
\DefMacro{table-numbers_of_programs_and_bugs-20230423-row1-exec-not_compile}{0}
\DefMacro{table-numbers_of_programs_and_bugs-20230423-row1-exec-timeout}{32}
\DefMacro{table-numbers_of_programs_and_bugs-20230423-row1-exec-ok}{5,837}
\DefMacro{table-numbers_of_programs_and_bugs-20230423-row1-exec-skipped_diff}{49}
\DefMacro{table-numbers_of_programs_and_bugs-20230423-row1-exec-skipped_crash}{0}
\DefMacro{table-numbers_of_programs_and_bugs-20230423-row1-exec-final_diff}{0}
\DefMacro{table-numbers_of_programs_and_bugs-20230423-row1-exec-final_crash}{0}
\DefMacro{table-numbers_of_programs_and_bugs-20230423-row2-project}{text}
\DefMacro{table-numbers_of_programs_and_bugs-20230423-row2-num_unit_tests}{1,724}
\DefMacro{table-numbers_of_programs_and_bugs-20230423-row2-num_unique_entry_methods}{477}
\DefMacro{table-numbers_of_programs_and_bugs-20230423-row2-num_unique_classes}{40}
\DefMacro{table-numbers_of_programs_and_bugs-20230423-row2-convert-fail}{17}
\DefMacro{table-numbers_of_programs_and_bugs-20230423-row2-convert-timeout}{0}
\DefMacro{table-numbers_of_programs_and_bugs-20230423-row2-convert-not_compile}{0}
\DefMacro{table-numbers_of_programs_and_bugs-20230423-row2-convert-no_original}{0}
\DefMacro{table-numbers_of_programs_and_bugs-20230423-row2-gen-fail}{7}
\DefMacro{table-numbers_of_programs_and_bugs-20230423-row2-gen-timeout}{47}
\DefMacro{table-numbers_of_programs_and_bugs-20230423-row2-gen-no_hole}{93}
\DefMacro{table-numbers_of_programs_and_bugs-20230423-row2-gen-no_gen}{481}
\DefMacro{table-numbers_of_programs_and_bugs-20230423-row2-num_working_templates}{1,133}
\DefMacro{table-numbers_of_programs_and_bugs-20230423-row2-num_unique_entry_methods_out_of_working_templates}{370}
\DefMacro{table-numbers_of_programs_and_bugs-20230423-row2-num_unique_classes_out_of_working_templates}{32}
\DefMacro{table-numbers_of_programs_and_bugs-20230423-row2-exec-total}{10,465}
\DefMacro{table-numbers_of_programs_and_bugs-20230423-row2-exec-not_compile}{0}
\DefMacro{table-numbers_of_programs_and_bugs-20230423-row2-exec-timeout}{52}
\DefMacro{table-numbers_of_programs_and_bugs-20230423-row2-exec-ok}{10,333}
\DefMacro{table-numbers_of_programs_and_bugs-20230423-row2-exec-skipped_diff}{71}
\DefMacro{table-numbers_of_programs_and_bugs-20230423-row2-exec-skipped_crash}{8}
\DefMacro{table-numbers_of_programs_and_bugs-20230423-row2-exec-final_diff}{1}
\DefMacro{table-numbers_of_programs_and_bugs-20230423-row2-exec-final_crash}{0}
\DefMacro{table-numbers_of_programs_and_bugs-20230423-row3-project}{numbers}
\DefMacro{table-numbers_of_programs_and_bugs-20230423-row3-num_unit_tests}{2,996}
\DefMacro{table-numbers_of_programs_and_bugs-20230423-row3-num_unique_entry_methods}{44}
\DefMacro{table-numbers_of_programs_and_bugs-20230423-row3-num_unique_classes}{4}
\DefMacro{table-numbers_of_programs_and_bugs-20230423-row3-convert-fail}{79}
\DefMacro{table-numbers_of_programs_and_bugs-20230423-row3-convert-timeout}{0}
\DefMacro{table-numbers_of_programs_and_bugs-20230423-row3-convert-not_compile}{0}
\DefMacro{table-numbers_of_programs_and_bugs-20230423-row3-convert-no_original}{0}
\DefMacro{table-numbers_of_programs_and_bugs-20230423-row3-gen-fail}{0}
\DefMacro{table-numbers_of_programs_and_bugs-20230423-row3-gen-timeout}{89}
\DefMacro{table-numbers_of_programs_and_bugs-20230423-row3-gen-no_hole}{0}
\DefMacro{table-numbers_of_programs_and_bugs-20230423-row3-gen-no_gen}{39}
\DefMacro{table-numbers_of_programs_and_bugs-20230423-row3-num_working_templates}{2,878}
\DefMacro{table-numbers_of_programs_and_bugs-20230423-row3-num_unique_entry_methods_out_of_working_templates}{42}
\DefMacro{table-numbers_of_programs_and_bugs-20230423-row3-num_unique_classes_out_of_working_templates}{3}
\DefMacro{table-numbers_of_programs_and_bugs-20230423-row3-exec-total}{28,272}
\DefMacro{table-numbers_of_programs_and_bugs-20230423-row3-exec-not_compile}{0}
\DefMacro{table-numbers_of_programs_and_bugs-20230423-row3-exec-timeout}{4}
\DefMacro{table-numbers_of_programs_and_bugs-20230423-row3-exec-ok}{28,268}
\DefMacro{table-numbers_of_programs_and_bugs-20230423-row3-exec-skipped_diff}{0}
\DefMacro{table-numbers_of_programs_and_bugs-20230423-row3-exec-skipped_crash}{0}
\DefMacro{table-numbers_of_programs_and_bugs-20230423-row3-exec-final_diff}{0}
\DefMacro{table-numbers_of_programs_and_bugs-20230423-row3-exec-final_crash}{0}
\DefMacro{table-numbers_of_programs_and_bugs-20230423-row4-project}{vectorz}
\DefMacro{table-numbers_of_programs_and_bugs-20230423-row4-num_unit_tests}{1,148}
\DefMacro{table-numbers_of_programs_and_bugs-20230423-row4-num_unique_entry_methods}{907}
\DefMacro{table-numbers_of_programs_and_bugs-20230423-row4-num_unique_classes}{156}
\DefMacro{table-numbers_of_programs_and_bugs-20230423-row4-convert-fail}{0}
\DefMacro{table-numbers_of_programs_and_bugs-20230423-row4-convert-timeout}{0}
\DefMacro{table-numbers_of_programs_and_bugs-20230423-row4-convert-not_compile}{0}
\DefMacro{table-numbers_of_programs_and_bugs-20230423-row4-convert-no_original}{0}
\DefMacro{table-numbers_of_programs_and_bugs-20230423-row4-gen-fail}{61}
\DefMacro{table-numbers_of_programs_and_bugs-20230423-row4-gen-timeout}{18}
\DefMacro{table-numbers_of_programs_and_bugs-20230423-row4-gen-no_hole}{88}
\DefMacro{table-numbers_of_programs_and_bugs-20230423-row4-gen-no_gen}{259}
\DefMacro{table-numbers_of_programs_and_bugs-20230423-row4-num_working_templates}{801}
\DefMacro{table-numbers_of_programs_and_bugs-20230423-row4-num_unique_entry_methods_out_of_working_templates}{639}
\DefMacro{table-numbers_of_programs_and_bugs-20230423-row4-num_unique_classes_out_of_working_templates}{139}
\DefMacro{table-numbers_of_programs_and_bugs-20230423-row4-exec-total}{6,135}
\DefMacro{table-numbers_of_programs_and_bugs-20230423-row4-exec-not_compile}{0}
\DefMacro{table-numbers_of_programs_and_bugs-20230423-row4-exec-timeout}{269}
\DefMacro{table-numbers_of_programs_and_bugs-20230423-row4-exec-ok}{5,818}
\DefMacro{table-numbers_of_programs_and_bugs-20230423-row4-exec-skipped_diff}{37}
\DefMacro{table-numbers_of_programs_and_bugs-20230423-row4-exec-skipped_crash}{3}
\DefMacro{table-numbers_of_programs_and_bugs-20230423-row4-exec-final_diff}{7}
\DefMacro{table-numbers_of_programs_and_bugs-20230423-row4-exec-final_crash}{1}
\DefMacro{table-numbers_of_programs_and_bugs-20230423-row5-project}{statistics}
\DefMacro{table-numbers_of_programs_and_bugs-20230423-row5-num_unit_tests}{1,898}
\DefMacro{table-numbers_of_programs_and_bugs-20230423-row5-num_unique_entry_methods}{373}
\DefMacro{table-numbers_of_programs_and_bugs-20230423-row5-num_unique_classes}{27}
\DefMacro{table-numbers_of_programs_and_bugs-20230423-row5-convert-fail}{173}
\DefMacro{table-numbers_of_programs_and_bugs-20230423-row5-convert-timeout}{0}
\DefMacro{table-numbers_of_programs_and_bugs-20230423-row5-convert-not_compile}{0}
\DefMacro{table-numbers_of_programs_and_bugs-20230423-row5-convert-no_original}{0}
\DefMacro{table-numbers_of_programs_and_bugs-20230423-row5-gen-fail}{178}
\DefMacro{table-numbers_of_programs_and_bugs-20230423-row5-gen-timeout}{17}
\DefMacro{table-numbers_of_programs_and_bugs-20230423-row5-gen-no_hole}{0}
\DefMacro{table-numbers_of_programs_and_bugs-20230423-row5-gen-no_gen}{637}
\DefMacro{table-numbers_of_programs_and_bugs-20230423-row5-num_working_templates}{1,088}
\DefMacro{table-numbers_of_programs_and_bugs-20230423-row5-num_unique_entry_methods_out_of_working_templates}{249}
\DefMacro{table-numbers_of_programs_and_bugs-20230423-row5-num_unique_classes_out_of_working_templates}{24}
\DefMacro{table-numbers_of_programs_and_bugs-20230423-row5-exec-total}{8,309}
\DefMacro{table-numbers_of_programs_and_bugs-20230423-row5-exec-not_compile}{0}
\DefMacro{table-numbers_of_programs_and_bugs-20230423-row5-exec-timeout}{26}
\DefMacro{table-numbers_of_programs_and_bugs-20230423-row5-exec-ok}{8,237}
\DefMacro{table-numbers_of_programs_and_bugs-20230423-row5-exec-skipped_diff}{46}
\DefMacro{table-numbers_of_programs_and_bugs-20230423-row5-exec-skipped_crash}{0}
\DefMacro{table-numbers_of_programs_and_bugs-20230423-row5-exec-final_diff}{0}
\DefMacro{table-numbers_of_programs_and_bugs-20230423-row5-exec-final_crash}{0}
\DefMacro{table-numbers_of_programs_and_bugs-20230423-row6-project}{codec}
\DefMacro{table-numbers_of_programs_and_bugs-20230423-row6-num_unit_tests}{1,479}
\DefMacro{table-numbers_of_programs_and_bugs-20230423-row6-num_unique_entry_methods}{388}
\DefMacro{table-numbers_of_programs_and_bugs-20230423-row6-num_unique_classes}{49}
\DefMacro{table-numbers_of_programs_and_bugs-20230423-row6-convert-fail}{26}
\DefMacro{table-numbers_of_programs_and_bugs-20230423-row6-convert-timeout}{0}
\DefMacro{table-numbers_of_programs_and_bugs-20230423-row6-convert-not_compile}{10}
\DefMacro{table-numbers_of_programs_and_bugs-20230423-row6-convert-no_original}{0}
\DefMacro{table-numbers_of_programs_and_bugs-20230423-row6-gen-fail}{85}
\DefMacro{table-numbers_of_programs_and_bugs-20230423-row6-gen-timeout}{51}
\DefMacro{table-numbers_of_programs_and_bugs-20230423-row6-gen-no_hole}{0}
\DefMacro{table-numbers_of_programs_and_bugs-20230423-row6-gen-no_gen}{379}
\DefMacro{table-numbers_of_programs_and_bugs-20230423-row6-num_working_templates}{1,064}
\DefMacro{table-numbers_of_programs_and_bugs-20230423-row6-num_unique_entry_methods_out_of_working_templates}{300}
\DefMacro{table-numbers_of_programs_and_bugs-20230423-row6-num_unique_classes_out_of_working_templates}{36}
\DefMacro{table-numbers_of_programs_and_bugs-20230423-row6-exec-total}{6,037}
\DefMacro{table-numbers_of_programs_and_bugs-20230423-row6-exec-not_compile}{0}
\DefMacro{table-numbers_of_programs_and_bugs-20230423-row6-exec-timeout}{145}
\DefMacro{table-numbers_of_programs_and_bugs-20230423-row6-exec-ok}{5,606}
\DefMacro{table-numbers_of_programs_and_bugs-20230423-row6-exec-skipped_diff}{237}
\DefMacro{table-numbers_of_programs_and_bugs-20230423-row6-exec-skipped_crash}{21}
\DefMacro{table-numbers_of_programs_and_bugs-20230423-row6-exec-final_diff}{28}
\DefMacro{table-numbers_of_programs_and_bugs-20230423-row6-exec-final_crash}{0}
\DefMacro{table-numbers_of_programs_and_bugs-20230423-row7-project}{zxing}
\DefMacro{table-numbers_of_programs_and_bugs-20230423-row7-num_unit_tests}{1,481}
\DefMacro{table-numbers_of_programs_and_bugs-20230423-row7-num_unique_entry_methods}{352}
\DefMacro{table-numbers_of_programs_and_bugs-20230423-row7-num_unique_classes}{91}
\DefMacro{table-numbers_of_programs_and_bugs-20230423-row7-convert-fail}{60}
\DefMacro{table-numbers_of_programs_and_bugs-20230423-row7-convert-timeout}{0}
\DefMacro{table-numbers_of_programs_and_bugs-20230423-row7-convert-not_compile}{12}
\DefMacro{table-numbers_of_programs_and_bugs-20230423-row7-convert-no_original}{0}
\DefMacro{table-numbers_of_programs_and_bugs-20230423-row7-gen-fail}{92}
\DefMacro{table-numbers_of_programs_and_bugs-20230423-row7-gen-timeout}{57}
\DefMacro{table-numbers_of_programs_and_bugs-20230423-row7-gen-no_hole}{19}
\DefMacro{table-numbers_of_programs_and_bugs-20230423-row7-gen-no_gen}{211}
\DefMacro{table-numbers_of_programs_and_bugs-20230423-row7-num_working_templates}{1,179}
\DefMacro{table-numbers_of_programs_and_bugs-20230423-row7-num_unique_entry_methods_out_of_working_templates}{288}
\DefMacro{table-numbers_of_programs_and_bugs-20230423-row7-num_unique_classes_out_of_working_templates}{73}
\DefMacro{table-numbers_of_programs_and_bugs-20230423-row7-exec-total}{10,366}
\DefMacro{table-numbers_of_programs_and_bugs-20230423-row7-exec-not_compile}{0}
\DefMacro{table-numbers_of_programs_and_bugs-20230423-row7-exec-timeout}{148}
\DefMacro{table-numbers_of_programs_and_bugs-20230423-row7-exec-ok}{9,320}
\DefMacro{table-numbers_of_programs_and_bugs-20230423-row7-exec-skipped_diff}{892}
\DefMacro{table-numbers_of_programs_and_bugs-20230423-row7-exec-skipped_crash}{0}
\DefMacro{table-numbers_of_programs_and_bugs-20230423-row7-exec-final_diff}{4}
\DefMacro{table-numbers_of_programs_and_bugs-20230423-row7-exec-final_crash}{2}
\DefMacro{table-numbers_of_programs_and_bugs-20230423-row8-project}{compress}
\DefMacro{table-numbers_of_programs_and_bugs-20230423-row8-num_unit_tests}{1,089}
\DefMacro{table-numbers_of_programs_and_bugs-20230423-row8-num_unique_entry_methods}{575}
\DefMacro{table-numbers_of_programs_and_bugs-20230423-row8-num_unique_classes}{110}
\DefMacro{table-numbers_of_programs_and_bugs-20230423-row8-convert-fail}{14}
\DefMacro{table-numbers_of_programs_and_bugs-20230423-row8-convert-timeout}{0}
\DefMacro{table-numbers_of_programs_and_bugs-20230423-row8-convert-not_compile}{20}
\DefMacro{table-numbers_of_programs_and_bugs-20230423-row8-convert-no_original}{0}
\DefMacro{table-numbers_of_programs_and_bugs-20230423-row8-gen-fail}{48}
\DefMacro{table-numbers_of_programs_and_bugs-20230423-row8-gen-timeout}{31}
\DefMacro{table-numbers_of_programs_and_bugs-20230423-row8-gen-no_hole}{17}
\DefMacro{table-numbers_of_programs_and_bugs-20230423-row8-gen-no_gen}{156}
\DefMacro{table-numbers_of_programs_and_bugs-20230423-row8-num_working_templates}{882}
\DefMacro{table-numbers_of_programs_and_bugs-20230423-row8-num_unique_entry_methods_out_of_working_templates}{479}
\DefMacro{table-numbers_of_programs_and_bugs-20230423-row8-num_unique_classes_out_of_working_templates}{97}
\DefMacro{table-numbers_of_programs_and_bugs-20230423-row8-exec-total}{7,646}
\DefMacro{table-numbers_of_programs_and_bugs-20230423-row8-exec-not_compile}{0}
\DefMacro{table-numbers_of_programs_and_bugs-20230423-row8-exec-timeout}{102}
\DefMacro{table-numbers_of_programs_and_bugs-20230423-row8-exec-ok}{6,759}
\DefMacro{table-numbers_of_programs_and_bugs-20230423-row8-exec-skipped_diff}{695}
\DefMacro{table-numbers_of_programs_and_bugs-20230423-row8-exec-skipped_crash}{1}
\DefMacro{table-numbers_of_programs_and_bugs-20230423-row8-exec-final_diff}{89}
\DefMacro{table-numbers_of_programs_and_bugs-20230423-row8-exec-final_crash}{0}
\DefMacro{table-numbers_of_programs_and_bugs-20230423-row9-project}{$\sum$}
\DefMacro{table-numbers_of_programs_and_bugs-20230423-row9-num_unit_tests}{14,770}
\DefMacro{table-numbers_of_programs_and_bugs-20230423-row9-num_unique_entry_methods}{5,004}
\DefMacro{table-numbers_of_programs_and_bugs-20230423-row9-num_unique_classes}{782}
\DefMacro{table-numbers_of_programs_and_bugs-20230423-row9-convert-fail}{491}
\DefMacro{table-numbers_of_programs_and_bugs-20230423-row9-convert-timeout}{0}
\DefMacro{table-numbers_of_programs_and_bugs-20230423-row9-convert-not_compile}{55}
\DefMacro{table-numbers_of_programs_and_bugs-20230423-row9-convert-no_original}{0}
\DefMacro{table-numbers_of_programs_and_bugs-20230423-row9-gen-fail}{687}
\DefMacro{table-numbers_of_programs_and_bugs-20230423-row9-gen-timeout}{393}
\DefMacro{table-numbers_of_programs_and_bugs-20230423-row9-gen-no_hole}{301}
\DefMacro{table-numbers_of_programs_and_bugs-20230423-row9-gen-no_gen}{2,723}
\DefMacro{table-numbers_of_programs_and_bugs-20230423-row9-num_working_templates}{11,200}
\DefMacro{table-numbers_of_programs_and_bugs-20230423-row9-num_unique_entry_methods_out_of_working_templates}{3,730}
\DefMacro{table-numbers_of_programs_and_bugs-20230423-row9-num_unique_classes_out_of_working_templates}{650}
\DefMacro{table-numbers_of_programs_and_bugs-20230423-row9-exec-total}{95,605}
\DefMacro{table-numbers_of_programs_and_bugs-20230423-row9-exec-not_compile}{0}
\DefMacro{table-numbers_of_programs_and_bugs-20230423-row9-exec-timeout}{909}
\DefMacro{table-numbers_of_programs_and_bugs-20230423-row9-exec-ok}{92,249}
\DefMacro{table-numbers_of_programs_and_bugs-20230423-row9-exec-skipped_diff}{2,245}
\DefMacro{table-numbers_of_programs_and_bugs-20230423-row9-exec-skipped_crash}{49}
\DefMacro{table-numbers_of_programs_and_bugs-20230423-row9-exec-final_diff}{140}
\DefMacro{table-numbers_of_programs_and_bugs-20230423-row9-exec-final_crash}{13}

\DefMacro{table-numbers_of_programs_and_bugs-20230409_20230411-row0-project}{lang}
\DefMacro{table-numbers_of_programs_and_bugs-20230409_20230411-row0-num_unit_tests}{1,971}
\DefMacro{table-numbers_of_programs_and_bugs-20230409_20230411-row0-num_unique_entry_methods}{1,148}
\DefMacro{table-numbers_of_programs_and_bugs-20230409_20230411-row0-num_unique_classes}{100}
\DefMacro{table-numbers_of_programs_and_bugs-20230409_20230411-row0-convert-fail}{98}
\DefMacro{table-numbers_of_programs_and_bugs-20230409_20230411-row0-convert-timeout}{0}
\DefMacro{table-numbers_of_programs_and_bugs-20230409_20230411-row0-convert-not_compile}{3}
\DefMacro{table-numbers_of_programs_and_bugs-20230409_20230411-row0-convert-no_original}{0}
\DefMacro{table-numbers_of_programs_and_bugs-20230409_20230411-row0-gen-fail}{46}
\DefMacro{table-numbers_of_programs_and_bugs-20230409_20230411-row0-gen-timeout}{96}
\DefMacro{table-numbers_of_programs_and_bugs-20230409_20230411-row0-gen-no_hole}{48}
\DefMacro{table-numbers_of_programs_and_bugs-20230409_20230411-row0-gen-no_gen}{323}
\DefMacro{table-numbers_of_programs_and_bugs-20230409_20230411-row0-num_working_templates}{1,499}
\DefMacro{table-numbers_of_programs_and_bugs-20230409_20230411-row0-num_unique_entry_methods_out_of_working_templates}{879}
\DefMacro{table-numbers_of_programs_and_bugs-20230409_20230411-row0-num_unique_classes_out_of_working_templates}{80}
\DefMacro{table-numbers_of_programs_and_bugs-20230409_20230411-row0-exec-total}{12,522}
\DefMacro{table-numbers_of_programs_and_bugs-20230409_20230411-row0-exec-not_compile}{0}
\DefMacro{table-numbers_of_programs_and_bugs-20230409_20230411-row0-exec-timeout}{114}
\DefMacro{table-numbers_of_programs_and_bugs-20230409_20230411-row0-exec-ok}{12,174}
\DefMacro{table-numbers_of_programs_and_bugs-20230409_20230411-row0-exec-skipped_diff}{213}
\DefMacro{table-numbers_of_programs_and_bugs-20230409_20230411-row0-exec-skipped_crash}{12}
\DefMacro{table-numbers_of_programs_and_bugs-20230409_20230411-row0-exec-final_diff}{0}
\DefMacro{table-numbers_of_programs_and_bugs-20230409_20230411-row0-exec-final_crash}{9}
\DefMacro{table-numbers_of_programs_and_bugs-20230409_20230411-row1-project}{math}
\DefMacro{table-numbers_of_programs_and_bugs-20230409_20230411-row1-num_unit_tests}{1,085}
\DefMacro{table-numbers_of_programs_and_bugs-20230409_20230411-row1-num_unique_entry_methods}{843}
\DefMacro{table-numbers_of_programs_and_bugs-20230409_20230411-row1-num_unique_classes}{240}
\DefMacro{table-numbers_of_programs_and_bugs-20230409_20230411-row1-convert-fail}{25}
\DefMacro{table-numbers_of_programs_and_bugs-20230409_20230411-row1-convert-timeout}{0}
\DefMacro{table-numbers_of_programs_and_bugs-20230409_20230411-row1-convert-not_compile}{19}
\DefMacro{table-numbers_of_programs_and_bugs-20230409_20230411-row1-convert-no_original}{0}
\DefMacro{table-numbers_of_programs_and_bugs-20230409_20230411-row1-gen-fail}{162}
\DefMacro{table-numbers_of_programs_and_bugs-20230409_20230411-row1-gen-timeout}{52}
\DefMacro{table-numbers_of_programs_and_bugs-20230409_20230411-row1-gen-no_hole}{40}
\DefMacro{table-numbers_of_programs_and_bugs-20230409_20230411-row1-gen-no_gen}{272}
\DefMacro{table-numbers_of_programs_and_bugs-20230409_20230411-row1-num_working_templates}{729}
\DefMacro{table-numbers_of_programs_and_bugs-20230409_20230411-row1-num_unique_entry_methods_out_of_working_templates}{583}
\DefMacro{table-numbers_of_programs_and_bugs-20230409_20230411-row1-num_unique_classes_out_of_working_templates}{188}
\DefMacro{table-numbers_of_programs_and_bugs-20230409_20230411-row1-exec-total}{5,961}
\DefMacro{table-numbers_of_programs_and_bugs-20230409_20230411-row1-exec-not_compile}{0}
\DefMacro{table-numbers_of_programs_and_bugs-20230409_20230411-row1-exec-timeout}{29}
\DefMacro{table-numbers_of_programs_and_bugs-20230409_20230411-row1-exec-ok}{5,826}
\DefMacro{table-numbers_of_programs_and_bugs-20230409_20230411-row1-exec-skipped_diff}{106}
\DefMacro{table-numbers_of_programs_and_bugs-20230409_20230411-row1-exec-skipped_crash}{0}
\DefMacro{table-numbers_of_programs_and_bugs-20230409_20230411-row1-exec-final_diff}{0}
\DefMacro{table-numbers_of_programs_and_bugs-20230409_20230411-row1-exec-final_crash}{0}
\DefMacro{table-numbers_of_programs_and_bugs-20230409_20230411-row2-project}{text}
\DefMacro{table-numbers_of_programs_and_bugs-20230409_20230411-row2-num_unit_tests}{1,624}
\DefMacro{table-numbers_of_programs_and_bugs-20230409_20230411-row2-num_unique_entry_methods}{525}
\DefMacro{table-numbers_of_programs_and_bugs-20230409_20230411-row2-num_unique_classes}{40}
\DefMacro{table-numbers_of_programs_and_bugs-20230409_20230411-row2-convert-fail}{18}
\DefMacro{table-numbers_of_programs_and_bugs-20230409_20230411-row2-convert-timeout}{0}
\DefMacro{table-numbers_of_programs_and_bugs-20230409_20230411-row2-convert-not_compile}{0}
\DefMacro{table-numbers_of_programs_and_bugs-20230409_20230411-row2-convert-no_original}{0}
\DefMacro{table-numbers_of_programs_and_bugs-20230409_20230411-row2-gen-fail}{10}
\DefMacro{table-numbers_of_programs_and_bugs-20230409_20230411-row2-gen-timeout}{44}
\DefMacro{table-numbers_of_programs_and_bugs-20230409_20230411-row2-gen-no_hole}{10}
\DefMacro{table-numbers_of_programs_and_bugs-20230409_20230411-row2-gen-no_gen}{330}
\DefMacro{table-numbers_of_programs_and_bugs-20230409_20230411-row2-num_working_templates}{1,266}
\DefMacro{table-numbers_of_programs_and_bugs-20230409_20230411-row2-num_unique_entry_methods_out_of_working_templates}{453}
\DefMacro{table-numbers_of_programs_and_bugs-20230409_20230411-row2-num_unique_classes_out_of_working_templates}{33}
\DefMacro{table-numbers_of_programs_and_bugs-20230409_20230411-row2-exec-total}{11,748}
\DefMacro{table-numbers_of_programs_and_bugs-20230409_20230411-row2-exec-not_compile}{0}
\DefMacro{table-numbers_of_programs_and_bugs-20230409_20230411-row2-exec-timeout}{68}
\DefMacro{table-numbers_of_programs_and_bugs-20230409_20230411-row2-exec-ok}{11,552}
\DefMacro{table-numbers_of_programs_and_bugs-20230409_20230411-row2-exec-skipped_diff}{125}
\DefMacro{table-numbers_of_programs_and_bugs-20230409_20230411-row2-exec-skipped_crash}{2}
\DefMacro{table-numbers_of_programs_and_bugs-20230409_20230411-row2-exec-final_diff}{0}
\DefMacro{table-numbers_of_programs_and_bugs-20230409_20230411-row2-exec-final_crash}{1}
\DefMacro{table-numbers_of_programs_and_bugs-20230409_20230411-row3-project}{numbers}
\DefMacro{table-numbers_of_programs_and_bugs-20230409_20230411-row3-num_unit_tests}{3,061}
\DefMacro{table-numbers_of_programs_and_bugs-20230409_20230411-row3-num_unique_entry_methods}{45}
\DefMacro{table-numbers_of_programs_and_bugs-20230409_20230411-row3-num_unique_classes}{4}
\DefMacro{table-numbers_of_programs_and_bugs-20230409_20230411-row3-convert-fail}{124}
\DefMacro{table-numbers_of_programs_and_bugs-20230409_20230411-row3-convert-timeout}{0}
\DefMacro{table-numbers_of_programs_and_bugs-20230409_20230411-row3-convert-not_compile}{0}
\DefMacro{table-numbers_of_programs_and_bugs-20230409_20230411-row3-convert-no_original}{0}
\DefMacro{table-numbers_of_programs_and_bugs-20230409_20230411-row3-gen-fail}{0}
\DefMacro{table-numbers_of_programs_and_bugs-20230409_20230411-row3-gen-timeout}{0}
\DefMacro{table-numbers_of_programs_and_bugs-20230409_20230411-row3-gen-no_hole}{0}
\DefMacro{table-numbers_of_programs_and_bugs-20230409_20230411-row3-gen-no_gen}{87}
\DefMacro{table-numbers_of_programs_and_bugs-20230409_20230411-row3-num_working_templates}{2,850}
\DefMacro{table-numbers_of_programs_and_bugs-20230409_20230411-row3-num_unique_entry_methods_out_of_working_templates}{41}
\DefMacro{table-numbers_of_programs_and_bugs-20230409_20230411-row3-num_unique_classes_out_of_working_templates}{3}
\DefMacro{table-numbers_of_programs_and_bugs-20230409_20230411-row3-exec-total}{28,500}
\DefMacro{table-numbers_of_programs_and_bugs-20230409_20230411-row3-exec-not_compile}{0}
\DefMacro{table-numbers_of_programs_and_bugs-20230409_20230411-row3-exec-timeout}{0}
\DefMacro{table-numbers_of_programs_and_bugs-20230409_20230411-row3-exec-ok}{28,500}
\DefMacro{table-numbers_of_programs_and_bugs-20230409_20230411-row3-exec-skipped_diff}{0}
\DefMacro{table-numbers_of_programs_and_bugs-20230409_20230411-row3-exec-skipped_crash}{0}
\DefMacro{table-numbers_of_programs_and_bugs-20230409_20230411-row3-exec-final_diff}{0}
\DefMacro{table-numbers_of_programs_and_bugs-20230409_20230411-row3-exec-final_crash}{0}
\DefMacro{table-numbers_of_programs_and_bugs-20230409_20230411-row4-project}{vectorz}
\DefMacro{table-numbers_of_programs_and_bugs-20230409_20230411-row4-num_unit_tests}{1,139}
\DefMacro{table-numbers_of_programs_and_bugs-20230409_20230411-row4-num_unique_entry_methods}{910}
\DefMacro{table-numbers_of_programs_and_bugs-20230409_20230411-row4-num_unique_classes}{160}
\DefMacro{table-numbers_of_programs_and_bugs-20230409_20230411-row4-convert-fail}{0}
\DefMacro{table-numbers_of_programs_and_bugs-20230409_20230411-row4-convert-timeout}{0}
\DefMacro{table-numbers_of_programs_and_bugs-20230409_20230411-row4-convert-not_compile}{0}
\DefMacro{table-numbers_of_programs_and_bugs-20230409_20230411-row4-convert-no_original}{0}
\DefMacro{table-numbers_of_programs_and_bugs-20230409_20230411-row4-gen-fail}{71}
\DefMacro{table-numbers_of_programs_and_bugs-20230409_20230411-row4-gen-timeout}{25}
\DefMacro{table-numbers_of_programs_and_bugs-20230409_20230411-row4-gen-no_hole}{123}
\DefMacro{table-numbers_of_programs_and_bugs-20230409_20230411-row4-gen-no_gen}{216}
\DefMacro{table-numbers_of_programs_and_bugs-20230409_20230411-row4-num_working_templates}{800}
\DefMacro{table-numbers_of_programs_and_bugs-20230409_20230411-row4-num_unique_entry_methods_out_of_working_templates}{658}
\DefMacro{table-numbers_of_programs_and_bugs-20230409_20230411-row4-num_unique_classes_out_of_working_templates}{143}
\DefMacro{table-numbers_of_programs_and_bugs-20230409_20230411-row4-exec-total}{5,967}
\DefMacro{table-numbers_of_programs_and_bugs-20230409_20230411-row4-exec-not_compile}{0}
\DefMacro{table-numbers_of_programs_and_bugs-20230409_20230411-row4-exec-timeout}{37}
\DefMacro{table-numbers_of_programs_and_bugs-20230409_20230411-row4-exec-ok}{5,775}
\DefMacro{table-numbers_of_programs_and_bugs-20230409_20230411-row4-exec-skipped_diff}{151}
\DefMacro{table-numbers_of_programs_and_bugs-20230409_20230411-row4-exec-skipped_crash}{0}
\DefMacro{table-numbers_of_programs_and_bugs-20230409_20230411-row4-exec-final_diff}{3}
\DefMacro{table-numbers_of_programs_and_bugs-20230409_20230411-row4-exec-final_crash}{1}
\DefMacro{table-numbers_of_programs_and_bugs-20230409_20230411-row5-project}{statistics}
\DefMacro{table-numbers_of_programs_and_bugs-20230409_20230411-row5-num_unit_tests}{1,952}
\DefMacro{table-numbers_of_programs_and_bugs-20230409_20230411-row5-num_unique_entry_methods}{379}
\DefMacro{table-numbers_of_programs_and_bugs-20230409_20230411-row5-num_unique_classes}{27}
\DefMacro{table-numbers_of_programs_and_bugs-20230409_20230411-row5-convert-fail}{197}
\DefMacro{table-numbers_of_programs_and_bugs-20230409_20230411-row5-convert-timeout}{0}
\DefMacro{table-numbers_of_programs_and_bugs-20230409_20230411-row5-convert-not_compile}{0}
\DefMacro{table-numbers_of_programs_and_bugs-20230409_20230411-row5-convert-no_original}{0}
\DefMacro{table-numbers_of_programs_and_bugs-20230409_20230411-row5-gen-fail}{140}
\DefMacro{table-numbers_of_programs_and_bugs-20230409_20230411-row5-gen-timeout}{30}
\DefMacro{table-numbers_of_programs_and_bugs-20230409_20230411-row5-gen-no_hole}{0}
\DefMacro{table-numbers_of_programs_and_bugs-20230409_20230411-row5-gen-no_gen}{602}
\DefMacro{table-numbers_of_programs_and_bugs-20230409_20230411-row5-num_working_templates}{1,153}
\DefMacro{table-numbers_of_programs_and_bugs-20230409_20230411-row5-num_unique_entry_methods_out_of_working_templates}{259}
\DefMacro{table-numbers_of_programs_and_bugs-20230409_20230411-row5-num_unique_classes_out_of_working_templates}{25}
\DefMacro{table-numbers_of_programs_and_bugs-20230409_20230411-row5-exec-total}{9,135}
\DefMacro{table-numbers_of_programs_and_bugs-20230409_20230411-row5-exec-not_compile}{0}
\DefMacro{table-numbers_of_programs_and_bugs-20230409_20230411-row5-exec-timeout}{16}
\DefMacro{table-numbers_of_programs_and_bugs-20230409_20230411-row5-exec-ok}{9,047}
\DefMacro{table-numbers_of_programs_and_bugs-20230409_20230411-row5-exec-skipped_diff}{72}
\DefMacro{table-numbers_of_programs_and_bugs-20230409_20230411-row5-exec-skipped_crash}{0}
\DefMacro{table-numbers_of_programs_and_bugs-20230409_20230411-row5-exec-final_diff}{0}
\DefMacro{table-numbers_of_programs_and_bugs-20230409_20230411-row5-exec-final_crash}{0}
\DefMacro{table-numbers_of_programs_and_bugs-20230409_20230411-row6-project}{codec}
\DefMacro{table-numbers_of_programs_and_bugs-20230409_20230411-row6-num_unit_tests}{1,349}
\DefMacro{table-numbers_of_programs_and_bugs-20230409_20230411-row6-num_unique_entry_methods}{340}
\DefMacro{table-numbers_of_programs_and_bugs-20230409_20230411-row6-num_unique_classes}{49}
\DefMacro{table-numbers_of_programs_and_bugs-20230409_20230411-row6-convert-fail}{16}
\DefMacro{table-numbers_of_programs_and_bugs-20230409_20230411-row6-convert-timeout}{0}
\DefMacro{table-numbers_of_programs_and_bugs-20230409_20230411-row6-convert-not_compile}{10}
\DefMacro{table-numbers_of_programs_and_bugs-20230409_20230411-row6-convert-no_original}{0}
\DefMacro{table-numbers_of_programs_and_bugs-20230409_20230411-row6-gen-fail}{78}
\DefMacro{table-numbers_of_programs_and_bugs-20230409_20230411-row6-gen-timeout}{27}
\DefMacro{table-numbers_of_programs_and_bugs-20230409_20230411-row6-gen-no_hole}{1}
\DefMacro{table-numbers_of_programs_and_bugs-20230409_20230411-row6-gen-no_gen}{192}
\DefMacro{table-numbers_of_programs_and_bugs-20230409_20230411-row6-num_working_templates}{1,130}
\DefMacro{table-numbers_of_programs_and_bugs-20230409_20230411-row6-num_unique_entry_methods_out_of_working_templates}{286}
\DefMacro{table-numbers_of_programs_and_bugs-20230409_20230411-row6-num_unique_classes_out_of_working_templates}{38}
\DefMacro{table-numbers_of_programs_and_bugs-20230409_20230411-row6-exec-total}{5,889}
\DefMacro{table-numbers_of_programs_and_bugs-20230409_20230411-row6-exec-not_compile}{0}
\DefMacro{table-numbers_of_programs_and_bugs-20230409_20230411-row6-exec-timeout}{106}
\DefMacro{table-numbers_of_programs_and_bugs-20230409_20230411-row6-exec-ok}{5,705}
\DefMacro{table-numbers_of_programs_and_bugs-20230409_20230411-row6-exec-skipped_diff}{74}
\DefMacro{table-numbers_of_programs_and_bugs-20230409_20230411-row6-exec-skipped_crash}{2}
\DefMacro{table-numbers_of_programs_and_bugs-20230409_20230411-row6-exec-final_diff}{2}
\DefMacro{table-numbers_of_programs_and_bugs-20230409_20230411-row6-exec-final_crash}{0}
\DefMacro{table-numbers_of_programs_and_bugs-20230409_20230411-row7-project}{jfreechart}
\DefMacro{table-numbers_of_programs_and_bugs-20230409_20230411-row7-num_unit_tests}{1,294}
\DefMacro{table-numbers_of_programs_and_bugs-20230409_20230411-row7-num_unique_entry_methods}{1,103}
\DefMacro{table-numbers_of_programs_and_bugs-20230409_20230411-row7-num_unique_classes}{269}
\DefMacro{table-numbers_of_programs_and_bugs-20230409_20230411-row7-convert-fail}{5}
\DefMacro{table-numbers_of_programs_and_bugs-20230409_20230411-row7-convert-timeout}{116}
\DefMacro{table-numbers_of_programs_and_bugs-20230409_20230411-row7-convert-not_compile}{11}
\DefMacro{table-numbers_of_programs_and_bugs-20230409_20230411-row7-convert-no_original}{0}
\DefMacro{table-numbers_of_programs_and_bugs-20230409_20230411-row7-gen-fail}{162}
\DefMacro{table-numbers_of_programs_and_bugs-20230409_20230411-row7-gen-timeout}{33}
\DefMacro{table-numbers_of_programs_and_bugs-20230409_20230411-row7-gen-no_hole}{23}
\DefMacro{table-numbers_of_programs_and_bugs-20230409_20230411-row7-gen-no_gen}{232}
\DefMacro{table-numbers_of_programs_and_bugs-20230409_20230411-row7-num_working_templates}{907}
\DefMacro{table-numbers_of_programs_and_bugs-20230409_20230411-row7-num_unique_entry_methods_out_of_working_templates}{775}
\DefMacro{table-numbers_of_programs_and_bugs-20230409_20230411-row7-num_unique_classes_out_of_working_templates}{225}
\DefMacro{table-numbers_of_programs_and_bugs-20230409_20230411-row7-exec-total}{7,660}
\DefMacro{table-numbers_of_programs_and_bugs-20230409_20230411-row7-exec-not_compile}{0}
\DefMacro{table-numbers_of_programs_and_bugs-20230409_20230411-row7-exec-timeout}{231}
\DefMacro{table-numbers_of_programs_and_bugs-20230409_20230411-row7-exec-ok}{7,382}
\DefMacro{table-numbers_of_programs_and_bugs-20230409_20230411-row7-exec-skipped_diff}{45}
\DefMacro{table-numbers_of_programs_and_bugs-20230409_20230411-row7-exec-skipped_crash}{1}
\DefMacro{table-numbers_of_programs_and_bugs-20230409_20230411-row7-exec-final_diff}{1}
\DefMacro{table-numbers_of_programs_and_bugs-20230409_20230411-row7-exec-final_crash}{0}
\DefMacro{table-numbers_of_programs_and_bugs-20230409_20230411-row8-project}{zxing}
\DefMacro{table-numbers_of_programs_and_bugs-20230409_20230411-row8-num_unit_tests}{1,420}
\DefMacro{table-numbers_of_programs_and_bugs-20230409_20230411-row8-num_unique_entry_methods}{327}
\DefMacro{table-numbers_of_programs_and_bugs-20230409_20230411-row8-num_unique_classes}{92}
\DefMacro{table-numbers_of_programs_and_bugs-20230409_20230411-row8-convert-fail}{56}
\DefMacro{table-numbers_of_programs_and_bugs-20230409_20230411-row8-convert-timeout}{0}
\DefMacro{table-numbers_of_programs_and_bugs-20230409_20230411-row8-convert-not_compile}{15}
\DefMacro{table-numbers_of_programs_and_bugs-20230409_20230411-row8-convert-no_original}{0}
\DefMacro{table-numbers_of_programs_and_bugs-20230409_20230411-row8-gen-fail}{75}
\DefMacro{table-numbers_of_programs_and_bugs-20230409_20230411-row8-gen-timeout}{87}
\DefMacro{table-numbers_of_programs_and_bugs-20230409_20230411-row8-gen-no_hole}{20}
\DefMacro{table-numbers_of_programs_and_bugs-20230409_20230411-row8-gen-no_gen}{161}
\DefMacro{table-numbers_of_programs_and_bugs-20230409_20230411-row8-num_working_templates}{1,168}
\DefMacro{table-numbers_of_programs_and_bugs-20230409_20230411-row8-num_unique_entry_methods_out_of_working_templates}{287}
\DefMacro{table-numbers_of_programs_and_bugs-20230409_20230411-row8-num_unique_classes_out_of_working_templates}{76}
\DefMacro{table-numbers_of_programs_and_bugs-20230409_20230411-row8-exec-total}{10,050}
\DefMacro{table-numbers_of_programs_and_bugs-20230409_20230411-row8-exec-not_compile}{0}
\DefMacro{table-numbers_of_programs_and_bugs-20230409_20230411-row8-exec-timeout}{153}
\DefMacro{table-numbers_of_programs_and_bugs-20230409_20230411-row8-exec-ok}{9,431}
\DefMacro{table-numbers_of_programs_and_bugs-20230409_20230411-row8-exec-skipped_diff}{457}
\DefMacro{table-numbers_of_programs_and_bugs-20230409_20230411-row8-exec-skipped_crash}{0}
\DefMacro{table-numbers_of_programs_and_bugs-20230409_20230411-row8-exec-final_diff}{6}
\DefMacro{table-numbers_of_programs_and_bugs-20230409_20230411-row8-exec-final_crash}{3}
\DefMacro{table-numbers_of_programs_and_bugs-20230409_20230411-row9-project}{$\sum$}
\DefMacro{table-numbers_of_programs_and_bugs-20230409_20230411-row9-num_unit_tests}{14,895}
\DefMacro{table-numbers_of_programs_and_bugs-20230409_20230411-row9-num_unique_entry_methods}{5,620}
\DefMacro{table-numbers_of_programs_and_bugs-20230409_20230411-row9-num_unique_classes}{981}
\DefMacro{table-numbers_of_programs_and_bugs-20230409_20230411-row9-convert-fail}{539}
\DefMacro{table-numbers_of_programs_and_bugs-20230409_20230411-row9-convert-timeout}{116}
\DefMacro{table-numbers_of_programs_and_bugs-20230409_20230411-row9-convert-not_compile}{58}
\DefMacro{table-numbers_of_programs_and_bugs-20230409_20230411-row9-convert-no_original}{0}
\DefMacro{table-numbers_of_programs_and_bugs-20230409_20230411-row9-gen-fail}{744}
\DefMacro{table-numbers_of_programs_and_bugs-20230409_20230411-row9-gen-timeout}{394}
\DefMacro{table-numbers_of_programs_and_bugs-20230409_20230411-row9-gen-no_hole}{265}
\DefMacro{table-numbers_of_programs_and_bugs-20230409_20230411-row9-gen-no_gen}{2,415}
\DefMacro{table-numbers_of_programs_and_bugs-20230409_20230411-row9-num_working_templates}{11,502}
\DefMacro{table-numbers_of_programs_and_bugs-20230409_20230411-row9-num_unique_entry_methods_out_of_working_templates}{4,221}
\DefMacro{table-numbers_of_programs_and_bugs-20230409_20230411-row9-num_unique_classes_out_of_working_templates}{811}
\DefMacro{table-numbers_of_programs_and_bugs-20230409_20230411-row9-exec-total}{97,432}
\DefMacro{table-numbers_of_programs_and_bugs-20230409_20230411-row9-exec-not_compile}{0}
\DefMacro{table-numbers_of_programs_and_bugs-20230409_20230411-row9-exec-timeout}{754}
\DefMacro{table-numbers_of_programs_and_bugs-20230409_20230411-row9-exec-ok}{95,392}
\DefMacro{table-numbers_of_programs_and_bugs-20230409_20230411-row9-exec-skipped_diff}{1,243}
\DefMacro{table-numbers_of_programs_and_bugs-20230409_20230411-row9-exec-skipped_crash}{17}
\DefMacro{table-numbers_of_programs_and_bugs-20230409_20230411-row9-exec-final_diff}{12}
\DefMacro{table-numbers_of_programs_and_bugs-20230409_20230411-row9-exec-final_crash}{14}

\DefMacro{table-numbers_of_programs_and_bugs-20231009-row0-project}{lang}
\DefMacro{table-numbers_of_programs_and_bugs-20231009-row0-num_unit_tests}{2,827}
\DefMacro{table-numbers_of_programs_and_bugs-20231009-row0-num_unique_entry_methods}{2,821}
\DefMacro{table-numbers_of_programs_and_bugs-20231009-row0-num_unique_classes}{131}
\DefMacro{table-numbers_of_programs_and_bugs-20231009-row0-convert-fail}{0}
\DefMacro{table-numbers_of_programs_and_bugs-20231009-row0-convert-timeout}{2}
\DefMacro{table-numbers_of_programs_and_bugs-20231009-row0-convert-not_compile}{202}
\DefMacro{table-numbers_of_programs_and_bugs-20231009-row0-convert-no_original}{0}
\DefMacro{table-numbers_of_programs_and_bugs-20231009-row0-gen-fail}{140}
\DefMacro{table-numbers_of_programs_and_bugs-20231009-row0-gen-timeout}{127}
\DefMacro{table-numbers_of_programs_and_bugs-20231009-row0-gen-no_hole}{241}
\DefMacro{table-numbers_of_programs_and_bugs-20231009-row0-gen-no_gen}{526}
\DefMacro{table-numbers_of_programs_and_bugs-20231009-row0-num_working_templates}{1,856}
\DefMacro{table-numbers_of_programs_and_bugs-20231009-row0-num_unique_entry_methods_out_of_working_templates}{1,856}
\DefMacro{table-numbers_of_programs_and_bugs-20231009-row0-num_unique_classes_out_of_working_templates}{104}
\DefMacro{table-numbers_of_programs_and_bugs-20231009-row0-exec-total}{15,433}
\DefMacro{table-numbers_of_programs_and_bugs-20231009-row0-exec-not_compile}{0}
\DefMacro{table-numbers_of_programs_and_bugs-20231009-row0-exec-timeout}{238}
\DefMacro{table-numbers_of_programs_and_bugs-20231009-row0-exec-ok}{14,598}
\DefMacro{table-numbers_of_programs_and_bugs-20231009-row0-exec-skipped_diff}{554}
\DefMacro{table-numbers_of_programs_and_bugs-20231009-row0-exec-skipped_crash}{23}
\DefMacro{table-numbers_of_programs_and_bugs-20231009-row0-exec-final_diff}{14}
\DefMacro{table-numbers_of_programs_and_bugs-20231009-row0-exec-final_crash}{6}
\DefMacro{table-numbers_of_programs_and_bugs-20231009-row1-project}{math}
\DefMacro{table-numbers_of_programs_and_bugs-20231009-row1-num_unit_tests}{5,296}
\DefMacro{table-numbers_of_programs_and_bugs-20231009-row1-num_unique_entry_methods}{5,296}
\DefMacro{table-numbers_of_programs_and_bugs-20231009-row1-num_unique_classes}{591}
\DefMacro{table-numbers_of_programs_and_bugs-20231009-row1-convert-fail}{8}
\DefMacro{table-numbers_of_programs_and_bugs-20231009-row1-convert-timeout}{7}
\DefMacro{table-numbers_of_programs_and_bugs-20231009-row1-convert-not_compile}{174}
\DefMacro{table-numbers_of_programs_and_bugs-20231009-row1-convert-no_original}{0}
\DefMacro{table-numbers_of_programs_and_bugs-20231009-row1-gen-fail}{367}
\DefMacro{table-numbers_of_programs_and_bugs-20231009-row1-gen-timeout}{261}
\DefMacro{table-numbers_of_programs_and_bugs-20231009-row1-gen-no_hole}{760}
\DefMacro{table-numbers_of_programs_and_bugs-20231009-row1-gen-no_gen}{986}
\DefMacro{table-numbers_of_programs_and_bugs-20231009-row1-num_working_templates}{3,361}
\DefMacro{table-numbers_of_programs_and_bugs-20231009-row1-num_unique_entry_methods_out_of_working_templates}{3,361}
\DefMacro{table-numbers_of_programs_and_bugs-20231009-row1-num_unique_classes_out_of_working_templates}{458}
\DefMacro{table-numbers_of_programs_and_bugs-20231009-row1-exec-total}{27,451}
\DefMacro{table-numbers_of_programs_and_bugs-20231009-row1-exec-not_compile}{0}
\DefMacro{table-numbers_of_programs_and_bugs-20231009-row1-exec-timeout}{625}
\DefMacro{table-numbers_of_programs_and_bugs-20231009-row1-exec-ok}{25,905}
\DefMacro{table-numbers_of_programs_and_bugs-20231009-row1-exec-skipped_diff}{878}
\DefMacro{table-numbers_of_programs_and_bugs-20231009-row1-exec-skipped_crash}{15}
\DefMacro{table-numbers_of_programs_and_bugs-20231009-row1-exec-final_diff}{0}
\DefMacro{table-numbers_of_programs_and_bugs-20231009-row1-exec-final_crash}{28}
\DefMacro{table-numbers_of_programs_and_bugs-20231009-row2-project}{text}
\DefMacro{table-numbers_of_programs_and_bugs-20231009-row2-num_unit_tests}{851}
\DefMacro{table-numbers_of_programs_and_bugs-20231009-row2-num_unique_entry_methods}{851}
\DefMacro{table-numbers_of_programs_and_bugs-20231009-row2-num_unique_classes}{77}
\DefMacro{table-numbers_of_programs_and_bugs-20231009-row2-convert-fail}{0}
\DefMacro{table-numbers_of_programs_and_bugs-20231009-row2-convert-timeout}{0}
\DefMacro{table-numbers_of_programs_and_bugs-20231009-row2-convert-not_compile}{4}
\DefMacro{table-numbers_of_programs_and_bugs-20231009-row2-convert-no_original}{0}
\DefMacro{table-numbers_of_programs_and_bugs-20231009-row2-gen-fail}{22}
\DefMacro{table-numbers_of_programs_and_bugs-20231009-row2-gen-timeout}{46}
\DefMacro{table-numbers_of_programs_and_bugs-20231009-row2-gen-no_hole}{54}
\DefMacro{table-numbers_of_programs_and_bugs-20231009-row2-gen-no_gen}{196}
\DefMacro{table-numbers_of_programs_and_bugs-20231009-row2-num_working_templates}{597}
\DefMacro{table-numbers_of_programs_and_bugs-20231009-row2-num_unique_entry_methods_out_of_working_templates}{597}
\DefMacro{table-numbers_of_programs_and_bugs-20231009-row2-num_unique_classes_out_of_working_templates}{52}
\DefMacro{table-numbers_of_programs_and_bugs-20231009-row2-exec-total}{5,483}
\DefMacro{table-numbers_of_programs_and_bugs-20231009-row2-exec-not_compile}{0}
\DefMacro{table-numbers_of_programs_and_bugs-20231009-row2-exec-timeout}{123}
\DefMacro{table-numbers_of_programs_and_bugs-20231009-row2-exec-ok}{5,256}
\DefMacro{table-numbers_of_programs_and_bugs-20231009-row2-exec-skipped_diff}{91}
\DefMacro{table-numbers_of_programs_and_bugs-20231009-row2-exec-skipped_crash}{2}
\DefMacro{table-numbers_of_programs_and_bugs-20231009-row2-exec-final_diff}{9}
\DefMacro{table-numbers_of_programs_and_bugs-20231009-row2-exec-final_crash}{2}
\DefMacro{table-numbers_of_programs_and_bugs-20231009-row3-project}{numbers}
\DefMacro{table-numbers_of_programs_and_bugs-20231009-row3-num_unit_tests}{51}
\DefMacro{table-numbers_of_programs_and_bugs-20231009-row3-num_unique_entry_methods}{51}
\DefMacro{table-numbers_of_programs_and_bugs-20231009-row3-num_unique_classes}{4}
\DefMacro{table-numbers_of_programs_and_bugs-20231009-row3-convert-fail}{0}
\DefMacro{table-numbers_of_programs_and_bugs-20231009-row3-convert-timeout}{0}
\DefMacro{table-numbers_of_programs_and_bugs-20231009-row3-convert-not_compile}{0}
\DefMacro{table-numbers_of_programs_and_bugs-20231009-row3-convert-no_original}{0}
\DefMacro{table-numbers_of_programs_and_bugs-20231009-row3-gen-fail}{0}
\DefMacro{table-numbers_of_programs_and_bugs-20231009-row3-gen-timeout}{3}
\DefMacro{table-numbers_of_programs_and_bugs-20231009-row3-gen-no_hole}{0}
\DefMacro{table-numbers_of_programs_and_bugs-20231009-row3-gen-no_gen}{3}
\DefMacro{table-numbers_of_programs_and_bugs-20231009-row3-num_working_templates}{48}
\DefMacro{table-numbers_of_programs_and_bugs-20231009-row3-num_unique_entry_methods_out_of_working_templates}{48}
\DefMacro{table-numbers_of_programs_and_bugs-20231009-row3-num_unique_classes_out_of_working_templates}{4}
\DefMacro{table-numbers_of_programs_and_bugs-20231009-row3-exec-total}{480}
\DefMacro{table-numbers_of_programs_and_bugs-20231009-row3-exec-not_compile}{0}
\DefMacro{table-numbers_of_programs_and_bugs-20231009-row3-exec-timeout}{0}
\DefMacro{table-numbers_of_programs_and_bugs-20231009-row3-exec-ok}{480}
\DefMacro{table-numbers_of_programs_and_bugs-20231009-row3-exec-skipped_diff}{0}
\DefMacro{table-numbers_of_programs_and_bugs-20231009-row3-exec-skipped_crash}{0}
\DefMacro{table-numbers_of_programs_and_bugs-20231009-row3-exec-final_diff}{0}
\DefMacro{table-numbers_of_programs_and_bugs-20231009-row3-exec-final_crash}{0}
\DefMacro{table-numbers_of_programs_and_bugs-20231009-row4-project}{vectorz}
\DefMacro{table-numbers_of_programs_and_bugs-20231009-row4-num_unit_tests}{5,582}
\DefMacro{table-numbers_of_programs_and_bugs-20231009-row4-num_unique_entry_methods}{5,582}
\DefMacro{table-numbers_of_programs_and_bugs-20231009-row4-num_unique_classes}{263}
\DefMacro{table-numbers_of_programs_and_bugs-20231009-row4-convert-fail}{0}
\DefMacro{table-numbers_of_programs_and_bugs-20231009-row4-convert-timeout}{153}
\DefMacro{table-numbers_of_programs_and_bugs-20231009-row4-convert-not_compile}{0}
\DefMacro{table-numbers_of_programs_and_bugs-20231009-row4-convert-no_original}{0}
\DefMacro{table-numbers_of_programs_and_bugs-20231009-row4-gen-fail}{415}
\DefMacro{table-numbers_of_programs_and_bugs-20231009-row4-gen-timeout}{126}
\DefMacro{table-numbers_of_programs_and_bugs-20231009-row4-gen-no_hole}{712}
\DefMacro{table-numbers_of_programs_and_bugs-20231009-row4-gen-no_gen}{1,287}
\DefMacro{table-numbers_of_programs_and_bugs-20231009-row4-num_working_templates}{3,430}
\DefMacro{table-numbers_of_programs_and_bugs-20231009-row4-num_unique_entry_methods_out_of_working_templates}{3,430}
\DefMacro{table-numbers_of_programs_and_bugs-20231009-row4-num_unique_classes_out_of_working_templates}{239}
\DefMacro{table-numbers_of_programs_and_bugs-20231009-row4-exec-total}{30,768}
\DefMacro{table-numbers_of_programs_and_bugs-20231009-row4-exec-not_compile}{0}
\DefMacro{table-numbers_of_programs_and_bugs-20231009-row4-exec-timeout}{310}
\DefMacro{table-numbers_of_programs_and_bugs-20231009-row4-exec-ok}{29,817}
\DefMacro{table-numbers_of_programs_and_bugs-20231009-row4-exec-skipped_diff}{613}
\DefMacro{table-numbers_of_programs_and_bugs-20231009-row4-exec-skipped_crash}{2}
\DefMacro{table-numbers_of_programs_and_bugs-20231009-row4-exec-final_diff}{4}
\DefMacro{table-numbers_of_programs_and_bugs-20231009-row4-exec-final_crash}{22}
\DefMacro{table-numbers_of_programs_and_bugs-20231009-row5-project}{statistics}
\DefMacro{table-numbers_of_programs_and_bugs-20231009-row5-num_unit_tests}{387}
\DefMacro{table-numbers_of_programs_and_bugs-20231009-row5-num_unique_entry_methods}{387}
\DefMacro{table-numbers_of_programs_and_bugs-20231009-row5-num_unique_classes}{32}
\DefMacro{table-numbers_of_programs_and_bugs-20231009-row5-convert-fail}{0}
\DefMacro{table-numbers_of_programs_and_bugs-20231009-row5-convert-timeout}{0}
\DefMacro{table-numbers_of_programs_and_bugs-20231009-row5-convert-not_compile}{0}
\DefMacro{table-numbers_of_programs_and_bugs-20231009-row5-convert-no_original}{0}
\DefMacro{table-numbers_of_programs_and_bugs-20231009-row5-gen-fail}{40}
\DefMacro{table-numbers_of_programs_and_bugs-20231009-row5-gen-timeout}{22}
\DefMacro{table-numbers_of_programs_and_bugs-20231009-row5-gen-no_hole}{1}
\DefMacro{table-numbers_of_programs_and_bugs-20231009-row5-gen-no_gen}{130}
\DefMacro{table-numbers_of_programs_and_bugs-20231009-row5-num_working_templates}{256}
\DefMacro{table-numbers_of_programs_and_bugs-20231009-row5-num_unique_entry_methods_out_of_working_templates}{256}
\DefMacro{table-numbers_of_programs_and_bugs-20231009-row5-num_unique_classes_out_of_working_templates}{30}
\DefMacro{table-numbers_of_programs_and_bugs-20231009-row5-exec-total}{2,161}
\DefMacro{table-numbers_of_programs_and_bugs-20231009-row5-exec-not_compile}{0}
\DefMacro{table-numbers_of_programs_and_bugs-20231009-row5-exec-timeout}{0}
\DefMacro{table-numbers_of_programs_and_bugs-20231009-row5-exec-ok}{2,158}
\DefMacro{table-numbers_of_programs_and_bugs-20231009-row5-exec-skipped_diff}{1}
\DefMacro{table-numbers_of_programs_and_bugs-20231009-row5-exec-skipped_crash}{2}
\DefMacro{table-numbers_of_programs_and_bugs-20231009-row5-exec-final_diff}{0}
\DefMacro{table-numbers_of_programs_and_bugs-20231009-row5-exec-final_crash}{0}
\DefMacro{table-numbers_of_programs_and_bugs-20231009-row6-project}{codec}
\DefMacro{table-numbers_of_programs_and_bugs-20231009-row6-num_unit_tests}{655}
\DefMacro{table-numbers_of_programs_and_bugs-20231009-row6-num_unique_entry_methods}{655}
\DefMacro{table-numbers_of_programs_and_bugs-20231009-row6-num_unique_classes}{53}
\DefMacro{table-numbers_of_programs_and_bugs-20231009-row6-convert-fail}{0}
\DefMacro{table-numbers_of_programs_and_bugs-20231009-row6-convert-timeout}{0}
\DefMacro{table-numbers_of_programs_and_bugs-20231009-row6-convert-not_compile}{0}
\DefMacro{table-numbers_of_programs_and_bugs-20231009-row6-convert-no_original}{0}
\DefMacro{table-numbers_of_programs_and_bugs-20231009-row6-gen-fail}{57}
\DefMacro{table-numbers_of_programs_and_bugs-20231009-row6-gen-timeout}{29}
\DefMacro{table-numbers_of_programs_and_bugs-20231009-row6-gen-no_hole}{13}
\DefMacro{table-numbers_of_programs_and_bugs-20231009-row6-gen-no_gen}{153}
\DefMacro{table-numbers_of_programs_and_bugs-20231009-row6-num_working_templates}{489}
\DefMacro{table-numbers_of_programs_and_bugs-20231009-row6-num_unique_entry_methods_out_of_working_templates}{489}
\DefMacro{table-numbers_of_programs_and_bugs-20231009-row6-num_unique_classes_out_of_working_templates}{42}
\DefMacro{table-numbers_of_programs_and_bugs-20231009-row6-exec-total}{3,360}
\DefMacro{table-numbers_of_programs_and_bugs-20231009-row6-exec-not_compile}{0}
\DefMacro{table-numbers_of_programs_and_bugs-20231009-row6-exec-timeout}{54}
\DefMacro{table-numbers_of_programs_and_bugs-20231009-row6-exec-ok}{3,147}
\DefMacro{table-numbers_of_programs_and_bugs-20231009-row6-exec-skipped_diff}{93}
\DefMacro{table-numbers_of_programs_and_bugs-20231009-row6-exec-skipped_crash}{57}
\DefMacro{table-numbers_of_programs_and_bugs-20231009-row6-exec-final_diff}{5}
\DefMacro{table-numbers_of_programs_and_bugs-20231009-row6-exec-final_crash}{4}
\DefMacro{table-numbers_of_programs_and_bugs-20231009-row7-project}{jfreechart}
\DefMacro{table-numbers_of_programs_and_bugs-20231009-row7-num_unit_tests}{7,137}
\DefMacro{table-numbers_of_programs_and_bugs-20231009-row7-num_unique_entry_methods}{7,137}
\DefMacro{table-numbers_of_programs_and_bugs-20231009-row7-num_unique_classes}{503}
\DefMacro{table-numbers_of_programs_and_bugs-20231009-row7-convert-fail}{6,626}
\DefMacro{table-numbers_of_programs_and_bugs-20231009-row7-convert-timeout}{418}
\DefMacro{table-numbers_of_programs_and_bugs-20231009-row7-convert-not_compile}{3}
\DefMacro{table-numbers_of_programs_and_bugs-20231009-row7-convert-no_original}{0}
\DefMacro{table-numbers_of_programs_and_bugs-20231009-row7-gen-fail}{10}
\DefMacro{table-numbers_of_programs_and_bugs-20231009-row7-gen-timeout}{1}
\DefMacro{table-numbers_of_programs_and_bugs-20231009-row7-gen-no_hole}{7}
\DefMacro{table-numbers_of_programs_and_bugs-20231009-row7-gen-no_gen}{13}
\DefMacro{table-numbers_of_programs_and_bugs-20231009-row7-num_working_templates}{70}
\DefMacro{table-numbers_of_programs_and_bugs-20231009-row7-num_unique_entry_methods_out_of_working_templates}{70}
\DefMacro{table-numbers_of_programs_and_bugs-20231009-row7-num_unique_classes_out_of_working_templates}{26}
\DefMacro{table-numbers_of_programs_and_bugs-20231009-row7-exec-total}{577}
\DefMacro{table-numbers_of_programs_and_bugs-20231009-row7-exec-not_compile}{0}
\DefMacro{table-numbers_of_programs_and_bugs-20231009-row7-exec-timeout}{3}
\DefMacro{table-numbers_of_programs_and_bugs-20231009-row7-exec-ok}{574}
\DefMacro{table-numbers_of_programs_and_bugs-20231009-row7-exec-skipped_diff}{0}
\DefMacro{table-numbers_of_programs_and_bugs-20231009-row7-exec-skipped_crash}{0}
\DefMacro{table-numbers_of_programs_and_bugs-20231009-row7-exec-final_diff}{0}
\DefMacro{table-numbers_of_programs_and_bugs-20231009-row7-exec-final_crash}{0}
\DefMacro{table-numbers_of_programs_and_bugs-20231009-row8-project}{zxing}
\DefMacro{table-numbers_of_programs_and_bugs-20231009-row8-num_unit_tests}{1,323}
\DefMacro{table-numbers_of_programs_and_bugs-20231009-row8-num_unique_entry_methods}{1,323}
\DefMacro{table-numbers_of_programs_and_bugs-20231009-row8-num_unique_classes}{213}
\DefMacro{table-numbers_of_programs_and_bugs-20231009-row8-convert-fail}{0}
\DefMacro{table-numbers_of_programs_and_bugs-20231009-row8-convert-timeout}{0}
\DefMacro{table-numbers_of_programs_and_bugs-20231009-row8-convert-not_compile}{56}
\DefMacro{table-numbers_of_programs_and_bugs-20231009-row8-convert-no_original}{0}
\DefMacro{table-numbers_of_programs_and_bugs-20231009-row8-gen-fail}{89}
\DefMacro{table-numbers_of_programs_and_bugs-20231009-row8-gen-timeout}{37}
\DefMacro{table-numbers_of_programs_and_bugs-20231009-row8-gen-no_hole}{241}
\DefMacro{table-numbers_of_programs_and_bugs-20231009-row8-gen-no_gen}{160}
\DefMacro{table-numbers_of_programs_and_bugs-20231009-row8-num_working_templates}{866}
\DefMacro{table-numbers_of_programs_and_bugs-20231009-row8-num_unique_entry_methods_out_of_working_templates}{866}
\DefMacro{table-numbers_of_programs_and_bugs-20231009-row8-num_unique_classes_out_of_working_templates}{174}
\DefMacro{table-numbers_of_programs_and_bugs-20231009-row8-exec-total}{7,467}
\DefMacro{table-numbers_of_programs_and_bugs-20231009-row8-exec-not_compile}{0}
\DefMacro{table-numbers_of_programs_and_bugs-20231009-row8-exec-timeout}{176}
\DefMacro{table-numbers_of_programs_and_bugs-20231009-row8-exec-ok}{7,067}
\DefMacro{table-numbers_of_programs_and_bugs-20231009-row8-exec-skipped_diff}{212}
\DefMacro{table-numbers_of_programs_and_bugs-20231009-row8-exec-skipped_crash}{10}
\DefMacro{table-numbers_of_programs_and_bugs-20231009-row8-exec-final_diff}{0}
\DefMacro{table-numbers_of_programs_and_bugs-20231009-row8-exec-final_crash}{2}
\DefMacro{table-numbers_of_programs_and_bugs-20231009-row9-project}{compress}
\DefMacro{table-numbers_of_programs_and_bugs-20231009-row9-num_unit_tests}{1,939}
\DefMacro{table-numbers_of_programs_and_bugs-20231009-row9-num_unique_entry_methods}{1,939}
\DefMacro{table-numbers_of_programs_and_bugs-20231009-row9-num_unique_classes}{177}
\DefMacro{table-numbers_of_programs_and_bugs-20231009-row9-convert-fail}{0}
\DefMacro{table-numbers_of_programs_and_bugs-20231009-row9-convert-timeout}{2}
\DefMacro{table-numbers_of_programs_and_bugs-20231009-row9-convert-not_compile}{60}
\DefMacro{table-numbers_of_programs_and_bugs-20231009-row9-convert-no_original}{0}
\DefMacro{table-numbers_of_programs_and_bugs-20231009-row9-gen-fail}{86}
\DefMacro{table-numbers_of_programs_and_bugs-20231009-row9-gen-timeout}{43}
\DefMacro{table-numbers_of_programs_and_bugs-20231009-row9-gen-no_hole}{368}
\DefMacro{table-numbers_of_programs_and_bugs-20231009-row9-gen-no_gen}{179}
\DefMacro{table-numbers_of_programs_and_bugs-20231009-row9-num_working_templates}{1,330}
\DefMacro{table-numbers_of_programs_and_bugs-20231009-row9-num_unique_entry_methods_out_of_working_templates}{1,330}
\DefMacro{table-numbers_of_programs_and_bugs-20231009-row9-num_unique_classes_out_of_working_templates}{147}
\DefMacro{table-numbers_of_programs_and_bugs-20231009-row9-exec-total}{11,416}
\DefMacro{table-numbers_of_programs_and_bugs-20231009-row9-exec-not_compile}{0}
\DefMacro{table-numbers_of_programs_and_bugs-20231009-row9-exec-timeout}{205}
\DefMacro{table-numbers_of_programs_and_bugs-20231009-row9-exec-ok}{9,839}
\DefMacro{table-numbers_of_programs_and_bugs-20231009-row9-exec-skipped_diff}{1,278}
\DefMacro{table-numbers_of_programs_and_bugs-20231009-row9-exec-skipped_crash}{1}
\DefMacro{table-numbers_of_programs_and_bugs-20231009-row9-exec-final_diff}{92}
\DefMacro{table-numbers_of_programs_and_bugs-20231009-row9-exec-final_crash}{1}
\DefMacro{table-numbers_of_programs_and_bugs-20231009-row10-project}{$\sum$}
\DefMacro{table-numbers_of_programs_and_bugs-20231009-row10-num_unit_tests}{26,048}
\DefMacro{table-numbers_of_programs_and_bugs-20231009-row10-num_unique_entry_methods}{26,042}
\DefMacro{table-numbers_of_programs_and_bugs-20231009-row10-num_unique_classes}{2,044}
\DefMacro{table-numbers_of_programs_and_bugs-20231009-row10-convert-fail}{6,634}
\DefMacro{table-numbers_of_programs_and_bugs-20231009-row10-convert-timeout}{582}
\DefMacro{table-numbers_of_programs_and_bugs-20231009-row10-convert-not_compile}{499}
\DefMacro{table-numbers_of_programs_and_bugs-20231009-row10-convert-no_original}{0}
\DefMacro{table-numbers_of_programs_and_bugs-20231009-row10-gen-fail}{1,226}
\DefMacro{table-numbers_of_programs_and_bugs-20231009-row10-gen-timeout}{695}
\DefMacro{table-numbers_of_programs_and_bugs-20231009-row10-gen-no_hole}{2,397}
\DefMacro{table-numbers_of_programs_and_bugs-20231009-row10-gen-no_gen}{3,633}
\DefMacro{table-numbers_of_programs_and_bugs-20231009-row10-num_working_templates}{12,303}
\DefMacro{table-numbers_of_programs_and_bugs-20231009-row10-num_unique_entry_methods_out_of_working_templates}{12,303}
\DefMacro{table-numbers_of_programs_and_bugs-20231009-row10-num_unique_classes_out_of_working_templates}{1,276}
\DefMacro{table-numbers_of_programs_and_bugs-20231009-row10-exec-total}{104,596}
\DefMacro{table-numbers_of_programs_and_bugs-20231009-row10-exec-not_compile}{0}
\DefMacro{table-numbers_of_programs_and_bugs-20231009-row10-exec-timeout}{1,734}
\DefMacro{table-numbers_of_programs_and_bugs-20231009-row10-exec-ok}{98,841}
\DefMacro{table-numbers_of_programs_and_bugs-20231009-row10-exec-skipped_diff}{3,720}
\DefMacro{table-numbers_of_programs_and_bugs-20231009-row10-exec-skipped_crash}{112}
\DefMacro{table-numbers_of_programs_and_bugs-20231009-row10-exec-final_diff}{124}
\DefMacro{table-numbers_of_programs_and_bugs-20231009-row10-exec-final_crash}{65}

\DefMacro{table-numbers_of_programs_and_bugs-20230910-row0-project}{lang}
\DefMacro{table-numbers_of_programs_and_bugs-20230910-row0-num_unit_tests}{2,827}
\DefMacro{table-numbers_of_programs_and_bugs-20230910-row0-num_unique_entry_methods}{2,821}
\DefMacro{table-numbers_of_programs_and_bugs-20230910-row0-num_unique_classes}{131}
\DefMacro{table-numbers_of_programs_and_bugs-20230910-row0-convert-fail}{0}
\DefMacro{table-numbers_of_programs_and_bugs-20230910-row0-convert-timeout}{0}
\DefMacro{table-numbers_of_programs_and_bugs-20230910-row0-convert-not_compile}{808}
\DefMacro{table-numbers_of_programs_and_bugs-20230910-row0-convert-no_original}{0}
\DefMacro{table-numbers_of_programs_and_bugs-20230910-row0-gen-fail}{98}
\DefMacro{table-numbers_of_programs_and_bugs-20230910-row0-gen-timeout}{68}
\DefMacro{table-numbers_of_programs_and_bugs-20230910-row0-gen-no_hole}{218}
\DefMacro{table-numbers_of_programs_and_bugs-20230910-row0-gen-no_gen}{433}
\DefMacro{table-numbers_of_programs_and_bugs-20230910-row0-num_working_templates}{1,368}
\DefMacro{table-numbers_of_programs_and_bugs-20230910-row0-num_unique_entry_methods_out_of_working_templates}{1,368}
\DefMacro{table-numbers_of_programs_and_bugs-20230910-row0-num_unique_classes_out_of_working_templates}{95}
\DefMacro{table-numbers_of_programs_and_bugs-20230910-row0-exec-total}{10,925}
\DefMacro{table-numbers_of_programs_and_bugs-20230910-row0-exec-not_compile}{0}
\DefMacro{table-numbers_of_programs_and_bugs-20230910-row0-exec-timeout}{86}
\DefMacro{table-numbers_of_programs_and_bugs-20230910-row0-exec-ok}{10,585}
\DefMacro{table-numbers_of_programs_and_bugs-20230910-row0-exec-skipped_diff}{238}
\DefMacro{table-numbers_of_programs_and_bugs-20230910-row0-exec-skipped_crash}{8}
\DefMacro{table-numbers_of_programs_and_bugs-20230910-row0-exec-final_diff}{0}
\DefMacro{table-numbers_of_programs_and_bugs-20230910-row0-exec-final_crash}{8}
\DefMacro{table-numbers_of_programs_and_bugs-20230910-row1-project}{math}
\DefMacro{table-numbers_of_programs_and_bugs-20230910-row1-num_unit_tests}{5,296}
\DefMacro{table-numbers_of_programs_and_bugs-20230910-row1-num_unique_entry_methods}{5,296}
\DefMacro{table-numbers_of_programs_and_bugs-20230910-row1-num_unique_classes}{591}
\DefMacro{table-numbers_of_programs_and_bugs-20230910-row1-convert-fail}{8}
\DefMacro{table-numbers_of_programs_and_bugs-20230910-row1-convert-timeout}{8}
\DefMacro{table-numbers_of_programs_and_bugs-20230910-row1-convert-not_compile}{177}
\DefMacro{table-numbers_of_programs_and_bugs-20230910-row1-convert-no_original}{0}
\DefMacro{table-numbers_of_programs_and_bugs-20230910-row1-gen-fail}{410}
\DefMacro{table-numbers_of_programs_and_bugs-20230910-row1-gen-timeout}{301}
\DefMacro{table-numbers_of_programs_and_bugs-20230910-row1-gen-no_hole}{846}
\DefMacro{table-numbers_of_programs_and_bugs-20230910-row1-gen-no_gen}{1,138}
\DefMacro{table-numbers_of_programs_and_bugs-20230910-row1-num_working_templates}{3,119}
\DefMacro{table-numbers_of_programs_and_bugs-20230910-row1-num_unique_entry_methods_out_of_working_templates}{3,119}
\DefMacro{table-numbers_of_programs_and_bugs-20230910-row1-num_unique_classes_out_of_working_templates}{450}
\DefMacro{table-numbers_of_programs_and_bugs-20230910-row1-exec-total}{25,457}
\DefMacro{table-numbers_of_programs_and_bugs-20230910-row1-exec-not_compile}{0}
\DefMacro{table-numbers_of_programs_and_bugs-20230910-row1-exec-timeout}{408}
\DefMacro{table-numbers_of_programs_and_bugs-20230910-row1-exec-ok}{24,153}
\DefMacro{table-numbers_of_programs_and_bugs-20230910-row1-exec-skipped_diff}{871}
\DefMacro{table-numbers_of_programs_and_bugs-20230910-row1-exec-skipped_crash}{11}
\DefMacro{table-numbers_of_programs_and_bugs-20230910-row1-exec-final_diff}{2}
\DefMacro{table-numbers_of_programs_and_bugs-20230910-row1-exec-final_crash}{12}
\DefMacro{table-numbers_of_programs_and_bugs-20230910-row2-project}{text}
\DefMacro{table-numbers_of_programs_and_bugs-20230910-row2-num_unit_tests}{850}
\DefMacro{table-numbers_of_programs_and_bugs-20230910-row2-num_unique_entry_methods}{850}
\DefMacro{table-numbers_of_programs_and_bugs-20230910-row2-num_unique_classes}{77}
\DefMacro{table-numbers_of_programs_and_bugs-20230910-row2-convert-fail}{0}
\DefMacro{table-numbers_of_programs_and_bugs-20230910-row2-convert-timeout}{0}
\DefMacro{table-numbers_of_programs_and_bugs-20230910-row2-convert-not_compile}{4}
\DefMacro{table-numbers_of_programs_and_bugs-20230910-row2-convert-no_original}{0}
\DefMacro{table-numbers_of_programs_and_bugs-20230910-row2-gen-fail}{17}
\DefMacro{table-numbers_of_programs_and_bugs-20230910-row2-gen-timeout}{31}
\DefMacro{table-numbers_of_programs_and_bugs-20230910-row2-gen-no_hole}{81}
\DefMacro{table-numbers_of_programs_and_bugs-20230910-row2-gen-no_gen}{191}
\DefMacro{table-numbers_of_programs_and_bugs-20230910-row2-num_working_templates}{574}
\DefMacro{table-numbers_of_programs_and_bugs-20230910-row2-num_unique_entry_methods_out_of_working_templates}{574}
\DefMacro{table-numbers_of_programs_and_bugs-20230910-row2-num_unique_classes_out_of_working_templates}{51}
\DefMacro{table-numbers_of_programs_and_bugs-20230910-row2-exec-total}{5,317}
\DefMacro{table-numbers_of_programs_and_bugs-20230910-row2-exec-not_compile}{0}
\DefMacro{table-numbers_of_programs_and_bugs-20230910-row2-exec-timeout}{103}
\DefMacro{table-numbers_of_programs_and_bugs-20230910-row2-exec-ok}{5,050}
\DefMacro{table-numbers_of_programs_and_bugs-20230910-row2-exec-skipped_diff}{164}
\DefMacro{table-numbers_of_programs_and_bugs-20230910-row2-exec-skipped_crash}{0}
\DefMacro{table-numbers_of_programs_and_bugs-20230910-row2-exec-final_diff}{0}
\DefMacro{table-numbers_of_programs_and_bugs-20230910-row2-exec-final_crash}{0}
\DefMacro{table-numbers_of_programs_and_bugs-20230910-row3-project}{numbers}
\DefMacro{table-numbers_of_programs_and_bugs-20230910-row3-num_unit_tests}{51}
\DefMacro{table-numbers_of_programs_and_bugs-20230910-row3-num_unique_entry_methods}{51}
\DefMacro{table-numbers_of_programs_and_bugs-20230910-row3-num_unique_classes}{4}
\DefMacro{table-numbers_of_programs_and_bugs-20230910-row3-convert-fail}{0}
\DefMacro{table-numbers_of_programs_and_bugs-20230910-row3-convert-timeout}{0}
\DefMacro{table-numbers_of_programs_and_bugs-20230910-row3-convert-not_compile}{0}
\DefMacro{table-numbers_of_programs_and_bugs-20230910-row3-convert-no_original}{0}
\DefMacro{table-numbers_of_programs_and_bugs-20230910-row3-gen-fail}{0}
\DefMacro{table-numbers_of_programs_and_bugs-20230910-row3-gen-timeout}{1}
\DefMacro{table-numbers_of_programs_and_bugs-20230910-row3-gen-no_hole}{0}
\DefMacro{table-numbers_of_programs_and_bugs-20230910-row3-gen-no_gen}{0}
\DefMacro{table-numbers_of_programs_and_bugs-20230910-row3-num_working_templates}{51}
\DefMacro{table-numbers_of_programs_and_bugs-20230910-row3-num_unique_entry_methods_out_of_working_templates}{51}
\DefMacro{table-numbers_of_programs_and_bugs-20230910-row3-num_unique_classes_out_of_working_templates}{4}
\DefMacro{table-numbers_of_programs_and_bugs-20230910-row3-exec-total}{501}
\DefMacro{table-numbers_of_programs_and_bugs-20230910-row3-exec-not_compile}{0}
\DefMacro{table-numbers_of_programs_and_bugs-20230910-row3-exec-timeout}{0}
\DefMacro{table-numbers_of_programs_and_bugs-20230910-row3-exec-ok}{501}
\DefMacro{table-numbers_of_programs_and_bugs-20230910-row3-exec-skipped_diff}{0}
\DefMacro{table-numbers_of_programs_and_bugs-20230910-row3-exec-skipped_crash}{0}
\DefMacro{table-numbers_of_programs_and_bugs-20230910-row3-exec-final_diff}{0}
\DefMacro{table-numbers_of_programs_and_bugs-20230910-row3-exec-final_crash}{0}
\DefMacro{table-numbers_of_programs_and_bugs-20230910-row4-project}{vectorz}
\DefMacro{table-numbers_of_programs_and_bugs-20230910-row4-num_unit_tests}{5,582}
\DefMacro{table-numbers_of_programs_and_bugs-20230910-row4-num_unique_entry_methods}{5,582}
\DefMacro{table-numbers_of_programs_and_bugs-20230910-row4-num_unique_classes}{263}
\DefMacro{table-numbers_of_programs_and_bugs-20230910-row4-convert-fail}{0}
\DefMacro{table-numbers_of_programs_and_bugs-20230910-row4-convert-timeout}{4}
\DefMacro{table-numbers_of_programs_and_bugs-20230910-row4-convert-not_compile}{0}
\DefMacro{table-numbers_of_programs_and_bugs-20230910-row4-convert-no_original}{0}
\DefMacro{table-numbers_of_programs_and_bugs-20230910-row4-gen-fail}{327}
\DefMacro{table-numbers_of_programs_and_bugs-20230910-row4-gen-timeout}{60}
\DefMacro{table-numbers_of_programs_and_bugs-20230910-row4-gen-no_hole}{677}
\DefMacro{table-numbers_of_programs_and_bugs-20230910-row4-gen-no_gen}{1,541}
\DefMacro{table-numbers_of_programs_and_bugs-20230910-row4-num_working_templates}{3,360}
\DefMacro{table-numbers_of_programs_and_bugs-20230910-row4-num_unique_entry_methods_out_of_working_templates}{3,360}
\DefMacro{table-numbers_of_programs_and_bugs-20230910-row4-num_unique_classes_out_of_working_templates}{234}
\DefMacro{table-numbers_of_programs_and_bugs-20230910-row4-exec-total}{28,317}
\DefMacro{table-numbers_of_programs_and_bugs-20230910-row4-exec-not_compile}{0}
\DefMacro{table-numbers_of_programs_and_bugs-20230910-row4-exec-timeout}{595}
\DefMacro{table-numbers_of_programs_and_bugs-20230910-row4-exec-ok}{26,016}
\DefMacro{table-numbers_of_programs_and_bugs-20230910-row4-exec-skipped_diff}{1,279}
\DefMacro{table-numbers_of_programs_and_bugs-20230910-row4-exec-skipped_crash}{389}
\DefMacro{table-numbers_of_programs_and_bugs-20230910-row4-exec-final_diff}{1}
\DefMacro{table-numbers_of_programs_and_bugs-20230910-row4-exec-final_crash}{37}
\DefMacro{table-numbers_of_programs_and_bugs-20230910-row5-project}{statistics}
\DefMacro{table-numbers_of_programs_and_bugs-20230910-row5-num_unit_tests}{387}
\DefMacro{table-numbers_of_programs_and_bugs-20230910-row5-num_unique_entry_methods}{387}
\DefMacro{table-numbers_of_programs_and_bugs-20230910-row5-num_unique_classes}{32}
\DefMacro{table-numbers_of_programs_and_bugs-20230910-row5-convert-fail}{0}
\DefMacro{table-numbers_of_programs_and_bugs-20230910-row5-convert-timeout}{0}
\DefMacro{table-numbers_of_programs_and_bugs-20230910-row5-convert-not_compile}{0}
\DefMacro{table-numbers_of_programs_and_bugs-20230910-row5-convert-no_original}{0}
\DefMacro{table-numbers_of_programs_and_bugs-20230910-row5-gen-fail}{40}
\DefMacro{table-numbers_of_programs_and_bugs-20230910-row5-gen-timeout}{11}
\DefMacro{table-numbers_of_programs_and_bugs-20230910-row5-gen-no_hole}{2}
\DefMacro{table-numbers_of_programs_and_bugs-20230910-row5-gen-no_gen}{145}
\DefMacro{table-numbers_of_programs_and_bugs-20230910-row5-num_working_templates}{240}
\DefMacro{table-numbers_of_programs_and_bugs-20230910-row5-num_unique_entry_methods_out_of_working_templates}{240}
\DefMacro{table-numbers_of_programs_and_bugs-20230910-row5-num_unique_classes_out_of_working_templates}{30}
\DefMacro{table-numbers_of_programs_and_bugs-20230910-row5-exec-total}{1,845}
\DefMacro{table-numbers_of_programs_and_bugs-20230910-row5-exec-not_compile}{0}
\DefMacro{table-numbers_of_programs_and_bugs-20230910-row5-exec-timeout}{0}
\DefMacro{table-numbers_of_programs_and_bugs-20230910-row5-exec-ok}{1,838}
\DefMacro{table-numbers_of_programs_and_bugs-20230910-row5-exec-skipped_diff}{6}
\DefMacro{table-numbers_of_programs_and_bugs-20230910-row5-exec-skipped_crash}{1}
\DefMacro{table-numbers_of_programs_and_bugs-20230910-row5-exec-final_diff}{0}
\DefMacro{table-numbers_of_programs_and_bugs-20230910-row5-exec-final_crash}{0}
\DefMacro{table-numbers_of_programs_and_bugs-20230910-row6-project}{codec}
\DefMacro{table-numbers_of_programs_and_bugs-20230910-row6-num_unit_tests}{655}
\DefMacro{table-numbers_of_programs_and_bugs-20230910-row6-num_unique_entry_methods}{655}
\DefMacro{table-numbers_of_programs_and_bugs-20230910-row6-num_unique_classes}{53}
\DefMacro{table-numbers_of_programs_and_bugs-20230910-row6-convert-fail}{0}
\DefMacro{table-numbers_of_programs_and_bugs-20230910-row6-convert-timeout}{0}
\DefMacro{table-numbers_of_programs_and_bugs-20230910-row6-convert-not_compile}{5}
\DefMacro{table-numbers_of_programs_and_bugs-20230910-row6-convert-no_original}{0}
\DefMacro{table-numbers_of_programs_and_bugs-20230910-row6-gen-fail}{48}
\DefMacro{table-numbers_of_programs_and_bugs-20230910-row6-gen-timeout}{14}
\DefMacro{table-numbers_of_programs_and_bugs-20230910-row6-gen-no_hole}{15}
\DefMacro{table-numbers_of_programs_and_bugs-20230910-row6-gen-no_gen}{158}
\DefMacro{table-numbers_of_programs_and_bugs-20230910-row6-num_working_templates}{477}
\DefMacro{table-numbers_of_programs_and_bugs-20230910-row6-num_unique_entry_methods_out_of_working_templates}{477}
\DefMacro{table-numbers_of_programs_and_bugs-20230910-row6-num_unique_classes_out_of_working_templates}{42}
\DefMacro{table-numbers_of_programs_and_bugs-20230910-row6-exec-total}{3,286}
\DefMacro{table-numbers_of_programs_and_bugs-20230910-row6-exec-not_compile}{0}
\DefMacro{table-numbers_of_programs_and_bugs-20230910-row6-exec-timeout}{78}
\DefMacro{table-numbers_of_programs_and_bugs-20230910-row6-exec-ok}{3,118}
\DefMacro{table-numbers_of_programs_and_bugs-20230910-row6-exec-skipped_diff}{34}
\DefMacro{table-numbers_of_programs_and_bugs-20230910-row6-exec-skipped_crash}{52}
\DefMacro{table-numbers_of_programs_and_bugs-20230910-row6-exec-final_diff}{1}
\DefMacro{table-numbers_of_programs_and_bugs-20230910-row6-exec-final_crash}{3}
\DefMacro{table-numbers_of_programs_and_bugs-20230910-row7-project}{jfreechart}
\DefMacro{table-numbers_of_programs_and_bugs-20230910-row7-num_unit_tests}{7,137}
\DefMacro{table-numbers_of_programs_and_bugs-20230910-row7-num_unique_entry_methods}{7,137}
\DefMacro{table-numbers_of_programs_and_bugs-20230910-row7-num_unique_classes}{503}
\DefMacro{table-numbers_of_programs_and_bugs-20230910-row7-convert-fail}{6,662}
\DefMacro{table-numbers_of_programs_and_bugs-20230910-row7-convert-timeout}{411}
\DefMacro{table-numbers_of_programs_and_bugs-20230910-row7-convert-not_compile}{3}
\DefMacro{table-numbers_of_programs_and_bugs-20230910-row7-convert-no_original}{0}
\DefMacro{table-numbers_of_programs_and_bugs-20230910-row7-gen-fail}{9}
\DefMacro{table-numbers_of_programs_and_bugs-20230910-row7-gen-timeout}{0}
\DefMacro{table-numbers_of_programs_and_bugs-20230910-row7-gen-no_hole}{7}
\DefMacro{table-numbers_of_programs_and_bugs-20230910-row7-gen-no_gen}{9}
\DefMacro{table-numbers_of_programs_and_bugs-20230910-row7-num_working_templates}{45}
\DefMacro{table-numbers_of_programs_and_bugs-20230910-row7-num_unique_entry_methods_out_of_working_templates}{45}
\DefMacro{table-numbers_of_programs_and_bugs-20230910-row7-num_unique_classes_out_of_working_templates}{17}
\DefMacro{table-numbers_of_programs_and_bugs-20230910-row7-exec-total}{345}
\DefMacro{table-numbers_of_programs_and_bugs-20230910-row7-exec-not_compile}{0}
\DefMacro{table-numbers_of_programs_and_bugs-20230910-row7-exec-timeout}{0}
\DefMacro{table-numbers_of_programs_and_bugs-20230910-row7-exec-ok}{345}
\DefMacro{table-numbers_of_programs_and_bugs-20230910-row7-exec-skipped_diff}{0}
\DefMacro{table-numbers_of_programs_and_bugs-20230910-row7-exec-skipped_crash}{0}
\DefMacro{table-numbers_of_programs_and_bugs-20230910-row7-exec-final_diff}{0}
\DefMacro{table-numbers_of_programs_and_bugs-20230910-row7-exec-final_crash}{0}
\DefMacro{table-numbers_of_programs_and_bugs-20230910-row8-project}{zxing}
\DefMacro{table-numbers_of_programs_and_bugs-20230910-row8-num_unit_tests}{1,323}
\DefMacro{table-numbers_of_programs_and_bugs-20230910-row8-num_unique_entry_methods}{1,323}
\DefMacro{table-numbers_of_programs_and_bugs-20230910-row8-num_unique_classes}{213}
\DefMacro{table-numbers_of_programs_and_bugs-20230910-row8-convert-fail}{0}
\DefMacro{table-numbers_of_programs_and_bugs-20230910-row8-convert-timeout}{0}
\DefMacro{table-numbers_of_programs_and_bugs-20230910-row8-convert-not_compile}{72}
\DefMacro{table-numbers_of_programs_and_bugs-20230910-row8-convert-no_original}{0}
\DefMacro{table-numbers_of_programs_and_bugs-20230910-row8-gen-fail}{73}
\DefMacro{table-numbers_of_programs_and_bugs-20230910-row8-gen-timeout}{45}
\DefMacro{table-numbers_of_programs_and_bugs-20230910-row8-gen-no_hole}{253}
\DefMacro{table-numbers_of_programs_and_bugs-20230910-row8-gen-no_gen}{120}
\DefMacro{table-numbers_of_programs_and_bugs-20230910-row8-num_working_templates}{878}
\DefMacro{table-numbers_of_programs_and_bugs-20230910-row8-num_unique_entry_methods_out_of_working_templates}{878}
\DefMacro{table-numbers_of_programs_and_bugs-20230910-row8-num_unique_classes_out_of_working_templates}{173}
\DefMacro{table-numbers_of_programs_and_bugs-20230910-row8-exec-total}{7,297}
\DefMacro{table-numbers_of_programs_and_bugs-20230910-row8-exec-not_compile}{0}
\DefMacro{table-numbers_of_programs_and_bugs-20230910-row8-exec-timeout}{153}
\DefMacro{table-numbers_of_programs_and_bugs-20230910-row8-exec-ok}{7,078}
\DefMacro{table-numbers_of_programs_and_bugs-20230910-row8-exec-skipped_diff}{55}
\DefMacro{table-numbers_of_programs_and_bugs-20230910-row8-exec-skipped_crash}{10}
\DefMacro{table-numbers_of_programs_and_bugs-20230910-row8-exec-final_diff}{0}
\DefMacro{table-numbers_of_programs_and_bugs-20230910-row8-exec-final_crash}{1}
\DefMacro{table-numbers_of_programs_and_bugs-20230910-row9-project}{compress}
\DefMacro{table-numbers_of_programs_and_bugs-20230910-row9-num_unit_tests}{1,939}
\DefMacro{table-numbers_of_programs_and_bugs-20230910-row9-num_unique_entry_methods}{1,939}
\DefMacro{table-numbers_of_programs_and_bugs-20230910-row9-num_unique_classes}{177}
\DefMacro{table-numbers_of_programs_and_bugs-20230910-row9-convert-fail}{0}
\DefMacro{table-numbers_of_programs_and_bugs-20230910-row9-convert-timeout}{2}
\DefMacro{table-numbers_of_programs_and_bugs-20230910-row9-convert-not_compile}{95}
\DefMacro{table-numbers_of_programs_and_bugs-20230910-row9-convert-no_original}{0}
\DefMacro{table-numbers_of_programs_and_bugs-20230910-row9-gen-fail}{108}
\DefMacro{table-numbers_of_programs_and_bugs-20230910-row9-gen-timeout}{55}
\DefMacro{table-numbers_of_programs_and_bugs-20230910-row9-gen-no_hole}{336}
\DefMacro{table-numbers_of_programs_and_bugs-20230910-row9-gen-no_gen}{262}
\DefMacro{table-numbers_of_programs_and_bugs-20230910-row9-num_working_templates}{1,244}
\DefMacro{table-numbers_of_programs_and_bugs-20230910-row9-num_unique_entry_methods_out_of_working_templates}{1,244}
\DefMacro{table-numbers_of_programs_and_bugs-20230910-row9-num_unique_classes_out_of_working_templates}{148}
\DefMacro{table-numbers_of_programs_and_bugs-20230910-row9-exec-total}{9,951}
\DefMacro{table-numbers_of_programs_and_bugs-20230910-row9-exec-not_compile}{0}
\DefMacro{table-numbers_of_programs_and_bugs-20230910-row9-exec-timeout}{185}
\DefMacro{table-numbers_of_programs_and_bugs-20230910-row9-exec-ok}{8,929}
\DefMacro{table-numbers_of_programs_and_bugs-20230910-row9-exec-skipped_diff}{827}
\DefMacro{table-numbers_of_programs_and_bugs-20230910-row9-exec-skipped_crash}{6}
\DefMacro{table-numbers_of_programs_and_bugs-20230910-row9-exec-final_diff}{2}
\DefMacro{table-numbers_of_programs_and_bugs-20230910-row9-exec-final_crash}{2}
\DefMacro{table-numbers_of_programs_and_bugs-20230910-row10-project}{$\sum$}
\DefMacro{table-numbers_of_programs_and_bugs-20230910-row10-num_unit_tests}{26,047}
\DefMacro{table-numbers_of_programs_and_bugs-20230910-row10-num_unique_entry_methods}{26,041}
\DefMacro{table-numbers_of_programs_and_bugs-20230910-row10-num_unique_classes}{2,044}
\DefMacro{table-numbers_of_programs_and_bugs-20230910-row10-convert-fail}{6,670}
\DefMacro{table-numbers_of_programs_and_bugs-20230910-row10-convert-timeout}{425}
\DefMacro{table-numbers_of_programs_and_bugs-20230910-row10-convert-not_compile}{1,164}
\DefMacro{table-numbers_of_programs_and_bugs-20230910-row10-convert-no_original}{0}
\DefMacro{table-numbers_of_programs_and_bugs-20230910-row10-gen-fail}{1,130}
\DefMacro{table-numbers_of_programs_and_bugs-20230910-row10-gen-timeout}{586}
\DefMacro{table-numbers_of_programs_and_bugs-20230910-row10-gen-no_hole}{2,435}
\DefMacro{table-numbers_of_programs_and_bugs-20230910-row10-gen-no_gen}{3,997}
\DefMacro{table-numbers_of_programs_and_bugs-20230910-row10-num_working_templates}{11,356}
\DefMacro{table-numbers_of_programs_and_bugs-20230910-row10-num_unique_entry_methods_out_of_working_templates}{11,356}
\DefMacro{table-numbers_of_programs_and_bugs-20230910-row10-num_unique_classes_out_of_working_templates}{1,244}
\DefMacro{table-numbers_of_programs_and_bugs-20230910-row10-exec-total}{93,241}
\DefMacro{table-numbers_of_programs_and_bugs-20230910-row10-exec-not_compile}{0}
\DefMacro{table-numbers_of_programs_and_bugs-20230910-row10-exec-timeout}{1,608}
\DefMacro{table-numbers_of_programs_and_bugs-20230910-row10-exec-ok}{87,613}
\DefMacro{table-numbers_of_programs_and_bugs-20230910-row10-exec-skipped_diff}{3,474}
\DefMacro{table-numbers_of_programs_and_bugs-20230910-row10-exec-skipped_crash}{477}
\DefMacro{table-numbers_of_programs_and_bugs-20230910-row10-exec-final_diff}{6}
\DefMacro{table-numbers_of_programs_and_bugs-20230910-row10-exec-final_crash}{63}

\DefMacro{table-numbers_of_programs_and_bugs-20230412_20230415-row0-project}{lang}
\DefMacro{table-numbers_of_programs_and_bugs-20230412_20230415-row0-num_unit_tests}{2,827}
\DefMacro{table-numbers_of_programs_and_bugs-20230412_20230415-row0-num_unique_entry_methods}{2,821}
\DefMacro{table-numbers_of_programs_and_bugs-20230412_20230415-row0-num_unique_classes}{131}
\DefMacro{table-numbers_of_programs_and_bugs-20230412_20230415-row0-convert-fail}{0}
\DefMacro{table-numbers_of_programs_and_bugs-20230412_20230415-row0-convert-timeout}{0}
\DefMacro{table-numbers_of_programs_and_bugs-20230412_20230415-row0-convert-not_compile}{167}
\DefMacro{table-numbers_of_programs_and_bugs-20230412_20230415-row0-convert-no_original}{0}
\DefMacro{table-numbers_of_programs_and_bugs-20230412_20230415-row0-gen-fail}{142}
\DefMacro{table-numbers_of_programs_and_bugs-20230412_20230415-row0-gen-timeout}{187}
\DefMacro{table-numbers_of_programs_and_bugs-20230412_20230415-row0-gen-no_hole}{250}
\DefMacro{table-numbers_of_programs_and_bugs-20230412_20230415-row0-gen-no_gen}{518}
\DefMacro{table-numbers_of_programs_and_bugs-20230412_20230415-row0-num_working_templates}{1,892}
\DefMacro{table-numbers_of_programs_and_bugs-20230412_20230415-row0-num_unique_entry_methods_out_of_working_templates}{1,892}
\DefMacro{table-numbers_of_programs_and_bugs-20230412_20230415-row0-num_unique_classes_out_of_working_templates}{103}
\DefMacro{table-numbers_of_programs_and_bugs-20230412_20230415-row0-exec-total}{15,978}
\DefMacro{table-numbers_of_programs_and_bugs-20230412_20230415-row0-exec-not_compile}{0}
\DefMacro{table-numbers_of_programs_and_bugs-20230412_20230415-row0-exec-timeout}{220}
\DefMacro{table-numbers_of_programs_and_bugs-20230412_20230415-row0-exec-ok}{15,422}
\DefMacro{table-numbers_of_programs_and_bugs-20230412_20230415-row0-exec-skipped_diff}{259}
\DefMacro{table-numbers_of_programs_and_bugs-20230412_20230415-row0-exec-skipped_crash}{71}
\DefMacro{table-numbers_of_programs_and_bugs-20230412_20230415-row0-exec-final_diff}{1}
\DefMacro{table-numbers_of_programs_and_bugs-20230412_20230415-row0-exec-final_crash}{5}
\DefMacro{table-numbers_of_programs_and_bugs-20230412_20230415-row1-project}{math}
\DefMacro{table-numbers_of_programs_and_bugs-20230412_20230415-row1-num_unit_tests}{5,296}
\DefMacro{table-numbers_of_programs_and_bugs-20230412_20230415-row1-num_unique_entry_methods}{5,296}
\DefMacro{table-numbers_of_programs_and_bugs-20230412_20230415-row1-num_unique_classes}{591}
\DefMacro{table-numbers_of_programs_and_bugs-20230412_20230415-row1-convert-fail}{8}
\DefMacro{table-numbers_of_programs_and_bugs-20230412_20230415-row1-convert-timeout}{2}
\DefMacro{table-numbers_of_programs_and_bugs-20230412_20230415-row1-convert-not_compile}{181}
\DefMacro{table-numbers_of_programs_and_bugs-20230412_20230415-row1-convert-no_original}{0}
\DefMacro{table-numbers_of_programs_and_bugs-20230412_20230415-row1-gen-fail}{435}
\DefMacro{table-numbers_of_programs_and_bugs-20230412_20230415-row1-gen-timeout}{259}
\DefMacro{table-numbers_of_programs_and_bugs-20230412_20230415-row1-gen-no_hole}{842}
\DefMacro{table-numbers_of_programs_and_bugs-20230412_20230415-row1-gen-no_gen}{1,192}
\DefMacro{table-numbers_of_programs_and_bugs-20230412_20230415-row1-num_working_templates}{3,071}
\DefMacro{table-numbers_of_programs_and_bugs-20230412_20230415-row1-num_unique_entry_methods_out_of_working_templates}{3,071}
\DefMacro{table-numbers_of_programs_and_bugs-20230412_20230415-row1-num_unique_classes_out_of_working_templates}{425}
\DefMacro{table-numbers_of_programs_and_bugs-20230412_20230415-row1-exec-total}{24,975}
\DefMacro{table-numbers_of_programs_and_bugs-20230412_20230415-row1-exec-not_compile}{0}
\DefMacro{table-numbers_of_programs_and_bugs-20230412_20230415-row1-exec-timeout}{411}
\DefMacro{table-numbers_of_programs_and_bugs-20230412_20230415-row1-exec-ok}{24,057}
\DefMacro{table-numbers_of_programs_and_bugs-20230412_20230415-row1-exec-skipped_diff}{474}
\DefMacro{table-numbers_of_programs_and_bugs-20230412_20230415-row1-exec-skipped_crash}{8}
\DefMacro{table-numbers_of_programs_and_bugs-20230412_20230415-row1-exec-final_diff}{2}
\DefMacro{table-numbers_of_programs_and_bugs-20230412_20230415-row1-exec-final_crash}{23}
\DefMacro{table-numbers_of_programs_and_bugs-20230412_20230415-row2-project}{text}
\DefMacro{table-numbers_of_programs_and_bugs-20230412_20230415-row2-num_unit_tests}{851}
\DefMacro{table-numbers_of_programs_and_bugs-20230412_20230415-row2-num_unique_entry_methods}{851}
\DefMacro{table-numbers_of_programs_and_bugs-20230412_20230415-row2-num_unique_classes}{77}
\DefMacro{table-numbers_of_programs_and_bugs-20230412_20230415-row2-convert-fail}{0}
\DefMacro{table-numbers_of_programs_and_bugs-20230412_20230415-row2-convert-timeout}{0}
\DefMacro{table-numbers_of_programs_and_bugs-20230412_20230415-row2-convert-not_compile}{59}
\DefMacro{table-numbers_of_programs_and_bugs-20230412_20230415-row2-convert-no_original}{0}
\DefMacro{table-numbers_of_programs_and_bugs-20230412_20230415-row2-gen-fail}{27}
\DefMacro{table-numbers_of_programs_and_bugs-20230412_20230415-row2-gen-timeout}{46}
\DefMacro{table-numbers_of_programs_and_bugs-20230412_20230415-row2-gen-no_hole}{64}
\DefMacro{table-numbers_of_programs_and_bugs-20230412_20230415-row2-gen-no_gen}{65}
\DefMacro{table-numbers_of_programs_and_bugs-20230412_20230415-row2-num_working_templates}{663}
\DefMacro{table-numbers_of_programs_and_bugs-20230412_20230415-row2-num_unique_entry_methods_out_of_working_templates}{663}
\DefMacro{table-numbers_of_programs_and_bugs-20230412_20230415-row2-num_unique_classes_out_of_working_templates}{50}
\DefMacro{table-numbers_of_programs_and_bugs-20230412_20230415-row2-exec-total}{6,134}
\DefMacro{table-numbers_of_programs_and_bugs-20230412_20230415-row2-exec-not_compile}{0}
\DefMacro{table-numbers_of_programs_and_bugs-20230412_20230415-row2-exec-timeout}{32}
\DefMacro{table-numbers_of_programs_and_bugs-20230412_20230415-row2-exec-ok}{6,077}
\DefMacro{table-numbers_of_programs_and_bugs-20230412_20230415-row2-exec-skipped_diff}{23}
\DefMacro{table-numbers_of_programs_and_bugs-20230412_20230415-row2-exec-skipped_crash}{1}
\DefMacro{table-numbers_of_programs_and_bugs-20230412_20230415-row2-exec-final_diff}{0}
\DefMacro{table-numbers_of_programs_and_bugs-20230412_20230415-row2-exec-final_crash}{1}
\DefMacro{table-numbers_of_programs_and_bugs-20230412_20230415-row3-project}{numbers}
\DefMacro{table-numbers_of_programs_and_bugs-20230412_20230415-row3-num_unit_tests}{51}
\DefMacro{table-numbers_of_programs_and_bugs-20230412_20230415-row3-num_unique_entry_methods}{51}
\DefMacro{table-numbers_of_programs_and_bugs-20230412_20230415-row3-num_unique_classes}{4}
\DefMacro{table-numbers_of_programs_and_bugs-20230412_20230415-row3-convert-fail}{0}
\DefMacro{table-numbers_of_programs_and_bugs-20230412_20230415-row3-convert-timeout}{0}
\DefMacro{table-numbers_of_programs_and_bugs-20230412_20230415-row3-convert-not_compile}{0}
\DefMacro{table-numbers_of_programs_and_bugs-20230412_20230415-row3-convert-no_original}{0}
\DefMacro{table-numbers_of_programs_and_bugs-20230412_20230415-row3-gen-fail}{0}
\DefMacro{table-numbers_of_programs_and_bugs-20230412_20230415-row3-gen-timeout}{0}
\DefMacro{table-numbers_of_programs_and_bugs-20230412_20230415-row3-gen-no_hole}{0}
\DefMacro{table-numbers_of_programs_and_bugs-20230412_20230415-row3-gen-no_gen}{0}
\DefMacro{table-numbers_of_programs_and_bugs-20230412_20230415-row3-num_working_templates}{51}
\DefMacro{table-numbers_of_programs_and_bugs-20230412_20230415-row3-num_unique_entry_methods_out_of_working_templates}{51}
\DefMacro{table-numbers_of_programs_and_bugs-20230412_20230415-row3-num_unique_classes_out_of_working_templates}{4}
\DefMacro{table-numbers_of_programs_and_bugs-20230412_20230415-row3-exec-total}{490}
\DefMacro{table-numbers_of_programs_and_bugs-20230412_20230415-row3-exec-not_compile}{0}
\DefMacro{table-numbers_of_programs_and_bugs-20230412_20230415-row3-exec-timeout}{1}
\DefMacro{table-numbers_of_programs_and_bugs-20230412_20230415-row3-exec-ok}{489}
\DefMacro{table-numbers_of_programs_and_bugs-20230412_20230415-row3-exec-skipped_diff}{0}
\DefMacro{table-numbers_of_programs_and_bugs-20230412_20230415-row3-exec-skipped_crash}{0}
\DefMacro{table-numbers_of_programs_and_bugs-20230412_20230415-row3-exec-final_diff}{0}
\DefMacro{table-numbers_of_programs_and_bugs-20230412_20230415-row3-exec-final_crash}{0}
\DefMacro{table-numbers_of_programs_and_bugs-20230412_20230415-row4-project}{vectorz}
\DefMacro{table-numbers_of_programs_and_bugs-20230412_20230415-row4-num_unit_tests}{5,582}
\DefMacro{table-numbers_of_programs_and_bugs-20230412_20230415-row4-num_unique_entry_methods}{5,582}
\DefMacro{table-numbers_of_programs_and_bugs-20230412_20230415-row4-num_unique_classes}{263}
\DefMacro{table-numbers_of_programs_and_bugs-20230412_20230415-row4-convert-fail}{0}
\DefMacro{table-numbers_of_programs_and_bugs-20230412_20230415-row4-convert-timeout}{3}
\DefMacro{table-numbers_of_programs_and_bugs-20230412_20230415-row4-convert-not_compile}{0}
\DefMacro{table-numbers_of_programs_and_bugs-20230412_20230415-row4-convert-no_original}{0}
\DefMacro{table-numbers_of_programs_and_bugs-20230412_20230415-row4-gen-fail}{433}
\DefMacro{table-numbers_of_programs_and_bugs-20230412_20230415-row4-gen-timeout}{126}
\DefMacro{table-numbers_of_programs_and_bugs-20230412_20230415-row4-gen-no_hole}{899}
\DefMacro{table-numbers_of_programs_and_bugs-20230412_20230415-row4-gen-no_gen}{1,329}
\DefMacro{table-numbers_of_programs_and_bugs-20230412_20230415-row4-num_working_templates}{3,351}
\DefMacro{table-numbers_of_programs_and_bugs-20230412_20230415-row4-num_unique_entry_methods_out_of_working_templates}{3,351}
\DefMacro{table-numbers_of_programs_and_bugs-20230412_20230415-row4-num_unique_classes_out_of_working_templates}{237}
\DefMacro{table-numbers_of_programs_and_bugs-20230412_20230415-row4-exec-total}{28,478}
\DefMacro{table-numbers_of_programs_and_bugs-20230412_20230415-row4-exec-not_compile}{0}
\DefMacro{table-numbers_of_programs_and_bugs-20230412_20230415-row4-exec-timeout}{189}
\DefMacro{table-numbers_of_programs_and_bugs-20230412_20230415-row4-exec-ok}{28,009}
\DefMacro{table-numbers_of_programs_and_bugs-20230412_20230415-row4-exec-skipped_diff}{259}
\DefMacro{table-numbers_of_programs_and_bugs-20230412_20230415-row4-exec-skipped_crash}{4}
\DefMacro{table-numbers_of_programs_and_bugs-20230412_20230415-row4-exec-final_diff}{0}
\DefMacro{table-numbers_of_programs_and_bugs-20230412_20230415-row4-exec-final_crash}{17}
\DefMacro{table-numbers_of_programs_and_bugs-20230412_20230415-row5-project}{statistics}
\DefMacro{table-numbers_of_programs_and_bugs-20230412_20230415-row5-num_unit_tests}{387}
\DefMacro{table-numbers_of_programs_and_bugs-20230412_20230415-row5-num_unique_entry_methods}{387}
\DefMacro{table-numbers_of_programs_and_bugs-20230412_20230415-row5-num_unique_classes}{32}
\DefMacro{table-numbers_of_programs_and_bugs-20230412_20230415-row5-convert-fail}{0}
\DefMacro{table-numbers_of_programs_and_bugs-20230412_20230415-row5-convert-timeout}{0}
\DefMacro{table-numbers_of_programs_and_bugs-20230412_20230415-row5-convert-not_compile}{0}
\DefMacro{table-numbers_of_programs_and_bugs-20230412_20230415-row5-convert-no_original}{0}
\DefMacro{table-numbers_of_programs_and_bugs-20230412_20230415-row5-gen-fail}{39}
\DefMacro{table-numbers_of_programs_and_bugs-20230412_20230415-row5-gen-timeout}{11}
\DefMacro{table-numbers_of_programs_and_bugs-20230412_20230415-row5-gen-no_hole}{2}
\DefMacro{table-numbers_of_programs_and_bugs-20230412_20230415-row5-gen-no_gen}{136}
\DefMacro{table-numbers_of_programs_and_bugs-20230412_20230415-row5-num_working_templates}{249}
\DefMacro{table-numbers_of_programs_and_bugs-20230412_20230415-row5-num_unique_entry_methods_out_of_working_templates}{249}
\DefMacro{table-numbers_of_programs_and_bugs-20230412_20230415-row5-num_unique_classes_out_of_working_templates}{30}
\DefMacro{table-numbers_of_programs_and_bugs-20230412_20230415-row5-exec-total}{2,096}
\DefMacro{table-numbers_of_programs_and_bugs-20230412_20230415-row5-exec-not_compile}{0}
\DefMacro{table-numbers_of_programs_and_bugs-20230412_20230415-row5-exec-timeout}{0}
\DefMacro{table-numbers_of_programs_and_bugs-20230412_20230415-row5-exec-ok}{2,091}
\DefMacro{table-numbers_of_programs_and_bugs-20230412_20230415-row5-exec-skipped_diff}{3}
\DefMacro{table-numbers_of_programs_and_bugs-20230412_20230415-row5-exec-skipped_crash}{2}
\DefMacro{table-numbers_of_programs_and_bugs-20230412_20230415-row5-exec-final_diff}{0}
\DefMacro{table-numbers_of_programs_and_bugs-20230412_20230415-row5-exec-final_crash}{0}
\DefMacro{table-numbers_of_programs_and_bugs-20230412_20230415-row6-project}{codec}
\DefMacro{table-numbers_of_programs_and_bugs-20230412_20230415-row6-num_unit_tests}{655}
\DefMacro{table-numbers_of_programs_and_bugs-20230412_20230415-row6-num_unique_entry_methods}{655}
\DefMacro{table-numbers_of_programs_and_bugs-20230412_20230415-row6-num_unique_classes}{53}
\DefMacro{table-numbers_of_programs_and_bugs-20230412_20230415-row6-convert-fail}{0}
\DefMacro{table-numbers_of_programs_and_bugs-20230412_20230415-row6-convert-timeout}{0}
\DefMacro{table-numbers_of_programs_and_bugs-20230412_20230415-row6-convert-not_compile}{5}
\DefMacro{table-numbers_of_programs_and_bugs-20230412_20230415-row6-convert-no_original}{0}
\DefMacro{table-numbers_of_programs_and_bugs-20230412_20230415-row6-gen-fail}{70}
\DefMacro{table-numbers_of_programs_and_bugs-20230412_20230415-row6-gen-timeout}{4}
\DefMacro{table-numbers_of_programs_and_bugs-20230412_20230415-row6-gen-no_hole}{9}
\DefMacro{table-numbers_of_programs_and_bugs-20230412_20230415-row6-gen-no_gen}{123}
\DefMacro{table-numbers_of_programs_and_bugs-20230412_20230415-row6-num_working_templates}{518}
\DefMacro{table-numbers_of_programs_and_bugs-20230412_20230415-row6-num_unique_entry_methods_out_of_working_templates}{518}
\DefMacro{table-numbers_of_programs_and_bugs-20230412_20230415-row6-num_unique_classes_out_of_working_templates}{43}
\DefMacro{table-numbers_of_programs_and_bugs-20230412_20230415-row6-exec-total}{3,532}
\DefMacro{table-numbers_of_programs_and_bugs-20230412_20230415-row6-exec-not_compile}{0}
\DefMacro{table-numbers_of_programs_and_bugs-20230412_20230415-row6-exec-timeout}{75}
\DefMacro{table-numbers_of_programs_and_bugs-20230412_20230415-row6-exec-ok}{3,439}
\DefMacro{table-numbers_of_programs_and_bugs-20230412_20230415-row6-exec-skipped_diff}{15}
\DefMacro{table-numbers_of_programs_and_bugs-20230412_20230415-row6-exec-skipped_crash}{3}
\DefMacro{table-numbers_of_programs_and_bugs-20230412_20230415-row6-exec-final_diff}{0}
\DefMacro{table-numbers_of_programs_and_bugs-20230412_20230415-row6-exec-final_crash}{0}
\DefMacro{table-numbers_of_programs_and_bugs-20230412_20230415-row7-project}{jfreechart}
\DefMacro{table-numbers_of_programs_and_bugs-20230412_20230415-row7-num_unit_tests}{7,137}
\DefMacro{table-numbers_of_programs_and_bugs-20230412_20230415-row7-num_unique_entry_methods}{7,137}
\DefMacro{table-numbers_of_programs_and_bugs-20230412_20230415-row7-num_unique_classes}{503}
\DefMacro{table-numbers_of_programs_and_bugs-20230412_20230415-row7-convert-fail}{7,006}
\DefMacro{table-numbers_of_programs_and_bugs-20230412_20230415-row7-convert-timeout}{67}
\DefMacro{table-numbers_of_programs_and_bugs-20230412_20230415-row7-convert-not_compile}{3}
\DefMacro{table-numbers_of_programs_and_bugs-20230412_20230415-row7-convert-no_original}{0}
\DefMacro{table-numbers_of_programs_and_bugs-20230412_20230415-row7-gen-fail}{9}
\DefMacro{table-numbers_of_programs_and_bugs-20230412_20230415-row7-gen-timeout}{0}
\DefMacro{table-numbers_of_programs_and_bugs-20230412_20230415-row7-gen-no_hole}{7}
\DefMacro{table-numbers_of_programs_and_bugs-20230412_20230415-row7-gen-no_gen}{9}
\DefMacro{table-numbers_of_programs_and_bugs-20230412_20230415-row7-num_working_templates}{45}
\DefMacro{table-numbers_of_programs_and_bugs-20230412_20230415-row7-num_unique_entry_methods_out_of_working_templates}{45}
\DefMacro{table-numbers_of_programs_and_bugs-20230412_20230415-row7-num_unique_classes_out_of_working_templates}{17}
\DefMacro{table-numbers_of_programs_and_bugs-20230412_20230415-row7-exec-total}{345}
\DefMacro{table-numbers_of_programs_and_bugs-20230412_20230415-row7-exec-not_compile}{0}
\DefMacro{table-numbers_of_programs_and_bugs-20230412_20230415-row7-exec-timeout}{0}
\DefMacro{table-numbers_of_programs_and_bugs-20230412_20230415-row7-exec-ok}{345}
\DefMacro{table-numbers_of_programs_and_bugs-20230412_20230415-row7-exec-skipped_diff}{0}
\DefMacro{table-numbers_of_programs_and_bugs-20230412_20230415-row7-exec-skipped_crash}{0}
\DefMacro{table-numbers_of_programs_and_bugs-20230412_20230415-row7-exec-final_diff}{0}
\DefMacro{table-numbers_of_programs_and_bugs-20230412_20230415-row7-exec-final_crash}{0}
\DefMacro{table-numbers_of_programs_and_bugs-20230412_20230415-row8-project}{zxing}
\DefMacro{table-numbers_of_programs_and_bugs-20230412_20230415-row8-num_unit_tests}{1,323}
\DefMacro{table-numbers_of_programs_and_bugs-20230412_20230415-row8-num_unique_entry_methods}{1,323}
\DefMacro{table-numbers_of_programs_and_bugs-20230412_20230415-row8-num_unique_classes}{213}
\DefMacro{table-numbers_of_programs_and_bugs-20230412_20230415-row8-convert-fail}{0}
\DefMacro{table-numbers_of_programs_and_bugs-20230412_20230415-row8-convert-timeout}{0}
\DefMacro{table-numbers_of_programs_and_bugs-20230412_20230415-row8-convert-not_compile}{72}
\DefMacro{table-numbers_of_programs_and_bugs-20230412_20230415-row8-convert-no_original}{0}
\DefMacro{table-numbers_of_programs_and_bugs-20230412_20230415-row8-gen-fail}{78}
\DefMacro{table-numbers_of_programs_and_bugs-20230412_20230415-row8-gen-timeout}{65}
\DefMacro{table-numbers_of_programs_and_bugs-20230412_20230415-row8-gen-no_hole}{234}
\DefMacro{table-numbers_of_programs_and_bugs-20230412_20230415-row8-gen-no_gen}{138}
\DefMacro{table-numbers_of_programs_and_bugs-20230412_20230415-row8-num_working_templates}{879}
\DefMacro{table-numbers_of_programs_and_bugs-20230412_20230415-row8-num_unique_entry_methods_out_of_working_templates}{879}
\DefMacro{table-numbers_of_programs_and_bugs-20230412_20230415-row8-num_unique_classes_out_of_working_templates}{172}
\DefMacro{table-numbers_of_programs_and_bugs-20230412_20230415-row8-exec-total}{7,364}
\DefMacro{table-numbers_of_programs_and_bugs-20230412_20230415-row8-exec-not_compile}{0}
\DefMacro{table-numbers_of_programs_and_bugs-20230412_20230415-row8-exec-timeout}{192}
\DefMacro{table-numbers_of_programs_and_bugs-20230412_20230415-row8-exec-ok}{7,099}
\DefMacro{table-numbers_of_programs_and_bugs-20230412_20230415-row8-exec-skipped_diff}{58}
\DefMacro{table-numbers_of_programs_and_bugs-20230412_20230415-row8-exec-skipped_crash}{11}
\DefMacro{table-numbers_of_programs_and_bugs-20230412_20230415-row8-exec-final_diff}{1}
\DefMacro{table-numbers_of_programs_and_bugs-20230412_20230415-row8-exec-final_crash}{3}
\DefMacro{table-numbers_of_programs_and_bugs-20230412_20230415-row9-project}{compress}
\DefMacro{table-numbers_of_programs_and_bugs-20230412_20230415-row9-num_unit_tests}{1,939}
\DefMacro{table-numbers_of_programs_and_bugs-20230412_20230415-row9-num_unique_entry_methods}{1,939}
\DefMacro{table-numbers_of_programs_and_bugs-20230412_20230415-row9-num_unique_classes}{177}
\DefMacro{table-numbers_of_programs_and_bugs-20230412_20230415-row9-convert-fail}{0}
\DefMacro{table-numbers_of_programs_and_bugs-20230412_20230415-row9-convert-timeout}{0}
\DefMacro{table-numbers_of_programs_and_bugs-20230412_20230415-row9-convert-not_compile}{95}
\DefMacro{table-numbers_of_programs_and_bugs-20230412_20230415-row9-convert-no_original}{0}
\DefMacro{table-numbers_of_programs_and_bugs-20230412_20230415-row9-gen-fail}{99}
\DefMacro{table-numbers_of_programs_and_bugs-20230412_20230415-row9-gen-timeout}{51}
\DefMacro{table-numbers_of_programs_and_bugs-20230412_20230415-row9-gen-no_hole}{337}
\DefMacro{table-numbers_of_programs_and_bugs-20230412_20230415-row9-gen-no_gen}{120}
\DefMacro{table-numbers_of_programs_and_bugs-20230412_20230415-row9-num_working_templates}{1,387}
\DefMacro{table-numbers_of_programs_and_bugs-20230412_20230415-row9-num_unique_entry_methods_out_of_working_templates}{1,387}
\DefMacro{table-numbers_of_programs_and_bugs-20230412_20230415-row9-num_unique_classes_out_of_working_templates}{149}
\DefMacro{table-numbers_of_programs_and_bugs-20230412_20230415-row9-exec-total}{11,705}
\DefMacro{table-numbers_of_programs_and_bugs-20230412_20230415-row9-exec-not_compile}{0}
\DefMacro{table-numbers_of_programs_and_bugs-20230412_20230415-row9-exec-timeout}{225}
\DefMacro{table-numbers_of_programs_and_bugs-20230412_20230415-row9-exec-ok}{10,656}
\DefMacro{table-numbers_of_programs_and_bugs-20230412_20230415-row9-exec-skipped_diff}{747}
\DefMacro{table-numbers_of_programs_and_bugs-20230412_20230415-row9-exec-skipped_crash}{1}
\DefMacro{table-numbers_of_programs_and_bugs-20230412_20230415-row9-exec-final_diff}{73}
\DefMacro{table-numbers_of_programs_and_bugs-20230412_20230415-row9-exec-final_crash}{3}
\DefMacro{table-numbers_of_programs_and_bugs-20230412_20230415-row10-project}{$\sum$}
\DefMacro{table-numbers_of_programs_and_bugs-20230412_20230415-row10-num_unit_tests}{26,048}
\DefMacro{table-numbers_of_programs_and_bugs-20230412_20230415-row10-num_unique_entry_methods}{26,042}
\DefMacro{table-numbers_of_programs_and_bugs-20230412_20230415-row10-num_unique_classes}{2,044}
\DefMacro{table-numbers_of_programs_and_bugs-20230412_20230415-row10-convert-fail}{7,014}
\DefMacro{table-numbers_of_programs_and_bugs-20230412_20230415-row10-convert-timeout}{72}
\DefMacro{table-numbers_of_programs_and_bugs-20230412_20230415-row10-convert-not_compile}{582}
\DefMacro{table-numbers_of_programs_and_bugs-20230412_20230415-row10-convert-no_original}{0}
\DefMacro{table-numbers_of_programs_and_bugs-20230412_20230415-row10-gen-fail}{1,332}
\DefMacro{table-numbers_of_programs_and_bugs-20230412_20230415-row10-gen-timeout}{749}
\DefMacro{table-numbers_of_programs_and_bugs-20230412_20230415-row10-gen-no_hole}{2,644}
\DefMacro{table-numbers_of_programs_and_bugs-20230412_20230415-row10-gen-no_gen}{3,630}
\DefMacro{table-numbers_of_programs_and_bugs-20230412_20230415-row10-num_working_templates}{12,106}
\DefMacro{table-numbers_of_programs_and_bugs-20230412_20230415-row10-num_unique_entry_methods_out_of_working_templates}{12,106}
\DefMacro{table-numbers_of_programs_and_bugs-20230412_20230415-row10-num_unique_classes_out_of_working_templates}{1,230}
\DefMacro{table-numbers_of_programs_and_bugs-20230412_20230415-row10-exec-total}{101,097}
\DefMacro{table-numbers_of_programs_and_bugs-20230412_20230415-row10-exec-not_compile}{0}
\DefMacro{table-numbers_of_programs_and_bugs-20230412_20230415-row10-exec-timeout}{1,345}
\DefMacro{table-numbers_of_programs_and_bugs-20230412_20230415-row10-exec-ok}{97,684}
\DefMacro{table-numbers_of_programs_and_bugs-20230412_20230415-row10-exec-skipped_diff}{1,838}
\DefMacro{table-numbers_of_programs_and_bugs-20230412_20230415-row10-exec-skipped_crash}{101}
\DefMacro{table-numbers_of_programs_and_bugs-20230412_20230415-row10-exec-final_diff}{77}
\DefMacro{table-numbers_of_programs_and_bugs-20230412_20230415-row10-exec-final_crash}{52}

\DefMacro{table-numbers_of_programs_and_bugs-20230419-row0-project}{lang}
\DefMacro{table-numbers_of_programs_and_bugs-20230419-row0-num_unit_tests}{1,911}
\DefMacro{table-numbers_of_programs_and_bugs-20230419-row0-num_unique_entry_methods}{1,132}
\DefMacro{table-numbers_of_programs_and_bugs-20230419-row0-num_unique_classes}{98}
\DefMacro{table-numbers_of_programs_and_bugs-20230419-row0-convert-fail}{110}
\DefMacro{table-numbers_of_programs_and_bugs-20230419-row0-convert-timeout}{0}
\DefMacro{table-numbers_of_programs_and_bugs-20230419-row0-convert-not_compile}{0}
\DefMacro{table-numbers_of_programs_and_bugs-20230419-row0-convert-no_original}{0}
\DefMacro{table-numbers_of_programs_and_bugs-20230419-row0-gen-fail}{0}
\DefMacro{table-numbers_of_programs_and_bugs-20230419-row0-gen-timeout}{0}
\DefMacro{table-numbers_of_programs_and_bugs-20230419-row0-gen-no_hole}{0}
\DefMacro{table-numbers_of_programs_and_bugs-20230419-row0-gen-no_gen}{0}
\DefMacro{table-numbers_of_programs_and_bugs-20230419-row0-num_working_templates}{1,801}
\DefMacro{table-numbers_of_programs_and_bugs-20230419-row0-num_unique_entry_methods_out_of_working_templates}{1,058}
\DefMacro{table-numbers_of_programs_and_bugs-20230419-row0-num_unique_classes_out_of_working_templates}{94}
\DefMacro{table-numbers_of_programs_and_bugs-20230419-row0-exec-total}{1,801}
\DefMacro{table-numbers_of_programs_and_bugs-20230419-row0-exec-not_compile}{0}
\DefMacro{table-numbers_of_programs_and_bugs-20230419-row0-exec-timeout}{27}
\DefMacro{table-numbers_of_programs_and_bugs-20230419-row0-exec-ok}{1,688}
\DefMacro{table-numbers_of_programs_and_bugs-20230419-row0-exec-skipped_diff}{86}
\DefMacro{table-numbers_of_programs_and_bugs-20230419-row0-exec-skipped_crash}{0}
\DefMacro{table-numbers_of_programs_and_bugs-20230419-row0-exec-final_diff}{0}
\DefMacro{table-numbers_of_programs_and_bugs-20230419-row0-exec-final_crash}{0}
\DefMacro{table-numbers_of_programs_and_bugs-20230419-row1-project}{math}
\DefMacro{table-numbers_of_programs_and_bugs-20230419-row1-num_unit_tests}{1,067}
\DefMacro{table-numbers_of_programs_and_bugs-20230419-row1-num_unique_entry_methods}{667}
\DefMacro{table-numbers_of_programs_and_bugs-20230419-row1-num_unique_classes}{177}
\DefMacro{table-numbers_of_programs_and_bugs-20230419-row1-convert-fail}{12}
\DefMacro{table-numbers_of_programs_and_bugs-20230419-row1-convert-timeout}{0}
\DefMacro{table-numbers_of_programs_and_bugs-20230419-row1-convert-not_compile}{0}
\DefMacro{table-numbers_of_programs_and_bugs-20230419-row1-convert-no_original}{0}
\DefMacro{table-numbers_of_programs_and_bugs-20230419-row1-gen-fail}{0}
\DefMacro{table-numbers_of_programs_and_bugs-20230419-row1-gen-timeout}{0}
\DefMacro{table-numbers_of_programs_and_bugs-20230419-row1-gen-no_hole}{0}
\DefMacro{table-numbers_of_programs_and_bugs-20230419-row1-gen-no_gen}{0}
\DefMacro{table-numbers_of_programs_and_bugs-20230419-row1-num_working_templates}{1,055}
\DefMacro{table-numbers_of_programs_and_bugs-20230419-row1-num_unique_entry_methods_out_of_working_templates}{657}
\DefMacro{table-numbers_of_programs_and_bugs-20230419-row1-num_unique_classes_out_of_working_templates}{173}
\DefMacro{table-numbers_of_programs_and_bugs-20230419-row1-exec-total}{1,055}
\DefMacro{table-numbers_of_programs_and_bugs-20230419-row1-exec-not_compile}{0}
\DefMacro{table-numbers_of_programs_and_bugs-20230419-row1-exec-timeout}{7}
\DefMacro{table-numbers_of_programs_and_bugs-20230419-row1-exec-ok}{1,020}
\DefMacro{table-numbers_of_programs_and_bugs-20230419-row1-exec-skipped_diff}{28}
\DefMacro{table-numbers_of_programs_and_bugs-20230419-row1-exec-skipped_crash}{0}
\DefMacro{table-numbers_of_programs_and_bugs-20230419-row1-exec-final_diff}{0}
\DefMacro{table-numbers_of_programs_and_bugs-20230419-row1-exec-final_crash}{0}
\DefMacro{table-numbers_of_programs_and_bugs-20230419-row2-project}{text}
\DefMacro{table-numbers_of_programs_and_bugs-20230419-row2-num_unit_tests}{1,726}
\DefMacro{table-numbers_of_programs_and_bugs-20230419-row2-num_unique_entry_methods}{501}
\DefMacro{table-numbers_of_programs_and_bugs-20230419-row2-num_unique_classes}{39}
\DefMacro{table-numbers_of_programs_and_bugs-20230419-row2-convert-fail}{23}
\DefMacro{table-numbers_of_programs_and_bugs-20230419-row2-convert-timeout}{0}
\DefMacro{table-numbers_of_programs_and_bugs-20230419-row2-convert-not_compile}{0}
\DefMacro{table-numbers_of_programs_and_bugs-20230419-row2-convert-no_original}{0}
\DefMacro{table-numbers_of_programs_and_bugs-20230419-row2-gen-fail}{0}
\DefMacro{table-numbers_of_programs_and_bugs-20230419-row2-gen-timeout}{0}
\DefMacro{table-numbers_of_programs_and_bugs-20230419-row2-gen-no_hole}{0}
\DefMacro{table-numbers_of_programs_and_bugs-20230419-row2-gen-no_gen}{0}
\DefMacro{table-numbers_of_programs_and_bugs-20230419-row2-num_working_templates}{1,703}
\DefMacro{table-numbers_of_programs_and_bugs-20230419-row2-num_unique_entry_methods_out_of_working_templates}{492}
\DefMacro{table-numbers_of_programs_and_bugs-20230419-row2-num_unique_classes_out_of_working_templates}{35}
\DefMacro{table-numbers_of_programs_and_bugs-20230419-row2-exec-total}{1,703}
\DefMacro{table-numbers_of_programs_and_bugs-20230419-row2-exec-not_compile}{0}
\DefMacro{table-numbers_of_programs_and_bugs-20230419-row2-exec-timeout}{165}
\DefMacro{table-numbers_of_programs_and_bugs-20230419-row2-exec-ok}{1,511}
\DefMacro{table-numbers_of_programs_and_bugs-20230419-row2-exec-skipped_diff}{27}
\DefMacro{table-numbers_of_programs_and_bugs-20230419-row2-exec-skipped_crash}{0}
\DefMacro{table-numbers_of_programs_and_bugs-20230419-row2-exec-final_diff}{0}
\DefMacro{table-numbers_of_programs_and_bugs-20230419-row2-exec-final_crash}{0}
\DefMacro{table-numbers_of_programs_and_bugs-20230419-row3-project}{numbers}
\DefMacro{table-numbers_of_programs_and_bugs-20230419-row3-num_unit_tests}{3,180}
\DefMacro{table-numbers_of_programs_and_bugs-20230419-row3-num_unique_entry_methods}{45}
\DefMacro{table-numbers_of_programs_and_bugs-20230419-row3-num_unique_classes}{4}
\DefMacro{table-numbers_of_programs_and_bugs-20230419-row3-convert-fail}{122}
\DefMacro{table-numbers_of_programs_and_bugs-20230419-row3-convert-timeout}{0}
\DefMacro{table-numbers_of_programs_and_bugs-20230419-row3-convert-not_compile}{0}
\DefMacro{table-numbers_of_programs_and_bugs-20230419-row3-convert-no_original}{0}
\DefMacro{table-numbers_of_programs_and_bugs-20230419-row3-gen-fail}{0}
\DefMacro{table-numbers_of_programs_and_bugs-20230419-row3-gen-timeout}{0}
\DefMacro{table-numbers_of_programs_and_bugs-20230419-row3-gen-no_hole}{0}
\DefMacro{table-numbers_of_programs_and_bugs-20230419-row3-gen-no_gen}{0}
\DefMacro{table-numbers_of_programs_and_bugs-20230419-row3-num_working_templates}{3,058}
\DefMacro{table-numbers_of_programs_and_bugs-20230419-row3-num_unique_entry_methods_out_of_working_templates}{42}
\DefMacro{table-numbers_of_programs_and_bugs-20230419-row3-num_unique_classes_out_of_working_templates}{3}
\DefMacro{table-numbers_of_programs_and_bugs-20230419-row3-exec-total}{3,058}
\DefMacro{table-numbers_of_programs_and_bugs-20230419-row3-exec-not_compile}{0}
\DefMacro{table-numbers_of_programs_and_bugs-20230419-row3-exec-timeout}{6}
\DefMacro{table-numbers_of_programs_and_bugs-20230419-row3-exec-ok}{3,052}
\DefMacro{table-numbers_of_programs_and_bugs-20230419-row3-exec-skipped_diff}{0}
\DefMacro{table-numbers_of_programs_and_bugs-20230419-row3-exec-skipped_crash}{0}
\DefMacro{table-numbers_of_programs_and_bugs-20230419-row3-exec-final_diff}{0}
\DefMacro{table-numbers_of_programs_and_bugs-20230419-row3-exec-final_crash}{0}
\DefMacro{table-numbers_of_programs_and_bugs-20230419-row4-project}{vectorz}
\DefMacro{table-numbers_of_programs_and_bugs-20230419-row4-num_unit_tests}{1,072}
\DefMacro{table-numbers_of_programs_and_bugs-20230419-row4-num_unique_entry_methods}{880}
\DefMacro{table-numbers_of_programs_and_bugs-20230419-row4-num_unique_classes}{153}
\DefMacro{table-numbers_of_programs_and_bugs-20230419-row4-convert-fail}{0}
\DefMacro{table-numbers_of_programs_and_bugs-20230419-row4-convert-timeout}{0}
\DefMacro{table-numbers_of_programs_and_bugs-20230419-row4-convert-not_compile}{0}
\DefMacro{table-numbers_of_programs_and_bugs-20230419-row4-convert-no_original}{0}
\DefMacro{table-numbers_of_programs_and_bugs-20230419-row4-gen-fail}{0}
\DefMacro{table-numbers_of_programs_and_bugs-20230419-row4-gen-timeout}{0}
\DefMacro{table-numbers_of_programs_and_bugs-20230419-row4-gen-no_hole}{0}
\DefMacro{table-numbers_of_programs_and_bugs-20230419-row4-gen-no_gen}{0}
\DefMacro{table-numbers_of_programs_and_bugs-20230419-row4-num_working_templates}{1,072}
\DefMacro{table-numbers_of_programs_and_bugs-20230419-row4-num_unique_entry_methods_out_of_working_templates}{880}
\DefMacro{table-numbers_of_programs_and_bugs-20230419-row4-num_unique_classes_out_of_working_templates}{153}
\DefMacro{table-numbers_of_programs_and_bugs-20230419-row4-exec-total}{1,072}
\DefMacro{table-numbers_of_programs_and_bugs-20230419-row4-exec-not_compile}{0}
\DefMacro{table-numbers_of_programs_and_bugs-20230419-row4-exec-timeout}{10}
\DefMacro{table-numbers_of_programs_and_bugs-20230419-row4-exec-ok}{1,051}
\DefMacro{table-numbers_of_programs_and_bugs-20230419-row4-exec-skipped_diff}{11}
\DefMacro{table-numbers_of_programs_and_bugs-20230419-row4-exec-skipped_crash}{0}
\DefMacro{table-numbers_of_programs_and_bugs-20230419-row4-exec-final_diff}{0}
\DefMacro{table-numbers_of_programs_and_bugs-20230419-row4-exec-final_crash}{0}
\DefMacro{table-numbers_of_programs_and_bugs-20230419-row5-project}{statistics}
\DefMacro{table-numbers_of_programs_and_bugs-20230419-row5-num_unit_tests}{1,909}
\DefMacro{table-numbers_of_programs_and_bugs-20230419-row5-num_unique_entry_methods}{373}
\DefMacro{table-numbers_of_programs_and_bugs-20230419-row5-num_unique_classes}{27}
\DefMacro{table-numbers_of_programs_and_bugs-20230419-row5-convert-fail}{191}
\DefMacro{table-numbers_of_programs_and_bugs-20230419-row5-convert-timeout}{0}
\DefMacro{table-numbers_of_programs_and_bugs-20230419-row5-convert-not_compile}{0}
\DefMacro{table-numbers_of_programs_and_bugs-20230419-row5-convert-no_original}{0}
\DefMacro{table-numbers_of_programs_and_bugs-20230419-row5-gen-fail}{0}
\DefMacro{table-numbers_of_programs_and_bugs-20230419-row5-gen-timeout}{0}
\DefMacro{table-numbers_of_programs_and_bugs-20230419-row5-gen-no_hole}{0}
\DefMacro{table-numbers_of_programs_and_bugs-20230419-row5-gen-no_gen}{0}
\DefMacro{table-numbers_of_programs_and_bugs-20230419-row5-num_working_templates}{1,718}
\DefMacro{table-numbers_of_programs_and_bugs-20230419-row5-num_unique_entry_methods_out_of_working_templates}{327}
\DefMacro{table-numbers_of_programs_and_bugs-20230419-row5-num_unique_classes_out_of_working_templates}{27}
\DefMacro{table-numbers_of_programs_and_bugs-20230419-row5-exec-total}{1,718}
\DefMacro{table-numbers_of_programs_and_bugs-20230419-row5-exec-not_compile}{0}
\DefMacro{table-numbers_of_programs_and_bugs-20230419-row5-exec-timeout}{1}
\DefMacro{table-numbers_of_programs_and_bugs-20230419-row5-exec-ok}{1,670}
\DefMacro{table-numbers_of_programs_and_bugs-20230419-row5-exec-skipped_diff}{47}
\DefMacro{table-numbers_of_programs_and_bugs-20230419-row5-exec-skipped_crash}{0}
\DefMacro{table-numbers_of_programs_and_bugs-20230419-row5-exec-final_diff}{0}
\DefMacro{table-numbers_of_programs_and_bugs-20230419-row5-exec-final_crash}{0}
\DefMacro{table-numbers_of_programs_and_bugs-20230419-row6-project}{codec}
\DefMacro{table-numbers_of_programs_and_bugs-20230419-row6-num_unit_tests}{2,247}
\DefMacro{table-numbers_of_programs_and_bugs-20230419-row6-num_unique_entry_methods}{291}
\DefMacro{table-numbers_of_programs_and_bugs-20230419-row6-num_unique_classes}{47}
\DefMacro{table-numbers_of_programs_and_bugs-20230419-row6-convert-fail}{9}
\DefMacro{table-numbers_of_programs_and_bugs-20230419-row6-convert-timeout}{0}
\DefMacro{table-numbers_of_programs_and_bugs-20230419-row6-convert-not_compile}{0}
\DefMacro{table-numbers_of_programs_and_bugs-20230419-row6-convert-no_original}{0}
\DefMacro{table-numbers_of_programs_and_bugs-20230419-row6-gen-fail}{0}
\DefMacro{table-numbers_of_programs_and_bugs-20230419-row6-gen-timeout}{0}
\DefMacro{table-numbers_of_programs_and_bugs-20230419-row6-gen-no_hole}{0}
\DefMacro{table-numbers_of_programs_and_bugs-20230419-row6-gen-no_gen}{0}
\DefMacro{table-numbers_of_programs_and_bugs-20230419-row6-num_working_templates}{2,238}
\DefMacro{table-numbers_of_programs_and_bugs-20230419-row6-num_unique_entry_methods_out_of_working_templates}{287}
\DefMacro{table-numbers_of_programs_and_bugs-20230419-row6-num_unique_classes_out_of_working_templates}{44}
\DefMacro{table-numbers_of_programs_and_bugs-20230419-row6-exec-total}{2,238}
\DefMacro{table-numbers_of_programs_and_bugs-20230419-row6-exec-not_compile}{0}
\DefMacro{table-numbers_of_programs_and_bugs-20230419-row6-exec-timeout}{64}
\DefMacro{table-numbers_of_programs_and_bugs-20230419-row6-exec-ok}{2,035}
\DefMacro{table-numbers_of_programs_and_bugs-20230419-row6-exec-skipped_diff}{132}
\DefMacro{table-numbers_of_programs_and_bugs-20230419-row6-exec-skipped_crash}{0}
\DefMacro{table-numbers_of_programs_and_bugs-20230419-row6-exec-final_diff}{7}
\DefMacro{table-numbers_of_programs_and_bugs-20230419-row6-exec-final_crash}{0}
\DefMacro{table-numbers_of_programs_and_bugs-20230419-row7-project}{jfreechart}
\DefMacro{table-numbers_of_programs_and_bugs-20230419-row7-num_unit_tests}{1,301}
\DefMacro{table-numbers_of_programs_and_bugs-20230419-row7-num_unique_entry_methods}{1,135}
\DefMacro{table-numbers_of_programs_and_bugs-20230419-row7-num_unique_classes}{282}
\DefMacro{table-numbers_of_programs_and_bugs-20230419-row7-convert-fail}{2}
\DefMacro{table-numbers_of_programs_and_bugs-20230419-row7-convert-timeout}{0}
\DefMacro{table-numbers_of_programs_and_bugs-20230419-row7-convert-not_compile}{0}
\DefMacro{table-numbers_of_programs_and_bugs-20230419-row7-convert-no_original}{0}
\DefMacro{table-numbers_of_programs_and_bugs-20230419-row7-gen-fail}{0}
\DefMacro{table-numbers_of_programs_and_bugs-20230419-row7-gen-timeout}{0}
\DefMacro{table-numbers_of_programs_and_bugs-20230419-row7-gen-no_hole}{0}
\DefMacro{table-numbers_of_programs_and_bugs-20230419-row7-gen-no_gen}{0}
\DefMacro{table-numbers_of_programs_and_bugs-20230419-row7-num_working_templates}{1,299}
\DefMacro{table-numbers_of_programs_and_bugs-20230419-row7-num_unique_entry_methods_out_of_working_templates}{1,133}
\DefMacro{table-numbers_of_programs_and_bugs-20230419-row7-num_unique_classes_out_of_working_templates}{280}
\DefMacro{table-numbers_of_programs_and_bugs-20230419-row7-exec-total}{1,299}
\DefMacro{table-numbers_of_programs_and_bugs-20230419-row7-exec-not_compile}{0}
\DefMacro{table-numbers_of_programs_and_bugs-20230419-row7-exec-timeout}{13}
\DefMacro{table-numbers_of_programs_and_bugs-20230419-row7-exec-ok}{1,277}
\DefMacro{table-numbers_of_programs_and_bugs-20230419-row7-exec-skipped_diff}{7}
\DefMacro{table-numbers_of_programs_and_bugs-20230419-row7-exec-skipped_crash}{0}
\DefMacro{table-numbers_of_programs_and_bugs-20230419-row7-exec-final_diff}{2}
\DefMacro{table-numbers_of_programs_and_bugs-20230419-row7-exec-final_crash}{0}
\DefMacro{table-numbers_of_programs_and_bugs-20230419-row8-project}{zxing}
\DefMacro{table-numbers_of_programs_and_bugs-20230419-row8-num_unit_tests}{1,406}
\DefMacro{table-numbers_of_programs_and_bugs-20230419-row8-num_unique_entry_methods}{335}
\DefMacro{table-numbers_of_programs_and_bugs-20230419-row8-num_unique_classes}{89}
\DefMacro{table-numbers_of_programs_and_bugs-20230419-row8-convert-fail}{48}
\DefMacro{table-numbers_of_programs_and_bugs-20230419-row8-convert-timeout}{0}
\DefMacro{table-numbers_of_programs_and_bugs-20230419-row8-convert-not_compile}{0}
\DefMacro{table-numbers_of_programs_and_bugs-20230419-row8-convert-no_original}{0}
\DefMacro{table-numbers_of_programs_and_bugs-20230419-row8-gen-fail}{0}
\DefMacro{table-numbers_of_programs_and_bugs-20230419-row8-gen-timeout}{0}
\DefMacro{table-numbers_of_programs_and_bugs-20230419-row8-gen-no_hole}{0}
\DefMacro{table-numbers_of_programs_and_bugs-20230419-row8-gen-no_gen}{0}
\DefMacro{table-numbers_of_programs_and_bugs-20230419-row8-num_working_templates}{1,358}
\DefMacro{table-numbers_of_programs_and_bugs-20230419-row8-num_unique_entry_methods_out_of_working_templates}{325}
\DefMacro{table-numbers_of_programs_and_bugs-20230419-row8-num_unique_classes_out_of_working_templates}{85}
\DefMacro{table-numbers_of_programs_and_bugs-20230419-row8-exec-total}{1,358}
\DefMacro{table-numbers_of_programs_and_bugs-20230419-row8-exec-not_compile}{0}
\DefMacro{table-numbers_of_programs_and_bugs-20230419-row8-exec-timeout}{1}
\DefMacro{table-numbers_of_programs_and_bugs-20230419-row8-exec-ok}{1,237}
\DefMacro{table-numbers_of_programs_and_bugs-20230419-row8-exec-skipped_diff}{100}
\DefMacro{table-numbers_of_programs_and_bugs-20230419-row8-exec-skipped_crash}{0}
\DefMacro{table-numbers_of_programs_and_bugs-20230419-row8-exec-final_diff}{0}
\DefMacro{table-numbers_of_programs_and_bugs-20230419-row8-exec-final_crash}{20}
\DefMacro{table-numbers_of_programs_and_bugs-20230419-row9-project}{compress}
\DefMacro{table-numbers_of_programs_and_bugs-20230419-row9-num_unit_tests}{1,073}
\DefMacro{table-numbers_of_programs_and_bugs-20230419-row9-num_unique_entry_methods}{586}
\DefMacro{table-numbers_of_programs_and_bugs-20230419-row9-num_unique_classes}{112}
\DefMacro{table-numbers_of_programs_and_bugs-20230419-row9-convert-fail}{9}
\DefMacro{table-numbers_of_programs_and_bugs-20230419-row9-convert-timeout}{0}
\DefMacro{table-numbers_of_programs_and_bugs-20230419-row9-convert-not_compile}{3}
\DefMacro{table-numbers_of_programs_and_bugs-20230419-row9-convert-no_original}{0}
\DefMacro{table-numbers_of_programs_and_bugs-20230419-row9-gen-fail}{0}
\DefMacro{table-numbers_of_programs_and_bugs-20230419-row9-gen-timeout}{0}
\DefMacro{table-numbers_of_programs_and_bugs-20230419-row9-gen-no_hole}{0}
\DefMacro{table-numbers_of_programs_and_bugs-20230419-row9-gen-no_gen}{0}
\DefMacro{table-numbers_of_programs_and_bugs-20230419-row9-num_working_templates}{1,061}
\DefMacro{table-numbers_of_programs_and_bugs-20230419-row9-num_unique_entry_methods_out_of_working_templates}{580}
\DefMacro{table-numbers_of_programs_and_bugs-20230419-row9-num_unique_classes_out_of_working_templates}{108}
\DefMacro{table-numbers_of_programs_and_bugs-20230419-row9-exec-total}{1,061}
\DefMacro{table-numbers_of_programs_and_bugs-20230419-row9-exec-not_compile}{0}
\DefMacro{table-numbers_of_programs_and_bugs-20230419-row9-exec-timeout}{3}
\DefMacro{table-numbers_of_programs_and_bugs-20230419-row9-exec-ok}{857}
\DefMacro{table-numbers_of_programs_and_bugs-20230419-row9-exec-skipped_diff}{200}
\DefMacro{table-numbers_of_programs_and_bugs-20230419-row9-exec-skipped_crash}{0}
\DefMacro{table-numbers_of_programs_and_bugs-20230419-row9-exec-final_diff}{1}
\DefMacro{table-numbers_of_programs_and_bugs-20230419-row9-exec-final_crash}{0}
\DefMacro{table-numbers_of_programs_and_bugs-20230419-row10-project}{$\sum$}
\DefMacro{table-numbers_of_programs_and_bugs-20230419-row10-num_unit_tests}{16,892}
\DefMacro{table-numbers_of_programs_and_bugs-20230419-row10-num_unique_entry_methods}{5,945}
\DefMacro{table-numbers_of_programs_and_bugs-20230419-row10-num_unique_classes}{1,028}
\DefMacro{table-numbers_of_programs_and_bugs-20230419-row10-convert-fail}{526}
\DefMacro{table-numbers_of_programs_and_bugs-20230419-row10-convert-timeout}{0}
\DefMacro{table-numbers_of_programs_and_bugs-20230419-row10-convert-not_compile}{3}
\DefMacro{table-numbers_of_programs_and_bugs-20230419-row10-convert-no_original}{0}
\DefMacro{table-numbers_of_programs_and_bugs-20230419-row10-gen-fail}{0}
\DefMacro{table-numbers_of_programs_and_bugs-20230419-row10-gen-timeout}{0}
\DefMacro{table-numbers_of_programs_and_bugs-20230419-row10-gen-no_hole}{0}
\DefMacro{table-numbers_of_programs_and_bugs-20230419-row10-gen-no_gen}{0}
\DefMacro{table-numbers_of_programs_and_bugs-20230419-row10-num_working_templates}{16,363}
\DefMacro{table-numbers_of_programs_and_bugs-20230419-row10-num_unique_entry_methods_out_of_working_templates}{5,781}
\DefMacro{table-numbers_of_programs_and_bugs-20230419-row10-num_unique_classes_out_of_working_templates}{1,002}
\DefMacro{table-numbers_of_programs_and_bugs-20230419-row10-exec-total}{16,363}
\DefMacro{table-numbers_of_programs_and_bugs-20230419-row10-exec-not_compile}{0}
\DefMacro{table-numbers_of_programs_and_bugs-20230419-row10-exec-timeout}{297}
\DefMacro{table-numbers_of_programs_and_bugs-20230419-row10-exec-ok}{15,398}
\DefMacro{table-numbers_of_programs_and_bugs-20230419-row10-exec-skipped_diff}{638}
\DefMacro{table-numbers_of_programs_and_bugs-20230419-row10-exec-skipped_crash}{0}
\DefMacro{table-numbers_of_programs_and_bugs-20230419-row10-exec-final_diff}{10}
\DefMacro{table-numbers_of_programs_and_bugs-20230419-row10-exec-final_crash}{20}

\DefMacro{table-numbers_of_programs_and_bugs-20230418-row0-project}{lang}
\DefMacro{table-numbers_of_programs_and_bugs-20230418-row0-num_unit_tests}{1,897}
\DefMacro{table-numbers_of_programs_and_bugs-20230418-row0-num_unique_entry_methods}{1,150}
\DefMacro{table-numbers_of_programs_and_bugs-20230418-row0-num_unique_classes}{95}
\DefMacro{table-numbers_of_programs_and_bugs-20230418-row0-convert-fail}{84}
\DefMacro{table-numbers_of_programs_and_bugs-20230418-row0-convert-timeout}{0}
\DefMacro{table-numbers_of_programs_and_bugs-20230418-row0-convert-not_compile}{0}
\DefMacro{table-numbers_of_programs_and_bugs-20230418-row0-convert-no_original}{0}
\DefMacro{table-numbers_of_programs_and_bugs-20230418-row0-gen-fail}{0}
\DefMacro{table-numbers_of_programs_and_bugs-20230418-row0-gen-timeout}{0}
\DefMacro{table-numbers_of_programs_and_bugs-20230418-row0-gen-no_hole}{0}
\DefMacro{table-numbers_of_programs_and_bugs-20230418-row0-gen-no_gen}{0}
\DefMacro{table-numbers_of_programs_and_bugs-20230418-row0-num_working_templates}{1,813}
\DefMacro{table-numbers_of_programs_and_bugs-20230418-row0-num_unique_entry_methods_out_of_working_templates}{1,088}
\DefMacro{table-numbers_of_programs_and_bugs-20230418-row0-num_unique_classes_out_of_working_templates}{92}
\DefMacro{table-numbers_of_programs_and_bugs-20230418-row0-exec-total}{1,813}
\DefMacro{table-numbers_of_programs_and_bugs-20230418-row0-exec-not_compile}{0}
\DefMacro{table-numbers_of_programs_and_bugs-20230418-row0-exec-timeout}{22}
\DefMacro{table-numbers_of_programs_and_bugs-20230418-row0-exec-ok}{1,706}
\DefMacro{table-numbers_of_programs_and_bugs-20230418-row0-exec-skipped_diff}{81}
\DefMacro{table-numbers_of_programs_and_bugs-20230418-row0-exec-skipped_crash}{1}
\DefMacro{table-numbers_of_programs_and_bugs-20230418-row0-exec-final_diff}{3}
\DefMacro{table-numbers_of_programs_and_bugs-20230418-row0-exec-final_crash}{0}
\DefMacro{table-numbers_of_programs_and_bugs-20230418-row1-project}{math}
\DefMacro{table-numbers_of_programs_and_bugs-20230418-row1-num_unit_tests}{1,042}
\DefMacro{table-numbers_of_programs_and_bugs-20230418-row1-num_unique_entry_methods}{543}
\DefMacro{table-numbers_of_programs_and_bugs-20230418-row1-num_unique_classes}{137}
\DefMacro{table-numbers_of_programs_and_bugs-20230418-row1-convert-fail}{17}
\DefMacro{table-numbers_of_programs_and_bugs-20230418-row1-convert-timeout}{0}
\DefMacro{table-numbers_of_programs_and_bugs-20230418-row1-convert-not_compile}{0}
\DefMacro{table-numbers_of_programs_and_bugs-20230418-row1-convert-no_original}{0}
\DefMacro{table-numbers_of_programs_and_bugs-20230418-row1-gen-fail}{0}
\DefMacro{table-numbers_of_programs_and_bugs-20230418-row1-gen-timeout}{0}
\DefMacro{table-numbers_of_programs_and_bugs-20230418-row1-gen-no_hole}{0}
\DefMacro{table-numbers_of_programs_and_bugs-20230418-row1-gen-no_gen}{0}
\DefMacro{table-numbers_of_programs_and_bugs-20230418-row1-num_working_templates}{1,025}
\DefMacro{table-numbers_of_programs_and_bugs-20230418-row1-num_unique_entry_methods_out_of_working_templates}{535}
\DefMacro{table-numbers_of_programs_and_bugs-20230418-row1-num_unique_classes_out_of_working_templates}{134}
\DefMacro{table-numbers_of_programs_and_bugs-20230418-row1-exec-total}{1,025}
\DefMacro{table-numbers_of_programs_and_bugs-20230418-row1-exec-not_compile}{0}
\DefMacro{table-numbers_of_programs_and_bugs-20230418-row1-exec-timeout}{3}
\DefMacro{table-numbers_of_programs_and_bugs-20230418-row1-exec-ok}{1,004}
\DefMacro{table-numbers_of_programs_and_bugs-20230418-row1-exec-skipped_diff}{18}
\DefMacro{table-numbers_of_programs_and_bugs-20230418-row1-exec-skipped_crash}{0}
\DefMacro{table-numbers_of_programs_and_bugs-20230418-row1-exec-final_diff}{0}
\DefMacro{table-numbers_of_programs_and_bugs-20230418-row1-exec-final_crash}{0}
\DefMacro{table-numbers_of_programs_and_bugs-20230418-row2-project}{text}
\DefMacro{table-numbers_of_programs_and_bugs-20230418-row2-num_unit_tests}{1,641}
\DefMacro{table-numbers_of_programs_and_bugs-20230418-row2-num_unique_entry_methods}{505}
\DefMacro{table-numbers_of_programs_and_bugs-20230418-row2-num_unique_classes}{39}
\DefMacro{table-numbers_of_programs_and_bugs-20230418-row2-convert-fail}{26}
\DefMacro{table-numbers_of_programs_and_bugs-20230418-row2-convert-timeout}{0}
\DefMacro{table-numbers_of_programs_and_bugs-20230418-row2-convert-not_compile}{0}
\DefMacro{table-numbers_of_programs_and_bugs-20230418-row2-convert-no_original}{0}
\DefMacro{table-numbers_of_programs_and_bugs-20230418-row2-gen-fail}{0}
\DefMacro{table-numbers_of_programs_and_bugs-20230418-row2-gen-timeout}{0}
\DefMacro{table-numbers_of_programs_and_bugs-20230418-row2-gen-no_hole}{0}
\DefMacro{table-numbers_of_programs_and_bugs-20230418-row2-gen-no_gen}{0}
\DefMacro{table-numbers_of_programs_and_bugs-20230418-row2-num_working_templates}{1,615}
\DefMacro{table-numbers_of_programs_and_bugs-20230418-row2-num_unique_entry_methods_out_of_working_templates}{496}
\DefMacro{table-numbers_of_programs_and_bugs-20230418-row2-num_unique_classes_out_of_working_templates}{35}
\DefMacro{table-numbers_of_programs_and_bugs-20230418-row2-exec-total}{1,615}
\DefMacro{table-numbers_of_programs_and_bugs-20230418-row2-exec-not_compile}{0}
\DefMacro{table-numbers_of_programs_and_bugs-20230418-row2-exec-timeout}{191}
\DefMacro{table-numbers_of_programs_and_bugs-20230418-row2-exec-ok}{1,379}
\DefMacro{table-numbers_of_programs_and_bugs-20230418-row2-exec-skipped_diff}{45}
\DefMacro{table-numbers_of_programs_and_bugs-20230418-row2-exec-skipped_crash}{0}
\DefMacro{table-numbers_of_programs_and_bugs-20230418-row2-exec-final_diff}{0}
\DefMacro{table-numbers_of_programs_and_bugs-20230418-row2-exec-final_crash}{0}
\DefMacro{table-numbers_of_programs_and_bugs-20230418-row3-project}{numbers}
\DefMacro{table-numbers_of_programs_and_bugs-20230418-row3-num_unit_tests}{3,063}
\DefMacro{table-numbers_of_programs_and_bugs-20230418-row3-num_unique_entry_methods}{45}
\DefMacro{table-numbers_of_programs_and_bugs-20230418-row3-num_unique_classes}{4}
\DefMacro{table-numbers_of_programs_and_bugs-20230418-row3-convert-fail}{110}
\DefMacro{table-numbers_of_programs_and_bugs-20230418-row3-convert-timeout}{0}
\DefMacro{table-numbers_of_programs_and_bugs-20230418-row3-convert-not_compile}{0}
\DefMacro{table-numbers_of_programs_and_bugs-20230418-row3-convert-no_original}{0}
\DefMacro{table-numbers_of_programs_and_bugs-20230418-row3-gen-fail}{0}
\DefMacro{table-numbers_of_programs_and_bugs-20230418-row3-gen-timeout}{0}
\DefMacro{table-numbers_of_programs_and_bugs-20230418-row3-gen-no_hole}{0}
\DefMacro{table-numbers_of_programs_and_bugs-20230418-row3-gen-no_gen}{0}
\DefMacro{table-numbers_of_programs_and_bugs-20230418-row3-num_working_templates}{2,953}
\DefMacro{table-numbers_of_programs_and_bugs-20230418-row3-num_unique_entry_methods_out_of_working_templates}{42}
\DefMacro{table-numbers_of_programs_and_bugs-20230418-row3-num_unique_classes_out_of_working_templates}{3}
\DefMacro{table-numbers_of_programs_and_bugs-20230418-row3-exec-total}{2,953}
\DefMacro{table-numbers_of_programs_and_bugs-20230418-row3-exec-not_compile}{0}
\DefMacro{table-numbers_of_programs_and_bugs-20230418-row3-exec-timeout}{8}
\DefMacro{table-numbers_of_programs_and_bugs-20230418-row3-exec-ok}{2,945}
\DefMacro{table-numbers_of_programs_and_bugs-20230418-row3-exec-skipped_diff}{0}
\DefMacro{table-numbers_of_programs_and_bugs-20230418-row3-exec-skipped_crash}{0}
\DefMacro{table-numbers_of_programs_and_bugs-20230418-row3-exec-final_diff}{0}
\DefMacro{table-numbers_of_programs_and_bugs-20230418-row3-exec-final_crash}{0}
\DefMacro{table-numbers_of_programs_and_bugs-20230418-row4-project}{vectorz}
\DefMacro{table-numbers_of_programs_and_bugs-20230418-row4-num_unit_tests}{1,144}
\DefMacro{table-numbers_of_programs_and_bugs-20230418-row4-num_unique_entry_methods}{893}
\DefMacro{table-numbers_of_programs_and_bugs-20230418-row4-num_unique_classes}{153}
\DefMacro{table-numbers_of_programs_and_bugs-20230418-row4-convert-fail}{0}
\DefMacro{table-numbers_of_programs_and_bugs-20230418-row4-convert-timeout}{0}
\DefMacro{table-numbers_of_programs_and_bugs-20230418-row4-convert-not_compile}{0}
\DefMacro{table-numbers_of_programs_and_bugs-20230418-row4-convert-no_original}{0}
\DefMacro{table-numbers_of_programs_and_bugs-20230418-row4-gen-fail}{0}
\DefMacro{table-numbers_of_programs_and_bugs-20230418-row4-gen-timeout}{0}
\DefMacro{table-numbers_of_programs_and_bugs-20230418-row4-gen-no_hole}{0}
\DefMacro{table-numbers_of_programs_and_bugs-20230418-row4-gen-no_gen}{0}
\DefMacro{table-numbers_of_programs_and_bugs-20230418-row4-num_working_templates}{1,144}
\DefMacro{table-numbers_of_programs_and_bugs-20230418-row4-num_unique_entry_methods_out_of_working_templates}{893}
\DefMacro{table-numbers_of_programs_and_bugs-20230418-row4-num_unique_classes_out_of_working_templates}{153}
\DefMacro{table-numbers_of_programs_and_bugs-20230418-row4-exec-total}{1,144}
\DefMacro{table-numbers_of_programs_and_bugs-20230418-row4-exec-not_compile}{0}
\DefMacro{table-numbers_of_programs_and_bugs-20230418-row4-exec-timeout}{5}
\DefMacro{table-numbers_of_programs_and_bugs-20230418-row4-exec-ok}{1,076}
\DefMacro{table-numbers_of_programs_and_bugs-20230418-row4-exec-skipped_diff}{63}
\DefMacro{table-numbers_of_programs_and_bugs-20230418-row4-exec-skipped_crash}{0}
\DefMacro{table-numbers_of_programs_and_bugs-20230418-row4-exec-final_diff}{0}
\DefMacro{table-numbers_of_programs_and_bugs-20230418-row4-exec-final_crash}{0}
\DefMacro{table-numbers_of_programs_and_bugs-20230418-row5-project}{statistics}
\DefMacro{table-numbers_of_programs_and_bugs-20230418-row5-num_unit_tests}{1,881}
\DefMacro{table-numbers_of_programs_and_bugs-20230418-row5-num_unique_entry_methods}{364}
\DefMacro{table-numbers_of_programs_and_bugs-20230418-row5-num_unique_classes}{26}
\DefMacro{table-numbers_of_programs_and_bugs-20230418-row5-convert-fail}{159}
\DefMacro{table-numbers_of_programs_and_bugs-20230418-row5-convert-timeout}{0}
\DefMacro{table-numbers_of_programs_and_bugs-20230418-row5-convert-not_compile}{0}
\DefMacro{table-numbers_of_programs_and_bugs-20230418-row5-convert-no_original}{0}
\DefMacro{table-numbers_of_programs_and_bugs-20230418-row5-gen-fail}{0}
\DefMacro{table-numbers_of_programs_and_bugs-20230418-row5-gen-timeout}{0}
\DefMacro{table-numbers_of_programs_and_bugs-20230418-row5-gen-no_hole}{0}
\DefMacro{table-numbers_of_programs_and_bugs-20230418-row5-gen-no_gen}{0}
\DefMacro{table-numbers_of_programs_and_bugs-20230418-row5-num_working_templates}{1,722}
\DefMacro{table-numbers_of_programs_and_bugs-20230418-row5-num_unique_entry_methods_out_of_working_templates}{320}
\DefMacro{table-numbers_of_programs_and_bugs-20230418-row5-num_unique_classes_out_of_working_templates}{26}
\DefMacro{table-numbers_of_programs_and_bugs-20230418-row5-exec-total}{1,722}
\DefMacro{table-numbers_of_programs_and_bugs-20230418-row5-exec-not_compile}{0}
\DefMacro{table-numbers_of_programs_and_bugs-20230418-row5-exec-timeout}{0}
\DefMacro{table-numbers_of_programs_and_bugs-20230418-row5-exec-ok}{1,634}
\DefMacro{table-numbers_of_programs_and_bugs-20230418-row5-exec-skipped_diff}{88}
\DefMacro{table-numbers_of_programs_and_bugs-20230418-row5-exec-skipped_crash}{0}
\DefMacro{table-numbers_of_programs_and_bugs-20230418-row5-exec-final_diff}{0}
\DefMacro{table-numbers_of_programs_and_bugs-20230418-row5-exec-final_crash}{0}
\DefMacro{table-numbers_of_programs_and_bugs-20230418-row6-project}{codec}
\DefMacro{table-numbers_of_programs_and_bugs-20230418-row6-num_unit_tests}{1,421}
\DefMacro{table-numbers_of_programs_and_bugs-20230418-row6-num_unique_entry_methods}{367}
\DefMacro{table-numbers_of_programs_and_bugs-20230418-row6-num_unique_classes}{49}
\DefMacro{table-numbers_of_programs_and_bugs-20230418-row6-convert-fail}{9}
\DefMacro{table-numbers_of_programs_and_bugs-20230418-row6-convert-timeout}{0}
\DefMacro{table-numbers_of_programs_and_bugs-20230418-row6-convert-not_compile}{0}
\DefMacro{table-numbers_of_programs_and_bugs-20230418-row6-convert-no_original}{0}
\DefMacro{table-numbers_of_programs_and_bugs-20230418-row6-gen-fail}{0}
\DefMacro{table-numbers_of_programs_and_bugs-20230418-row6-gen-timeout}{0}
\DefMacro{table-numbers_of_programs_and_bugs-20230418-row6-gen-no_hole}{0}
\DefMacro{table-numbers_of_programs_and_bugs-20230418-row6-gen-no_gen}{0}
\DefMacro{table-numbers_of_programs_and_bugs-20230418-row6-num_working_templates}{1,412}
\DefMacro{table-numbers_of_programs_and_bugs-20230418-row6-num_unique_entry_methods_out_of_working_templates}{364}
\DefMacro{table-numbers_of_programs_and_bugs-20230418-row6-num_unique_classes_out_of_working_templates}{46}
\DefMacro{table-numbers_of_programs_and_bugs-20230418-row6-exec-total}{1,412}
\DefMacro{table-numbers_of_programs_and_bugs-20230418-row6-exec-not_compile}{0}
\DefMacro{table-numbers_of_programs_and_bugs-20230418-row6-exec-timeout}{33}
\DefMacro{table-numbers_of_programs_and_bugs-20230418-row6-exec-ok}{1,309}
\DefMacro{table-numbers_of_programs_and_bugs-20230418-row6-exec-skipped_diff}{66}
\DefMacro{table-numbers_of_programs_and_bugs-20230418-row6-exec-skipped_crash}{0}
\DefMacro{table-numbers_of_programs_and_bugs-20230418-row6-exec-final_diff}{4}
\DefMacro{table-numbers_of_programs_and_bugs-20230418-row6-exec-final_crash}{0}
\DefMacro{table-numbers_of_programs_and_bugs-20230418-row7-project}{jfreechart}
\DefMacro{table-numbers_of_programs_and_bugs-20230418-row7-num_unit_tests}{1,240}
\DefMacro{table-numbers_of_programs_and_bugs-20230418-row7-num_unique_entry_methods}{1,047}
\DefMacro{table-numbers_of_programs_and_bugs-20230418-row7-num_unique_classes}{265}
\DefMacro{table-numbers_of_programs_and_bugs-20230418-row7-convert-fail}{3}
\DefMacro{table-numbers_of_programs_and_bugs-20230418-row7-convert-timeout}{0}
\DefMacro{table-numbers_of_programs_and_bugs-20230418-row7-convert-not_compile}{0}
\DefMacro{table-numbers_of_programs_and_bugs-20230418-row7-convert-no_original}{0}
\DefMacro{table-numbers_of_programs_and_bugs-20230418-row7-gen-fail}{0}
\DefMacro{table-numbers_of_programs_and_bugs-20230418-row7-gen-timeout}{0}
\DefMacro{table-numbers_of_programs_and_bugs-20230418-row7-gen-no_hole}{0}
\DefMacro{table-numbers_of_programs_and_bugs-20230418-row7-gen-no_gen}{0}
\DefMacro{table-numbers_of_programs_and_bugs-20230418-row7-num_working_templates}{1,237}
\DefMacro{table-numbers_of_programs_and_bugs-20230418-row7-num_unique_entry_methods_out_of_working_templates}{1,044}
\DefMacro{table-numbers_of_programs_and_bugs-20230418-row7-num_unique_classes_out_of_working_templates}{262}
\DefMacro{table-numbers_of_programs_and_bugs-20230418-row7-exec-total}{1,237}
\DefMacro{table-numbers_of_programs_and_bugs-20230418-row7-exec-not_compile}{0}
\DefMacro{table-numbers_of_programs_and_bugs-20230418-row7-exec-timeout}{26}
\DefMacro{table-numbers_of_programs_and_bugs-20230418-row7-exec-ok}{1,187}
\DefMacro{table-numbers_of_programs_and_bugs-20230418-row7-exec-skipped_diff}{24}
\DefMacro{table-numbers_of_programs_and_bugs-20230418-row7-exec-skipped_crash}{0}
\DefMacro{table-numbers_of_programs_and_bugs-20230418-row7-exec-final_diff}{0}
\DefMacro{table-numbers_of_programs_and_bugs-20230418-row7-exec-final_crash}{0}
\DefMacro{table-numbers_of_programs_and_bugs-20230418-row8-project}{zxing}
\DefMacro{table-numbers_of_programs_and_bugs-20230418-row8-num_unit_tests}{1,393}
\DefMacro{table-numbers_of_programs_and_bugs-20230418-row8-num_unique_entry_methods}{323}
\DefMacro{table-numbers_of_programs_and_bugs-20230418-row8-num_unique_classes}{90}
\DefMacro{table-numbers_of_programs_and_bugs-20230418-row8-convert-fail}{49}
\DefMacro{table-numbers_of_programs_and_bugs-20230418-row8-convert-timeout}{0}
\DefMacro{table-numbers_of_programs_and_bugs-20230418-row8-convert-not_compile}{0}
\DefMacro{table-numbers_of_programs_and_bugs-20230418-row8-convert-no_original}{0}
\DefMacro{table-numbers_of_programs_and_bugs-20230418-row8-gen-fail}{0}
\DefMacro{table-numbers_of_programs_and_bugs-20230418-row8-gen-timeout}{0}
\DefMacro{table-numbers_of_programs_and_bugs-20230418-row8-gen-no_hole}{0}
\DefMacro{table-numbers_of_programs_and_bugs-20230418-row8-gen-no_gen}{0}
\DefMacro{table-numbers_of_programs_and_bugs-20230418-row8-num_working_templates}{1,344}
\DefMacro{table-numbers_of_programs_and_bugs-20230418-row8-num_unique_entry_methods_out_of_working_templates}{312}
\DefMacro{table-numbers_of_programs_and_bugs-20230418-row8-num_unique_classes_out_of_working_templates}{86}
\DefMacro{table-numbers_of_programs_and_bugs-20230418-row8-exec-total}{1,344}
\DefMacro{table-numbers_of_programs_and_bugs-20230418-row8-exec-not_compile}{0}
\DefMacro{table-numbers_of_programs_and_bugs-20230418-row8-exec-timeout}{2}
\DefMacro{table-numbers_of_programs_and_bugs-20230418-row8-exec-ok}{1,318}
\DefMacro{table-numbers_of_programs_and_bugs-20230418-row8-exec-skipped_diff}{14}
\DefMacro{table-numbers_of_programs_and_bugs-20230418-row8-exec-skipped_crash}{0}
\DefMacro{table-numbers_of_programs_and_bugs-20230418-row8-exec-final_diff}{0}
\DefMacro{table-numbers_of_programs_and_bugs-20230418-row8-exec-final_crash}{10}
\DefMacro{table-numbers_of_programs_and_bugs-20230418-row9-project}{$\sum$}
\DefMacro{table-numbers_of_programs_and_bugs-20230418-row9-num_unit_tests}{14,722}
\DefMacro{table-numbers_of_programs_and_bugs-20230418-row9-num_unique_entry_methods}{5,237}
\DefMacro{table-numbers_of_programs_and_bugs-20230418-row9-num_unique_classes}{858}
\DefMacro{table-numbers_of_programs_and_bugs-20230418-row9-convert-fail}{457}
\DefMacro{table-numbers_of_programs_and_bugs-20230418-row9-convert-timeout}{0}
\DefMacro{table-numbers_of_programs_and_bugs-20230418-row9-convert-not_compile}{0}
\DefMacro{table-numbers_of_programs_and_bugs-20230418-row9-convert-no_original}{0}
\DefMacro{table-numbers_of_programs_and_bugs-20230418-row9-gen-fail}{0}
\DefMacro{table-numbers_of_programs_and_bugs-20230418-row9-gen-timeout}{0}
\DefMacro{table-numbers_of_programs_and_bugs-20230418-row9-gen-no_hole}{0}
\DefMacro{table-numbers_of_programs_and_bugs-20230418-row9-gen-no_gen}{0}
\DefMacro{table-numbers_of_programs_and_bugs-20230418-row9-num_working_templates}{14,265}
\DefMacro{table-numbers_of_programs_and_bugs-20230418-row9-num_unique_entry_methods_out_of_working_templates}{5,094}
\DefMacro{table-numbers_of_programs_and_bugs-20230418-row9-num_unique_classes_out_of_working_templates}{837}
\DefMacro{table-numbers_of_programs_and_bugs-20230418-row9-exec-total}{14,265}
\DefMacro{table-numbers_of_programs_and_bugs-20230418-row9-exec-not_compile}{0}
\DefMacro{table-numbers_of_programs_and_bugs-20230418-row9-exec-timeout}{290}
\DefMacro{table-numbers_of_programs_and_bugs-20230418-row9-exec-ok}{13,558}
\DefMacro{table-numbers_of_programs_and_bugs-20230418-row9-exec-skipped_diff}{399}
\DefMacro{table-numbers_of_programs_and_bugs-20230418-row9-exec-skipped_crash}{1}
\DefMacro{table-numbers_of_programs_and_bugs-20230418-row9-exec-final_diff}{7}
\DefMacro{table-numbers_of_programs_and_bugs-20230418-row9-exec-final_crash}{10}

\DefMacro{table-numbers_of_programs_and_bugs-20230417-row0-project}{lang}
\DefMacro{table-numbers_of_programs_and_bugs-20230417-row0-num_unit_tests}{1,822}
\DefMacro{table-numbers_of_programs_and_bugs-20230417-row0-num_unique_entry_methods}{1,087}
\DefMacro{table-numbers_of_programs_and_bugs-20230417-row0-num_unique_classes}{97}
\DefMacro{table-numbers_of_programs_and_bugs-20230417-row0-convert-fail}{90}
\DefMacro{table-numbers_of_programs_and_bugs-20230417-row0-convert-timeout}{0}
\DefMacro{table-numbers_of_programs_and_bugs-20230417-row0-convert-not_compile}{0}
\DefMacro{table-numbers_of_programs_and_bugs-20230417-row0-convert-no_original}{0}
\DefMacro{table-numbers_of_programs_and_bugs-20230417-row0-gen-fail}{0}
\DefMacro{table-numbers_of_programs_and_bugs-20230417-row0-gen-timeout}{0}
\DefMacro{table-numbers_of_programs_and_bugs-20230417-row0-gen-no_hole}{0}
\DefMacro{table-numbers_of_programs_and_bugs-20230417-row0-gen-no_gen}{0}
\DefMacro{table-numbers_of_programs_and_bugs-20230417-row0-num_working_templates}{1,732}
\DefMacro{table-numbers_of_programs_and_bugs-20230417-row0-num_unique_entry_methods_out_of_working_templates}{1,025}
\DefMacro{table-numbers_of_programs_and_bugs-20230417-row0-num_unique_classes_out_of_working_templates}{93}
\DefMacro{table-numbers_of_programs_and_bugs-20230417-row0-exec-total}{1,732}
\DefMacro{table-numbers_of_programs_and_bugs-20230417-row0-exec-not_compile}{0}
\DefMacro{table-numbers_of_programs_and_bugs-20230417-row0-exec-timeout}{29}
\DefMacro{table-numbers_of_programs_and_bugs-20230417-row0-exec-ok}{1,601}
\DefMacro{table-numbers_of_programs_and_bugs-20230417-row0-exec-skipped_diff}{99}
\DefMacro{table-numbers_of_programs_and_bugs-20230417-row0-exec-skipped_crash}{0}
\DefMacro{table-numbers_of_programs_and_bugs-20230417-row0-exec-final_diff}{3}
\DefMacro{table-numbers_of_programs_and_bugs-20230417-row0-exec-final_crash}{0}
\DefMacro{table-numbers_of_programs_and_bugs-20230417-row1-project}{math}
\DefMacro{table-numbers_of_programs_and_bugs-20230417-row1-num_unit_tests}{772}
\DefMacro{table-numbers_of_programs_and_bugs-20230417-row1-num_unique_entry_methods}{629}
\DefMacro{table-numbers_of_programs_and_bugs-20230417-row1-num_unique_classes}{196}
\DefMacro{table-numbers_of_programs_and_bugs-20230417-row1-convert-fail}{21}
\DefMacro{table-numbers_of_programs_and_bugs-20230417-row1-convert-timeout}{0}
\DefMacro{table-numbers_of_programs_and_bugs-20230417-row1-convert-not_compile}{0}
\DefMacro{table-numbers_of_programs_and_bugs-20230417-row1-convert-no_original}{0}
\DefMacro{table-numbers_of_programs_and_bugs-20230417-row1-gen-fail}{0}
\DefMacro{table-numbers_of_programs_and_bugs-20230417-row1-gen-timeout}{0}
\DefMacro{table-numbers_of_programs_and_bugs-20230417-row1-gen-no_hole}{0}
\DefMacro{table-numbers_of_programs_and_bugs-20230417-row1-gen-no_gen}{0}
\DefMacro{table-numbers_of_programs_and_bugs-20230417-row1-num_working_templates}{751}
\DefMacro{table-numbers_of_programs_and_bugs-20230417-row1-num_unique_entry_methods_out_of_working_templates}{612}
\DefMacro{table-numbers_of_programs_and_bugs-20230417-row1-num_unique_classes_out_of_working_templates}{193}
\DefMacro{table-numbers_of_programs_and_bugs-20230417-row1-exec-total}{751}
\DefMacro{table-numbers_of_programs_and_bugs-20230417-row1-exec-not_compile}{0}
\DefMacro{table-numbers_of_programs_and_bugs-20230417-row1-exec-timeout}{1}
\DefMacro{table-numbers_of_programs_and_bugs-20230417-row1-exec-ok}{738}
\DefMacro{table-numbers_of_programs_and_bugs-20230417-row1-exec-skipped_diff}{12}
\DefMacro{table-numbers_of_programs_and_bugs-20230417-row1-exec-skipped_crash}{0}
\DefMacro{table-numbers_of_programs_and_bugs-20230417-row1-exec-final_diff}{0}
\DefMacro{table-numbers_of_programs_and_bugs-20230417-row1-exec-final_crash}{0}
\DefMacro{table-numbers_of_programs_and_bugs-20230417-row2-project}{text}
\DefMacro{table-numbers_of_programs_and_bugs-20230417-row2-num_unit_tests}{1,604}
\DefMacro{table-numbers_of_programs_and_bugs-20230417-row2-num_unique_entry_methods}{502}
\DefMacro{table-numbers_of_programs_and_bugs-20230417-row2-num_unique_classes}{37}
\DefMacro{table-numbers_of_programs_and_bugs-20230417-row2-convert-fail}{19}
\DefMacro{table-numbers_of_programs_and_bugs-20230417-row2-convert-timeout}{0}
\DefMacro{table-numbers_of_programs_and_bugs-20230417-row2-convert-not_compile}{0}
\DefMacro{table-numbers_of_programs_and_bugs-20230417-row2-convert-no_original}{0}
\DefMacro{table-numbers_of_programs_and_bugs-20230417-row2-gen-fail}{0}
\DefMacro{table-numbers_of_programs_and_bugs-20230417-row2-gen-timeout}{0}
\DefMacro{table-numbers_of_programs_and_bugs-20230417-row2-gen-no_hole}{0}
\DefMacro{table-numbers_of_programs_and_bugs-20230417-row2-gen-no_gen}{0}
\DefMacro{table-numbers_of_programs_and_bugs-20230417-row2-num_working_templates}{1,585}
\DefMacro{table-numbers_of_programs_and_bugs-20230417-row2-num_unique_entry_methods_out_of_working_templates}{497}
\DefMacro{table-numbers_of_programs_and_bugs-20230417-row2-num_unique_classes_out_of_working_templates}{33}
\DefMacro{table-numbers_of_programs_and_bugs-20230417-row2-exec-total}{1,585}
\DefMacro{table-numbers_of_programs_and_bugs-20230417-row2-exec-not_compile}{0}
\DefMacro{table-numbers_of_programs_and_bugs-20230417-row2-exec-timeout}{185}
\DefMacro{table-numbers_of_programs_and_bugs-20230417-row2-exec-ok}{1,383}
\DefMacro{table-numbers_of_programs_and_bugs-20230417-row2-exec-skipped_diff}{17}
\DefMacro{table-numbers_of_programs_and_bugs-20230417-row2-exec-skipped_crash}{0}
\DefMacro{table-numbers_of_programs_and_bugs-20230417-row2-exec-final_diff}{0}
\DefMacro{table-numbers_of_programs_and_bugs-20230417-row2-exec-final_crash}{0}
\DefMacro{table-numbers_of_programs_and_bugs-20230417-row3-project}{numbers}
\DefMacro{table-numbers_of_programs_and_bugs-20230417-row3-num_unit_tests}{3,072}
\DefMacro{table-numbers_of_programs_and_bugs-20230417-row3-num_unique_entry_methods}{45}
\DefMacro{table-numbers_of_programs_and_bugs-20230417-row3-num_unique_classes}{4}
\DefMacro{table-numbers_of_programs_and_bugs-20230417-row3-convert-fail}{109}
\DefMacro{table-numbers_of_programs_and_bugs-20230417-row3-convert-timeout}{0}
\DefMacro{table-numbers_of_programs_and_bugs-20230417-row3-convert-not_compile}{0}
\DefMacro{table-numbers_of_programs_and_bugs-20230417-row3-convert-no_original}{0}
\DefMacro{table-numbers_of_programs_and_bugs-20230417-row3-gen-fail}{0}
\DefMacro{table-numbers_of_programs_and_bugs-20230417-row3-gen-timeout}{0}
\DefMacro{table-numbers_of_programs_and_bugs-20230417-row3-gen-no_hole}{0}
\DefMacro{table-numbers_of_programs_and_bugs-20230417-row3-gen-no_gen}{0}
\DefMacro{table-numbers_of_programs_and_bugs-20230417-row3-num_working_templates}{2,963}
\DefMacro{table-numbers_of_programs_and_bugs-20230417-row3-num_unique_entry_methods_out_of_working_templates}{42}
\DefMacro{table-numbers_of_programs_and_bugs-20230417-row3-num_unique_classes_out_of_working_templates}{3}
\DefMacro{table-numbers_of_programs_and_bugs-20230417-row3-exec-total}{2,963}
\DefMacro{table-numbers_of_programs_and_bugs-20230417-row3-exec-not_compile}{0}
\DefMacro{table-numbers_of_programs_and_bugs-20230417-row3-exec-timeout}{8}
\DefMacro{table-numbers_of_programs_and_bugs-20230417-row3-exec-ok}{2,955}
\DefMacro{table-numbers_of_programs_and_bugs-20230417-row3-exec-skipped_diff}{0}
\DefMacro{table-numbers_of_programs_and_bugs-20230417-row3-exec-skipped_crash}{0}
\DefMacro{table-numbers_of_programs_and_bugs-20230417-row3-exec-final_diff}{0}
\DefMacro{table-numbers_of_programs_and_bugs-20230417-row3-exec-final_crash}{0}
\DefMacro{table-numbers_of_programs_and_bugs-20230417-row4-project}{vectorz}
\DefMacro{table-numbers_of_programs_and_bugs-20230417-row4-num_unit_tests}{1,152}
\DefMacro{table-numbers_of_programs_and_bugs-20230417-row4-num_unique_entry_methods}{914}
\DefMacro{table-numbers_of_programs_and_bugs-20230417-row4-num_unique_classes}{163}
\DefMacro{table-numbers_of_programs_and_bugs-20230417-row4-convert-fail}{0}
\DefMacro{table-numbers_of_programs_and_bugs-20230417-row4-convert-timeout}{0}
\DefMacro{table-numbers_of_programs_and_bugs-20230417-row4-convert-not_compile}{0}
\DefMacro{table-numbers_of_programs_and_bugs-20230417-row4-convert-no_original}{0}
\DefMacro{table-numbers_of_programs_and_bugs-20230417-row4-gen-fail}{0}
\DefMacro{table-numbers_of_programs_and_bugs-20230417-row4-gen-timeout}{0}
\DefMacro{table-numbers_of_programs_and_bugs-20230417-row4-gen-no_hole}{0}
\DefMacro{table-numbers_of_programs_and_bugs-20230417-row4-gen-no_gen}{0}
\DefMacro{table-numbers_of_programs_and_bugs-20230417-row4-num_working_templates}{1,152}
\DefMacro{table-numbers_of_programs_and_bugs-20230417-row4-num_unique_entry_methods_out_of_working_templates}{914}
\DefMacro{table-numbers_of_programs_and_bugs-20230417-row4-num_unique_classes_out_of_working_templates}{163}
\DefMacro{table-numbers_of_programs_and_bugs-20230417-row4-exec-total}{1,152}
\DefMacro{table-numbers_of_programs_and_bugs-20230417-row4-exec-not_compile}{0}
\DefMacro{table-numbers_of_programs_and_bugs-20230417-row4-exec-timeout}{7}
\DefMacro{table-numbers_of_programs_and_bugs-20230417-row4-exec-ok}{1,139}
\DefMacro{table-numbers_of_programs_and_bugs-20230417-row4-exec-skipped_diff}{6}
\DefMacro{table-numbers_of_programs_and_bugs-20230417-row4-exec-skipped_crash}{0}
\DefMacro{table-numbers_of_programs_and_bugs-20230417-row4-exec-final_diff}{0}
\DefMacro{table-numbers_of_programs_and_bugs-20230417-row4-exec-final_crash}{0}
\DefMacro{table-numbers_of_programs_and_bugs-20230417-row5-project}{statistics}
\DefMacro{table-numbers_of_programs_and_bugs-20230417-row5-num_unit_tests}{1,890}
\DefMacro{table-numbers_of_programs_and_bugs-20230417-row5-num_unique_entry_methods}{358}
\DefMacro{table-numbers_of_programs_and_bugs-20230417-row5-num_unique_classes}{26}
\DefMacro{table-numbers_of_programs_and_bugs-20230417-row5-convert-fail}{154}
\DefMacro{table-numbers_of_programs_and_bugs-20230417-row5-convert-timeout}{0}
\DefMacro{table-numbers_of_programs_and_bugs-20230417-row5-convert-not_compile}{0}
\DefMacro{table-numbers_of_programs_and_bugs-20230417-row5-convert-no_original}{0}
\DefMacro{table-numbers_of_programs_and_bugs-20230417-row5-gen-fail}{0}
\DefMacro{table-numbers_of_programs_and_bugs-20230417-row5-gen-timeout}{0}
\DefMacro{table-numbers_of_programs_and_bugs-20230417-row5-gen-no_hole}{0}
\DefMacro{table-numbers_of_programs_and_bugs-20230417-row5-gen-no_gen}{0}
\DefMacro{table-numbers_of_programs_and_bugs-20230417-row5-num_working_templates}{1,736}
\DefMacro{table-numbers_of_programs_and_bugs-20230417-row5-num_unique_entry_methods_out_of_working_templates}{315}
\DefMacro{table-numbers_of_programs_and_bugs-20230417-row5-num_unique_classes_out_of_working_templates}{26}
\DefMacro{table-numbers_of_programs_and_bugs-20230417-row5-exec-total}{1,736}
\DefMacro{table-numbers_of_programs_and_bugs-20230417-row5-exec-not_compile}{0}
\DefMacro{table-numbers_of_programs_and_bugs-20230417-row5-exec-timeout}{0}
\DefMacro{table-numbers_of_programs_and_bugs-20230417-row5-exec-ok}{1,689}
\DefMacro{table-numbers_of_programs_and_bugs-20230417-row5-exec-skipped_diff}{47}
\DefMacro{table-numbers_of_programs_and_bugs-20230417-row5-exec-skipped_crash}{0}
\DefMacro{table-numbers_of_programs_and_bugs-20230417-row5-exec-final_diff}{0}
\DefMacro{table-numbers_of_programs_and_bugs-20230417-row5-exec-final_crash}{0}
\DefMacro{table-numbers_of_programs_and_bugs-20230417-row6-project}{codec}
\DefMacro{table-numbers_of_programs_and_bugs-20230417-row6-num_unit_tests}{1,382}
\DefMacro{table-numbers_of_programs_and_bugs-20230417-row6-num_unique_entry_methods}{369}
\DefMacro{table-numbers_of_programs_and_bugs-20230417-row6-num_unique_classes}{49}
\DefMacro{table-numbers_of_programs_and_bugs-20230417-row6-convert-fail}{25}
\DefMacro{table-numbers_of_programs_and_bugs-20230417-row6-convert-timeout}{0}
\DefMacro{table-numbers_of_programs_and_bugs-20230417-row6-convert-not_compile}{0}
\DefMacro{table-numbers_of_programs_and_bugs-20230417-row6-convert-no_original}{0}
\DefMacro{table-numbers_of_programs_and_bugs-20230417-row6-gen-fail}{0}
\DefMacro{table-numbers_of_programs_and_bugs-20230417-row6-gen-timeout}{0}
\DefMacro{table-numbers_of_programs_and_bugs-20230417-row6-gen-no_hole}{0}
\DefMacro{table-numbers_of_programs_and_bugs-20230417-row6-gen-no_gen}{0}
\DefMacro{table-numbers_of_programs_and_bugs-20230417-row6-num_working_templates}{1,357}
\DefMacro{table-numbers_of_programs_and_bugs-20230417-row6-num_unique_entry_methods_out_of_working_templates}{362}
\DefMacro{table-numbers_of_programs_and_bugs-20230417-row6-num_unique_classes_out_of_working_templates}{46}
\DefMacro{table-numbers_of_programs_and_bugs-20230417-row6-exec-total}{1,357}
\DefMacro{table-numbers_of_programs_and_bugs-20230417-row6-exec-not_compile}{0}
\DefMacro{table-numbers_of_programs_and_bugs-20230417-row6-exec-timeout}{26}
\DefMacro{table-numbers_of_programs_and_bugs-20230417-row6-exec-ok}{1,249}
\DefMacro{table-numbers_of_programs_and_bugs-20230417-row6-exec-skipped_diff}{77}
\DefMacro{table-numbers_of_programs_and_bugs-20230417-row6-exec-skipped_crash}{0}
\DefMacro{table-numbers_of_programs_and_bugs-20230417-row6-exec-final_diff}{5}
\DefMacro{table-numbers_of_programs_and_bugs-20230417-row6-exec-final_crash}{0}
\DefMacro{table-numbers_of_programs_and_bugs-20230417-row7-project}{jfreechart}
\DefMacro{table-numbers_of_programs_and_bugs-20230417-row7-num_unit_tests}{1,237}
\DefMacro{table-numbers_of_programs_and_bugs-20230417-row7-num_unique_entry_methods}{1,080}
\DefMacro{table-numbers_of_programs_and_bugs-20230417-row7-num_unique_classes}{281}
\DefMacro{table-numbers_of_programs_and_bugs-20230417-row7-convert-fail}{2}
\DefMacro{table-numbers_of_programs_and_bugs-20230417-row7-convert-timeout}{0}
\DefMacro{table-numbers_of_programs_and_bugs-20230417-row7-convert-not_compile}{0}
\DefMacro{table-numbers_of_programs_and_bugs-20230417-row7-convert-no_original}{0}
\DefMacro{table-numbers_of_programs_and_bugs-20230417-row7-gen-fail}{0}
\DefMacro{table-numbers_of_programs_and_bugs-20230417-row7-gen-timeout}{0}
\DefMacro{table-numbers_of_programs_and_bugs-20230417-row7-gen-no_hole}{0}
\DefMacro{table-numbers_of_programs_and_bugs-20230417-row7-gen-no_gen}{0}
\DefMacro{table-numbers_of_programs_and_bugs-20230417-row7-num_working_templates}{1,235}
\DefMacro{table-numbers_of_programs_and_bugs-20230417-row7-num_unique_entry_methods_out_of_working_templates}{1,078}
\DefMacro{table-numbers_of_programs_and_bugs-20230417-row7-num_unique_classes_out_of_working_templates}{279}
\DefMacro{table-numbers_of_programs_and_bugs-20230417-row7-exec-total}{1,235}
\DefMacro{table-numbers_of_programs_and_bugs-20230417-row7-exec-not_compile}{0}
\DefMacro{table-numbers_of_programs_and_bugs-20230417-row7-exec-timeout}{10}
\DefMacro{table-numbers_of_programs_and_bugs-20230417-row7-exec-ok}{1,205}
\DefMacro{table-numbers_of_programs_and_bugs-20230417-row7-exec-skipped_diff}{19}
\DefMacro{table-numbers_of_programs_and_bugs-20230417-row7-exec-skipped_crash}{0}
\DefMacro{table-numbers_of_programs_and_bugs-20230417-row7-exec-final_diff}{1}
\DefMacro{table-numbers_of_programs_and_bugs-20230417-row7-exec-final_crash}{0}
\DefMacro{table-numbers_of_programs_and_bugs-20230417-row8-project}{zxing}
\DefMacro{table-numbers_of_programs_and_bugs-20230417-row8-num_unit_tests}{1,481}
\DefMacro{table-numbers_of_programs_and_bugs-20230417-row8-num_unique_entry_methods}{353}
\DefMacro{table-numbers_of_programs_and_bugs-20230417-row8-num_unique_classes}{93}
\DefMacro{table-numbers_of_programs_and_bugs-20230417-row8-convert-fail}{57}
\DefMacro{table-numbers_of_programs_and_bugs-20230417-row8-convert-timeout}{0}
\DefMacro{table-numbers_of_programs_and_bugs-20230417-row8-convert-not_compile}{0}
\DefMacro{table-numbers_of_programs_and_bugs-20230417-row8-convert-no_original}{0}
\DefMacro{table-numbers_of_programs_and_bugs-20230417-row8-gen-fail}{0}
\DefMacro{table-numbers_of_programs_and_bugs-20230417-row8-gen-timeout}{0}
\DefMacro{table-numbers_of_programs_and_bugs-20230417-row8-gen-no_hole}{0}
\DefMacro{table-numbers_of_programs_and_bugs-20230417-row8-gen-no_gen}{0}
\DefMacro{table-numbers_of_programs_and_bugs-20230417-row8-num_working_templates}{1,424}
\DefMacro{table-numbers_of_programs_and_bugs-20230417-row8-num_unique_entry_methods_out_of_working_templates}{343}
\DefMacro{table-numbers_of_programs_and_bugs-20230417-row8-num_unique_classes_out_of_working_templates}{89}
\DefMacro{table-numbers_of_programs_and_bugs-20230417-row8-exec-total}{1,424}
\DefMacro{table-numbers_of_programs_and_bugs-20230417-row8-exec-not_compile}{77}
\DefMacro{table-numbers_of_programs_and_bugs-20230417-row8-exec-timeout}{5}
\DefMacro{table-numbers_of_programs_and_bugs-20230417-row8-exec-ok}{1,324}
\DefMacro{table-numbers_of_programs_and_bugs-20230417-row8-exec-skipped_diff}{0}
\DefMacro{table-numbers_of_programs_and_bugs-20230417-row8-exec-skipped_crash}{0}
\DefMacro{table-numbers_of_programs_and_bugs-20230417-row8-exec-final_diff}{0}
\DefMacro{table-numbers_of_programs_and_bugs-20230417-row8-exec-final_crash}{18}
\DefMacro{table-numbers_of_programs_and_bugs-20230417-row9-project}{$\sum$}
\DefMacro{table-numbers_of_programs_and_bugs-20230417-row9-num_unit_tests}{14,412}
\DefMacro{table-numbers_of_programs_and_bugs-20230417-row9-num_unique_entry_methods}{5,337}
\DefMacro{table-numbers_of_programs_and_bugs-20230417-row9-num_unique_classes}{946}
\DefMacro{table-numbers_of_programs_and_bugs-20230417-row9-convert-fail}{477}
\DefMacro{table-numbers_of_programs_and_bugs-20230417-row9-convert-timeout}{0}
\DefMacro{table-numbers_of_programs_and_bugs-20230417-row9-convert-not_compile}{0}
\DefMacro{table-numbers_of_programs_and_bugs-20230417-row9-convert-no_original}{0}
\DefMacro{table-numbers_of_programs_and_bugs-20230417-row9-gen-fail}{0}
\DefMacro{table-numbers_of_programs_and_bugs-20230417-row9-gen-timeout}{0}
\DefMacro{table-numbers_of_programs_and_bugs-20230417-row9-gen-no_hole}{0}
\DefMacro{table-numbers_of_programs_and_bugs-20230417-row9-gen-no_gen}{0}
\DefMacro{table-numbers_of_programs_and_bugs-20230417-row9-num_working_templates}{13,935}
\DefMacro{table-numbers_of_programs_and_bugs-20230417-row9-num_unique_entry_methods_out_of_working_templates}{5,188}
\DefMacro{table-numbers_of_programs_and_bugs-20230417-row9-num_unique_classes_out_of_working_templates}{925}
\DefMacro{table-numbers_of_programs_and_bugs-20230417-row9-exec-total}{13,935}
\DefMacro{table-numbers_of_programs_and_bugs-20230417-row9-exec-not_compile}{77}
\DefMacro{table-numbers_of_programs_and_bugs-20230417-row9-exec-timeout}{271}
\DefMacro{table-numbers_of_programs_and_bugs-20230417-row9-exec-ok}{13,283}
\DefMacro{table-numbers_of_programs_and_bugs-20230417-row9-exec-skipped_diff}{277}
\DefMacro{table-numbers_of_programs_and_bugs-20230417-row9-exec-skipped_crash}{0}
\DefMacro{table-numbers_of_programs_and_bugs-20230417-row9-exec-final_diff}{9}
\DefMacro{table-numbers_of_programs_and_bugs-20230417-row9-exec-final_crash}{18}

\DefMacro{table-numbers_of_programs_and_bugs-20231018-row0-project}{lang}
\DefMacro{table-numbers_of_programs_and_bugs-20231018-row0-num_unit_tests}{2,827}
\DefMacro{table-numbers_of_programs_and_bugs-20231018-row0-num_unique_entry_methods}{2,821}
\DefMacro{table-numbers_of_programs_and_bugs-20231018-row0-num_unique_classes}{131}
\DefMacro{table-numbers_of_programs_and_bugs-20231018-row0-convert-fail}{0}
\DefMacro{table-numbers_of_programs_and_bugs-20231018-row0-convert-timeout}{841}
\DefMacro{table-numbers_of_programs_and_bugs-20231018-row0-convert-not_compile}{147}
\DefMacro{table-numbers_of_programs_and_bugs-20231018-row0-convert-no_original}{0}
\DefMacro{table-numbers_of_programs_and_bugs-20231018-row0-gen-fail}{44}
\DefMacro{table-numbers_of_programs_and_bugs-20231018-row0-gen-timeout}{0}
\DefMacro{table-numbers_of_programs_and_bugs-20231018-row0-gen-no_hole}{525}
\DefMacro{table-numbers_of_programs_and_bugs-20231018-row0-gen-no_gen}{66}
\DefMacro{table-numbers_of_programs_and_bugs-20231018-row0-num_working_templates}{1,248}
\DefMacro{table-numbers_of_programs_and_bugs-20231018-row0-num_unique_entry_methods_out_of_working_templates}{1,248}
\DefMacro{table-numbers_of_programs_and_bugs-20231018-row0-num_unique_classes_out_of_working_templates}{83}
\DefMacro{table-numbers_of_programs_and_bugs-20231018-row0-exec-total}{10,916}
\DefMacro{table-numbers_of_programs_and_bugs-20231018-row0-exec-not_compile}{0}
\DefMacro{table-numbers_of_programs_and_bugs-20231018-row0-exec-timeout}{132}
\DefMacro{table-numbers_of_programs_and_bugs-20231018-row0-exec-ok}{10,696}
\DefMacro{table-numbers_of_programs_and_bugs-20231018-row0-exec-skipped_diff}{72}
\DefMacro{table-numbers_of_programs_and_bugs-20231018-row0-exec-skipped_crash}{8}
\DefMacro{table-numbers_of_programs_and_bugs-20231018-row0-exec-final_diff}{0}
\DefMacro{table-numbers_of_programs_and_bugs-20231018-row0-exec-final_crash}{8}
\DefMacro{table-numbers_of_programs_and_bugs-20231018-row1-project}{math}
\DefMacro{table-numbers_of_programs_and_bugs-20231018-row1-num_unit_tests}{5,296}
\DefMacro{table-numbers_of_programs_and_bugs-20231018-row1-num_unique_entry_methods}{5,296}
\DefMacro{table-numbers_of_programs_and_bugs-20231018-row1-num_unique_classes}{591}
\DefMacro{table-numbers_of_programs_and_bugs-20231018-row1-convert-fail}{8}
\DefMacro{table-numbers_of_programs_and_bugs-20231018-row1-convert-timeout}{207}
\DefMacro{table-numbers_of_programs_and_bugs-20231018-row1-convert-not_compile}{350}
\DefMacro{table-numbers_of_programs_and_bugs-20231018-row1-convert-no_original}{0}
\DefMacro{table-numbers_of_programs_and_bugs-20231018-row1-gen-fail}{262}
\DefMacro{table-numbers_of_programs_and_bugs-20231018-row1-gen-timeout}{135}
\DefMacro{table-numbers_of_programs_and_bugs-20231018-row1-gen-no_hole}{2,310}
\DefMacro{table-numbers_of_programs_and_bugs-20231018-row1-gen-no_gen}{392}
\DefMacro{table-numbers_of_programs_and_bugs-20231018-row1-num_working_templates}{2,029}
\DefMacro{table-numbers_of_programs_and_bugs-20231018-row1-num_unique_entry_methods_out_of_working_templates}{2,029}
\DefMacro{table-numbers_of_programs_and_bugs-20231018-row1-num_unique_classes_out_of_working_templates}{397}
\DefMacro{table-numbers_of_programs_and_bugs-20231018-row1-exec-total}{18,177}
\DefMacro{table-numbers_of_programs_and_bugs-20231018-row1-exec-not_compile}{0}
\DefMacro{table-numbers_of_programs_and_bugs-20231018-row1-exec-timeout}{248}
\DefMacro{table-numbers_of_programs_and_bugs-20231018-row1-exec-ok}{17,420}
\DefMacro{table-numbers_of_programs_and_bugs-20231018-row1-exec-skipped_diff}{477}
\DefMacro{table-numbers_of_programs_and_bugs-20231018-row1-exec-skipped_crash}{14}
\DefMacro{table-numbers_of_programs_and_bugs-20231018-row1-exec-final_diff}{0}
\DefMacro{table-numbers_of_programs_and_bugs-20231018-row1-exec-final_crash}{18}
\DefMacro{table-numbers_of_programs_and_bugs-20231018-row2-project}{text}
\DefMacro{table-numbers_of_programs_and_bugs-20231018-row2-num_unit_tests}{851}
\DefMacro{table-numbers_of_programs_and_bugs-20231018-row2-num_unique_entry_methods}{851}
\DefMacro{table-numbers_of_programs_and_bugs-20231018-row2-num_unique_classes}{77}
\DefMacro{table-numbers_of_programs_and_bugs-20231018-row2-convert-fail}{0}
\DefMacro{table-numbers_of_programs_and_bugs-20231018-row2-convert-timeout}{335}
\DefMacro{table-numbers_of_programs_and_bugs-20231018-row2-convert-not_compile}{8}
\DefMacro{table-numbers_of_programs_and_bugs-20231018-row2-convert-no_original}{0}
\DefMacro{table-numbers_of_programs_and_bugs-20231018-row2-gen-fail}{4}
\DefMacro{table-numbers_of_programs_and_bugs-20231018-row2-gen-timeout}{20}
\DefMacro{table-numbers_of_programs_and_bugs-20231018-row2-gen-no_hole}{117}
\DefMacro{table-numbers_of_programs_and_bugs-20231018-row2-gen-no_gen}{0}
\DefMacro{table-numbers_of_programs_and_bugs-20231018-row2-num_working_templates}{391}
\DefMacro{table-numbers_of_programs_and_bugs-20231018-row2-num_unique_entry_methods_out_of_working_templates}{391}
\DefMacro{table-numbers_of_programs_and_bugs-20231018-row2-num_unique_classes_out_of_working_templates}{56}
\DefMacro{table-numbers_of_programs_and_bugs-20231018-row2-exec-total}{3,572}
\DefMacro{table-numbers_of_programs_and_bugs-20231018-row2-exec-not_compile}{0}
\DefMacro{table-numbers_of_programs_and_bugs-20231018-row2-exec-timeout}{11}
\DefMacro{table-numbers_of_programs_and_bugs-20231018-row2-exec-ok}{3,543}
\DefMacro{table-numbers_of_programs_and_bugs-20231018-row2-exec-skipped_diff}{15}
\DefMacro{table-numbers_of_programs_and_bugs-20231018-row2-exec-skipped_crash}{2}
\DefMacro{table-numbers_of_programs_and_bugs-20231018-row2-exec-final_diff}{0}
\DefMacro{table-numbers_of_programs_and_bugs-20231018-row2-exec-final_crash}{1}
\DefMacro{table-numbers_of_programs_and_bugs-20231018-row3-project}{numbers}
\DefMacro{table-numbers_of_programs_and_bugs-20231018-row3-num_unit_tests}{51}
\DefMacro{table-numbers_of_programs_and_bugs-20231018-row3-num_unique_entry_methods}{51}
\DefMacro{table-numbers_of_programs_and_bugs-20231018-row3-num_unique_classes}{4}
\DefMacro{table-numbers_of_programs_and_bugs-20231018-row3-convert-fail}{0}
\DefMacro{table-numbers_of_programs_and_bugs-20231018-row3-convert-timeout}{0}
\DefMacro{table-numbers_of_programs_and_bugs-20231018-row3-convert-not_compile}{0}
\DefMacro{table-numbers_of_programs_and_bugs-20231018-row3-convert-no_original}{0}
\DefMacro{table-numbers_of_programs_and_bugs-20231018-row3-gen-fail}{0}
\DefMacro{table-numbers_of_programs_and_bugs-20231018-row3-gen-timeout}{0}
\DefMacro{table-numbers_of_programs_and_bugs-20231018-row3-gen-no_hole}{0}
\DefMacro{table-numbers_of_programs_and_bugs-20231018-row3-gen-no_gen}{0}
\DefMacro{table-numbers_of_programs_and_bugs-20231018-row3-num_working_templates}{51}
\DefMacro{table-numbers_of_programs_and_bugs-20231018-row3-num_unique_entry_methods_out_of_working_templates}{51}
\DefMacro{table-numbers_of_programs_and_bugs-20231018-row3-num_unique_classes_out_of_working_templates}{4}
\DefMacro{table-numbers_of_programs_and_bugs-20231018-row3-exec-total}{502}
\DefMacro{table-numbers_of_programs_and_bugs-20231018-row3-exec-not_compile}{0}
\DefMacro{table-numbers_of_programs_and_bugs-20231018-row3-exec-timeout}{0}
\DefMacro{table-numbers_of_programs_and_bugs-20231018-row3-exec-ok}{502}
\DefMacro{table-numbers_of_programs_and_bugs-20231018-row3-exec-skipped_diff}{0}
\DefMacro{table-numbers_of_programs_and_bugs-20231018-row3-exec-skipped_crash}{0}
\DefMacro{table-numbers_of_programs_and_bugs-20231018-row3-exec-final_diff}{0}
\DefMacro{table-numbers_of_programs_and_bugs-20231018-row3-exec-final_crash}{0}
\DefMacro{table-numbers_of_programs_and_bugs-20231018-row4-project}{vectorz}
\DefMacro{table-numbers_of_programs_and_bugs-20231018-row4-num_unit_tests}{5,582}
\DefMacro{table-numbers_of_programs_and_bugs-20231018-row4-num_unique_entry_methods}{5,582}
\DefMacro{table-numbers_of_programs_and_bugs-20231018-row4-num_unique_classes}{263}
\DefMacro{table-numbers_of_programs_and_bugs-20231018-row4-convert-fail}{0}
\DefMacro{table-numbers_of_programs_and_bugs-20231018-row4-convert-timeout}{1,166}
\DefMacro{table-numbers_of_programs_and_bugs-20231018-row4-convert-not_compile}{18}
\DefMacro{table-numbers_of_programs_and_bugs-20231018-row4-convert-no_original}{0}
\DefMacro{table-numbers_of_programs_and_bugs-20231018-row4-gen-fail}{360}
\DefMacro{table-numbers_of_programs_and_bugs-20231018-row4-gen-timeout}{304}
\DefMacro{table-numbers_of_programs_and_bugs-20231018-row4-gen-no_hole}{1,371}
\DefMacro{table-numbers_of_programs_and_bugs-20231018-row4-gen-no_gen}{695}
\DefMacro{table-numbers_of_programs_and_bugs-20231018-row4-num_working_templates}{2,332}
\DefMacro{table-numbers_of_programs_and_bugs-20231018-row4-num_unique_entry_methods_out_of_working_templates}{2,332}
\DefMacro{table-numbers_of_programs_and_bugs-20231018-row4-num_unique_classes_out_of_working_templates}{205}
\DefMacro{table-numbers_of_programs_and_bugs-20231018-row4-exec-total}{22,381}
\DefMacro{table-numbers_of_programs_and_bugs-20231018-row4-exec-not_compile}{0}
\DefMacro{table-numbers_of_programs_and_bugs-20231018-row4-exec-timeout}{245}
\DefMacro{table-numbers_of_programs_and_bugs-20231018-row4-exec-ok}{22,086}
\DefMacro{table-numbers_of_programs_and_bugs-20231018-row4-exec-skipped_diff}{38}
\DefMacro{table-numbers_of_programs_and_bugs-20231018-row4-exec-skipped_crash}{1}
\DefMacro{table-numbers_of_programs_and_bugs-20231018-row4-exec-final_diff}{0}
\DefMacro{table-numbers_of_programs_and_bugs-20231018-row4-exec-final_crash}{11}
\DefMacro{table-numbers_of_programs_and_bugs-20231018-row5-project}{statistics}
\DefMacro{table-numbers_of_programs_and_bugs-20231018-row5-num_unit_tests}{387}
\DefMacro{table-numbers_of_programs_and_bugs-20231018-row5-num_unique_entry_methods}{387}
\DefMacro{table-numbers_of_programs_and_bugs-20231018-row5-num_unique_classes}{32}
\DefMacro{table-numbers_of_programs_and_bugs-20231018-row5-convert-fail}{0}
\DefMacro{table-numbers_of_programs_and_bugs-20231018-row5-convert-timeout}{0}
\DefMacro{table-numbers_of_programs_and_bugs-20231018-row5-convert-not_compile}{0}
\DefMacro{table-numbers_of_programs_and_bugs-20231018-row5-convert-no_original}{0}
\DefMacro{table-numbers_of_programs_and_bugs-20231018-row5-gen-fail}{40}
\DefMacro{table-numbers_of_programs_and_bugs-20231018-row5-gen-timeout}{0}
\DefMacro{table-numbers_of_programs_and_bugs-20231018-row5-gen-no_hole}{5}
\DefMacro{table-numbers_of_programs_and_bugs-20231018-row5-gen-no_gen}{119}
\DefMacro{table-numbers_of_programs_and_bugs-20231018-row5-num_working_templates}{263}
\DefMacro{table-numbers_of_programs_and_bugs-20231018-row5-num_unique_entry_methods_out_of_working_templates}{263}
\DefMacro{table-numbers_of_programs_and_bugs-20231018-row5-num_unique_classes_out_of_working_templates}{30}
\DefMacro{table-numbers_of_programs_and_bugs-20231018-row5-exec-total}{2,158}
\DefMacro{table-numbers_of_programs_and_bugs-20231018-row5-exec-not_compile}{0}
\DefMacro{table-numbers_of_programs_and_bugs-20231018-row5-exec-timeout}{0}
\DefMacro{table-numbers_of_programs_and_bugs-20231018-row5-exec-ok}{2,154}
\DefMacro{table-numbers_of_programs_and_bugs-20231018-row5-exec-skipped_diff}{4}
\DefMacro{table-numbers_of_programs_and_bugs-20231018-row5-exec-skipped_crash}{0}
\DefMacro{table-numbers_of_programs_and_bugs-20231018-row5-exec-final_diff}{0}
\DefMacro{table-numbers_of_programs_and_bugs-20231018-row5-exec-final_crash}{0}
\DefMacro{table-numbers_of_programs_and_bugs-20231018-row6-project}{codec}
\DefMacro{table-numbers_of_programs_and_bugs-20231018-row6-num_unit_tests}{655}
\DefMacro{table-numbers_of_programs_and_bugs-20231018-row6-num_unique_entry_methods}{655}
\DefMacro{table-numbers_of_programs_and_bugs-20231018-row6-num_unique_classes}{53}
\DefMacro{table-numbers_of_programs_and_bugs-20231018-row6-convert-fail}{0}
\DefMacro{table-numbers_of_programs_and_bugs-20231018-row6-convert-timeout}{4}
\DefMacro{table-numbers_of_programs_and_bugs-20231018-row6-convert-not_compile}{0}
\DefMacro{table-numbers_of_programs_and_bugs-20231018-row6-convert-no_original}{0}
\DefMacro{table-numbers_of_programs_and_bugs-20231018-row6-gen-fail}{52}
\DefMacro{table-numbers_of_programs_and_bugs-20231018-row6-gen-timeout}{4}
\DefMacro{table-numbers_of_programs_and_bugs-20231018-row6-gen-no_hole}{164}
\DefMacro{table-numbers_of_programs_and_bugs-20231018-row6-gen-no_gen}{54}
\DefMacro{table-numbers_of_programs_and_bugs-20231018-row6-num_working_templates}{433}
\DefMacro{table-numbers_of_programs_and_bugs-20231018-row6-num_unique_entry_methods_out_of_working_templates}{433}
\DefMacro{table-numbers_of_programs_and_bugs-20231018-row6-num_unique_classes_out_of_working_templates}{42}
\DefMacro{table-numbers_of_programs_and_bugs-20231018-row6-exec-total}{3,409}
\DefMacro{table-numbers_of_programs_and_bugs-20231018-row6-exec-not_compile}{0}
\DefMacro{table-numbers_of_programs_and_bugs-20231018-row6-exec-timeout}{43}
\DefMacro{table-numbers_of_programs_and_bugs-20231018-row6-exec-ok}{3,305}
\DefMacro{table-numbers_of_programs_and_bugs-20231018-row6-exec-skipped_diff}{47}
\DefMacro{table-numbers_of_programs_and_bugs-20231018-row6-exec-skipped_crash}{0}
\DefMacro{table-numbers_of_programs_and_bugs-20231018-row6-exec-final_diff}{0}
\DefMacro{table-numbers_of_programs_and_bugs-20231018-row6-exec-final_crash}{14}
\DefMacro{table-numbers_of_programs_and_bugs-20231018-row7-project}{jfreechart}
\DefMacro{table-numbers_of_programs_and_bugs-20231018-row7-num_unit_tests}{7,137}
\DefMacro{table-numbers_of_programs_and_bugs-20231018-row7-num_unique_entry_methods}{7,137}
\DefMacro{table-numbers_of_programs_and_bugs-20231018-row7-num_unique_classes}{503}
\DefMacro{table-numbers_of_programs_and_bugs-20231018-row7-convert-fail}{13}
\DefMacro{table-numbers_of_programs_and_bugs-20231018-row7-convert-timeout}{869}
\DefMacro{table-numbers_of_programs_and_bugs-20231018-row7-convert-not_compile}{21}
\DefMacro{table-numbers_of_programs_and_bugs-20231018-row7-convert-no_original}{0}
\DefMacro{table-numbers_of_programs_and_bugs-20231018-row7-gen-fail}{626}
\DefMacro{table-numbers_of_programs_and_bugs-20231018-row7-gen-timeout}{1}
\DefMacro{table-numbers_of_programs_and_bugs-20231018-row7-gen-no_hole}{3,474}
\DefMacro{table-numbers_of_programs_and_bugs-20231018-row7-gen-no_gen}{634}
\DefMacro{table-numbers_of_programs_and_bugs-20231018-row7-num_working_templates}{2,126}
\DefMacro{table-numbers_of_programs_and_bugs-20231018-row7-num_unique_entry_methods_out_of_working_templates}{2,126}
\DefMacro{table-numbers_of_programs_and_bugs-20231018-row7-num_unique_classes_out_of_working_templates}{360}
\DefMacro{table-numbers_of_programs_and_bugs-20231018-row7-exec-total}{17,251}
\DefMacro{table-numbers_of_programs_and_bugs-20231018-row7-exec-not_compile}{0}
\DefMacro{table-numbers_of_programs_and_bugs-20231018-row7-exec-timeout}{90}
\DefMacro{table-numbers_of_programs_and_bugs-20231018-row7-exec-ok}{17,159}
\DefMacro{table-numbers_of_programs_and_bugs-20231018-row7-exec-skipped_diff}{0}
\DefMacro{table-numbers_of_programs_and_bugs-20231018-row7-exec-skipped_crash}{2}
\DefMacro{table-numbers_of_programs_and_bugs-20231018-row7-exec-final_diff}{0}
\DefMacro{table-numbers_of_programs_and_bugs-20231018-row7-exec-final_crash}{0}
\DefMacro{table-numbers_of_programs_and_bugs-20231018-row8-project}{zxing}
\DefMacro{table-numbers_of_programs_and_bugs-20231018-row8-num_unit_tests}{1,323}
\DefMacro{table-numbers_of_programs_and_bugs-20231018-row8-num_unique_entry_methods}{1,323}
\DefMacro{table-numbers_of_programs_and_bugs-20231018-row8-num_unique_classes}{213}
\DefMacro{table-numbers_of_programs_and_bugs-20231018-row8-convert-fail}{0}
\DefMacro{table-numbers_of_programs_and_bugs-20231018-row8-convert-timeout}{3}
\DefMacro{table-numbers_of_programs_and_bugs-20231018-row8-convert-not_compile}{57}
\DefMacro{table-numbers_of_programs_and_bugs-20231018-row8-convert-no_original}{0}
\DefMacro{table-numbers_of_programs_and_bugs-20231018-row8-gen-fail}{31}
\DefMacro{table-numbers_of_programs_and_bugs-20231018-row8-gen-timeout}{5}
\DefMacro{table-numbers_of_programs_and_bugs-20231018-row8-gen-no_hole}{610}
\DefMacro{table-numbers_of_programs_and_bugs-20231018-row8-gen-no_gen}{37}
\DefMacro{table-numbers_of_programs_and_bugs-20231018-row8-num_working_templates}{616}
\DefMacro{table-numbers_of_programs_and_bugs-20231018-row8-num_unique_entry_methods_out_of_working_templates}{616}
\DefMacro{table-numbers_of_programs_and_bugs-20231018-row8-num_unique_classes_out_of_working_templates}{142}
\DefMacro{table-numbers_of_programs_and_bugs-20231018-row8-exec-total}{5,525}
\DefMacro{table-numbers_of_programs_and_bugs-20231018-row8-exec-not_compile}{0}
\DefMacro{table-numbers_of_programs_and_bugs-20231018-row8-exec-timeout}{86}
\DefMacro{table-numbers_of_programs_and_bugs-20231018-row8-exec-ok}{5,413}
\DefMacro{table-numbers_of_programs_and_bugs-20231018-row8-exec-skipped_diff}{2}
\DefMacro{table-numbers_of_programs_and_bugs-20231018-row8-exec-skipped_crash}{21}
\DefMacro{table-numbers_of_programs_and_bugs-20231018-row8-exec-final_diff}{1}
\DefMacro{table-numbers_of_programs_and_bugs-20231018-row8-exec-final_crash}{2}
\DefMacro{table-numbers_of_programs_and_bugs-20231018-row9-project}{compress}
\DefMacro{table-numbers_of_programs_and_bugs-20231018-row9-num_unit_tests}{1,939}
\DefMacro{table-numbers_of_programs_and_bugs-20231018-row9-num_unique_entry_methods}{1,939}
\DefMacro{table-numbers_of_programs_and_bugs-20231018-row9-num_unique_classes}{177}
\DefMacro{table-numbers_of_programs_and_bugs-20231018-row9-convert-fail}{0}
\DefMacro{table-numbers_of_programs_and_bugs-20231018-row9-convert-timeout}{3}
\DefMacro{table-numbers_of_programs_and_bugs-20231018-row9-convert-not_compile}{62}
\DefMacro{table-numbers_of_programs_and_bugs-20231018-row9-convert-no_original}{0}
\DefMacro{table-numbers_of_programs_and_bugs-20231018-row9-gen-fail}{9}
\DefMacro{table-numbers_of_programs_and_bugs-20231018-row9-gen-timeout}{1}
\DefMacro{table-numbers_of_programs_and_bugs-20231018-row9-gen-no_hole}{1,074}
\DefMacro{table-numbers_of_programs_and_bugs-20231018-row9-gen-no_gen}{1}
\DefMacro{table-numbers_of_programs_and_bugs-20231018-row9-num_working_templates}{799}
\DefMacro{table-numbers_of_programs_and_bugs-20231018-row9-num_unique_entry_methods_out_of_working_templates}{799}
\DefMacro{table-numbers_of_programs_and_bugs-20231018-row9-num_unique_classes_out_of_working_templates}{134}
\DefMacro{table-numbers_of_programs_and_bugs-20231018-row9-exec-total}{6,943}
\DefMacro{table-numbers_of_programs_and_bugs-20231018-row9-exec-not_compile}{0}
\DefMacro{table-numbers_of_programs_and_bugs-20231018-row9-exec-timeout}{56}
\DefMacro{table-numbers_of_programs_and_bugs-20231018-row9-exec-ok}{6,886}
\DefMacro{table-numbers_of_programs_and_bugs-20231018-row9-exec-skipped_diff}{1}
\DefMacro{table-numbers_of_programs_and_bugs-20231018-row9-exec-skipped_crash}{0}
\DefMacro{table-numbers_of_programs_and_bugs-20231018-row9-exec-final_diff}{0}
\DefMacro{table-numbers_of_programs_and_bugs-20231018-row9-exec-final_crash}{0}
\DefMacro{table-numbers_of_programs_and_bugs-20231018-row10-project}{$\sum$}
\DefMacro{table-numbers_of_programs_and_bugs-20231018-row10-num_unit_tests}{26,048}
\DefMacro{table-numbers_of_programs_and_bugs-20231018-row10-num_unique_entry_methods}{26,042}
\DefMacro{table-numbers_of_programs_and_bugs-20231018-row10-num_unique_classes}{2,044}
\DefMacro{table-numbers_of_programs_and_bugs-20231018-row10-convert-fail}{21}
\DefMacro{table-numbers_of_programs_and_bugs-20231018-row10-convert-timeout}{3,428}
\DefMacro{table-numbers_of_programs_and_bugs-20231018-row10-convert-not_compile}{663}
\DefMacro{table-numbers_of_programs_and_bugs-20231018-row10-convert-no_original}{0}
\DefMacro{table-numbers_of_programs_and_bugs-20231018-row10-gen-fail}{1,428}
\DefMacro{table-numbers_of_programs_and_bugs-20231018-row10-gen-timeout}{470}
\DefMacro{table-numbers_of_programs_and_bugs-20231018-row10-gen-no_hole}{9,650}
\DefMacro{table-numbers_of_programs_and_bugs-20231018-row10-gen-no_gen}{1,998}
\DefMacro{table-numbers_of_programs_and_bugs-20231018-row10-num_working_templates}{10,288}
\DefMacro{table-numbers_of_programs_and_bugs-20231018-row10-num_unique_entry_methods_out_of_working_templates}{10,288}
\DefMacro{table-numbers_of_programs_and_bugs-20231018-row10-num_unique_classes_out_of_working_templates}{1,453}
\DefMacro{table-numbers_of_programs_and_bugs-20231018-row10-exec-total}{90,834}
\DefMacro{table-numbers_of_programs_and_bugs-20231018-row10-exec-not_compile}{0}
\DefMacro{table-numbers_of_programs_and_bugs-20231018-row10-exec-timeout}{911}
\DefMacro{table-numbers_of_programs_and_bugs-20231018-row10-exec-ok}{89,164}
\DefMacro{table-numbers_of_programs_and_bugs-20231018-row10-exec-skipped_diff}{656}
\DefMacro{table-numbers_of_programs_and_bugs-20231018-row10-exec-skipped_crash}{48}
\DefMacro{table-numbers_of_programs_and_bugs-20231018-row10-exec-final_diff}{1}
\DefMacro{table-numbers_of_programs_and_bugs-20231018-row10-exec-final_crash}{54}

\DefMacro{table-numbers_of_programs_and_bugs-20231013-row0-project}{lang}
\DefMacro{table-numbers_of_programs_and_bugs-20231013-row0-num_unit_tests}{2,827}
\DefMacro{table-numbers_of_programs_and_bugs-20231013-row0-num_unique_entry_methods}{2,821}
\DefMacro{table-numbers_of_programs_and_bugs-20231013-row0-num_unique_classes}{131}
\DefMacro{table-numbers_of_programs_and_bugs-20231013-row0-convert-fail}{0}
\DefMacro{table-numbers_of_programs_and_bugs-20231013-row0-convert-timeout}{844}
\DefMacro{table-numbers_of_programs_and_bugs-20231013-row0-convert-not_compile}{147}
\DefMacro{table-numbers_of_programs_and_bugs-20231013-row0-convert-no_original}{0}
\DefMacro{table-numbers_of_programs_and_bugs-20231013-row0-gen-fail}{46}
\DefMacro{table-numbers_of_programs_and_bugs-20231013-row0-gen-timeout}{1}
\DefMacro{table-numbers_of_programs_and_bugs-20231013-row0-gen-no_hole}{525}
\DefMacro{table-numbers_of_programs_and_bugs-20231013-row0-gen-no_gen}{38}
\DefMacro{table-numbers_of_programs_and_bugs-20231013-row0-num_working_templates}{1,273}
\DefMacro{table-numbers_of_programs_and_bugs-20231013-row0-num_unique_entry_methods_out_of_working_templates}{1,273}
\DefMacro{table-numbers_of_programs_and_bugs-20231013-row0-num_unique_classes_out_of_working_templates}{84}
\DefMacro{table-numbers_of_programs_and_bugs-20231013-row0-exec-total}{11,068}
\DefMacro{table-numbers_of_programs_and_bugs-20231013-row0-exec-not_compile}{0}
\DefMacro{table-numbers_of_programs_and_bugs-20231013-row0-exec-timeout}{104}
\DefMacro{table-numbers_of_programs_and_bugs-20231013-row0-exec-ok}{10,685}
\DefMacro{table-numbers_of_programs_and_bugs-20231013-row0-exec-skipped_diff}{259}
\DefMacro{table-numbers_of_programs_and_bugs-20231013-row0-exec-skipped_crash}{8}
\DefMacro{table-numbers_of_programs_and_bugs-20231013-row0-exec-final_diff}{0}
\DefMacro{table-numbers_of_programs_and_bugs-20231013-row0-exec-final_crash}{12}
\DefMacro{table-numbers_of_programs_and_bugs-20231013-row1-project}{math}
\DefMacro{table-numbers_of_programs_and_bugs-20231013-row1-num_unit_tests}{5,296}
\DefMacro{table-numbers_of_programs_and_bugs-20231013-row1-num_unique_entry_methods}{5,296}
\DefMacro{table-numbers_of_programs_and_bugs-20231013-row1-num_unique_classes}{591}
\DefMacro{table-numbers_of_programs_and_bugs-20231013-row1-convert-fail}{8}
\DefMacro{table-numbers_of_programs_and_bugs-20231013-row1-convert-timeout}{207}
\DefMacro{table-numbers_of_programs_and_bugs-20231013-row1-convert-not_compile}{350}
\DefMacro{table-numbers_of_programs_and_bugs-20231013-row1-convert-no_original}{0}
\DefMacro{table-numbers_of_programs_and_bugs-20231013-row1-gen-fail}{264}
\DefMacro{table-numbers_of_programs_and_bugs-20231013-row1-gen-timeout}{133}
\DefMacro{table-numbers_of_programs_and_bugs-20231013-row1-gen-no_hole}{2,310}
\DefMacro{table-numbers_of_programs_and_bugs-20231013-row1-gen-no_gen}{400}
\DefMacro{table-numbers_of_programs_and_bugs-20231013-row1-num_working_templates}{2,021}
\DefMacro{table-numbers_of_programs_and_bugs-20231013-row1-num_unique_entry_methods_out_of_working_templates}{2,021}
\DefMacro{table-numbers_of_programs_and_bugs-20231013-row1-num_unique_classes_out_of_working_templates}{397}
\DefMacro{table-numbers_of_programs_and_bugs-20231013-row1-exec-total}{18,231}
\DefMacro{table-numbers_of_programs_and_bugs-20231013-row1-exec-not_compile}{0}
\DefMacro{table-numbers_of_programs_and_bugs-20231013-row1-exec-timeout}{166}
\DefMacro{table-numbers_of_programs_and_bugs-20231013-row1-exec-ok}{17,561}
\DefMacro{table-numbers_of_programs_and_bugs-20231013-row1-exec-skipped_diff}{474}
\DefMacro{table-numbers_of_programs_and_bugs-20231013-row1-exec-skipped_crash}{11}
\DefMacro{table-numbers_of_programs_and_bugs-20231013-row1-exec-final_diff}{0}
\DefMacro{table-numbers_of_programs_and_bugs-20231013-row1-exec-final_crash}{19}
\DefMacro{table-numbers_of_programs_and_bugs-20231013-row2-project}{text}
\DefMacro{table-numbers_of_programs_and_bugs-20231013-row2-num_unit_tests}{851}
\DefMacro{table-numbers_of_programs_and_bugs-20231013-row2-num_unique_entry_methods}{851}
\DefMacro{table-numbers_of_programs_and_bugs-20231013-row2-num_unique_classes}{77}
\DefMacro{table-numbers_of_programs_and_bugs-20231013-row2-convert-fail}{0}
\DefMacro{table-numbers_of_programs_and_bugs-20231013-row2-convert-timeout}{336}
\DefMacro{table-numbers_of_programs_and_bugs-20231013-row2-convert-not_compile}{8}
\DefMacro{table-numbers_of_programs_and_bugs-20231013-row2-convert-no_original}{0}
\DefMacro{table-numbers_of_programs_and_bugs-20231013-row2-gen-fail}{0}
\DefMacro{table-numbers_of_programs_and_bugs-20231013-row2-gen-timeout}{26}
\DefMacro{table-numbers_of_programs_and_bugs-20231013-row2-gen-no_hole}{117}
\DefMacro{table-numbers_of_programs_and_bugs-20231013-row2-gen-no_gen}{26}
\DefMacro{table-numbers_of_programs_and_bugs-20231013-row2-num_working_templates}{364}
\DefMacro{table-numbers_of_programs_and_bugs-20231013-row2-num_unique_entry_methods_out_of_working_templates}{364}
\DefMacro{table-numbers_of_programs_and_bugs-20231013-row2-num_unique_classes_out_of_working_templates}{56}
\DefMacro{table-numbers_of_programs_and_bugs-20231013-row2-exec-total}{3,482}
\DefMacro{table-numbers_of_programs_and_bugs-20231013-row2-exec-not_compile}{0}
\DefMacro{table-numbers_of_programs_and_bugs-20231013-row2-exec-timeout}{18}
\DefMacro{table-numbers_of_programs_and_bugs-20231013-row2-exec-ok}{3,452}
\DefMacro{table-numbers_of_programs_and_bugs-20231013-row2-exec-skipped_diff}{11}
\DefMacro{table-numbers_of_programs_and_bugs-20231013-row2-exec-skipped_crash}{0}
\DefMacro{table-numbers_of_programs_and_bugs-20231013-row2-exec-final_diff}{0}
\DefMacro{table-numbers_of_programs_and_bugs-20231013-row2-exec-final_crash}{1}
\DefMacro{table-numbers_of_programs_and_bugs-20231013-row3-project}{numbers}
\DefMacro{table-numbers_of_programs_and_bugs-20231013-row3-num_unit_tests}{51}
\DefMacro{table-numbers_of_programs_and_bugs-20231013-row3-num_unique_entry_methods}{51}
\DefMacro{table-numbers_of_programs_and_bugs-20231013-row3-num_unique_classes}{4}
\DefMacro{table-numbers_of_programs_and_bugs-20231013-row3-convert-fail}{0}
\DefMacro{table-numbers_of_programs_and_bugs-20231013-row3-convert-timeout}{0}
\DefMacro{table-numbers_of_programs_and_bugs-20231013-row3-convert-not_compile}{0}
\DefMacro{table-numbers_of_programs_and_bugs-20231013-row3-convert-no_original}{0}
\DefMacro{table-numbers_of_programs_and_bugs-20231013-row3-gen-fail}{0}
\DefMacro{table-numbers_of_programs_and_bugs-20231013-row3-gen-timeout}{0}
\DefMacro{table-numbers_of_programs_and_bugs-20231013-row3-gen-no_hole}{0}
\DefMacro{table-numbers_of_programs_and_bugs-20231013-row3-gen-no_gen}{0}
\DefMacro{table-numbers_of_programs_and_bugs-20231013-row3-num_working_templates}{51}
\DefMacro{table-numbers_of_programs_and_bugs-20231013-row3-num_unique_entry_methods_out_of_working_templates}{51}
\DefMacro{table-numbers_of_programs_and_bugs-20231013-row3-num_unique_classes_out_of_working_templates}{4}
\DefMacro{table-numbers_of_programs_and_bugs-20231013-row3-exec-total}{502}
\DefMacro{table-numbers_of_programs_and_bugs-20231013-row3-exec-not_compile}{0}
\DefMacro{table-numbers_of_programs_and_bugs-20231013-row3-exec-timeout}{0}
\DefMacro{table-numbers_of_programs_and_bugs-20231013-row3-exec-ok}{502}
\DefMacro{table-numbers_of_programs_and_bugs-20231013-row3-exec-skipped_diff}{0}
\DefMacro{table-numbers_of_programs_and_bugs-20231013-row3-exec-skipped_crash}{0}
\DefMacro{table-numbers_of_programs_and_bugs-20231013-row3-exec-final_diff}{0}
\DefMacro{table-numbers_of_programs_and_bugs-20231013-row3-exec-final_crash}{0}
\DefMacro{table-numbers_of_programs_and_bugs-20231013-row4-project}{vectorz}
\DefMacro{table-numbers_of_programs_and_bugs-20231013-row4-num_unit_tests}{5,582}
\DefMacro{table-numbers_of_programs_and_bugs-20231013-row4-num_unique_entry_methods}{5,582}
\DefMacro{table-numbers_of_programs_and_bugs-20231013-row4-num_unique_classes}{263}
\DefMacro{table-numbers_of_programs_and_bugs-20231013-row4-convert-fail}{0}
\DefMacro{table-numbers_of_programs_and_bugs-20231013-row4-convert-timeout}{1,165}
\DefMacro{table-numbers_of_programs_and_bugs-20231013-row4-convert-not_compile}{3}
\DefMacro{table-numbers_of_programs_and_bugs-20231013-row4-convert-no_original}{0}
\DefMacro{table-numbers_of_programs_and_bugs-20231013-row4-gen-fail}{521}
\DefMacro{table-numbers_of_programs_and_bugs-20231013-row4-gen-timeout}{96}
\DefMacro{table-numbers_of_programs_and_bugs-20231013-row4-gen-no_hole}{1,423}
\DefMacro{table-numbers_of_programs_and_bugs-20231013-row4-gen-no_gen}{634}
\DefMacro{table-numbers_of_programs_and_bugs-20231013-row4-num_working_templates}{2,357}
\DefMacro{table-numbers_of_programs_and_bugs-20231013-row4-num_unique_entry_methods_out_of_working_templates}{2,357}
\DefMacro{table-numbers_of_programs_and_bugs-20231013-row4-num_unique_classes_out_of_working_templates}{225}
\DefMacro{table-numbers_of_programs_and_bugs-20231013-row4-exec-total}{22,473}
\DefMacro{table-numbers_of_programs_and_bugs-20231013-row4-exec-not_compile}{0}
\DefMacro{table-numbers_of_programs_and_bugs-20231013-row4-exec-timeout}{257}
\DefMacro{table-numbers_of_programs_and_bugs-20231013-row4-exec-ok}{22,172}
\DefMacro{table-numbers_of_programs_and_bugs-20231013-row4-exec-skipped_diff}{32}
\DefMacro{table-numbers_of_programs_and_bugs-20231013-row4-exec-skipped_crash}{3}
\DefMacro{table-numbers_of_programs_and_bugs-20231013-row4-exec-final_diff}{0}
\DefMacro{table-numbers_of_programs_and_bugs-20231013-row4-exec-final_crash}{9}
\DefMacro{table-numbers_of_programs_and_bugs-20231013-row5-project}{statistics}
\DefMacro{table-numbers_of_programs_and_bugs-20231013-row5-num_unit_tests}{387}
\DefMacro{table-numbers_of_programs_and_bugs-20231013-row5-num_unique_entry_methods}{387}
\DefMacro{table-numbers_of_programs_and_bugs-20231013-row5-num_unique_classes}{32}
\DefMacro{table-numbers_of_programs_and_bugs-20231013-row5-convert-fail}{0}
\DefMacro{table-numbers_of_programs_and_bugs-20231013-row5-convert-timeout}{0}
\DefMacro{table-numbers_of_programs_and_bugs-20231013-row5-convert-not_compile}{0}
\DefMacro{table-numbers_of_programs_and_bugs-20231013-row5-convert-no_original}{0}
\DefMacro{table-numbers_of_programs_and_bugs-20231013-row5-gen-fail}{40}
\DefMacro{table-numbers_of_programs_and_bugs-20231013-row5-gen-timeout}{0}
\DefMacro{table-numbers_of_programs_and_bugs-20231013-row5-gen-no_hole}{5}
\DefMacro{table-numbers_of_programs_and_bugs-20231013-row5-gen-no_gen}{140}
\DefMacro{table-numbers_of_programs_and_bugs-20231013-row5-num_working_templates}{242}
\DefMacro{table-numbers_of_programs_and_bugs-20231013-row5-num_unique_entry_methods_out_of_working_templates}{242}
\DefMacro{table-numbers_of_programs_and_bugs-20231013-row5-num_unique_classes_out_of_working_templates}{30}
\DefMacro{table-numbers_of_programs_and_bugs-20231013-row5-exec-total}{2,219}
\DefMacro{table-numbers_of_programs_and_bugs-20231013-row5-exec-not_compile}{0}
\DefMacro{table-numbers_of_programs_and_bugs-20231013-row5-exec-timeout}{0}
\DefMacro{table-numbers_of_programs_and_bugs-20231013-row5-exec-ok}{2,202}
\DefMacro{table-numbers_of_programs_and_bugs-20231013-row5-exec-skipped_diff}{13}
\DefMacro{table-numbers_of_programs_and_bugs-20231013-row5-exec-skipped_crash}{3}
\DefMacro{table-numbers_of_programs_and_bugs-20231013-row5-exec-final_diff}{1}
\DefMacro{table-numbers_of_programs_and_bugs-20231013-row5-exec-final_crash}{0}
\DefMacro{table-numbers_of_programs_and_bugs-20231013-row6-project}{codec}
\DefMacro{table-numbers_of_programs_and_bugs-20231013-row6-num_unit_tests}{655}
\DefMacro{table-numbers_of_programs_and_bugs-20231013-row6-num_unique_entry_methods}{655}
\DefMacro{table-numbers_of_programs_and_bugs-20231013-row6-num_unique_classes}{53}
\DefMacro{table-numbers_of_programs_and_bugs-20231013-row6-convert-fail}{0}
\DefMacro{table-numbers_of_programs_and_bugs-20231013-row6-convert-timeout}{6}
\DefMacro{table-numbers_of_programs_and_bugs-20231013-row6-convert-not_compile}{0}
\DefMacro{table-numbers_of_programs_and_bugs-20231013-row6-convert-no_original}{0}
\DefMacro{table-numbers_of_programs_and_bugs-20231013-row6-gen-fail}{52}
\DefMacro{table-numbers_of_programs_and_bugs-20231013-row6-gen-timeout}{3}
\DefMacro{table-numbers_of_programs_and_bugs-20231013-row6-gen-no_hole}{163}
\DefMacro{table-numbers_of_programs_and_bugs-20231013-row6-gen-no_gen}{52}
\DefMacro{table-numbers_of_programs_and_bugs-20231013-row6-num_working_templates}{434}
\DefMacro{table-numbers_of_programs_and_bugs-20231013-row6-num_unique_entry_methods_out_of_working_templates}{434}
\DefMacro{table-numbers_of_programs_and_bugs-20231013-row6-num_unique_classes_out_of_working_templates}{42}
\DefMacro{table-numbers_of_programs_and_bugs-20231013-row6-exec-total}{3,425}
\DefMacro{table-numbers_of_programs_and_bugs-20231013-row6-exec-not_compile}{0}
\DefMacro{table-numbers_of_programs_and_bugs-20231013-row6-exec-timeout}{54}
\DefMacro{table-numbers_of_programs_and_bugs-20231013-row6-exec-ok}{3,301}
\DefMacro{table-numbers_of_programs_and_bugs-20231013-row6-exec-skipped_diff}{51}
\DefMacro{table-numbers_of_programs_and_bugs-20231013-row6-exec-skipped_crash}{0}
\DefMacro{table-numbers_of_programs_and_bugs-20231013-row6-exec-final_diff}{0}
\DefMacro{table-numbers_of_programs_and_bugs-20231013-row6-exec-final_crash}{19}
\DefMacro{table-numbers_of_programs_and_bugs-20231013-row7-project}{jfreechart}
\DefMacro{table-numbers_of_programs_and_bugs-20231013-row7-num_unit_tests}{7,137}
\DefMacro{table-numbers_of_programs_and_bugs-20231013-row7-num_unique_entry_methods}{7,137}
\DefMacro{table-numbers_of_programs_and_bugs-20231013-row7-num_unique_classes}{503}
\DefMacro{table-numbers_of_programs_and_bugs-20231013-row7-convert-fail}{13}
\DefMacro{table-numbers_of_programs_and_bugs-20231013-row7-convert-timeout}{866}
\DefMacro{table-numbers_of_programs_and_bugs-20231013-row7-convert-not_compile}{21}
\DefMacro{table-numbers_of_programs_and_bugs-20231013-row7-convert-no_original}{0}
\DefMacro{table-numbers_of_programs_and_bugs-20231013-row7-gen-fail}{625}
\DefMacro{table-numbers_of_programs_and_bugs-20231013-row7-gen-timeout}{0}
\DefMacro{table-numbers_of_programs_and_bugs-20231013-row7-gen-no_hole}{3,537}
\DefMacro{table-numbers_of_programs_and_bugs-20231013-row7-gen-no_gen}{657}
\DefMacro{table-numbers_of_programs_and_bugs-20231013-row7-num_working_templates}{2,043}
\DefMacro{table-numbers_of_programs_and_bugs-20231013-row7-num_unique_entry_methods_out_of_working_templates}{2,043}
\DefMacro{table-numbers_of_programs_and_bugs-20231013-row7-num_unique_classes_out_of_working_templates}{346}
\DefMacro{table-numbers_of_programs_and_bugs-20231013-row7-exec-total}{16,131}
\DefMacro{table-numbers_of_programs_and_bugs-20231013-row7-exec-not_compile}{0}
\DefMacro{table-numbers_of_programs_and_bugs-20231013-row7-exec-timeout}{66}
\DefMacro{table-numbers_of_programs_and_bugs-20231013-row7-exec-ok}{16,065}
\DefMacro{table-numbers_of_programs_and_bugs-20231013-row7-exec-skipped_diff}{0}
\DefMacro{table-numbers_of_programs_and_bugs-20231013-row7-exec-skipped_crash}{0}
\DefMacro{table-numbers_of_programs_and_bugs-20231013-row7-exec-final_diff}{0}
\DefMacro{table-numbers_of_programs_and_bugs-20231013-row7-exec-final_crash}{0}
\DefMacro{table-numbers_of_programs_and_bugs-20231013-row8-project}{zxing}
\DefMacro{table-numbers_of_programs_and_bugs-20231013-row8-num_unit_tests}{1,323}
\DefMacro{table-numbers_of_programs_and_bugs-20231013-row8-num_unique_entry_methods}{1,323}
\DefMacro{table-numbers_of_programs_and_bugs-20231013-row8-num_unique_classes}{213}
\DefMacro{table-numbers_of_programs_and_bugs-20231013-row8-convert-fail}{0}
\DefMacro{table-numbers_of_programs_and_bugs-20231013-row8-convert-timeout}{3}
\DefMacro{table-numbers_of_programs_and_bugs-20231013-row8-convert-not_compile}{57}
\DefMacro{table-numbers_of_programs_and_bugs-20231013-row8-convert-no_original}{0}
\DefMacro{table-numbers_of_programs_and_bugs-20231013-row8-gen-fail}{29}
\DefMacro{table-numbers_of_programs_and_bugs-20231013-row8-gen-timeout}{5}
\DefMacro{table-numbers_of_programs_and_bugs-20231013-row8-gen-no_hole}{610}
\DefMacro{table-numbers_of_programs_and_bugs-20231013-row8-gen-no_gen}{39}
\DefMacro{table-numbers_of_programs_and_bugs-20231013-row8-num_working_templates}{614}
\DefMacro{table-numbers_of_programs_and_bugs-20231013-row8-num_unique_entry_methods_out_of_working_templates}{614}
\DefMacro{table-numbers_of_programs_and_bugs-20231013-row8-num_unique_classes_out_of_working_templates}{142}
\DefMacro{table-numbers_of_programs_and_bugs-20231013-row8-exec-total}{5,536}
\DefMacro{table-numbers_of_programs_and_bugs-20231013-row8-exec-not_compile}{0}
\DefMacro{table-numbers_of_programs_and_bugs-20231013-row8-exec-timeout}{91}
\DefMacro{table-numbers_of_programs_and_bugs-20231013-row8-exec-ok}{5,421}
\DefMacro{table-numbers_of_programs_and_bugs-20231013-row8-exec-skipped_diff}{3}
\DefMacro{table-numbers_of_programs_and_bugs-20231013-row8-exec-skipped_crash}{21}
\DefMacro{table-numbers_of_programs_and_bugs-20231013-row8-exec-final_diff}{0}
\DefMacro{table-numbers_of_programs_and_bugs-20231013-row8-exec-final_crash}{0}
\DefMacro{table-numbers_of_programs_and_bugs-20231013-row9-project}{compress}
\DefMacro{table-numbers_of_programs_and_bugs-20231013-row9-num_unit_tests}{1,939}
\DefMacro{table-numbers_of_programs_and_bugs-20231013-row9-num_unique_entry_methods}{1,939}
\DefMacro{table-numbers_of_programs_and_bugs-20231013-row9-num_unique_classes}{177}
\DefMacro{table-numbers_of_programs_and_bugs-20231013-row9-convert-fail}{0}
\DefMacro{table-numbers_of_programs_and_bugs-20231013-row9-convert-timeout}{3}
\DefMacro{table-numbers_of_programs_and_bugs-20231013-row9-convert-not_compile}{62}
\DefMacro{table-numbers_of_programs_and_bugs-20231013-row9-convert-no_original}{0}
\DefMacro{table-numbers_of_programs_and_bugs-20231013-row9-gen-fail}{9}
\DefMacro{table-numbers_of_programs_and_bugs-20231013-row9-gen-timeout}{1}
\DefMacro{table-numbers_of_programs_and_bugs-20231013-row9-gen-no_hole}{1,074}
\DefMacro{table-numbers_of_programs_and_bugs-20231013-row9-gen-no_gen}{9}
\DefMacro{table-numbers_of_programs_and_bugs-20231013-row9-num_working_templates}{791}
\DefMacro{table-numbers_of_programs_and_bugs-20231013-row9-num_unique_entry_methods_out_of_working_templates}{791}
\DefMacro{table-numbers_of_programs_and_bugs-20231013-row9-num_unique_classes_out_of_working_templates}{133}
\DefMacro{table-numbers_of_programs_and_bugs-20231013-row9-exec-total}{6,932}
\DefMacro{table-numbers_of_programs_and_bugs-20231013-row9-exec-not_compile}{0}
\DefMacro{table-numbers_of_programs_and_bugs-20231013-row9-exec-timeout}{52}
\DefMacro{table-numbers_of_programs_and_bugs-20231013-row9-exec-ok}{6,880}
\DefMacro{table-numbers_of_programs_and_bugs-20231013-row9-exec-skipped_diff}{0}
\DefMacro{table-numbers_of_programs_and_bugs-20231013-row9-exec-skipped_crash}{0}
\DefMacro{table-numbers_of_programs_and_bugs-20231013-row9-exec-final_diff}{0}
\DefMacro{table-numbers_of_programs_and_bugs-20231013-row9-exec-final_crash}{0}
\DefMacro{table-numbers_of_programs_and_bugs-20231013-row10-project}{$\sum$}
\DefMacro{table-numbers_of_programs_and_bugs-20231013-row10-num_unit_tests}{26,048}
\DefMacro{table-numbers_of_programs_and_bugs-20231013-row10-num_unique_entry_methods}{26,042}
\DefMacro{table-numbers_of_programs_and_bugs-20231013-row10-num_unique_classes}{2,044}
\DefMacro{table-numbers_of_programs_and_bugs-20231013-row10-convert-fail}{21}
\DefMacro{table-numbers_of_programs_and_bugs-20231013-row10-convert-timeout}{3,430}
\DefMacro{table-numbers_of_programs_and_bugs-20231013-row10-convert-not_compile}{648}
\DefMacro{table-numbers_of_programs_and_bugs-20231013-row10-convert-no_original}{0}
\DefMacro{table-numbers_of_programs_and_bugs-20231013-row10-gen-fail}{1,586}
\DefMacro{table-numbers_of_programs_and_bugs-20231013-row10-gen-timeout}{265}
\DefMacro{table-numbers_of_programs_and_bugs-20231013-row10-gen-no_hole}{9,764}
\DefMacro{table-numbers_of_programs_and_bugs-20231013-row10-gen-no_gen}{1,995}
\DefMacro{table-numbers_of_programs_and_bugs-20231013-row10-num_working_templates}{10,190}
\DefMacro{table-numbers_of_programs_and_bugs-20231013-row10-num_unique_entry_methods_out_of_working_templates}{10,190}
\DefMacro{table-numbers_of_programs_and_bugs-20231013-row10-num_unique_classes_out_of_working_templates}{1,459}
\DefMacro{table-numbers_of_programs_and_bugs-20231013-row10-exec-total}{89,999}
\DefMacro{table-numbers_of_programs_and_bugs-20231013-row10-exec-not_compile}{0}
\DefMacro{table-numbers_of_programs_and_bugs-20231013-row10-exec-timeout}{808}
\DefMacro{table-numbers_of_programs_and_bugs-20231013-row10-exec-ok}{88,241}
\DefMacro{table-numbers_of_programs_and_bugs-20231013-row10-exec-skipped_diff}{843}
\DefMacro{table-numbers_of_programs_and_bugs-20231013-row10-exec-skipped_crash}{46}
\DefMacro{table-numbers_of_programs_and_bugs-20231013-row10-exec-final_diff}{1}
\DefMacro{table-numbers_of_programs_and_bugs-20231013-row10-exec-final_crash}{60}

\DefMacro{table-numbers_of_programs_and_bugs-20230421-row0-project}{lang}
\DefMacro{table-numbers_of_programs_and_bugs-20230421-row0-num_unit_tests}{2,827}
\DefMacro{table-numbers_of_programs_and_bugs-20230421-row0-num_unique_entry_methods}{2,821}
\DefMacro{table-numbers_of_programs_and_bugs-20230421-row0-num_unique_classes}{131}
\DefMacro{table-numbers_of_programs_and_bugs-20230421-row0-convert-fail}{0}
\DefMacro{table-numbers_of_programs_and_bugs-20230421-row0-convert-timeout}{845}
\DefMacro{table-numbers_of_programs_and_bugs-20230421-row0-convert-not_compile}{147}
\DefMacro{table-numbers_of_programs_and_bugs-20230421-row0-convert-no_original}{0}
\DefMacro{table-numbers_of_programs_and_bugs-20230421-row0-gen-fail}{45}
\DefMacro{table-numbers_of_programs_and_bugs-20230421-row0-gen-timeout}{1}
\DefMacro{table-numbers_of_programs_and_bugs-20230421-row0-gen-no_hole}{525}
\DefMacro{table-numbers_of_programs_and_bugs-20230421-row0-gen-no_gen}{49}
\DefMacro{table-numbers_of_programs_and_bugs-20230421-row0-num_working_templates}{1,261}
\DefMacro{table-numbers_of_programs_and_bugs-20230421-row0-num_unique_entry_methods_out_of_working_templates}{1,261}
\DefMacro{table-numbers_of_programs_and_bugs-20230421-row0-num_unique_classes_out_of_working_templates}{83}
\DefMacro{table-numbers_of_programs_and_bugs-20230421-row0-exec-total}{11,023}
\DefMacro{table-numbers_of_programs_and_bugs-20230421-row0-exec-not_compile}{0}
\DefMacro{table-numbers_of_programs_and_bugs-20230421-row0-exec-timeout}{103}
\DefMacro{table-numbers_of_programs_and_bugs-20230421-row0-exec-ok}{10,678}
\DefMacro{table-numbers_of_programs_and_bugs-20230421-row0-exec-skipped_diff}{224}
\DefMacro{table-numbers_of_programs_and_bugs-20230421-row0-exec-skipped_crash}{7}
\DefMacro{table-numbers_of_programs_and_bugs-20230421-row0-exec-final_diff}{1}
\DefMacro{table-numbers_of_programs_and_bugs-20230421-row0-exec-final_crash}{10}
\DefMacro{table-numbers_of_programs_and_bugs-20230421-row1-project}{math}
\DefMacro{table-numbers_of_programs_and_bugs-20230421-row1-num_unit_tests}{5,296}
\DefMacro{table-numbers_of_programs_and_bugs-20230421-row1-num_unique_entry_methods}{5,296}
\DefMacro{table-numbers_of_programs_and_bugs-20230421-row1-num_unique_classes}{591}
\DefMacro{table-numbers_of_programs_and_bugs-20230421-row1-convert-fail}{8}
\DefMacro{table-numbers_of_programs_and_bugs-20230421-row1-convert-timeout}{209}
\DefMacro{table-numbers_of_programs_and_bugs-20230421-row1-convert-not_compile}{350}
\DefMacro{table-numbers_of_programs_and_bugs-20230421-row1-convert-no_original}{0}
\DefMacro{table-numbers_of_programs_and_bugs-20230421-row1-gen-fail}{258}
\DefMacro{table-numbers_of_programs_and_bugs-20230421-row1-gen-timeout}{134}
\DefMacro{table-numbers_of_programs_and_bugs-20230421-row1-gen-no_hole}{2,310}
\DefMacro{table-numbers_of_programs_and_bugs-20230421-row1-gen-no_gen}{393}
\DefMacro{table-numbers_of_programs_and_bugs-20230421-row1-num_working_templates}{2,026}
\DefMacro{table-numbers_of_programs_and_bugs-20230421-row1-num_unique_entry_methods_out_of_working_templates}{2,026}
\DefMacro{table-numbers_of_programs_and_bugs-20230421-row1-num_unique_classes_out_of_working_templates}{397}
\DefMacro{table-numbers_of_programs_and_bugs-20230421-row1-exec-total}{18,216}
\DefMacro{table-numbers_of_programs_and_bugs-20230421-row1-exec-not_compile}{0}
\DefMacro{table-numbers_of_programs_and_bugs-20230421-row1-exec-timeout}{210}
\DefMacro{table-numbers_of_programs_and_bugs-20230421-row1-exec-ok}{17,503}
\DefMacro{table-numbers_of_programs_and_bugs-20230421-row1-exec-skipped_diff}{477}
\DefMacro{table-numbers_of_programs_and_bugs-20230421-row1-exec-skipped_crash}{16}
\DefMacro{table-numbers_of_programs_and_bugs-20230421-row1-exec-final_diff}{0}
\DefMacro{table-numbers_of_programs_and_bugs-20230421-row1-exec-final_crash}{10}
\DefMacro{table-numbers_of_programs_and_bugs-20230421-row2-project}{text}
\DefMacro{table-numbers_of_programs_and_bugs-20230421-row2-num_unit_tests}{851}
\DefMacro{table-numbers_of_programs_and_bugs-20230421-row2-num_unique_entry_methods}{851}
\DefMacro{table-numbers_of_programs_and_bugs-20230421-row2-num_unique_classes}{77}
\DefMacro{table-numbers_of_programs_and_bugs-20230421-row2-convert-fail}{0}
\DefMacro{table-numbers_of_programs_and_bugs-20230421-row2-convert-timeout}{336}
\DefMacro{table-numbers_of_programs_and_bugs-20230421-row2-convert-not_compile}{8}
\DefMacro{table-numbers_of_programs_and_bugs-20230421-row2-convert-no_original}{0}
\DefMacro{table-numbers_of_programs_and_bugs-20230421-row2-gen-fail}{4}
\DefMacro{table-numbers_of_programs_and_bugs-20230421-row2-gen-timeout}{22}
\DefMacro{table-numbers_of_programs_and_bugs-20230421-row2-gen-no_hole}{117}
\DefMacro{table-numbers_of_programs_and_bugs-20230421-row2-gen-no_gen}{26}
\DefMacro{table-numbers_of_programs_and_bugs-20230421-row2-num_working_templates}{364}
\DefMacro{table-numbers_of_programs_and_bugs-20230421-row2-num_unique_entry_methods_out_of_working_templates}{364}
\DefMacro{table-numbers_of_programs_and_bugs-20230421-row2-num_unique_classes_out_of_working_templates}{56}
\DefMacro{table-numbers_of_programs_and_bugs-20230421-row2-exec-total}{3,472}
\DefMacro{table-numbers_of_programs_and_bugs-20230421-row2-exec-not_compile}{0}
\DefMacro{table-numbers_of_programs_and_bugs-20230421-row2-exec-timeout}{19}
\DefMacro{table-numbers_of_programs_and_bugs-20230421-row2-exec-ok}{3,428}
\DefMacro{table-numbers_of_programs_and_bugs-20230421-row2-exec-skipped_diff}{24}
\DefMacro{table-numbers_of_programs_and_bugs-20230421-row2-exec-skipped_crash}{1}
\DefMacro{table-numbers_of_programs_and_bugs-20230421-row2-exec-final_diff}{0}
\DefMacro{table-numbers_of_programs_and_bugs-20230421-row2-exec-final_crash}{0}
\DefMacro{table-numbers_of_programs_and_bugs-20230421-row3-project}{numbers}
\DefMacro{table-numbers_of_programs_and_bugs-20230421-row3-num_unit_tests}{51}
\DefMacro{table-numbers_of_programs_and_bugs-20230421-row3-num_unique_entry_methods}{51}
\DefMacro{table-numbers_of_programs_and_bugs-20230421-row3-num_unique_classes}{4}
\DefMacro{table-numbers_of_programs_and_bugs-20230421-row3-convert-fail}{0}
\DefMacro{table-numbers_of_programs_and_bugs-20230421-row3-convert-timeout}{0}
\DefMacro{table-numbers_of_programs_and_bugs-20230421-row3-convert-not_compile}{0}
\DefMacro{table-numbers_of_programs_and_bugs-20230421-row3-convert-no_original}{0}
\DefMacro{table-numbers_of_programs_and_bugs-20230421-row3-gen-fail}{0}
\DefMacro{table-numbers_of_programs_and_bugs-20230421-row3-gen-timeout}{0}
\DefMacro{table-numbers_of_programs_and_bugs-20230421-row3-gen-no_hole}{0}
\DefMacro{table-numbers_of_programs_and_bugs-20230421-row3-gen-no_gen}{0}
\DefMacro{table-numbers_of_programs_and_bugs-20230421-row3-num_working_templates}{51}
\DefMacro{table-numbers_of_programs_and_bugs-20230421-row3-num_unique_entry_methods_out_of_working_templates}{51}
\DefMacro{table-numbers_of_programs_and_bugs-20230421-row3-num_unique_classes_out_of_working_templates}{4}
\DefMacro{table-numbers_of_programs_and_bugs-20230421-row3-exec-total}{502}
\DefMacro{table-numbers_of_programs_and_bugs-20230421-row3-exec-not_compile}{0}
\DefMacro{table-numbers_of_programs_and_bugs-20230421-row3-exec-timeout}{0}
\DefMacro{table-numbers_of_programs_and_bugs-20230421-row3-exec-ok}{502}
\DefMacro{table-numbers_of_programs_and_bugs-20230421-row3-exec-skipped_diff}{0}
\DefMacro{table-numbers_of_programs_and_bugs-20230421-row3-exec-skipped_crash}{0}
\DefMacro{table-numbers_of_programs_and_bugs-20230421-row3-exec-final_diff}{0}
\DefMacro{table-numbers_of_programs_and_bugs-20230421-row3-exec-final_crash}{0}
\DefMacro{table-numbers_of_programs_and_bugs-20230421-row4-project}{vectorz}
\DefMacro{table-numbers_of_programs_and_bugs-20230421-row4-num_unit_tests}{5,582}
\DefMacro{table-numbers_of_programs_and_bugs-20230421-row4-num_unique_entry_methods}{5,582}
\DefMacro{table-numbers_of_programs_and_bugs-20230421-row4-num_unique_classes}{263}
\DefMacro{table-numbers_of_programs_and_bugs-20230421-row4-convert-fail}{0}
\DefMacro{table-numbers_of_programs_and_bugs-20230421-row4-convert-timeout}{1,154}
\DefMacro{table-numbers_of_programs_and_bugs-20230421-row4-convert-not_compile}{18}
\DefMacro{table-numbers_of_programs_and_bugs-20230421-row4-convert-no_original}{0}
\DefMacro{table-numbers_of_programs_and_bugs-20230421-row4-gen-fail}{509}
\DefMacro{table-numbers_of_programs_and_bugs-20230421-row4-gen-timeout}{179}
\DefMacro{table-numbers_of_programs_and_bugs-20230421-row4-gen-no_hole}{1,373}
\DefMacro{table-numbers_of_programs_and_bugs-20230421-row4-gen-no_gen}{696}
\DefMacro{table-numbers_of_programs_and_bugs-20230421-row4-num_working_templates}{2,341}
\DefMacro{table-numbers_of_programs_and_bugs-20230421-row4-num_unique_entry_methods_out_of_working_templates}{2,341}
\DefMacro{table-numbers_of_programs_and_bugs-20230421-row4-num_unique_classes_out_of_working_templates}{204}
\DefMacro{table-numbers_of_programs_and_bugs-20230421-row4-exec-total}{22,310}
\DefMacro{table-numbers_of_programs_and_bugs-20230421-row4-exec-not_compile}{0}
\DefMacro{table-numbers_of_programs_and_bugs-20230421-row4-exec-timeout}{180}
\DefMacro{table-numbers_of_programs_and_bugs-20230421-row4-exec-ok}{22,061}
\DefMacro{table-numbers_of_programs_and_bugs-20230421-row4-exec-skipped_diff}{57}
\DefMacro{table-numbers_of_programs_and_bugs-20230421-row4-exec-skipped_crash}{1}
\DefMacro{table-numbers_of_programs_and_bugs-20230421-row4-exec-final_diff}{2}
\DefMacro{table-numbers_of_programs_and_bugs-20230421-row4-exec-final_crash}{9}
\DefMacro{table-numbers_of_programs_and_bugs-20230421-row5-project}{statistics}
\DefMacro{table-numbers_of_programs_and_bugs-20230421-row5-num_unit_tests}{387}
\DefMacro{table-numbers_of_programs_and_bugs-20230421-row5-num_unique_entry_methods}{387}
\DefMacro{table-numbers_of_programs_and_bugs-20230421-row5-num_unique_classes}{32}
\DefMacro{table-numbers_of_programs_and_bugs-20230421-row5-convert-fail}{0}
\DefMacro{table-numbers_of_programs_and_bugs-20230421-row5-convert-timeout}{0}
\DefMacro{table-numbers_of_programs_and_bugs-20230421-row5-convert-not_compile}{0}
\DefMacro{table-numbers_of_programs_and_bugs-20230421-row5-convert-no_original}{0}
\DefMacro{table-numbers_of_programs_and_bugs-20230421-row5-gen-fail}{41}
\DefMacro{table-numbers_of_programs_and_bugs-20230421-row5-gen-timeout}{0}
\DefMacro{table-numbers_of_programs_and_bugs-20230421-row5-gen-no_hole}{5}
\DefMacro{table-numbers_of_programs_and_bugs-20230421-row5-gen-no_gen}{143}
\DefMacro{table-numbers_of_programs_and_bugs-20230421-row5-num_working_templates}{239}
\DefMacro{table-numbers_of_programs_and_bugs-20230421-row5-num_unique_entry_methods_out_of_working_templates}{239}
\DefMacro{table-numbers_of_programs_and_bugs-20230421-row5-num_unique_classes_out_of_working_templates}{30}
\DefMacro{table-numbers_of_programs_and_bugs-20230421-row5-exec-total}{2,218}
\DefMacro{table-numbers_of_programs_and_bugs-20230421-row5-exec-not_compile}{0}
\DefMacro{table-numbers_of_programs_and_bugs-20230421-row5-exec-timeout}{0}
\DefMacro{table-numbers_of_programs_and_bugs-20230421-row5-exec-ok}{2,215}
\DefMacro{table-numbers_of_programs_and_bugs-20230421-row5-exec-skipped_diff}{3}
\DefMacro{table-numbers_of_programs_and_bugs-20230421-row5-exec-skipped_crash}{0}
\DefMacro{table-numbers_of_programs_and_bugs-20230421-row5-exec-final_diff}{0}
\DefMacro{table-numbers_of_programs_and_bugs-20230421-row5-exec-final_crash}{0}
\DefMacro{table-numbers_of_programs_and_bugs-20230421-row6-project}{codec}
\DefMacro{table-numbers_of_programs_and_bugs-20230421-row6-num_unit_tests}{655}
\DefMacro{table-numbers_of_programs_and_bugs-20230421-row6-num_unique_entry_methods}{655}
\DefMacro{table-numbers_of_programs_and_bugs-20230421-row6-num_unique_classes}{53}
\DefMacro{table-numbers_of_programs_and_bugs-20230421-row6-convert-fail}{0}
\DefMacro{table-numbers_of_programs_and_bugs-20230421-row6-convert-timeout}{3}
\DefMacro{table-numbers_of_programs_and_bugs-20230421-row6-convert-not_compile}{5}
\DefMacro{table-numbers_of_programs_and_bugs-20230421-row6-convert-no_original}{0}
\DefMacro{table-numbers_of_programs_and_bugs-20230421-row6-gen-fail}{53}
\DefMacro{table-numbers_of_programs_and_bugs-20230421-row6-gen-timeout}{5}
\DefMacro{table-numbers_of_programs_and_bugs-20230421-row6-gen-no_hole}{162}
\DefMacro{table-numbers_of_programs_and_bugs-20230421-row6-gen-no_gen}{55}
\DefMacro{table-numbers_of_programs_and_bugs-20230421-row6-num_working_templates}{430}
\DefMacro{table-numbers_of_programs_and_bugs-20230421-row6-num_unique_entry_methods_out_of_working_templates}{430}
\DefMacro{table-numbers_of_programs_and_bugs-20230421-row6-num_unique_classes_out_of_working_templates}{41}
\DefMacro{table-numbers_of_programs_and_bugs-20230421-row6-exec-total}{3,373}
\DefMacro{table-numbers_of_programs_and_bugs-20230421-row6-exec-not_compile}{0}
\DefMacro{table-numbers_of_programs_and_bugs-20230421-row6-exec-timeout}{52}
\DefMacro{table-numbers_of_programs_and_bugs-20230421-row6-exec-ok}{3,286}
\DefMacro{table-numbers_of_programs_and_bugs-20230421-row6-exec-skipped_diff}{17}
\DefMacro{table-numbers_of_programs_and_bugs-20230421-row6-exec-skipped_crash}{0}
\DefMacro{table-numbers_of_programs_and_bugs-20230421-row6-exec-final_diff}{0}
\DefMacro{table-numbers_of_programs_and_bugs-20230421-row6-exec-final_crash}{18}
\DefMacro{table-numbers_of_programs_and_bugs-20230421-row7-project}{jfreechart}
\DefMacro{table-numbers_of_programs_and_bugs-20230421-row7-num_unit_tests}{7,137}
\DefMacro{table-numbers_of_programs_and_bugs-20230421-row7-num_unique_entry_methods}{7,137}
\DefMacro{table-numbers_of_programs_and_bugs-20230421-row7-num_unique_classes}{503}
\DefMacro{table-numbers_of_programs_and_bugs-20230421-row7-convert-fail}{13}
\DefMacro{table-numbers_of_programs_and_bugs-20230421-row7-convert-timeout}{805}
\DefMacro{table-numbers_of_programs_and_bugs-20230421-row7-convert-not_compile}{33}
\DefMacro{table-numbers_of_programs_and_bugs-20230421-row7-convert-no_original}{0}
\DefMacro{table-numbers_of_programs_and_bugs-20230421-row7-gen-fail}{639}
\DefMacro{table-numbers_of_programs_and_bugs-20230421-row7-gen-timeout}{1}
\DefMacro{table-numbers_of_programs_and_bugs-20230421-row7-gen-no_hole}{3,491}
\DefMacro{table-numbers_of_programs_and_bugs-20230421-row7-gen-no_gen}{802}
\DefMacro{table-numbers_of_programs_and_bugs-20230421-row7-num_working_templates}{1,993}
\DefMacro{table-numbers_of_programs_and_bugs-20230421-row7-num_unique_entry_methods_out_of_working_templates}{1,993}
\DefMacro{table-numbers_of_programs_and_bugs-20230421-row7-num_unique_classes_out_of_working_templates}{332}
\DefMacro{table-numbers_of_programs_and_bugs-20230421-row7-exec-total}{15,803}
\DefMacro{table-numbers_of_programs_and_bugs-20230421-row7-exec-not_compile}{0}
\DefMacro{table-numbers_of_programs_and_bugs-20230421-row7-exec-timeout}{15}
\DefMacro{table-numbers_of_programs_and_bugs-20230421-row7-exec-ok}{15,788}
\DefMacro{table-numbers_of_programs_and_bugs-20230421-row7-exec-skipped_diff}{0}
\DefMacro{table-numbers_of_programs_and_bugs-20230421-row7-exec-skipped_crash}{0}
\DefMacro{table-numbers_of_programs_and_bugs-20230421-row7-exec-final_diff}{0}
\DefMacro{table-numbers_of_programs_and_bugs-20230421-row7-exec-final_crash}{0}
\DefMacro{table-numbers_of_programs_and_bugs-20230421-row8-project}{zxing}
\DefMacro{table-numbers_of_programs_and_bugs-20230421-row8-num_unit_tests}{1,323}
\DefMacro{table-numbers_of_programs_and_bugs-20230421-row8-num_unique_entry_methods}{1,323}
\DefMacro{table-numbers_of_programs_and_bugs-20230421-row8-num_unique_classes}{213}
\DefMacro{table-numbers_of_programs_and_bugs-20230421-row8-convert-fail}{0}
\DefMacro{table-numbers_of_programs_and_bugs-20230421-row8-convert-timeout}{3}
\DefMacro{table-numbers_of_programs_and_bugs-20230421-row8-convert-not_compile}{73}
\DefMacro{table-numbers_of_programs_and_bugs-20230421-row8-convert-no_original}{0}
\DefMacro{table-numbers_of_programs_and_bugs-20230421-row8-gen-fail}{29}
\DefMacro{table-numbers_of_programs_and_bugs-20230421-row8-gen-timeout}{4}
\DefMacro{table-numbers_of_programs_and_bugs-20230421-row8-gen-no_hole}{607}
\DefMacro{table-numbers_of_programs_and_bugs-20230421-row8-gen-no_gen}{41}
\DefMacro{table-numbers_of_programs_and_bugs-20230421-row8-num_working_templates}{599}
\DefMacro{table-numbers_of_programs_and_bugs-20230421-row8-num_unique_entry_methods_out_of_working_templates}{599}
\DefMacro{table-numbers_of_programs_and_bugs-20230421-row8-num_unique_classes_out_of_working_templates}{139}
\DefMacro{table-numbers_of_programs_and_bugs-20230421-row8-exec-total}{5,405}
\DefMacro{table-numbers_of_programs_and_bugs-20230421-row8-exec-not_compile}{0}
\DefMacro{table-numbers_of_programs_and_bugs-20230421-row8-exec-timeout}{70}
\DefMacro{table-numbers_of_programs_and_bugs-20230421-row8-exec-ok}{5,320}
\DefMacro{table-numbers_of_programs_and_bugs-20230421-row8-exec-skipped_diff}{3}
\DefMacro{table-numbers_of_programs_and_bugs-20230421-row8-exec-skipped_crash}{10}
\DefMacro{table-numbers_of_programs_and_bugs-20230421-row8-exec-final_diff}{1}
\DefMacro{table-numbers_of_programs_and_bugs-20230421-row8-exec-final_crash}{1}
\DefMacro{table-numbers_of_programs_and_bugs-20230421-row9-project}{compress}
\DefMacro{table-numbers_of_programs_and_bugs-20230421-row9-num_unit_tests}{1,939}
\DefMacro{table-numbers_of_programs_and_bugs-20230421-row9-num_unique_entry_methods}{1,939}
\DefMacro{table-numbers_of_programs_and_bugs-20230421-row9-num_unique_classes}{177}
\DefMacro{table-numbers_of_programs_and_bugs-20230421-row9-convert-fail}{0}
\DefMacro{table-numbers_of_programs_and_bugs-20230421-row9-convert-timeout}{3}
\DefMacro{table-numbers_of_programs_and_bugs-20230421-row9-convert-not_compile}{97}
\DefMacro{table-numbers_of_programs_and_bugs-20230421-row9-convert-no_original}{0}
\DefMacro{table-numbers_of_programs_and_bugs-20230421-row9-gen-fail}{9}
\DefMacro{table-numbers_of_programs_and_bugs-20230421-row9-gen-timeout}{0}
\DefMacro{table-numbers_of_programs_and_bugs-20230421-row9-gen-no_hole}{1,055}
\DefMacro{table-numbers_of_programs_and_bugs-20230421-row9-gen-no_gen}{9}
\DefMacro{table-numbers_of_programs_and_bugs-20230421-row9-num_working_templates}{775}
\DefMacro{table-numbers_of_programs_and_bugs-20230421-row9-num_unique_entry_methods_out_of_working_templates}{775}
\DefMacro{table-numbers_of_programs_and_bugs-20230421-row9-num_unique_classes_out_of_working_templates}{132}
\DefMacro{table-numbers_of_programs_and_bugs-20230421-row9-exec-total}{6,777}
\DefMacro{table-numbers_of_programs_and_bugs-20230421-row9-exec-not_compile}{0}
\DefMacro{table-numbers_of_programs_and_bugs-20230421-row9-exec-timeout}{63}
\DefMacro{table-numbers_of_programs_and_bugs-20230421-row9-exec-ok}{6,714}
\DefMacro{table-numbers_of_programs_and_bugs-20230421-row9-exec-skipped_diff}{0}
\DefMacro{table-numbers_of_programs_and_bugs-20230421-row9-exec-skipped_crash}{0}
\DefMacro{table-numbers_of_programs_and_bugs-20230421-row9-exec-final_diff}{0}
\DefMacro{table-numbers_of_programs_and_bugs-20230421-row9-exec-final_crash}{0}
\DefMacro{table-numbers_of_programs_and_bugs-20230421-row10-project}{$\sum$}
\DefMacro{table-numbers_of_programs_and_bugs-20230421-row10-num_unit_tests}{26,048}
\DefMacro{table-numbers_of_programs_and_bugs-20230421-row10-num_unique_entry_methods}{26,042}
\DefMacro{table-numbers_of_programs_and_bugs-20230421-row10-num_unique_classes}{2,044}
\DefMacro{table-numbers_of_programs_and_bugs-20230421-row10-convert-fail}{21}
\DefMacro{table-numbers_of_programs_and_bugs-20230421-row10-convert-timeout}{3,358}
\DefMacro{table-numbers_of_programs_and_bugs-20230421-row10-convert-not_compile}{731}
\DefMacro{table-numbers_of_programs_and_bugs-20230421-row10-convert-no_original}{0}
\DefMacro{table-numbers_of_programs_and_bugs-20230421-row10-gen-fail}{1,587}
\DefMacro{table-numbers_of_programs_and_bugs-20230421-row10-gen-timeout}{346}
\DefMacro{table-numbers_of_programs_and_bugs-20230421-row10-gen-no_hole}{9,645}
\DefMacro{table-numbers_of_programs_and_bugs-20230421-row10-gen-no_gen}{2,214}
\DefMacro{table-numbers_of_programs_and_bugs-20230421-row10-num_working_templates}{10,079}
\DefMacro{table-numbers_of_programs_and_bugs-20230421-row10-num_unique_entry_methods_out_of_working_templates}{10,079}
\DefMacro{table-numbers_of_programs_and_bugs-20230421-row10-num_unique_classes_out_of_working_templates}{1,418}
\DefMacro{table-numbers_of_programs_and_bugs-20230421-row10-exec-total}{89,099}
\DefMacro{table-numbers_of_programs_and_bugs-20230421-row10-exec-not_compile}{0}
\DefMacro{table-numbers_of_programs_and_bugs-20230421-row10-exec-timeout}{712}
\DefMacro{table-numbers_of_programs_and_bugs-20230421-row10-exec-ok}{87,495}
\DefMacro{table-numbers_of_programs_and_bugs-20230421-row10-exec-skipped_diff}{805}
\DefMacro{table-numbers_of_programs_and_bugs-20230421-row10-exec-skipped_crash}{35}
\DefMacro{table-numbers_of_programs_and_bugs-20230421-row10-exec-final_diff}{4}
\DefMacro{table-numbers_of_programs_and_bugs-20230421-row10-exec-final_crash}{48}

\DefMacro{table-numbers_of_programs_and_bugs-20231002-row0-project}{lang}
\DefMacro{table-numbers_of_programs_and_bugs-20231002-row0-num_unit_tests}{1,912}
\DefMacro{table-numbers_of_programs_and_bugs-20231002-row0-num_unique_entry_methods}{1,155}
\DefMacro{table-numbers_of_programs_and_bugs-20231002-row0-num_unique_classes}{97}
\DefMacro{table-numbers_of_programs_and_bugs-20231002-row0-convert-fail}{91}
\DefMacro{table-numbers_of_programs_and_bugs-20231002-row0-convert-timeout}{0}
\DefMacro{table-numbers_of_programs_and_bugs-20231002-row0-convert-not_compile}{1}
\DefMacro{table-numbers_of_programs_and_bugs-20231002-row0-convert-no_original}{0}
\DefMacro{table-numbers_of_programs_and_bugs-20231002-row0-gen-fail}{22}
\DefMacro{table-numbers_of_programs_and_bugs-20231002-row0-gen-timeout}{77}
\DefMacro{table-numbers_of_programs_and_bugs-20231002-row0-gen-no_hole}{51}
\DefMacro{table-numbers_of_programs_and_bugs-20231002-row0-gen-no_gen}{85}
\DefMacro{table-numbers_of_programs_and_bugs-20231002-row0-num_working_templates}{1,684}
\DefMacro{table-numbers_of_programs_and_bugs-20231002-row0-num_unique_entry_methods_out_of_working_templates}{1,001}
\DefMacro{table-numbers_of_programs_and_bugs-20231002-row0-num_unique_classes_out_of_working_templates}{80}
\DefMacro{table-numbers_of_programs_and_bugs-20231002-row0-exec-total}{4,627}
\DefMacro{table-numbers_of_programs_and_bugs-20231002-row0-exec-not_compile}{0}
\DefMacro{table-numbers_of_programs_and_bugs-20231002-row0-exec-timeout}{1}
\DefMacro{table-numbers_of_programs_and_bugs-20231002-row0-exec-ok}{4,336}
\DefMacro{table-numbers_of_programs_and_bugs-20231002-row0-exec-skipped_diff}{276}
\DefMacro{table-numbers_of_programs_and_bugs-20231002-row0-exec-skipped_crash}{10}
\DefMacro{table-numbers_of_programs_and_bugs-20231002-row0-exec-final_diff}{3}
\DefMacro{table-numbers_of_programs_and_bugs-20231002-row0-exec-final_crash}{1}
\DefMacro{table-numbers_of_programs_and_bugs-20231002-row1-project}{math}
\DefMacro{table-numbers_of_programs_and_bugs-20231002-row1-num_unit_tests}{1,076}
\DefMacro{table-numbers_of_programs_and_bugs-20231002-row1-num_unique_entry_methods}{800}
\DefMacro{table-numbers_of_programs_and_bugs-20231002-row1-num_unique_classes}{202}
\DefMacro{table-numbers_of_programs_and_bugs-20231002-row1-convert-fail}{21}
\DefMacro{table-numbers_of_programs_and_bugs-20231002-row1-convert-timeout}{0}
\DefMacro{table-numbers_of_programs_and_bugs-20231002-row1-convert-not_compile}{26}
\DefMacro{table-numbers_of_programs_and_bugs-20231002-row1-convert-no_original}{0}
\DefMacro{table-numbers_of_programs_and_bugs-20231002-row1-gen-fail}{129}
\DefMacro{table-numbers_of_programs_and_bugs-20231002-row1-gen-timeout}{67}
\DefMacro{table-numbers_of_programs_and_bugs-20231002-row1-gen-no_hole}{38}
\DefMacro{table-numbers_of_programs_and_bugs-20231002-row1-gen-no_gen}{190}
\DefMacro{table-numbers_of_programs_and_bugs-20231002-row1-num_working_templates}{801}
\DefMacro{table-numbers_of_programs_and_bugs-20231002-row1-num_unique_entry_methods_out_of_working_templates}{620}
\DefMacro{table-numbers_of_programs_and_bugs-20231002-row1-num_unique_classes_out_of_working_templates}{169}
\DefMacro{table-numbers_of_programs_and_bugs-20231002-row1-exec-total}{2,793}
\DefMacro{table-numbers_of_programs_and_bugs-20231002-row1-exec-not_compile}{0}
\DefMacro{table-numbers_of_programs_and_bugs-20231002-row1-exec-timeout}{9}
\DefMacro{table-numbers_of_programs_and_bugs-20231002-row1-exec-ok}{2,733}
\DefMacro{table-numbers_of_programs_and_bugs-20231002-row1-exec-skipped_diff}{47}
\DefMacro{table-numbers_of_programs_and_bugs-20231002-row1-exec-skipped_crash}{1}
\DefMacro{table-numbers_of_programs_and_bugs-20231002-row1-exec-final_diff}{0}
\DefMacro{table-numbers_of_programs_and_bugs-20231002-row1-exec-final_crash}{3}
\DefMacro{table-numbers_of_programs_and_bugs-20231002-row2-project}{text}
\DefMacro{table-numbers_of_programs_and_bugs-20231002-row2-num_unit_tests}{1,711}
\DefMacro{table-numbers_of_programs_and_bugs-20231002-row2-num_unique_entry_methods}{516}
\DefMacro{table-numbers_of_programs_and_bugs-20231002-row2-num_unique_classes}{38}
\DefMacro{table-numbers_of_programs_and_bugs-20231002-row2-convert-fail}{18}
\DefMacro{table-numbers_of_programs_and_bugs-20231002-row2-convert-timeout}{0}
\DefMacro{table-numbers_of_programs_and_bugs-20231002-row2-convert-not_compile}{0}
\DefMacro{table-numbers_of_programs_and_bugs-20231002-row2-convert-no_original}{0}
\DefMacro{table-numbers_of_programs_and_bugs-20231002-row2-gen-fail}{0}
\DefMacro{table-numbers_of_programs_and_bugs-20231002-row2-gen-timeout}{230}
\DefMacro{table-numbers_of_programs_and_bugs-20231002-row2-gen-no_hole}{10}
\DefMacro{table-numbers_of_programs_and_bugs-20231002-row2-gen-no_gen}{152}
\DefMacro{table-numbers_of_programs_and_bugs-20231002-row2-num_working_templates}{1,531}
\DefMacro{table-numbers_of_programs_and_bugs-20231002-row2-num_unique_entry_methods_out_of_working_templates}{484}
\DefMacro{table-numbers_of_programs_and_bugs-20231002-row2-num_unique_classes_out_of_working_templates}{31}
\DefMacro{table-numbers_of_programs_and_bugs-20231002-row2-exec-total}{7,200}
\DefMacro{table-numbers_of_programs_and_bugs-20231002-row2-exec-not_compile}{0}
\DefMacro{table-numbers_of_programs_and_bugs-20231002-row2-exec-timeout}{30}
\DefMacro{table-numbers_of_programs_and_bugs-20231002-row2-exec-ok}{7,031}
\DefMacro{table-numbers_of_programs_and_bugs-20231002-row2-exec-skipped_diff}{133}
\DefMacro{table-numbers_of_programs_and_bugs-20231002-row2-exec-skipped_crash}{4}
\DefMacro{table-numbers_of_programs_and_bugs-20231002-row2-exec-final_diff}{2}
\DefMacro{table-numbers_of_programs_and_bugs-20231002-row2-exec-final_crash}{0}
\DefMacro{table-numbers_of_programs_and_bugs-20231002-row3-project}{numbers}
\DefMacro{table-numbers_of_programs_and_bugs-20231002-row3-num_unit_tests}{3,064}
\DefMacro{table-numbers_of_programs_and_bugs-20231002-row3-num_unique_entry_methods}{44}
\DefMacro{table-numbers_of_programs_and_bugs-20231002-row3-num_unique_classes}{4}
\DefMacro{table-numbers_of_programs_and_bugs-20231002-row3-convert-fail}{121}
\DefMacro{table-numbers_of_programs_and_bugs-20231002-row3-convert-timeout}{0}
\DefMacro{table-numbers_of_programs_and_bugs-20231002-row3-convert-not_compile}{0}
\DefMacro{table-numbers_of_programs_and_bugs-20231002-row3-convert-no_original}{0}
\DefMacro{table-numbers_of_programs_and_bugs-20231002-row3-gen-fail}{0}
\DefMacro{table-numbers_of_programs_and_bugs-20231002-row3-gen-timeout}{4}
\DefMacro{table-numbers_of_programs_and_bugs-20231002-row3-gen-no_hole}{0}
\DefMacro{table-numbers_of_programs_and_bugs-20231002-row3-gen-no_gen}{0}
\DefMacro{table-numbers_of_programs_and_bugs-20231002-row3-num_working_templates}{2,943}
\DefMacro{table-numbers_of_programs_and_bugs-20231002-row3-num_unique_entry_methods_out_of_working_templates}{41}
\DefMacro{table-numbers_of_programs_and_bugs-20231002-row3-num_unique_classes_out_of_working_templates}{3}
\DefMacro{table-numbers_of_programs_and_bugs-20231002-row3-exec-total}{19,405}
\DefMacro{table-numbers_of_programs_and_bugs-20231002-row3-exec-not_compile}{0}
\DefMacro{table-numbers_of_programs_and_bugs-20231002-row3-exec-timeout}{7}
\DefMacro{table-numbers_of_programs_and_bugs-20231002-row3-exec-ok}{19,398}
\DefMacro{table-numbers_of_programs_and_bugs-20231002-row3-exec-skipped_diff}{0}
\DefMacro{table-numbers_of_programs_and_bugs-20231002-row3-exec-skipped_crash}{0}
\DefMacro{table-numbers_of_programs_and_bugs-20231002-row3-exec-final_diff}{0}
\DefMacro{table-numbers_of_programs_and_bugs-20231002-row3-exec-final_crash}{0}
\DefMacro{table-numbers_of_programs_and_bugs-20231002-row4-project}{vectorz}
\DefMacro{table-numbers_of_programs_and_bugs-20231002-row4-num_unit_tests}{1,093}
\DefMacro{table-numbers_of_programs_and_bugs-20231002-row4-num_unique_entry_methods}{868}
\DefMacro{table-numbers_of_programs_and_bugs-20231002-row4-num_unique_classes}{152}
\DefMacro{table-numbers_of_programs_and_bugs-20231002-row4-convert-fail}{10}
\DefMacro{table-numbers_of_programs_and_bugs-20231002-row4-convert-timeout}{0}
\DefMacro{table-numbers_of_programs_and_bugs-20231002-row4-convert-not_compile}{0}
\DefMacro{table-numbers_of_programs_and_bugs-20231002-row4-convert-no_original}{0}
\DefMacro{table-numbers_of_programs_and_bugs-20231002-row4-gen-fail}{29}
\DefMacro{table-numbers_of_programs_and_bugs-20231002-row4-gen-timeout}{15}
\DefMacro{table-numbers_of_programs_and_bugs-20231002-row4-gen-no_hole}{94}
\DefMacro{table-numbers_of_programs_and_bugs-20231002-row4-gen-no_gen}{182}
\DefMacro{table-numbers_of_programs_and_bugs-20231002-row4-num_working_templates}{807}
\DefMacro{table-numbers_of_programs_and_bugs-20231002-row4-num_unique_entry_methods_out_of_working_templates}{635}
\DefMacro{table-numbers_of_programs_and_bugs-20231002-row4-num_unique_classes_out_of_working_templates}{140}
\DefMacro{table-numbers_of_programs_and_bugs-20231002-row4-exec-total}{1,838}
\DefMacro{table-numbers_of_programs_and_bugs-20231002-row4-exec-not_compile}{0}
\DefMacro{table-numbers_of_programs_and_bugs-20231002-row4-exec-timeout}{37}
\DefMacro{table-numbers_of_programs_and_bugs-20231002-row4-exec-ok}{1,758}
\DefMacro{table-numbers_of_programs_and_bugs-20231002-row4-exec-skipped_diff}{30}
\DefMacro{table-numbers_of_programs_and_bugs-20231002-row4-exec-skipped_crash}{12}
\DefMacro{table-numbers_of_programs_and_bugs-20231002-row4-exec-final_diff}{0}
\DefMacro{table-numbers_of_programs_and_bugs-20231002-row4-exec-final_crash}{1}
\DefMacro{table-numbers_of_programs_and_bugs-20231002-row5-project}{statistics}
\DefMacro{table-numbers_of_programs_and_bugs-20231002-row5-num_unit_tests}{1,924}
\DefMacro{table-numbers_of_programs_and_bugs-20231002-row5-num_unique_entry_methods}{368}
\DefMacro{table-numbers_of_programs_and_bugs-20231002-row5-num_unique_classes}{26}
\DefMacro{table-numbers_of_programs_and_bugs-20231002-row5-convert-fail}{191}
\DefMacro{table-numbers_of_programs_and_bugs-20231002-row5-convert-timeout}{0}
\DefMacro{table-numbers_of_programs_and_bugs-20231002-row5-convert-not_compile}{0}
\DefMacro{table-numbers_of_programs_and_bugs-20231002-row5-convert-no_original}{0}
\DefMacro{table-numbers_of_programs_and_bugs-20231002-row5-gen-fail}{150}
\DefMacro{table-numbers_of_programs_and_bugs-20231002-row5-gen-timeout}{25}
\DefMacro{table-numbers_of_programs_and_bugs-20231002-row5-gen-no_hole}{0}
\DefMacro{table-numbers_of_programs_and_bugs-20231002-row5-gen-no_gen}{151}
\DefMacro{table-numbers_of_programs_and_bugs-20231002-row5-num_working_templates}{1,582}
\DefMacro{table-numbers_of_programs_and_bugs-20231002-row5-num_unique_entry_methods_out_of_working_templates}{295}
\DefMacro{table-numbers_of_programs_and_bugs-20231002-row5-num_unique_classes_out_of_working_templates}{24}
\DefMacro{table-numbers_of_programs_and_bugs-20231002-row5-exec-total}{12,800}
\DefMacro{table-numbers_of_programs_and_bugs-20231002-row5-exec-not_compile}{0}
\DefMacro{table-numbers_of_programs_and_bugs-20231002-row5-exec-timeout}{0}
\DefMacro{table-numbers_of_programs_and_bugs-20231002-row5-exec-ok}{12,467}
\DefMacro{table-numbers_of_programs_and_bugs-20231002-row5-exec-skipped_diff}{333}
\DefMacro{table-numbers_of_programs_and_bugs-20231002-row5-exec-skipped_crash}{0}
\DefMacro{table-numbers_of_programs_and_bugs-20231002-row5-exec-final_diff}{0}
\DefMacro{table-numbers_of_programs_and_bugs-20231002-row5-exec-final_crash}{0}
\DefMacro{table-numbers_of_programs_and_bugs-20231002-row6-project}{codec}
\DefMacro{table-numbers_of_programs_and_bugs-20231002-row6-num_unit_tests}{1,497}
\DefMacro{table-numbers_of_programs_and_bugs-20231002-row6-num_unique_entry_methods}{394}
\DefMacro{table-numbers_of_programs_and_bugs-20231002-row6-num_unique_classes}{49}
\DefMacro{table-numbers_of_programs_and_bugs-20231002-row6-convert-fail}{19}
\DefMacro{table-numbers_of_programs_and_bugs-20231002-row6-convert-timeout}{0}
\DefMacro{table-numbers_of_programs_and_bugs-20231002-row6-convert-not_compile}{0}
\DefMacro{table-numbers_of_programs_and_bugs-20231002-row6-convert-no_original}{0}
\DefMacro{table-numbers_of_programs_and_bugs-20231002-row6-gen-fail}{62}
\DefMacro{table-numbers_of_programs_and_bugs-20231002-row6-gen-timeout}{3}
\DefMacro{table-numbers_of_programs_and_bugs-20231002-row6-gen-no_hole}{2}
\DefMacro{table-numbers_of_programs_and_bugs-20231002-row6-gen-no_gen}{129}
\DefMacro{table-numbers_of_programs_and_bugs-20231002-row6-num_working_templates}{1,347}
\DefMacro{table-numbers_of_programs_and_bugs-20231002-row6-num_unique_entry_methods_out_of_working_templates}{359}
\DefMacro{table-numbers_of_programs_and_bugs-20231002-row6-num_unique_classes_out_of_working_templates}{42}
\DefMacro{table-numbers_of_programs_and_bugs-20231002-row6-exec-total}{4,973}
\DefMacro{table-numbers_of_programs_and_bugs-20231002-row6-exec-not_compile}{0}
\DefMacro{table-numbers_of_programs_and_bugs-20231002-row6-exec-timeout}{110}
\DefMacro{table-numbers_of_programs_and_bugs-20231002-row6-exec-ok}{4,615}
\DefMacro{table-numbers_of_programs_and_bugs-20231002-row6-exec-skipped_diff}{205}
\DefMacro{table-numbers_of_programs_and_bugs-20231002-row6-exec-skipped_crash}{20}
\DefMacro{table-numbers_of_programs_and_bugs-20231002-row6-exec-final_diff}{23}
\DefMacro{table-numbers_of_programs_and_bugs-20231002-row6-exec-final_crash}{0}
\DefMacro{table-numbers_of_programs_and_bugs-20231002-row7-project}{jfreechart}
\DefMacro{table-numbers_of_programs_and_bugs-20231002-row7-num_unit_tests}{1,244}
\DefMacro{table-numbers_of_programs_and_bugs-20231002-row7-num_unique_entry_methods}{1,065}
\DefMacro{table-numbers_of_programs_and_bugs-20231002-row7-num_unique_classes}{275}
\DefMacro{table-numbers_of_programs_and_bugs-20231002-row7-convert-fail}{5}
\DefMacro{table-numbers_of_programs_and_bugs-20231002-row7-convert-timeout}{0}
\DefMacro{table-numbers_of_programs_and_bugs-20231002-row7-convert-not_compile}{8}
\DefMacro{table-numbers_of_programs_and_bugs-20231002-row7-convert-no_original}{0}
\DefMacro{table-numbers_of_programs_and_bugs-20231002-row7-gen-fail}{236}
\DefMacro{table-numbers_of_programs_and_bugs-20231002-row7-gen-timeout}{8}
\DefMacro{table-numbers_of_programs_and_bugs-20231002-row7-gen-no_hole}{21}
\DefMacro{table-numbers_of_programs_and_bugs-20231002-row7-gen-no_gen}{327}
\DefMacro{table-numbers_of_programs_and_bugs-20231002-row7-num_working_templates}{883}
\DefMacro{table-numbers_of_programs_and_bugs-20231002-row7-num_unique_entry_methods_out_of_working_templates}{762}
\DefMacro{table-numbers_of_programs_and_bugs-20231002-row7-num_unique_classes_out_of_working_templates}{236}
\DefMacro{table-numbers_of_programs_and_bugs-20231002-row7-exec-total}{1,411}
\DefMacro{table-numbers_of_programs_and_bugs-20231002-row7-exec-not_compile}{0}
\DefMacro{table-numbers_of_programs_and_bugs-20231002-row7-exec-timeout}{18}
\DefMacro{table-numbers_of_programs_and_bugs-20231002-row7-exec-ok}{1,347}
\DefMacro{table-numbers_of_programs_and_bugs-20231002-row7-exec-skipped_diff}{43}
\DefMacro{table-numbers_of_programs_and_bugs-20231002-row7-exec-skipped_crash}{0}
\DefMacro{table-numbers_of_programs_and_bugs-20231002-row7-exec-final_diff}{3}
\DefMacro{table-numbers_of_programs_and_bugs-20231002-row7-exec-final_crash}{0}
\DefMacro{table-numbers_of_programs_and_bugs-20231002-row8-project}{zxing}
\DefMacro{table-numbers_of_programs_and_bugs-20231002-row8-num_unit_tests}{1,432}
\DefMacro{table-numbers_of_programs_and_bugs-20231002-row8-num_unique_entry_methods}{318}
\DefMacro{table-numbers_of_programs_and_bugs-20231002-row8-num_unique_classes}{88}
\DefMacro{table-numbers_of_programs_and_bugs-20231002-row8-convert-fail}{58}
\DefMacro{table-numbers_of_programs_and_bugs-20231002-row8-convert-timeout}{0}
\DefMacro{table-numbers_of_programs_and_bugs-20231002-row8-convert-not_compile}{29}
\DefMacro{table-numbers_of_programs_and_bugs-20231002-row8-convert-no_original}{0}
\DefMacro{table-numbers_of_programs_and_bugs-20231002-row8-gen-fail}{71}
\DefMacro{table-numbers_of_programs_and_bugs-20231002-row8-gen-timeout}{11}
\DefMacro{table-numbers_of_programs_and_bugs-20231002-row8-gen-no_hole}{21}
\DefMacro{table-numbers_of_programs_and_bugs-20231002-row8-gen-no_gen}{77}
\DefMacro{table-numbers_of_programs_and_bugs-20231002-row8-num_working_templates}{1,247}
\DefMacro{table-numbers_of_programs_and_bugs-20231002-row8-num_unique_entry_methods_out_of_working_templates}{286}
\DefMacro{table-numbers_of_programs_and_bugs-20231002-row8-num_unique_classes_out_of_working_templates}{74}
\DefMacro{table-numbers_of_programs_and_bugs-20231002-row8-exec-total}{3,441}
\DefMacro{table-numbers_of_programs_and_bugs-20231002-row8-exec-not_compile}{0}
\DefMacro{table-numbers_of_programs_and_bugs-20231002-row8-exec-timeout}{16}
\DefMacro{table-numbers_of_programs_and_bugs-20231002-row8-exec-ok}{3,387}
\DefMacro{table-numbers_of_programs_and_bugs-20231002-row8-exec-skipped_diff}{36}
\DefMacro{table-numbers_of_programs_and_bugs-20231002-row8-exec-skipped_crash}{0}
\DefMacro{table-numbers_of_programs_and_bugs-20231002-row8-exec-final_diff}{1}
\DefMacro{table-numbers_of_programs_and_bugs-20231002-row8-exec-final_crash}{1}
\DefMacro{table-numbers_of_programs_and_bugs-20231002-row9-project}{compress}
\DefMacro{table-numbers_of_programs_and_bugs-20231002-row9-num_unit_tests}{1,059}
\DefMacro{table-numbers_of_programs_and_bugs-20231002-row9-num_unique_entry_methods}{565}
\DefMacro{table-numbers_of_programs_and_bugs-20231002-row9-num_unique_classes}{112}
\DefMacro{table-numbers_of_programs_and_bugs-20231002-row9-convert-fail}{10}
\DefMacro{table-numbers_of_programs_and_bugs-20231002-row9-convert-timeout}{0}
\DefMacro{table-numbers_of_programs_and_bugs-20231002-row9-convert-not_compile}{11}
\DefMacro{table-numbers_of_programs_and_bugs-20231002-row9-convert-no_original}{0}
\DefMacro{table-numbers_of_programs_and_bugs-20231002-row9-gen-fail}{14}
\DefMacro{table-numbers_of_programs_and_bugs-20231002-row9-gen-timeout}{12}
\DefMacro{table-numbers_of_programs_and_bugs-20231002-row9-gen-no_hole}{58}
\DefMacro{table-numbers_of_programs_and_bugs-20231002-row9-gen-no_gen}{199}
\DefMacro{table-numbers_of_programs_and_bugs-20231002-row9-num_working_templates}{781}
\DefMacro{table-numbers_of_programs_and_bugs-20231002-row9-num_unique_entry_methods_out_of_working_templates}{436}
\DefMacro{table-numbers_of_programs_and_bugs-20231002-row9-num_unique_classes_out_of_working_templates}{101}
\DefMacro{table-numbers_of_programs_and_bugs-20231002-row9-exec-total}{1,429}
\DefMacro{table-numbers_of_programs_and_bugs-20231002-row9-exec-not_compile}{0}
\DefMacro{table-numbers_of_programs_and_bugs-20231002-row9-exec-timeout}{0}
\DefMacro{table-numbers_of_programs_and_bugs-20231002-row9-exec-ok}{1,317}
\DefMacro{table-numbers_of_programs_and_bugs-20231002-row9-exec-skipped_diff}{111}
\DefMacro{table-numbers_of_programs_and_bugs-20231002-row9-exec-skipped_crash}{0}
\DefMacro{table-numbers_of_programs_and_bugs-20231002-row9-exec-final_diff}{1}
\DefMacro{table-numbers_of_programs_and_bugs-20231002-row9-exec-final_crash}{0}
\DefMacro{table-numbers_of_programs_and_bugs-20231002-row10-project}{$\sum$}
\DefMacro{table-numbers_of_programs_and_bugs-20231002-row10-num_unit_tests}{16,012}
\DefMacro{table-numbers_of_programs_and_bugs-20231002-row10-num_unique_entry_methods}{6,093}
\DefMacro{table-numbers_of_programs_and_bugs-20231002-row10-num_unique_classes}{1,043}
\DefMacro{table-numbers_of_programs_and_bugs-20231002-row10-convert-fail}{544}
\DefMacro{table-numbers_of_programs_and_bugs-20231002-row10-convert-timeout}{0}
\DefMacro{table-numbers_of_programs_and_bugs-20231002-row10-convert-not_compile}{75}
\DefMacro{table-numbers_of_programs_and_bugs-20231002-row10-convert-no_original}{0}
\DefMacro{table-numbers_of_programs_and_bugs-20231002-row10-gen-fail}{713}
\DefMacro{table-numbers_of_programs_and_bugs-20231002-row10-gen-timeout}{452}
\DefMacro{table-numbers_of_programs_and_bugs-20231002-row10-gen-no_hole}{295}
\DefMacro{table-numbers_of_programs_and_bugs-20231002-row10-gen-no_gen}{1,492}
\DefMacro{table-numbers_of_programs_and_bugs-20231002-row10-num_working_templates}{13,606}
\DefMacro{table-numbers_of_programs_and_bugs-20231002-row10-num_unique_entry_methods_out_of_working_templates}{4,919}
\DefMacro{table-numbers_of_programs_and_bugs-20231002-row10-num_unique_classes_out_of_working_templates}{900}
\DefMacro{table-numbers_of_programs_and_bugs-20231002-row10-exec-total}{59,917}
\DefMacro{table-numbers_of_programs_and_bugs-20231002-row10-exec-not_compile}{0}
\DefMacro{table-numbers_of_programs_and_bugs-20231002-row10-exec-timeout}{228}
\DefMacro{table-numbers_of_programs_and_bugs-20231002-row10-exec-ok}{58,389}
\DefMacro{table-numbers_of_programs_and_bugs-20231002-row10-exec-skipped_diff}{1,214}
\DefMacro{table-numbers_of_programs_and_bugs-20231002-row10-exec-skipped_crash}{47}
\DefMacro{table-numbers_of_programs_and_bugs-20231002-row10-exec-final_diff}{33}
\DefMacro{table-numbers_of_programs_and_bugs-20231002-row10-exec-final_crash}{6}

\DefMacro{table-numbers_of_programs_and_bugs-20230930-row0-project}{lang}
\DefMacro{table-numbers_of_programs_and_bugs-20230930-row0-num_unit_tests}{1,912}
\DefMacro{table-numbers_of_programs_and_bugs-20230930-row0-num_unique_entry_methods}{1,155}
\DefMacro{table-numbers_of_programs_and_bugs-20230930-row0-num_unique_classes}{97}
\DefMacro{table-numbers_of_programs_and_bugs-20230930-row0-convert-fail}{91}
\DefMacro{table-numbers_of_programs_and_bugs-20230930-row0-convert-timeout}{0}
\DefMacro{table-numbers_of_programs_and_bugs-20230930-row0-convert-not_compile}{1}
\DefMacro{table-numbers_of_programs_and_bugs-20230930-row0-convert-no_original}{0}
\DefMacro{table-numbers_of_programs_and_bugs-20230930-row0-gen-fail}{54}
\DefMacro{table-numbers_of_programs_and_bugs-20230930-row0-gen-timeout}{60}
\DefMacro{table-numbers_of_programs_and_bugs-20230930-row0-gen-no_hole}{53}
\DefMacro{table-numbers_of_programs_and_bugs-20230930-row0-gen-no_gen}{295}
\DefMacro{table-numbers_of_programs_and_bugs-20230930-row0-num_working_templates}{1,472}
\DefMacro{table-numbers_of_programs_and_bugs-20230930-row0-num_unique_entry_methods_out_of_working_templates}{869}
\DefMacro{table-numbers_of_programs_and_bugs-20230930-row0-num_unique_classes_out_of_working_templates}{79}
\DefMacro{table-numbers_of_programs_and_bugs-20230930-row0-exec-total}{12,368}
\DefMacro{table-numbers_of_programs_and_bugs-20230930-row0-exec-not_compile}{0}
\DefMacro{table-numbers_of_programs_and_bugs-20230930-row0-exec-timeout}{96}
\DefMacro{table-numbers_of_programs_and_bugs-20230930-row0-exec-ok}{12,090}
\DefMacro{table-numbers_of_programs_and_bugs-20230930-row0-exec-skipped_diff}{157}
\DefMacro{table-numbers_of_programs_and_bugs-20230930-row0-exec-skipped_crash}{13}
\DefMacro{table-numbers_of_programs_and_bugs-20230930-row0-exec-final_diff}{3}
\DefMacro{table-numbers_of_programs_and_bugs-20230930-row0-exec-final_crash}{9}
\DefMacro{table-numbers_of_programs_and_bugs-20230930-row1-project}{math}
\DefMacro{table-numbers_of_programs_and_bugs-20230930-row1-num_unit_tests}{1,076}
\DefMacro{table-numbers_of_programs_and_bugs-20230930-row1-num_unique_entry_methods}{800}
\DefMacro{table-numbers_of_programs_and_bugs-20230930-row1-num_unique_classes}{202}
\DefMacro{table-numbers_of_programs_and_bugs-20230930-row1-convert-fail}{21}
\DefMacro{table-numbers_of_programs_and_bugs-20230930-row1-convert-timeout}{0}
\DefMacro{table-numbers_of_programs_and_bugs-20230930-row1-convert-not_compile}{26}
\DefMacro{table-numbers_of_programs_and_bugs-20230930-row1-convert-no_original}{0}
\DefMacro{table-numbers_of_programs_and_bugs-20230930-row1-gen-fail}{169}
\DefMacro{table-numbers_of_programs_and_bugs-20230930-row1-gen-timeout}{63}
\DefMacro{table-numbers_of_programs_and_bugs-20230930-row1-gen-no_hole}{38}
\DefMacro{table-numbers_of_programs_and_bugs-20230930-row1-gen-no_gen}{335}
\DefMacro{table-numbers_of_programs_and_bugs-20230930-row1-num_working_templates}{656}
\DefMacro{table-numbers_of_programs_and_bugs-20230930-row1-num_unique_entry_methods_out_of_working_templates}{502}
\DefMacro{table-numbers_of_programs_and_bugs-20230930-row1-num_unique_classes_out_of_working_templates}{155}
\DefMacro{table-numbers_of_programs_and_bugs-20230930-row1-exec-total}{5,387}
\DefMacro{table-numbers_of_programs_and_bugs-20230930-row1-exec-not_compile}{0}
\DefMacro{table-numbers_of_programs_and_bugs-20230930-row1-exec-timeout}{52}
\DefMacro{table-numbers_of_programs_and_bugs-20230930-row1-exec-ok}{5,228}
\DefMacro{table-numbers_of_programs_and_bugs-20230930-row1-exec-skipped_diff}{100}
\DefMacro{table-numbers_of_programs_and_bugs-20230930-row1-exec-skipped_crash}{0}
\DefMacro{table-numbers_of_programs_and_bugs-20230930-row1-exec-final_diff}{0}
\DefMacro{table-numbers_of_programs_and_bugs-20230930-row1-exec-final_crash}{7}
\DefMacro{table-numbers_of_programs_and_bugs-20230930-row2-project}{text}
\DefMacro{table-numbers_of_programs_and_bugs-20230930-row2-num_unit_tests}{1,711}
\DefMacro{table-numbers_of_programs_and_bugs-20230930-row2-num_unique_entry_methods}{516}
\DefMacro{table-numbers_of_programs_and_bugs-20230930-row2-num_unique_classes}{38}
\DefMacro{table-numbers_of_programs_and_bugs-20230930-row2-convert-fail}{18}
\DefMacro{table-numbers_of_programs_and_bugs-20230930-row2-convert-timeout}{0}
\DefMacro{table-numbers_of_programs_and_bugs-20230930-row2-convert-not_compile}{0}
\DefMacro{table-numbers_of_programs_and_bugs-20230930-row2-convert-no_original}{0}
\DefMacro{table-numbers_of_programs_and_bugs-20230930-row2-gen-fail}{5}
\DefMacro{table-numbers_of_programs_and_bugs-20230930-row2-gen-timeout}{62}
\DefMacro{table-numbers_of_programs_and_bugs-20230930-row2-gen-no_hole}{61}
\DefMacro{table-numbers_of_programs_and_bugs-20230930-row2-gen-no_gen}{404}
\DefMacro{table-numbers_of_programs_and_bugs-20230930-row2-num_working_templates}{1,228}
\DefMacro{table-numbers_of_programs_and_bugs-20230930-row2-num_unique_entry_methods_out_of_working_templates}{422}
\DefMacro{table-numbers_of_programs_and_bugs-20230930-row2-num_unique_classes_out_of_working_templates}{30}
\DefMacro{table-numbers_of_programs_and_bugs-20230930-row2-exec-total}{10,712}
\DefMacro{table-numbers_of_programs_and_bugs-20230930-row2-exec-not_compile}{0}
\DefMacro{table-numbers_of_programs_and_bugs-20230930-row2-exec-timeout}{75}
\DefMacro{table-numbers_of_programs_and_bugs-20230930-row2-exec-ok}{10,567}
\DefMacro{table-numbers_of_programs_and_bugs-20230930-row2-exec-skipped_diff}{70}
\DefMacro{table-numbers_of_programs_and_bugs-20230930-row2-exec-skipped_crash}{0}
\DefMacro{table-numbers_of_programs_and_bugs-20230930-row2-exec-final_diff}{0}
\DefMacro{table-numbers_of_programs_and_bugs-20230930-row2-exec-final_crash}{0}
\DefMacro{table-numbers_of_programs_and_bugs-20230930-row3-project}{numbers}
\DefMacro{table-numbers_of_programs_and_bugs-20230930-row3-num_unit_tests}{3,064}
\DefMacro{table-numbers_of_programs_and_bugs-20230930-row3-num_unique_entry_methods}{44}
\DefMacro{table-numbers_of_programs_and_bugs-20230930-row3-num_unique_classes}{4}
\DefMacro{table-numbers_of_programs_and_bugs-20230930-row3-convert-fail}{121}
\DefMacro{table-numbers_of_programs_and_bugs-20230930-row3-convert-timeout}{0}
\DefMacro{table-numbers_of_programs_and_bugs-20230930-row3-convert-not_compile}{0}
\DefMacro{table-numbers_of_programs_and_bugs-20230930-row3-convert-no_original}{0}
\DefMacro{table-numbers_of_programs_and_bugs-20230930-row3-gen-fail}{0}
\DefMacro{table-numbers_of_programs_and_bugs-20230930-row3-gen-timeout}{9}
\DefMacro{table-numbers_of_programs_and_bugs-20230930-row3-gen-no_hole}{0}
\DefMacro{table-numbers_of_programs_and_bugs-20230930-row3-gen-no_gen}{77}
\DefMacro{table-numbers_of_programs_and_bugs-20230930-row3-num_working_templates}{2,866}
\DefMacro{table-numbers_of_programs_and_bugs-20230930-row3-num_unique_entry_methods_out_of_working_templates}{41}
\DefMacro{table-numbers_of_programs_and_bugs-20230930-row3-num_unique_classes_out_of_working_templates}{3}
\DefMacro{table-numbers_of_programs_and_bugs-20230930-row3-exec-total}{28,647}
\DefMacro{table-numbers_of_programs_and_bugs-20230930-row3-exec-not_compile}{0}
\DefMacro{table-numbers_of_programs_and_bugs-20230930-row3-exec-timeout}{16}
\DefMacro{table-numbers_of_programs_and_bugs-20230930-row3-exec-ok}{28,631}
\DefMacro{table-numbers_of_programs_and_bugs-20230930-row3-exec-skipped_diff}{0}
\DefMacro{table-numbers_of_programs_and_bugs-20230930-row3-exec-skipped_crash}{0}
\DefMacro{table-numbers_of_programs_and_bugs-20230930-row3-exec-final_diff}{0}
\DefMacro{table-numbers_of_programs_and_bugs-20230930-row3-exec-final_crash}{0}
\DefMacro{table-numbers_of_programs_and_bugs-20230930-row4-project}{vectorz}
\DefMacro{table-numbers_of_programs_and_bugs-20230930-row4-num_unit_tests}{1,093}
\DefMacro{table-numbers_of_programs_and_bugs-20230930-row4-num_unique_entry_methods}{868}
\DefMacro{table-numbers_of_programs_and_bugs-20230930-row4-num_unique_classes}{152}
\DefMacro{table-numbers_of_programs_and_bugs-20230930-row4-convert-fail}{10}
\DefMacro{table-numbers_of_programs_and_bugs-20230930-row4-convert-timeout}{0}
\DefMacro{table-numbers_of_programs_and_bugs-20230930-row4-convert-not_compile}{0}
\DefMacro{table-numbers_of_programs_and_bugs-20230930-row4-convert-no_original}{0}
\DefMacro{table-numbers_of_programs_and_bugs-20230930-row4-gen-fail}{49}
\DefMacro{table-numbers_of_programs_and_bugs-20230930-row4-gen-timeout}{20}
\DefMacro{table-numbers_of_programs_and_bugs-20230930-row4-gen-no_hole}{94}
\DefMacro{table-numbers_of_programs_and_bugs-20230930-row4-gen-no_gen}{204}
\DefMacro{table-numbers_of_programs_and_bugs-20230930-row4-num_working_templates}{785}
\DefMacro{table-numbers_of_programs_and_bugs-20230930-row4-num_unique_entry_methods_out_of_working_templates}{620}
\DefMacro{table-numbers_of_programs_and_bugs-20230930-row4-num_unique_classes_out_of_working_templates}{135}
\DefMacro{table-numbers_of_programs_and_bugs-20230930-row4-exec-total}{6,255}
\DefMacro{table-numbers_of_programs_and_bugs-20230930-row4-exec-not_compile}{0}
\DefMacro{table-numbers_of_programs_and_bugs-20230930-row4-exec-timeout}{185}
\DefMacro{table-numbers_of_programs_and_bugs-20230930-row4-exec-ok}{5,989}
\DefMacro{table-numbers_of_programs_and_bugs-20230930-row4-exec-skipped_diff}{74}
\DefMacro{table-numbers_of_programs_and_bugs-20230930-row4-exec-skipped_crash}{0}
\DefMacro{table-numbers_of_programs_and_bugs-20230930-row4-exec-final_diff}{5}
\DefMacro{table-numbers_of_programs_and_bugs-20230930-row4-exec-final_crash}{2}
\DefMacro{table-numbers_of_programs_and_bugs-20230930-row5-project}{statistics}
\DefMacro{table-numbers_of_programs_and_bugs-20230930-row5-num_unit_tests}{1,924}
\DefMacro{table-numbers_of_programs_and_bugs-20230930-row5-num_unique_entry_methods}{368}
\DefMacro{table-numbers_of_programs_and_bugs-20230930-row5-num_unique_classes}{26}
\DefMacro{table-numbers_of_programs_and_bugs-20230930-row5-convert-fail}{191}
\DefMacro{table-numbers_of_programs_and_bugs-20230930-row5-convert-timeout}{0}
\DefMacro{table-numbers_of_programs_and_bugs-20230930-row5-convert-not_compile}{0}
\DefMacro{table-numbers_of_programs_and_bugs-20230930-row5-convert-no_original}{0}
\DefMacro{table-numbers_of_programs_and_bugs-20230930-row5-gen-fail}{150}
\DefMacro{table-numbers_of_programs_and_bugs-20230930-row5-gen-timeout}{59}
\DefMacro{table-numbers_of_programs_and_bugs-20230930-row5-gen-no_hole}{0}
\DefMacro{table-numbers_of_programs_and_bugs-20230930-row5-gen-no_gen}{671}
\DefMacro{table-numbers_of_programs_and_bugs-20230930-row5-num_working_templates}{1,062}
\DefMacro{table-numbers_of_programs_and_bugs-20230930-row5-num_unique_entry_methods_out_of_working_templates}{232}
\DefMacro{table-numbers_of_programs_and_bugs-20230930-row5-num_unique_classes_out_of_working_templates}{24}
\DefMacro{table-numbers_of_programs_and_bugs-20230930-row5-exec-total}{8,567}
\DefMacro{table-numbers_of_programs_and_bugs-20230930-row5-exec-not_compile}{0}
\DefMacro{table-numbers_of_programs_and_bugs-20230930-row5-exec-timeout}{21}
\DefMacro{table-numbers_of_programs_and_bugs-20230930-row5-exec-ok}{8,485}
\DefMacro{table-numbers_of_programs_and_bugs-20230930-row5-exec-skipped_diff}{61}
\DefMacro{table-numbers_of_programs_and_bugs-20230930-row5-exec-skipped_crash}{0}
\DefMacro{table-numbers_of_programs_and_bugs-20230930-row5-exec-final_diff}{0}
\DefMacro{table-numbers_of_programs_and_bugs-20230930-row5-exec-final_crash}{0}
\DefMacro{table-numbers_of_programs_and_bugs-20230930-row6-project}{codec}
\DefMacro{table-numbers_of_programs_and_bugs-20230930-row6-num_unit_tests}{1,497}
\DefMacro{table-numbers_of_programs_and_bugs-20230930-row6-num_unique_entry_methods}{394}
\DefMacro{table-numbers_of_programs_and_bugs-20230930-row6-num_unique_classes}{49}
\DefMacro{table-numbers_of_programs_and_bugs-20230930-row6-convert-fail}{19}
\DefMacro{table-numbers_of_programs_and_bugs-20230930-row6-convert-timeout}{0}
\DefMacro{table-numbers_of_programs_and_bugs-20230930-row6-convert-not_compile}{0}
\DefMacro{table-numbers_of_programs_and_bugs-20230930-row6-convert-no_original}{0}
\DefMacro{table-numbers_of_programs_and_bugs-20230930-row6-gen-fail}{85}
\DefMacro{table-numbers_of_programs_and_bugs-20230930-row6-gen-timeout}{34}
\DefMacro{table-numbers_of_programs_and_bugs-20230930-row6-gen-no_hole}{1}
\DefMacro{table-numbers_of_programs_and_bugs-20230930-row6-gen-no_gen}{320}
\DefMacro{table-numbers_of_programs_and_bugs-20230930-row6-num_working_templates}{1,157}
\DefMacro{table-numbers_of_programs_and_bugs-20230930-row6-num_unique_entry_methods_out_of_working_templates}{304}
\DefMacro{table-numbers_of_programs_and_bugs-20230930-row6-num_unique_classes_out_of_working_templates}{37}
\DefMacro{table-numbers_of_programs_and_bugs-20230930-row6-exec-total}{6,355}
\DefMacro{table-numbers_of_programs_and_bugs-20230930-row6-exec-not_compile}{0}
\DefMacro{table-numbers_of_programs_and_bugs-20230930-row6-exec-timeout}{209}
\DefMacro{table-numbers_of_programs_and_bugs-20230930-row6-exec-ok}{5,936}
\DefMacro{table-numbers_of_programs_and_bugs-20230930-row6-exec-skipped_diff}{171}
\DefMacro{table-numbers_of_programs_and_bugs-20230930-row6-exec-skipped_crash}{22}
\DefMacro{table-numbers_of_programs_and_bugs-20230930-row6-exec-final_diff}{17}
\DefMacro{table-numbers_of_programs_and_bugs-20230930-row6-exec-final_crash}{0}
\DefMacro{table-numbers_of_programs_and_bugs-20230930-row7-project}{jfreechart}
\DefMacro{table-numbers_of_programs_and_bugs-20230930-row7-num_unit_tests}{1,244}
\DefMacro{table-numbers_of_programs_and_bugs-20230930-row7-num_unique_entry_methods}{1,065}
\DefMacro{table-numbers_of_programs_and_bugs-20230930-row7-num_unique_classes}{275}
\DefMacro{table-numbers_of_programs_and_bugs-20230930-row7-convert-fail}{5}
\DefMacro{table-numbers_of_programs_and_bugs-20230930-row7-convert-timeout}{6}
\DefMacro{table-numbers_of_programs_and_bugs-20230930-row7-convert-not_compile}{8}
\DefMacro{table-numbers_of_programs_and_bugs-20230930-row7-convert-no_original}{0}
\DefMacro{table-numbers_of_programs_and_bugs-20230930-row7-gen-fail}{252}
\DefMacro{table-numbers_of_programs_and_bugs-20230930-row7-gen-timeout}{37}
\DefMacro{table-numbers_of_programs_and_bugs-20230930-row7-gen-no_hole}{21}
\DefMacro{table-numbers_of_programs_and_bugs-20230930-row7-gen-no_gen}{343}
\DefMacro{table-numbers_of_programs_and_bugs-20230930-row7-num_working_templates}{861}
\DefMacro{table-numbers_of_programs_and_bugs-20230930-row7-num_unique_entry_methods_out_of_working_templates}{740}
\DefMacro{table-numbers_of_programs_and_bugs-20230930-row7-num_unique_classes_out_of_working_templates}{226}
\DefMacro{table-numbers_of_programs_and_bugs-20230930-row7-exec-total}{7,471}
\DefMacro{table-numbers_of_programs_and_bugs-20230930-row7-exec-not_compile}{0}
\DefMacro{table-numbers_of_programs_and_bugs-20230930-row7-exec-timeout}{196}
\DefMacro{table-numbers_of_programs_and_bugs-20230930-row7-exec-ok}{7,110}
\DefMacro{table-numbers_of_programs_and_bugs-20230930-row7-exec-skipped_diff}{153}
\DefMacro{table-numbers_of_programs_and_bugs-20230930-row7-exec-skipped_crash}{1}
\DefMacro{table-numbers_of_programs_and_bugs-20230930-row7-exec-final_diff}{11}
\DefMacro{table-numbers_of_programs_and_bugs-20230930-row7-exec-final_crash}{0}
\DefMacro{table-numbers_of_programs_and_bugs-20230930-row8-project}{zxing}
\DefMacro{table-numbers_of_programs_and_bugs-20230930-row8-num_unit_tests}{1,432}
\DefMacro{table-numbers_of_programs_and_bugs-20230930-row8-num_unique_entry_methods}{318}
\DefMacro{table-numbers_of_programs_and_bugs-20230930-row8-num_unique_classes}{88}
\DefMacro{table-numbers_of_programs_and_bugs-20230930-row8-convert-fail}{58}
\DefMacro{table-numbers_of_programs_and_bugs-20230930-row8-convert-timeout}{0}
\DefMacro{table-numbers_of_programs_and_bugs-20230930-row8-convert-not_compile}{29}
\DefMacro{table-numbers_of_programs_and_bugs-20230930-row8-convert-no_original}{0}
\DefMacro{table-numbers_of_programs_and_bugs-20230930-row8-gen-fail}{71}
\DefMacro{table-numbers_of_programs_and_bugs-20230930-row8-gen-timeout}{71}
\DefMacro{table-numbers_of_programs_and_bugs-20230930-row8-gen-no_hole}{21}
\DefMacro{table-numbers_of_programs_and_bugs-20230930-row8-gen-no_gen}{142}
\DefMacro{table-numbers_of_programs_and_bugs-20230930-row8-num_working_templates}{1,182}
\DefMacro{table-numbers_of_programs_and_bugs-20230930-row8-num_unique_entry_methods_out_of_working_templates}{263}
\DefMacro{table-numbers_of_programs_and_bugs-20230930-row8-num_unique_classes_out_of_working_templates}{70}
\DefMacro{table-numbers_of_programs_and_bugs-20230930-row8-exec-total}{10,326}
\DefMacro{table-numbers_of_programs_and_bugs-20230930-row8-exec-not_compile}{0}
\DefMacro{table-numbers_of_programs_and_bugs-20230930-row8-exec-timeout}{126}
\DefMacro{table-numbers_of_programs_and_bugs-20230930-row8-exec-ok}{9,751}
\DefMacro{table-numbers_of_programs_and_bugs-20230930-row8-exec-skipped_diff}{444}
\DefMacro{table-numbers_of_programs_and_bugs-20230930-row8-exec-skipped_crash}{0}
\DefMacro{table-numbers_of_programs_and_bugs-20230930-row8-exec-final_diff}{5}
\DefMacro{table-numbers_of_programs_and_bugs-20230930-row8-exec-final_crash}{0}
\DefMacro{table-numbers_of_programs_and_bugs-20230930-row9-project}{compress}
\DefMacro{table-numbers_of_programs_and_bugs-20230930-row9-num_unit_tests}{1,059}
\DefMacro{table-numbers_of_programs_and_bugs-20230930-row9-num_unique_entry_methods}{565}
\DefMacro{table-numbers_of_programs_and_bugs-20230930-row9-num_unique_classes}{112}
\DefMacro{table-numbers_of_programs_and_bugs-20230930-row9-convert-fail}{10}
\DefMacro{table-numbers_of_programs_and_bugs-20230930-row9-convert-timeout}{0}
\DefMacro{table-numbers_of_programs_and_bugs-20230930-row9-convert-not_compile}{10}
\DefMacro{table-numbers_of_programs_and_bugs-20230930-row9-convert-no_original}{0}
\DefMacro{table-numbers_of_programs_and_bugs-20230930-row9-gen-fail}{51}
\DefMacro{table-numbers_of_programs_and_bugs-20230930-row9-gen-timeout}{11}
\DefMacro{table-numbers_of_programs_and_bugs-20230930-row9-gen-no_hole}{58}
\DefMacro{table-numbers_of_programs_and_bugs-20230930-row9-gen-no_gen}{189}
\DefMacro{table-numbers_of_programs_and_bugs-20230930-row9-num_working_templates}{792}
\DefMacro{table-numbers_of_programs_and_bugs-20230930-row9-num_unique_entry_methods_out_of_working_templates}{446}
\DefMacro{table-numbers_of_programs_and_bugs-20230930-row9-num_unique_classes_out_of_working_templates}{101}
\DefMacro{table-numbers_of_programs_and_bugs-20230930-row9-exec-total}{6,582}
\DefMacro{table-numbers_of_programs_and_bugs-20230930-row9-exec-not_compile}{0}
\DefMacro{table-numbers_of_programs_and_bugs-20230930-row9-exec-timeout}{107}
\DefMacro{table-numbers_of_programs_and_bugs-20230930-row9-exec-ok}{5,402}
\DefMacro{table-numbers_of_programs_and_bugs-20230930-row9-exec-skipped_diff}{1,000}
\DefMacro{table-numbers_of_programs_and_bugs-20230930-row9-exec-skipped_crash}{1}
\DefMacro{table-numbers_of_programs_and_bugs-20230930-row9-exec-final_diff}{72}
\DefMacro{table-numbers_of_programs_and_bugs-20230930-row9-exec-final_crash}{0}
\DefMacro{table-numbers_of_programs_and_bugs-20230930-row10-project}{$\sum$}
\DefMacro{table-numbers_of_programs_and_bugs-20230930-row10-num_unit_tests}{16,012}
\DefMacro{table-numbers_of_programs_and_bugs-20230930-row10-num_unique_entry_methods}{6,093}
\DefMacro{table-numbers_of_programs_and_bugs-20230930-row10-num_unique_classes}{1,043}
\DefMacro{table-numbers_of_programs_and_bugs-20230930-row10-convert-fail}{544}
\DefMacro{table-numbers_of_programs_and_bugs-20230930-row10-convert-timeout}{6}
\DefMacro{table-numbers_of_programs_and_bugs-20230930-row10-convert-not_compile}{74}
\DefMacro{table-numbers_of_programs_and_bugs-20230930-row10-convert-no_original}{0}
\DefMacro{table-numbers_of_programs_and_bugs-20230930-row10-gen-fail}{886}
\DefMacro{table-numbers_of_programs_and_bugs-20230930-row10-gen-timeout}{426}
\DefMacro{table-numbers_of_programs_and_bugs-20230930-row10-gen-no_hole}{347}
\DefMacro{table-numbers_of_programs_and_bugs-20230930-row10-gen-no_gen}{2,980}
\DefMacro{table-numbers_of_programs_and_bugs-20230930-row10-num_working_templates}{12,061}
\DefMacro{table-numbers_of_programs_and_bugs-20230930-row10-num_unique_entry_methods_out_of_working_templates}{4,439}
\DefMacro{table-numbers_of_programs_and_bugs-20230930-row10-num_unique_classes_out_of_working_templates}{860}
\DefMacro{table-numbers_of_programs_and_bugs-20230930-row10-exec-total}{102,670}
\DefMacro{table-numbers_of_programs_and_bugs-20230930-row10-exec-not_compile}{0}
\DefMacro{table-numbers_of_programs_and_bugs-20230930-row10-exec-timeout}{1,083}
\DefMacro{table-numbers_of_programs_and_bugs-20230930-row10-exec-ok}{99,189}
\DefMacro{table-numbers_of_programs_and_bugs-20230930-row10-exec-skipped_diff}{2,230}
\DefMacro{table-numbers_of_programs_and_bugs-20230930-row10-exec-skipped_crash}{37}
\DefMacro{table-numbers_of_programs_and_bugs-20230930-row10-exec-final_diff}{113}
\DefMacro{table-numbers_of_programs_and_bugs-20230930-row10-exec-final_crash}{18}

\DefMacro{table-numbers_of_programs_and_bugs-20230928_2-row0-project}{lang}
\DefMacro{table-numbers_of_programs_and_bugs-20230928_2-row0-num_unit_tests}{1,912}
\DefMacro{table-numbers_of_programs_and_bugs-20230928_2-row0-num_unique_entry_methods}{1,155}
\DefMacro{table-numbers_of_programs_and_bugs-20230928_2-row0-num_unique_classes}{97}
\DefMacro{table-numbers_of_programs_and_bugs-20230928_2-row0-convert-fail}{91}
\DefMacro{table-numbers_of_programs_and_bugs-20230928_2-row0-convert-timeout}{0}
\DefMacro{table-numbers_of_programs_and_bugs-20230928_2-row0-convert-not_compile}{1}
\DefMacro{table-numbers_of_programs_and_bugs-20230928_2-row0-convert-no_original}{0}
\DefMacro{table-numbers_of_programs_and_bugs-20230928_2-row0-gen-fail}{32}
\DefMacro{table-numbers_of_programs_and_bugs-20230928_2-row0-gen-timeout}{67}
\DefMacro{table-numbers_of_programs_and_bugs-20230928_2-row0-gen-no_hole}{51}
\DefMacro{table-numbers_of_programs_and_bugs-20230928_2-row0-gen-no_gen}{66}
\DefMacro{table-numbers_of_programs_and_bugs-20230928_2-row0-num_working_templates}{1,703}
\DefMacro{table-numbers_of_programs_and_bugs-20230928_2-row0-num_unique_entry_methods_out_of_working_templates}{1,010}
\DefMacro{table-numbers_of_programs_and_bugs-20230928_2-row0-num_unique_classes_out_of_working_templates}{81}
\DefMacro{table-numbers_of_programs_and_bugs-20230928_2-row0-exec-total}{8,128}
\DefMacro{table-numbers_of_programs_and_bugs-20230928_2-row0-exec-not_compile}{0}
\DefMacro{table-numbers_of_programs_and_bugs-20230928_2-row0-exec-timeout}{13}
\DefMacro{table-numbers_of_programs_and_bugs-20230928_2-row0-exec-ok}{7,887}
\DefMacro{table-numbers_of_programs_and_bugs-20230928_2-row0-exec-skipped_diff}{184}
\DefMacro{table-numbers_of_programs_and_bugs-20230928_2-row0-exec-skipped_crash}{14}
\DefMacro{table-numbers_of_programs_and_bugs-20230928_2-row0-exec-final_diff}{5}
\DefMacro{table-numbers_of_programs_and_bugs-20230928_2-row0-exec-final_crash}{25}
\DefMacro{table-numbers_of_programs_and_bugs-20230928_2-row1-project}{math}
\DefMacro{table-numbers_of_programs_and_bugs-20230928_2-row1-num_unit_tests}{1,076}
\DefMacro{table-numbers_of_programs_and_bugs-20230928_2-row1-num_unique_entry_methods}{800}
\DefMacro{table-numbers_of_programs_and_bugs-20230928_2-row1-num_unique_classes}{202}
\DefMacro{table-numbers_of_programs_and_bugs-20230928_2-row1-convert-fail}{21}
\DefMacro{table-numbers_of_programs_and_bugs-20230928_2-row1-convert-timeout}{0}
\DefMacro{table-numbers_of_programs_and_bugs-20230928_2-row1-convert-not_compile}{26}
\DefMacro{table-numbers_of_programs_and_bugs-20230928_2-row1-convert-no_original}{0}
\DefMacro{table-numbers_of_programs_and_bugs-20230928_2-row1-gen-fail}{143}
\DefMacro{table-numbers_of_programs_and_bugs-20230928_2-row1-gen-timeout}{53}
\DefMacro{table-numbers_of_programs_and_bugs-20230928_2-row1-gen-no_hole}{38}
\DefMacro{table-numbers_of_programs_and_bugs-20230928_2-row1-gen-no_gen}{181}
\DefMacro{table-numbers_of_programs_and_bugs-20230928_2-row1-num_working_templates}{810}
\DefMacro{table-numbers_of_programs_and_bugs-20230928_2-row1-num_unique_entry_methods_out_of_working_templates}{627}
\DefMacro{table-numbers_of_programs_and_bugs-20230928_2-row1-num_unique_classes_out_of_working_templates}{171}
\DefMacro{table-numbers_of_programs_and_bugs-20230928_2-row1-exec-total}{3,440}
\DefMacro{table-numbers_of_programs_and_bugs-20230928_2-row1-exec-not_compile}{0}
\DefMacro{table-numbers_of_programs_and_bugs-20230928_2-row1-exec-timeout}{7}
\DefMacro{table-numbers_of_programs_and_bugs-20230928_2-row1-exec-ok}{3,382}
\DefMacro{table-numbers_of_programs_and_bugs-20230928_2-row1-exec-skipped_diff}{47}
\DefMacro{table-numbers_of_programs_and_bugs-20230928_2-row1-exec-skipped_crash}{4}
\DefMacro{table-numbers_of_programs_and_bugs-20230928_2-row1-exec-final_diff}{0}
\DefMacro{table-numbers_of_programs_and_bugs-20230928_2-row1-exec-final_crash}{0}
\DefMacro{table-numbers_of_programs_and_bugs-20230928_2-row2-project}{text}
\DefMacro{table-numbers_of_programs_and_bugs-20230928_2-row2-num_unit_tests}{1,711}
\DefMacro{table-numbers_of_programs_and_bugs-20230928_2-row2-num_unique_entry_methods}{516}
\DefMacro{table-numbers_of_programs_and_bugs-20230928_2-row2-num_unique_classes}{38}
\DefMacro{table-numbers_of_programs_and_bugs-20230928_2-row2-convert-fail}{18}
\DefMacro{table-numbers_of_programs_and_bugs-20230928_2-row2-convert-timeout}{0}
\DefMacro{table-numbers_of_programs_and_bugs-20230928_2-row2-convert-not_compile}{0}
\DefMacro{table-numbers_of_programs_and_bugs-20230928_2-row2-convert-no_original}{0}
\DefMacro{table-numbers_of_programs_and_bugs-20230928_2-row2-gen-fail}{1}
\DefMacro{table-numbers_of_programs_and_bugs-20230928_2-row2-gen-timeout}{200}
\DefMacro{table-numbers_of_programs_and_bugs-20230928_2-row2-gen-no_hole}{10}
\DefMacro{table-numbers_of_programs_and_bugs-20230928_2-row2-gen-no_gen}{118}
\DefMacro{table-numbers_of_programs_and_bugs-20230928_2-row2-num_working_templates}{1,565}
\DefMacro{table-numbers_of_programs_and_bugs-20230928_2-row2-num_unique_entry_methods_out_of_working_templates}{495}
\DefMacro{table-numbers_of_programs_and_bugs-20230928_2-row2-num_unique_classes_out_of_working_templates}{32}
\DefMacro{table-numbers_of_programs_and_bugs-20230928_2-row2-exec-total}{7,649}
\DefMacro{table-numbers_of_programs_and_bugs-20230928_2-row2-exec-not_compile}{0}
\DefMacro{table-numbers_of_programs_and_bugs-20230928_2-row2-exec-timeout}{36}
\DefMacro{table-numbers_of_programs_and_bugs-20230928_2-row2-exec-ok}{7,541}
\DefMacro{table-numbers_of_programs_and_bugs-20230928_2-row2-exec-skipped_diff}{71}
\DefMacro{table-numbers_of_programs_and_bugs-20230928_2-row2-exec-skipped_crash}{0}
\DefMacro{table-numbers_of_programs_and_bugs-20230928_2-row2-exec-final_diff}{0}
\DefMacro{table-numbers_of_programs_and_bugs-20230928_2-row2-exec-final_crash}{1}
\DefMacro{table-numbers_of_programs_and_bugs-20230928_2-row3-project}{numbers}
\DefMacro{table-numbers_of_programs_and_bugs-20230928_2-row3-num_unit_tests}{3,064}
\DefMacro{table-numbers_of_programs_and_bugs-20230928_2-row3-num_unique_entry_methods}{44}
\DefMacro{table-numbers_of_programs_and_bugs-20230928_2-row3-num_unique_classes}{4}
\DefMacro{table-numbers_of_programs_and_bugs-20230928_2-row3-convert-fail}{121}
\DefMacro{table-numbers_of_programs_and_bugs-20230928_2-row3-convert-timeout}{0}
\DefMacro{table-numbers_of_programs_and_bugs-20230928_2-row3-convert-not_compile}{0}
\DefMacro{table-numbers_of_programs_and_bugs-20230928_2-row3-convert-no_original}{0}
\DefMacro{table-numbers_of_programs_and_bugs-20230928_2-row3-gen-fail}{0}
\DefMacro{table-numbers_of_programs_and_bugs-20230928_2-row3-gen-timeout}{33}
\DefMacro{table-numbers_of_programs_and_bugs-20230928_2-row3-gen-no_hole}{0}
\DefMacro{table-numbers_of_programs_and_bugs-20230928_2-row3-gen-no_gen}{0}
\DefMacro{table-numbers_of_programs_and_bugs-20230928_2-row3-num_working_templates}{2,943}
\DefMacro{table-numbers_of_programs_and_bugs-20230928_2-row3-num_unique_entry_methods_out_of_working_templates}{41}
\DefMacro{table-numbers_of_programs_and_bugs-20230928_2-row3-num_unique_classes_out_of_working_templates}{3}
\DefMacro{table-numbers_of_programs_and_bugs-20230928_2-row3-exec-total}{14,853}
\DefMacro{table-numbers_of_programs_and_bugs-20230928_2-row3-exec-not_compile}{0}
\DefMacro{table-numbers_of_programs_and_bugs-20230928_2-row3-exec-timeout}{49}
\DefMacro{table-numbers_of_programs_and_bugs-20230928_2-row3-exec-ok}{14,804}
\DefMacro{table-numbers_of_programs_and_bugs-20230928_2-row3-exec-skipped_diff}{0}
\DefMacro{table-numbers_of_programs_and_bugs-20230928_2-row3-exec-skipped_crash}{0}
\DefMacro{table-numbers_of_programs_and_bugs-20230928_2-row3-exec-final_diff}{0}
\DefMacro{table-numbers_of_programs_and_bugs-20230928_2-row3-exec-final_crash}{0}
\DefMacro{table-numbers_of_programs_and_bugs-20230928_2-row4-project}{vectorz}
\DefMacro{table-numbers_of_programs_and_bugs-20230928_2-row4-num_unit_tests}{1,093}
\DefMacro{table-numbers_of_programs_and_bugs-20230928_2-row4-num_unique_entry_methods}{868}
\DefMacro{table-numbers_of_programs_and_bugs-20230928_2-row4-num_unique_classes}{152}
\DefMacro{table-numbers_of_programs_and_bugs-20230928_2-row4-convert-fail}{10}
\DefMacro{table-numbers_of_programs_and_bugs-20230928_2-row4-convert-timeout}{0}
\DefMacro{table-numbers_of_programs_and_bugs-20230928_2-row4-convert-not_compile}{0}
\DefMacro{table-numbers_of_programs_and_bugs-20230928_2-row4-convert-no_original}{0}
\DefMacro{table-numbers_of_programs_and_bugs-20230928_2-row4-gen-fail}{42}
\DefMacro{table-numbers_of_programs_and_bugs-20230928_2-row4-gen-timeout}{7}
\DefMacro{table-numbers_of_programs_and_bugs-20230928_2-row4-gen-no_hole}{94}
\DefMacro{table-numbers_of_programs_and_bugs-20230928_2-row4-gen-no_gen}{179}
\DefMacro{table-numbers_of_programs_and_bugs-20230928_2-row4-num_working_templates}{810}
\DefMacro{table-numbers_of_programs_and_bugs-20230928_2-row4-num_unique_entry_methods_out_of_working_templates}{638}
\DefMacro{table-numbers_of_programs_and_bugs-20230928_2-row4-num_unique_classes_out_of_working_templates}{141}
\DefMacro{table-numbers_of_programs_and_bugs-20230928_2-row4-exec-total}{3,951}
\DefMacro{table-numbers_of_programs_and_bugs-20230928_2-row4-exec-not_compile}{0}
\DefMacro{table-numbers_of_programs_and_bugs-20230928_2-row4-exec-timeout}{40}
\DefMacro{table-numbers_of_programs_and_bugs-20230928_2-row4-exec-ok}{3,842}
\DefMacro{table-numbers_of_programs_and_bugs-20230928_2-row4-exec-skipped_diff}{52}
\DefMacro{table-numbers_of_programs_and_bugs-20230928_2-row4-exec-skipped_crash}{12}
\DefMacro{table-numbers_of_programs_and_bugs-20230928_2-row4-exec-final_diff}{0}
\DefMacro{table-numbers_of_programs_and_bugs-20230928_2-row4-exec-final_crash}{5}
\DefMacro{table-numbers_of_programs_and_bugs-20230928_2-row5-project}{statistics}
\DefMacro{table-numbers_of_programs_and_bugs-20230928_2-row5-num_unit_tests}{1,924}
\DefMacro{table-numbers_of_programs_and_bugs-20230928_2-row5-num_unique_entry_methods}{368}
\DefMacro{table-numbers_of_programs_and_bugs-20230928_2-row5-num_unique_classes}{26}
\DefMacro{table-numbers_of_programs_and_bugs-20230928_2-row5-convert-fail}{191}
\DefMacro{table-numbers_of_programs_and_bugs-20230928_2-row5-convert-timeout}{0}
\DefMacro{table-numbers_of_programs_and_bugs-20230928_2-row5-convert-not_compile}{0}
\DefMacro{table-numbers_of_programs_and_bugs-20230928_2-row5-convert-no_original}{0}
\DefMacro{table-numbers_of_programs_and_bugs-20230928_2-row5-gen-fail}{150}
\DefMacro{table-numbers_of_programs_and_bugs-20230928_2-row5-gen-timeout}{34}
\DefMacro{table-numbers_of_programs_and_bugs-20230928_2-row5-gen-no_hole}{0}
\DefMacro{table-numbers_of_programs_and_bugs-20230928_2-row5-gen-no_gen}{156}
\DefMacro{table-numbers_of_programs_and_bugs-20230928_2-row5-num_working_templates}{1,577}
\DefMacro{table-numbers_of_programs_and_bugs-20230928_2-row5-num_unique_entry_methods_out_of_working_templates}{295}
\DefMacro{table-numbers_of_programs_and_bugs-20230928_2-row5-num_unique_classes_out_of_working_templates}{24}
\DefMacro{table-numbers_of_programs_and_bugs-20230928_2-row5-exec-total}{13,388}
\DefMacro{table-numbers_of_programs_and_bugs-20230928_2-row5-exec-not_compile}{0}
\DefMacro{table-numbers_of_programs_and_bugs-20230928_2-row5-exec-timeout}{0}
\DefMacro{table-numbers_of_programs_and_bugs-20230928_2-row5-exec-ok}{13,237}
\DefMacro{table-numbers_of_programs_and_bugs-20230928_2-row5-exec-skipped_diff}{151}
\DefMacro{table-numbers_of_programs_and_bugs-20230928_2-row5-exec-skipped_crash}{0}
\DefMacro{table-numbers_of_programs_and_bugs-20230928_2-row5-exec-final_diff}{0}
\DefMacro{table-numbers_of_programs_and_bugs-20230928_2-row5-exec-final_crash}{0}
\DefMacro{table-numbers_of_programs_and_bugs-20230928_2-row6-project}{codec}
\DefMacro{table-numbers_of_programs_and_bugs-20230928_2-row6-num_unit_tests}{1,497}
\DefMacro{table-numbers_of_programs_and_bugs-20230928_2-row6-num_unique_entry_methods}{394}
\DefMacro{table-numbers_of_programs_and_bugs-20230928_2-row6-num_unique_classes}{49}
\DefMacro{table-numbers_of_programs_and_bugs-20230928_2-row6-convert-fail}{19}
\DefMacro{table-numbers_of_programs_and_bugs-20230928_2-row6-convert-timeout}{0}
\DefMacro{table-numbers_of_programs_and_bugs-20230928_2-row6-convert-not_compile}{0}
\DefMacro{table-numbers_of_programs_and_bugs-20230928_2-row6-convert-no_original}{0}
\DefMacro{table-numbers_of_programs_and_bugs-20230928_2-row6-gen-fail}{85}
\DefMacro{table-numbers_of_programs_and_bugs-20230928_2-row6-gen-timeout}{11}
\DefMacro{table-numbers_of_programs_and_bugs-20230928_2-row6-gen-no_hole}{2}
\DefMacro{table-numbers_of_programs_and_bugs-20230928_2-row6-gen-no_gen}{149}
\DefMacro{table-numbers_of_programs_and_bugs-20230928_2-row6-num_working_templates}{1,327}
\DefMacro{table-numbers_of_programs_and_bugs-20230928_2-row6-num_unique_entry_methods_out_of_working_templates}{353}
\DefMacro{table-numbers_of_programs_and_bugs-20230928_2-row6-num_unique_classes_out_of_working_templates}{40}
\DefMacro{table-numbers_of_programs_and_bugs-20230928_2-row6-exec-total}{5,175}
\DefMacro{table-numbers_of_programs_and_bugs-20230928_2-row6-exec-not_compile}{0}
\DefMacro{table-numbers_of_programs_and_bugs-20230928_2-row6-exec-timeout}{65}
\DefMacro{table-numbers_of_programs_and_bugs-20230928_2-row6-exec-ok}{4,930}
\DefMacro{table-numbers_of_programs_and_bugs-20230928_2-row6-exec-skipped_diff}{135}
\DefMacro{table-numbers_of_programs_and_bugs-20230928_2-row6-exec-skipped_crash}{20}
\DefMacro{table-numbers_of_programs_and_bugs-20230928_2-row6-exec-final_diff}{0}
\DefMacro{table-numbers_of_programs_and_bugs-20230928_2-row6-exec-final_crash}{25}
\DefMacro{table-numbers_of_programs_and_bugs-20230928_2-row7-project}{jfreechart}
\DefMacro{table-numbers_of_programs_and_bugs-20230928_2-row7-num_unit_tests}{1,244}
\DefMacro{table-numbers_of_programs_and_bugs-20230928_2-row7-num_unique_entry_methods}{1,065}
\DefMacro{table-numbers_of_programs_and_bugs-20230928_2-row7-num_unique_classes}{275}
\DefMacro{table-numbers_of_programs_and_bugs-20230928_2-row7-convert-fail}{5}
\DefMacro{table-numbers_of_programs_and_bugs-20230928_2-row7-convert-timeout}{0}
\DefMacro{table-numbers_of_programs_and_bugs-20230928_2-row7-convert-not_compile}{8}
\DefMacro{table-numbers_of_programs_and_bugs-20230928_2-row7-convert-no_original}{0}
\DefMacro{table-numbers_of_programs_and_bugs-20230928_2-row7-gen-fail}{237}
\DefMacro{table-numbers_of_programs_and_bugs-20230928_2-row7-gen-timeout}{9}
\DefMacro{table-numbers_of_programs_and_bugs-20230928_2-row7-gen-no_hole}{21}
\DefMacro{table-numbers_of_programs_and_bugs-20230928_2-row7-gen-no_gen}{312}
\DefMacro{table-numbers_of_programs_and_bugs-20230928_2-row7-num_working_templates}{898}
\DefMacro{table-numbers_of_programs_and_bugs-20230928_2-row7-num_unique_entry_methods_out_of_working_templates}{777}
\DefMacro{table-numbers_of_programs_and_bugs-20230928_2-row7-num_unique_classes_out_of_working_templates}{237}
\DefMacro{table-numbers_of_programs_and_bugs-20230928_2-row7-exec-total}{2,311}
\DefMacro{table-numbers_of_programs_and_bugs-20230928_2-row7-exec-not_compile}{0}
\DefMacro{table-numbers_of_programs_and_bugs-20230928_2-row7-exec-timeout}{2}
\DefMacro{table-numbers_of_programs_and_bugs-20230928_2-row7-exec-ok}{2,269}
\DefMacro{table-numbers_of_programs_and_bugs-20230928_2-row7-exec-skipped_diff}{38}
\DefMacro{table-numbers_of_programs_and_bugs-20230928_2-row7-exec-skipped_crash}{0}
\DefMacro{table-numbers_of_programs_and_bugs-20230928_2-row7-exec-final_diff}{2}
\DefMacro{table-numbers_of_programs_and_bugs-20230928_2-row7-exec-final_crash}{0}
\DefMacro{table-numbers_of_programs_and_bugs-20230928_2-row8-project}{zxing}
\DefMacro{table-numbers_of_programs_and_bugs-20230928_2-row8-num_unit_tests}{1,432}
\DefMacro{table-numbers_of_programs_and_bugs-20230928_2-row8-num_unique_entry_methods}{318}
\DefMacro{table-numbers_of_programs_and_bugs-20230928_2-row8-num_unique_classes}{88}
\DefMacro{table-numbers_of_programs_and_bugs-20230928_2-row8-convert-fail}{58}
\DefMacro{table-numbers_of_programs_and_bugs-20230928_2-row8-convert-timeout}{0}
\DefMacro{table-numbers_of_programs_and_bugs-20230928_2-row8-convert-not_compile}{29}
\DefMacro{table-numbers_of_programs_and_bugs-20230928_2-row8-convert-no_original}{0}
\DefMacro{table-numbers_of_programs_and_bugs-20230928_2-row8-gen-fail}{71}
\DefMacro{table-numbers_of_programs_and_bugs-20230928_2-row8-gen-timeout}{10}
\DefMacro{table-numbers_of_programs_and_bugs-20230928_2-row8-gen-no_hole}{21}
\DefMacro{table-numbers_of_programs_and_bugs-20230928_2-row8-gen-no_gen}{55}
\DefMacro{table-numbers_of_programs_and_bugs-20230928_2-row8-num_working_templates}{1,269}
\DefMacro{table-numbers_of_programs_and_bugs-20230928_2-row8-num_unique_entry_methods_out_of_working_templates}{287}
\DefMacro{table-numbers_of_programs_and_bugs-20230928_2-row8-num_unique_classes_out_of_working_templates}{74}
\DefMacro{table-numbers_of_programs_and_bugs-20230928_2-row8-exec-total}{4,114}
\DefMacro{table-numbers_of_programs_and_bugs-20230928_2-row8-exec-not_compile}{0}
\DefMacro{table-numbers_of_programs_and_bugs-20230928_2-row8-exec-timeout}{16}
\DefMacro{table-numbers_of_programs_and_bugs-20230928_2-row8-exec-ok}{4,055}
\DefMacro{table-numbers_of_programs_and_bugs-20230928_2-row8-exec-skipped_diff}{36}
\DefMacro{table-numbers_of_programs_and_bugs-20230928_2-row8-exec-skipped_crash}{0}
\DefMacro{table-numbers_of_programs_and_bugs-20230928_2-row8-exec-final_diff}{0}
\DefMacro{table-numbers_of_programs_and_bugs-20230928_2-row8-exec-final_crash}{7}
\DefMacro{table-numbers_of_programs_and_bugs-20230928_2-row9-project}{compress}
\DefMacro{table-numbers_of_programs_and_bugs-20230928_2-row9-num_unit_tests}{1,059}
\DefMacro{table-numbers_of_programs_and_bugs-20230928_2-row9-num_unique_entry_methods}{565}
\DefMacro{table-numbers_of_programs_and_bugs-20230928_2-row9-num_unique_classes}{112}
\DefMacro{table-numbers_of_programs_and_bugs-20230928_2-row9-convert-fail}{10}
\DefMacro{table-numbers_of_programs_and_bugs-20230928_2-row9-convert-timeout}{0}
\DefMacro{table-numbers_of_programs_and_bugs-20230928_2-row9-convert-not_compile}{11}
\DefMacro{table-numbers_of_programs_and_bugs-20230928_2-row9-convert-no_original}{0}
\DefMacro{table-numbers_of_programs_and_bugs-20230928_2-row9-gen-fail}{14}
\DefMacro{table-numbers_of_programs_and_bugs-20230928_2-row9-gen-timeout}{11}
\DefMacro{table-numbers_of_programs_and_bugs-20230928_2-row9-gen-no_hole}{58}
\DefMacro{table-numbers_of_programs_and_bugs-20230928_2-row9-gen-no_gen}{200}
\DefMacro{table-numbers_of_programs_and_bugs-20230928_2-row9-num_working_templates}{780}
\DefMacro{table-numbers_of_programs_and_bugs-20230928_2-row9-num_unique_entry_methods_out_of_working_templates}{435}
\DefMacro{table-numbers_of_programs_and_bugs-20230928_2-row9-num_unique_classes_out_of_working_templates}{101}
\DefMacro{table-numbers_of_programs_and_bugs-20230928_2-row9-exec-total}{1,897}
\DefMacro{table-numbers_of_programs_and_bugs-20230928_2-row9-exec-not_compile}{0}
\DefMacro{table-numbers_of_programs_and_bugs-20230928_2-row9-exec-timeout}{7}
\DefMacro{table-numbers_of_programs_and_bugs-20230928_2-row9-exec-ok}{1,703}
\DefMacro{table-numbers_of_programs_and_bugs-20230928_2-row9-exec-skipped_diff}{187}
\DefMacro{table-numbers_of_programs_and_bugs-20230928_2-row9-exec-skipped_crash}{0}
\DefMacro{table-numbers_of_programs_and_bugs-20230928_2-row9-exec-final_diff}{0}
\DefMacro{table-numbers_of_programs_and_bugs-20230928_2-row9-exec-final_crash}{0}
\DefMacro{table-numbers_of_programs_and_bugs-20230928_2-row10-project}{$\sum$}
\DefMacro{table-numbers_of_programs_and_bugs-20230928_2-row10-num_unit_tests}{16,012}
\DefMacro{table-numbers_of_programs_and_bugs-20230928_2-row10-num_unique_entry_methods}{6,093}
\DefMacro{table-numbers_of_programs_and_bugs-20230928_2-row10-num_unique_classes}{1,043}
\DefMacro{table-numbers_of_programs_and_bugs-20230928_2-row10-convert-fail}{544}
\DefMacro{table-numbers_of_programs_and_bugs-20230928_2-row10-convert-timeout}{0}
\DefMacro{table-numbers_of_programs_and_bugs-20230928_2-row10-convert-not_compile}{75}
\DefMacro{table-numbers_of_programs_and_bugs-20230928_2-row10-convert-no_original}{0}
\DefMacro{table-numbers_of_programs_and_bugs-20230928_2-row10-gen-fail}{775}
\DefMacro{table-numbers_of_programs_and_bugs-20230928_2-row10-gen-timeout}{435}
\DefMacro{table-numbers_of_programs_and_bugs-20230928_2-row10-gen-no_hole}{295}
\DefMacro{table-numbers_of_programs_and_bugs-20230928_2-row10-gen-no_gen}{1,416}
\DefMacro{table-numbers_of_programs_and_bugs-20230928_2-row10-num_working_templates}{13,682}
\DefMacro{table-numbers_of_programs_and_bugs-20230928_2-row10-num_unique_entry_methods_out_of_working_templates}{4,958}
\DefMacro{table-numbers_of_programs_and_bugs-20230928_2-row10-num_unique_classes_out_of_working_templates}{904}
\DefMacro{table-numbers_of_programs_and_bugs-20230928_2-row10-exec-total}{64,906}
\DefMacro{table-numbers_of_programs_and_bugs-20230928_2-row10-exec-not_compile}{0}
\DefMacro{table-numbers_of_programs_and_bugs-20230928_2-row10-exec-timeout}{235}
\DefMacro{table-numbers_of_programs_and_bugs-20230928_2-row10-exec-ok}{63,650}
\DefMacro{table-numbers_of_programs_and_bugs-20230928_2-row10-exec-skipped_diff}{901}
\DefMacro{table-numbers_of_programs_and_bugs-20230928_2-row10-exec-skipped_crash}{50}
\DefMacro{table-numbers_of_programs_and_bugs-20230928_2-row10-exec-final_diff}{7}
\DefMacro{table-numbers_of_programs_and_bugs-20230928_2-row10-exec-final_crash}{63}

\DefMacro{table-numbers_of_programs_and_bugs-20230928-row0-project}{lang}
\DefMacro{table-numbers_of_programs_and_bugs-20230928-row0-num_unit_tests}{1,912}
\DefMacro{table-numbers_of_programs_and_bugs-20230928-row0-num_unique_entry_methods}{1,155}
\DefMacro{table-numbers_of_programs_and_bugs-20230928-row0-num_unique_classes}{97}
\DefMacro{table-numbers_of_programs_and_bugs-20230928-row0-convert-fail}{91}
\DefMacro{table-numbers_of_programs_and_bugs-20230928-row0-convert-timeout}{0}
\DefMacro{table-numbers_of_programs_and_bugs-20230928-row0-convert-not_compile}{1}
\DefMacro{table-numbers_of_programs_and_bugs-20230928-row0-convert-no_original}{0}
\DefMacro{table-numbers_of_programs_and_bugs-20230928-row0-gen-fail}{32}
\DefMacro{table-numbers_of_programs_and_bugs-20230928-row0-gen-timeout}{65}
\DefMacro{table-numbers_of_programs_and_bugs-20230928-row0-gen-no_hole}{53}
\DefMacro{table-numbers_of_programs_and_bugs-20230928-row0-gen-no_gen}{310}
\DefMacro{table-numbers_of_programs_and_bugs-20230928-row0-num_working_templates}{1,457}
\DefMacro{table-numbers_of_programs_and_bugs-20230928-row0-num_unique_entry_methods_out_of_working_templates}{870}
\DefMacro{table-numbers_of_programs_and_bugs-20230928-row0-num_unique_classes_out_of_working_templates}{79}
\DefMacro{table-numbers_of_programs_and_bugs-20230928-row0-exec-total}{8,907}
\DefMacro{table-numbers_of_programs_and_bugs-20230928-row0-exec-not_compile}{0}
\DefMacro{table-numbers_of_programs_and_bugs-20230928-row0-exec-timeout}{177}
\DefMacro{table-numbers_of_programs_and_bugs-20230928-row0-exec-ok}{8,551}
\DefMacro{table-numbers_of_programs_and_bugs-20230928-row0-exec-skipped_diff}{160}
\DefMacro{table-numbers_of_programs_and_bugs-20230928-row0-exec-skipped_crash}{0}
\DefMacro{table-numbers_of_programs_and_bugs-20230928-row0-exec-final_diff}{0}
\DefMacro{table-numbers_of_programs_and_bugs-20230928-row0-exec-final_crash}{19}
\DefMacro{table-numbers_of_programs_and_bugs-20230928-row1-project}{math}
\DefMacro{table-numbers_of_programs_and_bugs-20230928-row1-num_unit_tests}{1,076}
\DefMacro{table-numbers_of_programs_and_bugs-20230928-row1-num_unique_entry_methods}{800}
\DefMacro{table-numbers_of_programs_and_bugs-20230928-row1-num_unique_classes}{202}
\DefMacro{table-numbers_of_programs_and_bugs-20230928-row1-convert-fail}{21}
\DefMacro{table-numbers_of_programs_and_bugs-20230928-row1-convert-timeout}{0}
\DefMacro{table-numbers_of_programs_and_bugs-20230928-row1-convert-not_compile}{26}
\DefMacro{table-numbers_of_programs_and_bugs-20230928-row1-convert-no_original}{0}
\DefMacro{table-numbers_of_programs_and_bugs-20230928-row1-gen-fail}{136}
\DefMacro{table-numbers_of_programs_and_bugs-20230928-row1-gen-timeout}{34}
\DefMacro{table-numbers_of_programs_and_bugs-20230928-row1-gen-no_hole}{38}
\DefMacro{table-numbers_of_programs_and_bugs-20230928-row1-gen-no_gen}{257}
\DefMacro{table-numbers_of_programs_and_bugs-20230928-row1-num_working_templates}{734}
\DefMacro{table-numbers_of_programs_and_bugs-20230928-row1-num_unique_entry_methods_out_of_working_templates}{564}
\DefMacro{table-numbers_of_programs_and_bugs-20230928-row1-num_unique_classes_out_of_working_templates}{162}
\DefMacro{table-numbers_of_programs_and_bugs-20230928-row1-exec-total}{4,674}
\DefMacro{table-numbers_of_programs_and_bugs-20230928-row1-exec-not_compile}{0}
\DefMacro{table-numbers_of_programs_and_bugs-20230928-row1-exec-timeout}{51}
\DefMacro{table-numbers_of_programs_and_bugs-20230928-row1-exec-ok}{4,548}
\DefMacro{table-numbers_of_programs_and_bugs-20230928-row1-exec-skipped_diff}{75}
\DefMacro{table-numbers_of_programs_and_bugs-20230928-row1-exec-skipped_crash}{0}
\DefMacro{table-numbers_of_programs_and_bugs-20230928-row1-exec-final_diff}{0}
\DefMacro{table-numbers_of_programs_and_bugs-20230928-row1-exec-final_crash}{0}
\DefMacro{table-numbers_of_programs_and_bugs-20230928-row2-project}{text}
\DefMacro{table-numbers_of_programs_and_bugs-20230928-row2-num_unit_tests}{1,711}
\DefMacro{table-numbers_of_programs_and_bugs-20230928-row2-num_unique_entry_methods}{516}
\DefMacro{table-numbers_of_programs_and_bugs-20230928-row2-num_unique_classes}{38}
\DefMacro{table-numbers_of_programs_and_bugs-20230928-row2-convert-fail}{18}
\DefMacro{table-numbers_of_programs_and_bugs-20230928-row2-convert-timeout}{0}
\DefMacro{table-numbers_of_programs_and_bugs-20230928-row2-convert-not_compile}{0}
\DefMacro{table-numbers_of_programs_and_bugs-20230928-row2-convert-no_original}{0}
\DefMacro{table-numbers_of_programs_and_bugs-20230928-row2-gen-fail}{5}
\DefMacro{table-numbers_of_programs_and_bugs-20230928-row2-gen-timeout}{61}
\DefMacro{table-numbers_of_programs_and_bugs-20230928-row2-gen-no_hole}{10}
\DefMacro{table-numbers_of_programs_and_bugs-20230928-row2-gen-no_gen}{600}
\DefMacro{table-numbers_of_programs_and_bugs-20230928-row2-num_working_templates}{1,083}
\DefMacro{table-numbers_of_programs_and_bugs-20230928-row2-num_unique_entry_methods_out_of_working_templates}{387}
\DefMacro{table-numbers_of_programs_and_bugs-20230928-row2-num_unique_classes_out_of_working_templates}{31}
\DefMacro{table-numbers_of_programs_and_bugs-20230928-row2-exec-total}{6,682}
\DefMacro{table-numbers_of_programs_and_bugs-20230928-row2-exec-not_compile}{0}
\DefMacro{table-numbers_of_programs_and_bugs-20230928-row2-exec-timeout}{99}
\DefMacro{table-numbers_of_programs_and_bugs-20230928-row2-exec-ok}{6,578}
\DefMacro{table-numbers_of_programs_and_bugs-20230928-row2-exec-skipped_diff}{3}
\DefMacro{table-numbers_of_programs_and_bugs-20230928-row2-exec-skipped_crash}{2}
\DefMacro{table-numbers_of_programs_and_bugs-20230928-row2-exec-final_diff}{0}
\DefMacro{table-numbers_of_programs_and_bugs-20230928-row2-exec-final_crash}{0}
\DefMacro{table-numbers_of_programs_and_bugs-20230928-row3-project}{numbers}
\DefMacro{table-numbers_of_programs_and_bugs-20230928-row3-num_unit_tests}{3,064}
\DefMacro{table-numbers_of_programs_and_bugs-20230928-row3-num_unique_entry_methods}{44}
\DefMacro{table-numbers_of_programs_and_bugs-20230928-row3-num_unique_classes}{4}
\DefMacro{table-numbers_of_programs_and_bugs-20230928-row3-convert-fail}{121}
\DefMacro{table-numbers_of_programs_and_bugs-20230928-row3-convert-timeout}{0}
\DefMacro{table-numbers_of_programs_and_bugs-20230928-row3-convert-not_compile}{0}
\DefMacro{table-numbers_of_programs_and_bugs-20230928-row3-convert-no_original}{0}
\DefMacro{table-numbers_of_programs_and_bugs-20230928-row3-gen-fail}{0}
\DefMacro{table-numbers_of_programs_and_bugs-20230928-row3-gen-timeout}{2}
\DefMacro{table-numbers_of_programs_and_bugs-20230928-row3-gen-no_hole}{0}
\DefMacro{table-numbers_of_programs_and_bugs-20230928-row3-gen-no_gen}{101}
\DefMacro{table-numbers_of_programs_and_bugs-20230928-row3-num_working_templates}{2,842}
\DefMacro{table-numbers_of_programs_and_bugs-20230928-row3-num_unique_entry_methods_out_of_working_templates}{38}
\DefMacro{table-numbers_of_programs_and_bugs-20230928-row3-num_unique_classes_out_of_working_templates}{3}
\DefMacro{table-numbers_of_programs_and_bugs-20230928-row3-exec-total}{26,692}
\DefMacro{table-numbers_of_programs_and_bugs-20230928-row3-exec-not_compile}{0}
\DefMacro{table-numbers_of_programs_and_bugs-20230928-row3-exec-timeout}{7}
\DefMacro{table-numbers_of_programs_and_bugs-20230928-row3-exec-ok}{26,685}
\DefMacro{table-numbers_of_programs_and_bugs-20230928-row3-exec-skipped_diff}{0}
\DefMacro{table-numbers_of_programs_and_bugs-20230928-row3-exec-skipped_crash}{0}
\DefMacro{table-numbers_of_programs_and_bugs-20230928-row3-exec-final_diff}{0}
\DefMacro{table-numbers_of_programs_and_bugs-20230928-row3-exec-final_crash}{0}
\DefMacro{table-numbers_of_programs_and_bugs-20230928-row4-project}{vectorz}
\DefMacro{table-numbers_of_programs_and_bugs-20230928-row4-num_unit_tests}{1,093}
\DefMacro{table-numbers_of_programs_and_bugs-20230928-row4-num_unique_entry_methods}{868}
\DefMacro{table-numbers_of_programs_and_bugs-20230928-row4-num_unique_classes}{152}
\DefMacro{table-numbers_of_programs_and_bugs-20230928-row4-convert-fail}{10}
\DefMacro{table-numbers_of_programs_and_bugs-20230928-row4-convert-timeout}{0}
\DefMacro{table-numbers_of_programs_and_bugs-20230928-row4-convert-not_compile}{0}
\DefMacro{table-numbers_of_programs_and_bugs-20230928-row4-convert-no_original}{0}
\DefMacro{table-numbers_of_programs_and_bugs-20230928-row4-gen-fail}{39}
\DefMacro{table-numbers_of_programs_and_bugs-20230928-row4-gen-timeout}{11}
\DefMacro{table-numbers_of_programs_and_bugs-20230928-row4-gen-no_hole}{94}
\DefMacro{table-numbers_of_programs_and_bugs-20230928-row4-gen-no_gen}{276}
\DefMacro{table-numbers_of_programs_and_bugs-20230928-row4-num_working_templates}{713}
\DefMacro{table-numbers_of_programs_and_bugs-20230928-row4-num_unique_entry_methods_out_of_working_templates}{562}
\DefMacro{table-numbers_of_programs_and_bugs-20230928-row4-num_unique_classes_out_of_working_templates}{134}
\DefMacro{table-numbers_of_programs_and_bugs-20230928-row4-exec-total}{4,751}
\DefMacro{table-numbers_of_programs_and_bugs-20230928-row4-exec-not_compile}{0}
\DefMacro{table-numbers_of_programs_and_bugs-20230928-row4-exec-timeout}{156}
\DefMacro{table-numbers_of_programs_and_bugs-20230928-row4-exec-ok}{4,539}
\DefMacro{table-numbers_of_programs_and_bugs-20230928-row4-exec-skipped_diff}{54}
\DefMacro{table-numbers_of_programs_and_bugs-20230928-row4-exec-skipped_crash}{2}
\DefMacro{table-numbers_of_programs_and_bugs-20230928-row4-exec-final_diff}{0}
\DefMacro{table-numbers_of_programs_and_bugs-20230928-row4-exec-final_crash}{0}
\DefMacro{table-numbers_of_programs_and_bugs-20230928-row5-project}{statistics}
\DefMacro{table-numbers_of_programs_and_bugs-20230928-row5-num_unit_tests}{1,924}
\DefMacro{table-numbers_of_programs_and_bugs-20230928-row5-num_unique_entry_methods}{368}
\DefMacro{table-numbers_of_programs_and_bugs-20230928-row5-num_unique_classes}{26}
\DefMacro{table-numbers_of_programs_and_bugs-20230928-row5-convert-fail}{191}
\DefMacro{table-numbers_of_programs_and_bugs-20230928-row5-convert-timeout}{0}
\DefMacro{table-numbers_of_programs_and_bugs-20230928-row5-convert-not_compile}{0}
\DefMacro{table-numbers_of_programs_and_bugs-20230928-row5-convert-no_original}{0}
\DefMacro{table-numbers_of_programs_and_bugs-20230928-row5-gen-fail}{150}
\DefMacro{table-numbers_of_programs_and_bugs-20230928-row5-gen-timeout}{43}
\DefMacro{table-numbers_of_programs_and_bugs-20230928-row5-gen-no_hole}{0}
\DefMacro{table-numbers_of_programs_and_bugs-20230928-row5-gen-no_gen}{189}
\DefMacro{table-numbers_of_programs_and_bugs-20230928-row5-num_working_templates}{1,544}
\DefMacro{table-numbers_of_programs_and_bugs-20230928-row5-num_unique_entry_methods_out_of_working_templates}{295}
\DefMacro{table-numbers_of_programs_and_bugs-20230928-row5-num_unique_classes_out_of_working_templates}{24}
\DefMacro{table-numbers_of_programs_and_bugs-20230928-row5-exec-total}{15,231}
\DefMacro{table-numbers_of_programs_and_bugs-20230928-row5-exec-not_compile}{0}
\DefMacro{table-numbers_of_programs_and_bugs-20230928-row5-exec-timeout}{0}
\DefMacro{table-numbers_of_programs_and_bugs-20230928-row5-exec-ok}{15,014}
\DefMacro{table-numbers_of_programs_and_bugs-20230928-row5-exec-skipped_diff}{217}
\DefMacro{table-numbers_of_programs_and_bugs-20230928-row5-exec-skipped_crash}{0}
\DefMacro{table-numbers_of_programs_and_bugs-20230928-row5-exec-final_diff}{0}
\DefMacro{table-numbers_of_programs_and_bugs-20230928-row5-exec-final_crash}{0}
\DefMacro{table-numbers_of_programs_and_bugs-20230928-row6-project}{codec}
\DefMacro{table-numbers_of_programs_and_bugs-20230928-row6-num_unit_tests}{1,497}
\DefMacro{table-numbers_of_programs_and_bugs-20230928-row6-num_unique_entry_methods}{394}
\DefMacro{table-numbers_of_programs_and_bugs-20230928-row6-num_unique_classes}{49}
\DefMacro{table-numbers_of_programs_and_bugs-20230928-row6-convert-fail}{19}
\DefMacro{table-numbers_of_programs_and_bugs-20230928-row6-convert-timeout}{0}
\DefMacro{table-numbers_of_programs_and_bugs-20230928-row6-convert-not_compile}{0}
\DefMacro{table-numbers_of_programs_and_bugs-20230928-row6-convert-no_original}{0}
\DefMacro{table-numbers_of_programs_and_bugs-20230928-row6-gen-fail}{92}
\DefMacro{table-numbers_of_programs_and_bugs-20230928-row6-gen-timeout}{23}
\DefMacro{table-numbers_of_programs_and_bugs-20230928-row6-gen-no_hole}{2}
\DefMacro{table-numbers_of_programs_and_bugs-20230928-row6-gen-no_gen}{375}
\DefMacro{table-numbers_of_programs_and_bugs-20230928-row6-num_working_templates}{1,101}
\DefMacro{table-numbers_of_programs_and_bugs-20230928-row6-num_unique_entry_methods_out_of_working_templates}{297}
\DefMacro{table-numbers_of_programs_and_bugs-20230928-row6-num_unique_classes_out_of_working_templates}{37}
\DefMacro{table-numbers_of_programs_and_bugs-20230928-row6-exec-total}{4,918}
\DefMacro{table-numbers_of_programs_and_bugs-20230928-row6-exec-not_compile}{0}
\DefMacro{table-numbers_of_programs_and_bugs-20230928-row6-exec-timeout}{377}
\DefMacro{table-numbers_of_programs_and_bugs-20230928-row6-exec-ok}{4,425}
\DefMacro{table-numbers_of_programs_and_bugs-20230928-row6-exec-skipped_diff}{95}
\DefMacro{table-numbers_of_programs_and_bugs-20230928-row6-exec-skipped_crash}{20}
\DefMacro{table-numbers_of_programs_and_bugs-20230928-row6-exec-final_diff}{0}
\DefMacro{table-numbers_of_programs_and_bugs-20230928-row6-exec-final_crash}{1}
\DefMacro{table-numbers_of_programs_and_bugs-20230928-row7-project}{jfreechart}
\DefMacro{table-numbers_of_programs_and_bugs-20230928-row7-num_unit_tests}{1,244}
\DefMacro{table-numbers_of_programs_and_bugs-20230928-row7-num_unique_entry_methods}{1,065}
\DefMacro{table-numbers_of_programs_and_bugs-20230928-row7-num_unique_classes}{275}
\DefMacro{table-numbers_of_programs_and_bugs-20230928-row7-convert-fail}{5}
\DefMacro{table-numbers_of_programs_and_bugs-20230928-row7-convert-timeout}{0}
\DefMacro{table-numbers_of_programs_and_bugs-20230928-row7-convert-not_compile}{8}
\DefMacro{table-numbers_of_programs_and_bugs-20230928-row7-convert-no_original}{0}
\DefMacro{table-numbers_of_programs_and_bugs-20230928-row7-gen-fail}{236}
\DefMacro{table-numbers_of_programs_and_bugs-20230928-row7-gen-timeout}{24}
\DefMacro{table-numbers_of_programs_and_bugs-20230928-row7-gen-no_hole}{22}
\DefMacro{table-numbers_of_programs_and_bugs-20230928-row7-gen-no_gen}{420}
\DefMacro{table-numbers_of_programs_and_bugs-20230928-row7-num_working_templates}{789}
\DefMacro{table-numbers_of_programs_and_bugs-20230928-row7-num_unique_entry_methods_out_of_working_templates}{683}
\DefMacro{table-numbers_of_programs_and_bugs-20230928-row7-num_unique_classes_out_of_working_templates}{223}
\DefMacro{table-numbers_of_programs_and_bugs-20230928-row7-exec-total}{5,325}
\DefMacro{table-numbers_of_programs_and_bugs-20230928-row7-exec-not_compile}{0}
\DefMacro{table-numbers_of_programs_and_bugs-20230928-row7-exec-timeout}{158}
\DefMacro{table-numbers_of_programs_and_bugs-20230928-row7-exec-ok}{5,041}
\DefMacro{table-numbers_of_programs_and_bugs-20230928-row7-exec-skipped_diff}{118}
\DefMacro{table-numbers_of_programs_and_bugs-20230928-row7-exec-skipped_crash}{0}
\DefMacro{table-numbers_of_programs_and_bugs-20230928-row7-exec-final_diff}{8}
\DefMacro{table-numbers_of_programs_and_bugs-20230928-row7-exec-final_crash}{0}
\DefMacro{table-numbers_of_programs_and_bugs-20230928-row8-project}{zxing}
\DefMacro{table-numbers_of_programs_and_bugs-20230928-row8-num_unit_tests}{1,432}
\DefMacro{table-numbers_of_programs_and_bugs-20230928-row8-num_unique_entry_methods}{318}
\DefMacro{table-numbers_of_programs_and_bugs-20230928-row8-num_unique_classes}{88}
\DefMacro{table-numbers_of_programs_and_bugs-20230928-row8-convert-fail}{58}
\DefMacro{table-numbers_of_programs_and_bugs-20230928-row8-convert-timeout}{0}
\DefMacro{table-numbers_of_programs_and_bugs-20230928-row8-convert-not_compile}{29}
\DefMacro{table-numbers_of_programs_and_bugs-20230928-row8-convert-no_original}{0}
\DefMacro{table-numbers_of_programs_and_bugs-20230928-row8-gen-fail}{21}
\DefMacro{table-numbers_of_programs_and_bugs-20230928-row8-gen-timeout}{15}
\DefMacro{table-numbers_of_programs_and_bugs-20230928-row8-gen-no_hole}{21}
\DefMacro{table-numbers_of_programs_and_bugs-20230928-row8-gen-no_gen}{93}
\DefMacro{table-numbers_of_programs_and_bugs-20230928-row8-num_working_templates}{1,231}
\DefMacro{table-numbers_of_programs_and_bugs-20230928-row8-num_unique_entry_methods_out_of_working_templates}{274}
\DefMacro{table-numbers_of_programs_and_bugs-20230928-row8-num_unique_classes_out_of_working_templates}{71}
\DefMacro{table-numbers_of_programs_and_bugs-20230928-row8-exec-total}{6,490}
\DefMacro{table-numbers_of_programs_and_bugs-20230928-row8-exec-not_compile}{0}
\DefMacro{table-numbers_of_programs_and_bugs-20230928-row8-exec-timeout}{163}
\DefMacro{table-numbers_of_programs_and_bugs-20230928-row8-exec-ok}{5,854}
\DefMacro{table-numbers_of_programs_and_bugs-20230928-row8-exec-skipped_diff}{473}
\DefMacro{table-numbers_of_programs_and_bugs-20230928-row8-exec-skipped_crash}{0}
\DefMacro{table-numbers_of_programs_and_bugs-20230928-row8-exec-final_diff}{0}
\DefMacro{table-numbers_of_programs_and_bugs-20230928-row8-exec-final_crash}{0}
\DefMacro{table-numbers_of_programs_and_bugs-20230928-row9-project}{compress}
\DefMacro{table-numbers_of_programs_and_bugs-20230928-row9-num_unit_tests}{1,059}
\DefMacro{table-numbers_of_programs_and_bugs-20230928-row9-num_unique_entry_methods}{565}
\DefMacro{table-numbers_of_programs_and_bugs-20230928-row9-num_unique_classes}{112}
\DefMacro{table-numbers_of_programs_and_bugs-20230928-row9-convert-fail}{10}
\DefMacro{table-numbers_of_programs_and_bugs-20230928-row9-convert-timeout}{0}
\DefMacro{table-numbers_of_programs_and_bugs-20230928-row9-convert-not_compile}{10}
\DefMacro{table-numbers_of_programs_and_bugs-20230928-row9-convert-no_original}{0}
\DefMacro{table-numbers_of_programs_and_bugs-20230928-row9-gen-fail}{34}
\DefMacro{table-numbers_of_programs_and_bugs-20230928-row9-gen-timeout}{13}
\DefMacro{table-numbers_of_programs_and_bugs-20230928-row9-gen-no_hole}{58}
\DefMacro{table-numbers_of_programs_and_bugs-20230928-row9-gen-no_gen}{336}
\DefMacro{table-numbers_of_programs_and_bugs-20230928-row9-num_working_templates}{645}
\DefMacro{table-numbers_of_programs_and_bugs-20230928-row9-num_unique_entry_methods_out_of_working_templates}{374}
\DefMacro{table-numbers_of_programs_and_bugs-20230928-row9-num_unique_classes_out_of_working_templates}{97}
\DefMacro{table-numbers_of_programs_and_bugs-20230928-row9-exec-total}{3,560}
\DefMacro{table-numbers_of_programs_and_bugs-20230928-row9-exec-not_compile}{0}
\DefMacro{table-numbers_of_programs_and_bugs-20230928-row9-exec-timeout}{78}
\DefMacro{table-numbers_of_programs_and_bugs-20230928-row9-exec-ok}{3,184}
\DefMacro{table-numbers_of_programs_and_bugs-20230928-row9-exec-skipped_diff}{297}
\DefMacro{table-numbers_of_programs_and_bugs-20230928-row9-exec-skipped_crash}{0}
\DefMacro{table-numbers_of_programs_and_bugs-20230928-row9-exec-final_diff}{0}
\DefMacro{table-numbers_of_programs_and_bugs-20230928-row9-exec-final_crash}{1}
\DefMacro{table-numbers_of_programs_and_bugs-20230928-row10-project}{$\sum$}
\DefMacro{table-numbers_of_programs_and_bugs-20230928-row10-num_unit_tests}{16,012}
\DefMacro{table-numbers_of_programs_and_bugs-20230928-row10-num_unique_entry_methods}{6,093}
\DefMacro{table-numbers_of_programs_and_bugs-20230928-row10-num_unique_classes}{1,043}
\DefMacro{table-numbers_of_programs_and_bugs-20230928-row10-convert-fail}{544}
\DefMacro{table-numbers_of_programs_and_bugs-20230928-row10-convert-timeout}{0}
\DefMacro{table-numbers_of_programs_and_bugs-20230928-row10-convert-not_compile}{74}
\DefMacro{table-numbers_of_programs_and_bugs-20230928-row10-convert-no_original}{0}
\DefMacro{table-numbers_of_programs_and_bugs-20230928-row10-gen-fail}{745}
\DefMacro{table-numbers_of_programs_and_bugs-20230928-row10-gen-timeout}{291}
\DefMacro{table-numbers_of_programs_and_bugs-20230928-row10-gen-no_hole}{298}
\DefMacro{table-numbers_of_programs_and_bugs-20230928-row10-gen-no_gen}{2,957}
\DefMacro{table-numbers_of_programs_and_bugs-20230928-row10-num_working_templates}{12,139}
\DefMacro{table-numbers_of_programs_and_bugs-20230928-row10-num_unique_entry_methods_out_of_working_templates}{4,344}
\DefMacro{table-numbers_of_programs_and_bugs-20230928-row10-num_unique_classes_out_of_working_templates}{861}
\DefMacro{table-numbers_of_programs_and_bugs-20230928-row10-exec-total}{87,230}
\DefMacro{table-numbers_of_programs_and_bugs-20230928-row10-exec-not_compile}{0}
\DefMacro{table-numbers_of_programs_and_bugs-20230928-row10-exec-timeout}{1,266}
\DefMacro{table-numbers_of_programs_and_bugs-20230928-row10-exec-ok}{84,419}
\DefMacro{table-numbers_of_programs_and_bugs-20230928-row10-exec-skipped_diff}{1,492}
\DefMacro{table-numbers_of_programs_and_bugs-20230928-row10-exec-skipped_crash}{24}
\DefMacro{table-numbers_of_programs_and_bugs-20230928-row10-exec-final_diff}{8}
\DefMacro{table-numbers_of_programs_and_bugs-20230928-row10-exec-final_crash}{21}

\DefMacro{table-numbers_of_programs_and_bugs-20230924-row0-project}{lang}
\DefMacro{table-numbers_of_programs_and_bugs-20230924-row0-num_unit_tests}{1,912}
\DefMacro{table-numbers_of_programs_and_bugs-20230924-row0-num_unique_entry_methods}{1,155}
\DefMacro{table-numbers_of_programs_and_bugs-20230924-row0-num_unique_classes}{97}
\DefMacro{table-numbers_of_programs_and_bugs-20230924-row0-convert-fail}{91}
\DefMacro{table-numbers_of_programs_and_bugs-20230924-row0-convert-timeout}{0}
\DefMacro{table-numbers_of_programs_and_bugs-20230924-row0-convert-not_compile}{1}
\DefMacro{table-numbers_of_programs_and_bugs-20230924-row0-convert-no_original}{0}
\DefMacro{table-numbers_of_programs_and_bugs-20230924-row0-gen-fail}{54}
\DefMacro{table-numbers_of_programs_and_bugs-20230924-row0-gen-timeout}{73}
\DefMacro{table-numbers_of_programs_and_bugs-20230924-row0-gen-no_hole}{53}
\DefMacro{table-numbers_of_programs_and_bugs-20230924-row0-gen-no_gen}{160}
\DefMacro{table-numbers_of_programs_and_bugs-20230924-row0-num_working_templates}{1,607}
\DefMacro{table-numbers_of_programs_and_bugs-20230924-row0-num_unique_entry_methods_out_of_working_templates}{957}
\DefMacro{table-numbers_of_programs_and_bugs-20230924-row0-num_unique_classes_out_of_working_templates}{79}
\DefMacro{table-numbers_of_programs_and_bugs-20230924-row0-exec-total}{12,828}
\DefMacro{table-numbers_of_programs_and_bugs-20230924-row0-exec-not_compile}{0}
\DefMacro{table-numbers_of_programs_and_bugs-20230924-row0-exec-timeout}{19}
\DefMacro{table-numbers_of_programs_and_bugs-20230924-row0-exec-ok}{12,567}
\DefMacro{table-numbers_of_programs_and_bugs-20230924-row0-exec-skipped_diff}{226}
\DefMacro{table-numbers_of_programs_and_bugs-20230924-row0-exec-skipped_crash}{14}
\DefMacro{table-numbers_of_programs_and_bugs-20230924-row0-exec-final_diff}{0}
\DefMacro{table-numbers_of_programs_and_bugs-20230924-row0-exec-final_crash}{2}
\DefMacro{table-numbers_of_programs_and_bugs-20230924-row1-project}{math}
\DefMacro{table-numbers_of_programs_and_bugs-20230924-row1-num_unit_tests}{1,076}
\DefMacro{table-numbers_of_programs_and_bugs-20230924-row1-num_unique_entry_methods}{800}
\DefMacro{table-numbers_of_programs_and_bugs-20230924-row1-num_unique_classes}{202}
\DefMacro{table-numbers_of_programs_and_bugs-20230924-row1-convert-fail}{21}
\DefMacro{table-numbers_of_programs_and_bugs-20230924-row1-convert-timeout}{0}
\DefMacro{table-numbers_of_programs_and_bugs-20230924-row1-convert-not_compile}{26}
\DefMacro{table-numbers_of_programs_and_bugs-20230924-row1-convert-no_original}{0}
\DefMacro{table-numbers_of_programs_and_bugs-20230924-row1-gen-fail}{168}
\DefMacro{table-numbers_of_programs_and_bugs-20230924-row1-gen-timeout}{41}
\DefMacro{table-numbers_of_programs_and_bugs-20230924-row1-gen-no_hole}{38}
\DefMacro{table-numbers_of_programs_and_bugs-20230924-row1-gen-no_gen}{283}
\DefMacro{table-numbers_of_programs_and_bugs-20230924-row1-num_working_templates}{708}
\DefMacro{table-numbers_of_programs_and_bugs-20230924-row1-num_unique_entry_methods_out_of_working_templates}{547}
\DefMacro{table-numbers_of_programs_and_bugs-20230924-row1-num_unique_classes_out_of_working_templates}{161}
\DefMacro{table-numbers_of_programs_and_bugs-20230924-row1-exec-total}{5,356}
\DefMacro{table-numbers_of_programs_and_bugs-20230924-row1-exec-not_compile}{0}
\DefMacro{table-numbers_of_programs_and_bugs-20230924-row1-exec-timeout}{11}
\DefMacro{table-numbers_of_programs_and_bugs-20230924-row1-exec-ok}{5,222}
\DefMacro{table-numbers_of_programs_and_bugs-20230924-row1-exec-skipped_diff}{113}
\DefMacro{table-numbers_of_programs_and_bugs-20230924-row1-exec-skipped_crash}{1}
\DefMacro{table-numbers_of_programs_and_bugs-20230924-row1-exec-final_diff}{1}
\DefMacro{table-numbers_of_programs_and_bugs-20230924-row1-exec-final_crash}{8}
\DefMacro{table-numbers_of_programs_and_bugs-20230924-row2-project}{text}
\DefMacro{table-numbers_of_programs_and_bugs-20230924-row2-num_unit_tests}{1,711}
\DefMacro{table-numbers_of_programs_and_bugs-20230924-row2-num_unique_entry_methods}{516}
\DefMacro{table-numbers_of_programs_and_bugs-20230924-row2-num_unique_classes}{38}
\DefMacro{table-numbers_of_programs_and_bugs-20230924-row2-convert-fail}{18}
\DefMacro{table-numbers_of_programs_and_bugs-20230924-row2-convert-timeout}{0}
\DefMacro{table-numbers_of_programs_and_bugs-20230924-row2-convert-not_compile}{0}
\DefMacro{table-numbers_of_programs_and_bugs-20230924-row2-convert-no_original}{0}
\DefMacro{table-numbers_of_programs_and_bugs-20230924-row2-gen-fail}{5}
\DefMacro{table-numbers_of_programs_and_bugs-20230924-row2-gen-timeout}{57}
\DefMacro{table-numbers_of_programs_and_bugs-20230924-row2-gen-no_hole}{10}
\DefMacro{table-numbers_of_programs_and_bugs-20230924-row2-gen-no_gen}{306}
\DefMacro{table-numbers_of_programs_and_bugs-20230924-row2-num_working_templates}{1,377}
\DefMacro{table-numbers_of_programs_and_bugs-20230924-row2-num_unique_entry_methods_out_of_working_templates}{463}
\DefMacro{table-numbers_of_programs_and_bugs-20230924-row2-num_unique_classes_out_of_working_templates}{32}
\DefMacro{table-numbers_of_programs_and_bugs-20230924-row2-exec-total}{11,700}
\DefMacro{table-numbers_of_programs_and_bugs-20230924-row2-exec-not_compile}{0}
\DefMacro{table-numbers_of_programs_and_bugs-20230924-row2-exec-timeout}{0}
\DefMacro{table-numbers_of_programs_and_bugs-20230924-row2-exec-ok}{11,609}
\DefMacro{table-numbers_of_programs_and_bugs-20230924-row2-exec-skipped_diff}{88}
\DefMacro{table-numbers_of_programs_and_bugs-20230924-row2-exec-skipped_crash}{1}
\DefMacro{table-numbers_of_programs_and_bugs-20230924-row2-exec-final_diff}{0}
\DefMacro{table-numbers_of_programs_and_bugs-20230924-row2-exec-final_crash}{2}
\DefMacro{table-numbers_of_programs_and_bugs-20230924-row3-project}{numbers}
\DefMacro{table-numbers_of_programs_and_bugs-20230924-row3-num_unit_tests}{3,064}
\DefMacro{table-numbers_of_programs_and_bugs-20230924-row3-num_unique_entry_methods}{44}
\DefMacro{table-numbers_of_programs_and_bugs-20230924-row3-num_unique_classes}{4}
\DefMacro{table-numbers_of_programs_and_bugs-20230924-row3-convert-fail}{121}
\DefMacro{table-numbers_of_programs_and_bugs-20230924-row3-convert-timeout}{0}
\DefMacro{table-numbers_of_programs_and_bugs-20230924-row3-convert-not_compile}{0}
\DefMacro{table-numbers_of_programs_and_bugs-20230924-row3-convert-no_original}{0}
\DefMacro{table-numbers_of_programs_and_bugs-20230924-row3-gen-fail}{0}
\DefMacro{table-numbers_of_programs_and_bugs-20230924-row3-gen-timeout}{15}
\DefMacro{table-numbers_of_programs_and_bugs-20230924-row3-gen-no_hole}{0}
\DefMacro{table-numbers_of_programs_and_bugs-20230924-row3-gen-no_gen}{0}
\DefMacro{table-numbers_of_programs_and_bugs-20230924-row3-num_working_templates}{2,943}
\DefMacro{table-numbers_of_programs_and_bugs-20230924-row3-num_unique_entry_methods_out_of_working_templates}{41}
\DefMacro{table-numbers_of_programs_and_bugs-20230924-row3-num_unique_classes_out_of_working_templates}{3}
\DefMacro{table-numbers_of_programs_and_bugs-20230924-row3-exec-total}{28,252}
\DefMacro{table-numbers_of_programs_and_bugs-20230924-row3-exec-not_compile}{0}
\DefMacro{table-numbers_of_programs_and_bugs-20230924-row3-exec-timeout}{344}
\DefMacro{table-numbers_of_programs_and_bugs-20230924-row3-exec-ok}{27,908}
\DefMacro{table-numbers_of_programs_and_bugs-20230924-row3-exec-skipped_diff}{0}
\DefMacro{table-numbers_of_programs_and_bugs-20230924-row3-exec-skipped_crash}{0}
\DefMacro{table-numbers_of_programs_and_bugs-20230924-row3-exec-final_diff}{0}
\DefMacro{table-numbers_of_programs_and_bugs-20230924-row3-exec-final_crash}{0}
\DefMacro{table-numbers_of_programs_and_bugs-20230924-row4-project}{vectorz}
\DefMacro{table-numbers_of_programs_and_bugs-20230924-row4-num_unit_tests}{1,093}
\DefMacro{table-numbers_of_programs_and_bugs-20230924-row4-num_unique_entry_methods}{868}
\DefMacro{table-numbers_of_programs_and_bugs-20230924-row4-num_unique_classes}{152}
\DefMacro{table-numbers_of_programs_and_bugs-20230924-row4-convert-fail}{10}
\DefMacro{table-numbers_of_programs_and_bugs-20230924-row4-convert-timeout}{0}
\DefMacro{table-numbers_of_programs_and_bugs-20230924-row4-convert-not_compile}{0}
\DefMacro{table-numbers_of_programs_and_bugs-20230924-row4-convert-no_original}{0}
\DefMacro{table-numbers_of_programs_and_bugs-20230924-row4-gen-fail}{34}
\DefMacro{table-numbers_of_programs_and_bugs-20230924-row4-gen-timeout}{11}
\DefMacro{table-numbers_of_programs_and_bugs-20230924-row4-gen-no_hole}{94}
\DefMacro{table-numbers_of_programs_and_bugs-20230924-row4-gen-no_gen}{134}
\DefMacro{table-numbers_of_programs_and_bugs-20230924-row4-num_working_templates}{855}
\DefMacro{table-numbers_of_programs_and_bugs-20230924-row4-num_unique_entry_methods_out_of_working_templates}{678}
\DefMacro{table-numbers_of_programs_and_bugs-20230924-row4-num_unique_classes_out_of_working_templates}{142}
\DefMacro{table-numbers_of_programs_and_bugs-20230924-row4-exec-total}{5,476}
\DefMacro{table-numbers_of_programs_and_bugs-20230924-row4-exec-not_compile}{0}
\DefMacro{table-numbers_of_programs_and_bugs-20230924-row4-exec-timeout}{138}
\DefMacro{table-numbers_of_programs_and_bugs-20230924-row4-exec-ok}{5,239}
\DefMacro{table-numbers_of_programs_and_bugs-20230924-row4-exec-skipped_diff}{82}
\DefMacro{table-numbers_of_programs_and_bugs-20230924-row4-exec-skipped_crash}{13}
\DefMacro{table-numbers_of_programs_and_bugs-20230924-row4-exec-final_diff}{2}
\DefMacro{table-numbers_of_programs_and_bugs-20230924-row4-exec-final_crash}{2}
\DefMacro{table-numbers_of_programs_and_bugs-20230924-row5-project}{statistics}
\DefMacro{table-numbers_of_programs_and_bugs-20230924-row5-num_unit_tests}{1,924}
\DefMacro{table-numbers_of_programs_and_bugs-20230924-row5-num_unique_entry_methods}{368}
\DefMacro{table-numbers_of_programs_and_bugs-20230924-row5-num_unique_classes}{26}
\DefMacro{table-numbers_of_programs_and_bugs-20230924-row5-convert-fail}{191}
\DefMacro{table-numbers_of_programs_and_bugs-20230924-row5-convert-timeout}{0}
\DefMacro{table-numbers_of_programs_and_bugs-20230924-row5-convert-not_compile}{0}
\DefMacro{table-numbers_of_programs_and_bugs-20230924-row5-convert-no_original}{0}
\DefMacro{table-numbers_of_programs_and_bugs-20230924-row5-gen-fail}{152}
\DefMacro{table-numbers_of_programs_and_bugs-20230924-row5-gen-timeout}{93}
\DefMacro{table-numbers_of_programs_and_bugs-20230924-row5-gen-no_hole}{0}
\DefMacro{table-numbers_of_programs_and_bugs-20230924-row5-gen-no_gen}{556}
\DefMacro{table-numbers_of_programs_and_bugs-20230924-row5-num_working_templates}{1,177}
\DefMacro{table-numbers_of_programs_and_bugs-20230924-row5-num_unique_entry_methods_out_of_working_templates}{249}
\DefMacro{table-numbers_of_programs_and_bugs-20230924-row5-num_unique_classes_out_of_working_templates}{24}
\DefMacro{table-numbers_of_programs_and_bugs-20230924-row5-exec-total}{9,629}
\DefMacro{table-numbers_of_programs_and_bugs-20230924-row5-exec-not_compile}{0}
\DefMacro{table-numbers_of_programs_and_bugs-20230924-row5-exec-timeout}{0}
\DefMacro{table-numbers_of_programs_and_bugs-20230924-row5-exec-ok}{9,496}
\DefMacro{table-numbers_of_programs_and_bugs-20230924-row5-exec-skipped_diff}{133}
\DefMacro{table-numbers_of_programs_and_bugs-20230924-row5-exec-skipped_crash}{0}
\DefMacro{table-numbers_of_programs_and_bugs-20230924-row5-exec-final_diff}{0}
\DefMacro{table-numbers_of_programs_and_bugs-20230924-row5-exec-final_crash}{0}
\DefMacro{table-numbers_of_programs_and_bugs-20230924-row6-project}{codec}
\DefMacro{table-numbers_of_programs_and_bugs-20230924-row6-num_unit_tests}{1,497}
\DefMacro{table-numbers_of_programs_and_bugs-20230924-row6-num_unique_entry_methods}{394}
\DefMacro{table-numbers_of_programs_and_bugs-20230924-row6-num_unique_classes}{49}
\DefMacro{table-numbers_of_programs_and_bugs-20230924-row6-convert-fail}{19}
\DefMacro{table-numbers_of_programs_and_bugs-20230924-row6-convert-timeout}{0}
\DefMacro{table-numbers_of_programs_and_bugs-20230924-row6-convert-not_compile}{0}
\DefMacro{table-numbers_of_programs_and_bugs-20230924-row6-convert-no_original}{0}
\DefMacro{table-numbers_of_programs_and_bugs-20230924-row6-gen-fail}{81}
\DefMacro{table-numbers_of_programs_and_bugs-20230924-row6-gen-timeout}{11}
\DefMacro{table-numbers_of_programs_and_bugs-20230924-row6-gen-no_hole}{1}
\DefMacro{table-numbers_of_programs_and_bugs-20230924-row6-gen-no_gen}{161}
\DefMacro{table-numbers_of_programs_and_bugs-20230924-row6-num_working_templates}{1,316}
\DefMacro{table-numbers_of_programs_and_bugs-20230924-row6-num_unique_entry_methods_out_of_working_templates}{350}
\DefMacro{table-numbers_of_programs_and_bugs-20230924-row6-num_unique_classes_out_of_working_templates}{39}
\DefMacro{table-numbers_of_programs_and_bugs-20230924-row6-exec-total}{7,808}
\DefMacro{table-numbers_of_programs_and_bugs-20230924-row6-exec-not_compile}{0}
\DefMacro{table-numbers_of_programs_and_bugs-20230924-row6-exec-timeout}{607}
\DefMacro{table-numbers_of_programs_and_bugs-20230924-row6-exec-ok}{7,133}
\DefMacro{table-numbers_of_programs_and_bugs-20230924-row6-exec-skipped_diff}{50}
\DefMacro{table-numbers_of_programs_and_bugs-20230924-row6-exec-skipped_crash}{18}
\DefMacro{table-numbers_of_programs_and_bugs-20230924-row6-exec-final_diff}{0}
\DefMacro{table-numbers_of_programs_and_bugs-20230924-row6-exec-final_crash}{0}
\DefMacro{table-numbers_of_programs_and_bugs-20230924-row7-project}{jfreechart}
\DefMacro{table-numbers_of_programs_and_bugs-20230924-row7-num_unit_tests}{1,244}
\DefMacro{table-numbers_of_programs_and_bugs-20230924-row7-num_unique_entry_methods}{1,065}
\DefMacro{table-numbers_of_programs_and_bugs-20230924-row7-num_unique_classes}{275}
\DefMacro{table-numbers_of_programs_and_bugs-20230924-row7-convert-fail}{5}
\DefMacro{table-numbers_of_programs_and_bugs-20230924-row7-convert-timeout}{0}
\DefMacro{table-numbers_of_programs_and_bugs-20230924-row7-convert-not_compile}{8}
\DefMacro{table-numbers_of_programs_and_bugs-20230924-row7-convert-no_original}{0}
\DefMacro{table-numbers_of_programs_and_bugs-20230924-row7-gen-fail}{243}
\DefMacro{table-numbers_of_programs_and_bugs-20230924-row7-gen-timeout}{14}
\DefMacro{table-numbers_of_programs_and_bugs-20230924-row7-gen-no_hole}{21}
\DefMacro{table-numbers_of_programs_and_bugs-20230924-row7-gen-no_gen}{248}
\DefMacro{table-numbers_of_programs_and_bugs-20230924-row7-num_working_templates}{962}
\DefMacro{table-numbers_of_programs_and_bugs-20230924-row7-num_unique_entry_methods_out_of_working_templates}{829}
\DefMacro{table-numbers_of_programs_and_bugs-20230924-row7-num_unique_classes_out_of_working_templates}{238}
\DefMacro{table-numbers_of_programs_and_bugs-20230924-row7-exec-total}{7,822}
\DefMacro{table-numbers_of_programs_and_bugs-20230924-row7-exec-not_compile}{0}
\DefMacro{table-numbers_of_programs_and_bugs-20230924-row7-exec-timeout}{138}
\DefMacro{table-numbers_of_programs_and_bugs-20230924-row7-exec-ok}{7,544}
\DefMacro{table-numbers_of_programs_and_bugs-20230924-row7-exec-skipped_diff}{131}
\DefMacro{table-numbers_of_programs_and_bugs-20230924-row7-exec-skipped_crash}{1}
\DefMacro{table-numbers_of_programs_and_bugs-20230924-row7-exec-final_diff}{8}
\DefMacro{table-numbers_of_programs_and_bugs-20230924-row7-exec-final_crash}{0}
\DefMacro{table-numbers_of_programs_and_bugs-20230924-row8-project}{zxing}
\DefMacro{table-numbers_of_programs_and_bugs-20230924-row8-num_unit_tests}{1,432}
\DefMacro{table-numbers_of_programs_and_bugs-20230924-row8-num_unique_entry_methods}{318}
\DefMacro{table-numbers_of_programs_and_bugs-20230924-row8-num_unique_classes}{88}
\DefMacro{table-numbers_of_programs_and_bugs-20230924-row8-convert-fail}{58}
\DefMacro{table-numbers_of_programs_and_bugs-20230924-row8-convert-timeout}{0}
\DefMacro{table-numbers_of_programs_and_bugs-20230924-row8-convert-not_compile}{29}
\DefMacro{table-numbers_of_programs_and_bugs-20230924-row8-convert-no_original}{0}
\DefMacro{table-numbers_of_programs_and_bugs-20230924-row8-gen-fail}{63}
\DefMacro{table-numbers_of_programs_and_bugs-20230924-row8-gen-timeout}{9}
\DefMacro{table-numbers_of_programs_and_bugs-20230924-row8-gen-no_hole}{21}
\DefMacro{table-numbers_of_programs_and_bugs-20230924-row8-gen-no_gen}{69}
\DefMacro{table-numbers_of_programs_and_bugs-20230924-row8-num_working_templates}{1,255}
\DefMacro{table-numbers_of_programs_and_bugs-20230924-row8-num_unique_entry_methods_out_of_working_templates}{284}
\DefMacro{table-numbers_of_programs_and_bugs-20230924-row8-num_unique_classes_out_of_working_templates}{72}
\DefMacro{table-numbers_of_programs_and_bugs-20230924-row8-exec-total}{10,478}
\DefMacro{table-numbers_of_programs_and_bugs-20230924-row8-exec-not_compile}{0}
\DefMacro{table-numbers_of_programs_and_bugs-20230924-row8-exec-timeout}{28}
\DefMacro{table-numbers_of_programs_and_bugs-20230924-row8-exec-ok}{10,346}
\DefMacro{table-numbers_of_programs_and_bugs-20230924-row8-exec-skipped_diff}{102}
\DefMacro{table-numbers_of_programs_and_bugs-20230924-row8-exec-skipped_crash}{0}
\DefMacro{table-numbers_of_programs_and_bugs-20230924-row8-exec-final_diff}{0}
\DefMacro{table-numbers_of_programs_and_bugs-20230924-row8-exec-final_crash}{2}
\DefMacro{table-numbers_of_programs_and_bugs-20230924-row9-project}{compress}
\DefMacro{table-numbers_of_programs_and_bugs-20230924-row9-num_unit_tests}{1,059}
\DefMacro{table-numbers_of_programs_and_bugs-20230924-row9-num_unique_entry_methods}{565}
\DefMacro{table-numbers_of_programs_and_bugs-20230924-row9-num_unique_classes}{112}
\DefMacro{table-numbers_of_programs_and_bugs-20230924-row9-convert-fail}{10}
\DefMacro{table-numbers_of_programs_and_bugs-20230924-row9-convert-timeout}{0}
\DefMacro{table-numbers_of_programs_and_bugs-20230924-row9-convert-not_compile}{11}
\DefMacro{table-numbers_of_programs_and_bugs-20230924-row9-convert-no_original}{0}
\DefMacro{table-numbers_of_programs_and_bugs-20230924-row9-gen-fail}{30}
\DefMacro{table-numbers_of_programs_and_bugs-20230924-row9-gen-timeout}{8}
\DefMacro{table-numbers_of_programs_and_bugs-20230924-row9-gen-no_hole}{58}
\DefMacro{table-numbers_of_programs_and_bugs-20230924-row9-gen-no_gen}{90}
\DefMacro{table-numbers_of_programs_and_bugs-20230924-row9-num_working_templates}{890}
\DefMacro{table-numbers_of_programs_and_bugs-20230924-row9-num_unique_entry_methods_out_of_working_templates}{492}
\DefMacro{table-numbers_of_programs_and_bugs-20230924-row9-num_unique_classes_out_of_working_templates}{102}
\DefMacro{table-numbers_of_programs_and_bugs-20230924-row9-exec-total}{6,410}
\DefMacro{table-numbers_of_programs_and_bugs-20230924-row9-exec-not_compile}{0}
\DefMacro{table-numbers_of_programs_and_bugs-20230924-row9-exec-timeout}{9}
\DefMacro{table-numbers_of_programs_and_bugs-20230924-row9-exec-ok}{5,421}
\DefMacro{table-numbers_of_programs_and_bugs-20230924-row9-exec-skipped_diff}{901}
\DefMacro{table-numbers_of_programs_and_bugs-20230924-row9-exec-skipped_crash}{0}
\DefMacro{table-numbers_of_programs_and_bugs-20230924-row9-exec-final_diff}{73}
\DefMacro{table-numbers_of_programs_and_bugs-20230924-row9-exec-final_crash}{6}
\DefMacro{table-numbers_of_programs_and_bugs-20230924-row10-project}{$\sum$}
\DefMacro{table-numbers_of_programs_and_bugs-20230924-row10-num_unit_tests}{16,012}
\DefMacro{table-numbers_of_programs_and_bugs-20230924-row10-num_unique_entry_methods}{6,093}
\DefMacro{table-numbers_of_programs_and_bugs-20230924-row10-num_unique_classes}{1,043}
\DefMacro{table-numbers_of_programs_and_bugs-20230924-row10-convert-fail}{544}
\DefMacro{table-numbers_of_programs_and_bugs-20230924-row10-convert-timeout}{0}
\DefMacro{table-numbers_of_programs_and_bugs-20230924-row10-convert-not_compile}{75}
\DefMacro{table-numbers_of_programs_and_bugs-20230924-row10-convert-no_original}{0}
\DefMacro{table-numbers_of_programs_and_bugs-20230924-row10-gen-fail}{830}
\DefMacro{table-numbers_of_programs_and_bugs-20230924-row10-gen-timeout}{332}
\DefMacro{table-numbers_of_programs_and_bugs-20230924-row10-gen-no_hole}{296}
\DefMacro{table-numbers_of_programs_and_bugs-20230924-row10-gen-no_gen}{2,007}
\DefMacro{table-numbers_of_programs_and_bugs-20230924-row10-num_working_templates}{13,090}
\DefMacro{table-numbers_of_programs_and_bugs-20230924-row10-num_unique_entry_methods_out_of_working_templates}{4,890}
\DefMacro{table-numbers_of_programs_and_bugs-20230924-row10-num_unique_classes_out_of_working_templates}{892}
\DefMacro{table-numbers_of_programs_and_bugs-20230924-row10-exec-total}{105,759}
\DefMacro{table-numbers_of_programs_and_bugs-20230924-row10-exec-not_compile}{0}
\DefMacro{table-numbers_of_programs_and_bugs-20230924-row10-exec-timeout}{1,294}
\DefMacro{table-numbers_of_programs_and_bugs-20230924-row10-exec-ok}{102,485}
\DefMacro{table-numbers_of_programs_and_bugs-20230924-row10-exec-skipped_diff}{1,826}
\DefMacro{table-numbers_of_programs_and_bugs-20230924-row10-exec-skipped_crash}{48}
\DefMacro{table-numbers_of_programs_and_bugs-20230924-row10-exec-final_diff}{84}
\DefMacro{table-numbers_of_programs_and_bugs-20230924-row10-exec-final_crash}{22}

\DefMacro{table-numbers_of_programs_and_bugs-20231115-row0-project}{lang}
\DefMacro{table-numbers_of_programs_and_bugs-20231115-row0-num_unit_tests}{2,516}
\DefMacro{table-numbers_of_programs_and_bugs-20231115-row0-num_unique_entry_methods}{1,185}
\DefMacro{table-numbers_of_programs_and_bugs-20231115-row0-num_unique_classes}{91}
\DefMacro{table-numbers_of_programs_and_bugs-20231115-row0-convert-fail}{152}
\DefMacro{table-numbers_of_programs_and_bugs-20231115-row0-convert-timeout}{0}
\DefMacro{table-numbers_of_programs_and_bugs-20231115-row0-convert-not_compile}{1}
\DefMacro{table-numbers_of_programs_and_bugs-20231115-row0-convert-no_original}{0}
\DefMacro{table-numbers_of_programs_and_bugs-20231115-row0-gen-fail}{64}
\DefMacro{table-numbers_of_programs_and_bugs-20231115-row0-gen-timeout}{23}
\DefMacro{table-numbers_of_programs_and_bugs-20231115-row0-gen-no_hole}{59}
\DefMacro{table-numbers_of_programs_and_bugs-20231115-row0-gen-no_gen}{161}
\DefMacro{table-numbers_of_programs_and_bugs-20231115-row0-num_working_templates}{2,143}
\DefMacro{table-numbers_of_programs_and_bugs-20231115-row0-num_unique_entry_methods_out_of_working_templates}{1,004}
\DefMacro{table-numbers_of_programs_and_bugs-20231115-row0-num_unique_classes_out_of_working_templates}{69}
\DefMacro{table-numbers_of_programs_and_bugs-20231115-row0-exec-total}{18,921}
\DefMacro{table-numbers_of_programs_and_bugs-20231115-row0-exec-not_compile}{0}
\DefMacro{table-numbers_of_programs_and_bugs-20231115-row0-exec-timeout}{588}
\DefMacro{table-numbers_of_programs_and_bugs-20231115-row0-exec-ok}{18,249}
\DefMacro{table-numbers_of_programs_and_bugs-20231115-row0-exec-skipped_diff}{71}
\DefMacro{table-numbers_of_programs_and_bugs-20231115-row0-exec-skipped_crash}{13}
\DefMacro{table-numbers_of_programs_and_bugs-20231115-row0-exec-final_diff}{0}
\DefMacro{table-numbers_of_programs_and_bugs-20231115-row0-exec-final_crash}{0}
\DefMacro{table-numbers_of_programs_and_bugs-20231115-row1-project}{math}
\DefMacro{table-numbers_of_programs_and_bugs-20231115-row1-num_unit_tests}{1,092}
\DefMacro{table-numbers_of_programs_and_bugs-20231115-row1-num_unique_entry_methods}{589}
\DefMacro{table-numbers_of_programs_and_bugs-20231115-row1-num_unique_classes}{148}
\DefMacro{table-numbers_of_programs_and_bugs-20231115-row1-convert-fail}{20}
\DefMacro{table-numbers_of_programs_and_bugs-20231115-row1-convert-timeout}{0}
\DefMacro{table-numbers_of_programs_and_bugs-20231115-row1-convert-not_compile}{7}
\DefMacro{table-numbers_of_programs_and_bugs-20231115-row1-convert-no_original}{0}
\DefMacro{table-numbers_of_programs_and_bugs-20231115-row1-gen-fail}{250}
\DefMacro{table-numbers_of_programs_and_bugs-20231115-row1-gen-timeout}{41}
\DefMacro{table-numbers_of_programs_and_bugs-20231115-row1-gen-no_hole}{46}
\DefMacro{table-numbers_of_programs_and_bugs-20231115-row1-gen-no_gen}{289}
\DefMacro{table-numbers_of_programs_and_bugs-20231115-row1-num_working_templates}{730}
\DefMacro{table-numbers_of_programs_and_bugs-20231115-row1-num_unique_entry_methods_out_of_working_templates}{412}
\DefMacro{table-numbers_of_programs_and_bugs-20231115-row1-num_unique_classes_out_of_working_templates}{116}
\DefMacro{table-numbers_of_programs_and_bugs-20231115-row1-exec-total}{6,193}
\DefMacro{table-numbers_of_programs_and_bugs-20231115-row1-exec-not_compile}{0}
\DefMacro{table-numbers_of_programs_and_bugs-20231115-row1-exec-timeout}{284}
\DefMacro{table-numbers_of_programs_and_bugs-20231115-row1-exec-ok}{5,845}
\DefMacro{table-numbers_of_programs_and_bugs-20231115-row1-exec-skipped_diff}{64}
\DefMacro{table-numbers_of_programs_and_bugs-20231115-row1-exec-skipped_crash}{0}
\DefMacro{table-numbers_of_programs_and_bugs-20231115-row1-exec-final_diff}{0}
\DefMacro{table-numbers_of_programs_and_bugs-20231115-row1-exec-final_crash}{0}
\DefMacro{table-numbers_of_programs_and_bugs-20231115-row2-project}{text}
\DefMacro{table-numbers_of_programs_and_bugs-20231115-row2-num_unit_tests}{1,624}
\DefMacro{table-numbers_of_programs_and_bugs-20231115-row2-num_unique_entry_methods}{462}
\DefMacro{table-numbers_of_programs_and_bugs-20231115-row2-num_unique_classes}{39}
\DefMacro{table-numbers_of_programs_and_bugs-20231115-row2-convert-fail}{33}
\DefMacro{table-numbers_of_programs_and_bugs-20231115-row2-convert-timeout}{0}
\DefMacro{table-numbers_of_programs_and_bugs-20231115-row2-convert-not_compile}{0}
\DefMacro{table-numbers_of_programs_and_bugs-20231115-row2-convert-no_original}{0}
\DefMacro{table-numbers_of_programs_and_bugs-20231115-row2-gen-fail}{9}
\DefMacro{table-numbers_of_programs_and_bugs-20231115-row2-gen-timeout}{49}
\DefMacro{table-numbers_of_programs_and_bugs-20231115-row2-gen-no_hole}{13}
\DefMacro{table-numbers_of_programs_and_bugs-20231115-row2-gen-no_gen}{402}
\DefMacro{table-numbers_of_programs_and_bugs-20231115-row2-num_working_templates}{1,176}
\DefMacro{table-numbers_of_programs_and_bugs-20231115-row2-num_unique_entry_methods_out_of_working_templates}{374}
\DefMacro{table-numbers_of_programs_and_bugs-20231115-row2-num_unique_classes_out_of_working_templates}{30}
\DefMacro{table-numbers_of_programs_and_bugs-20231115-row2-exec-total}{10,746}
\DefMacro{table-numbers_of_programs_and_bugs-20231115-row2-exec-not_compile}{0}
\DefMacro{table-numbers_of_programs_and_bugs-20231115-row2-exec-timeout}{1,314}
\DefMacro{table-numbers_of_programs_and_bugs-20231115-row2-exec-ok}{9,410}
\DefMacro{table-numbers_of_programs_and_bugs-20231115-row2-exec-skipped_diff}{20}
\DefMacro{table-numbers_of_programs_and_bugs-20231115-row2-exec-skipped_crash}{2}
\DefMacro{table-numbers_of_programs_and_bugs-20231115-row2-exec-final_diff}{0}
\DefMacro{table-numbers_of_programs_and_bugs-20231115-row2-exec-final_crash}{0}
\DefMacro{table-numbers_of_programs_and_bugs-20231115-row3-project}{numbers}
\DefMacro{table-numbers_of_programs_and_bugs-20231115-row3-num_unit_tests}{3,101}
\DefMacro{table-numbers_of_programs_and_bugs-20231115-row3-num_unique_entry_methods}{45}
\DefMacro{table-numbers_of_programs_and_bugs-20231115-row3-num_unique_classes}{4}
\DefMacro{table-numbers_of_programs_and_bugs-20231115-row3-convert-fail}{121}
\DefMacro{table-numbers_of_programs_and_bugs-20231115-row3-convert-timeout}{0}
\DefMacro{table-numbers_of_programs_and_bugs-20231115-row3-convert-not_compile}{0}
\DefMacro{table-numbers_of_programs_and_bugs-20231115-row3-convert-no_original}{0}
\DefMacro{table-numbers_of_programs_and_bugs-20231115-row3-gen-fail}{0}
\DefMacro{table-numbers_of_programs_and_bugs-20231115-row3-gen-timeout}{59}
\DefMacro{table-numbers_of_programs_and_bugs-20231115-row3-gen-no_hole}{0}
\DefMacro{table-numbers_of_programs_and_bugs-20231115-row3-gen-no_gen}{0}
\DefMacro{table-numbers_of_programs_and_bugs-20231115-row3-num_working_templates}{2,980}
\DefMacro{table-numbers_of_programs_and_bugs-20231115-row3-num_unique_entry_methods_out_of_working_templates}{42}
\DefMacro{table-numbers_of_programs_and_bugs-20231115-row3-num_unique_classes_out_of_working_templates}{3}
\DefMacro{table-numbers_of_programs_and_bugs-20231115-row3-exec-total}{29,114}
\DefMacro{table-numbers_of_programs_and_bugs-20231115-row3-exec-not_compile}{0}
\DefMacro{table-numbers_of_programs_and_bugs-20231115-row3-exec-timeout}{107}
\DefMacro{table-numbers_of_programs_and_bugs-20231115-row3-exec-ok}{29,007}
\DefMacro{table-numbers_of_programs_and_bugs-20231115-row3-exec-skipped_diff}{0}
\DefMacro{table-numbers_of_programs_and_bugs-20231115-row3-exec-skipped_crash}{0}
\DefMacro{table-numbers_of_programs_and_bugs-20231115-row3-exec-final_diff}{0}
\DefMacro{table-numbers_of_programs_and_bugs-20231115-row3-exec-final_crash}{0}
\DefMacro{table-numbers_of_programs_and_bugs-20231115-row4-project}{vectorz}
\DefMacro{table-numbers_of_programs_and_bugs-20231115-row4-num_unit_tests}{1,183}
\DefMacro{table-numbers_of_programs_and_bugs-20231115-row4-num_unique_entry_methods}{914}
\DefMacro{table-numbers_of_programs_and_bugs-20231115-row4-num_unique_classes}{154}
\DefMacro{table-numbers_of_programs_and_bugs-20231115-row4-convert-fail}{0}
\DefMacro{table-numbers_of_programs_and_bugs-20231115-row4-convert-timeout}{0}
\DefMacro{table-numbers_of_programs_and_bugs-20231115-row4-convert-not_compile}{0}
\DefMacro{table-numbers_of_programs_and_bugs-20231115-row4-convert-no_original}{0}
\DefMacro{table-numbers_of_programs_and_bugs-20231115-row4-gen-fail}{51}
\DefMacro{table-numbers_of_programs_and_bugs-20231115-row4-gen-timeout}{43}
\DefMacro{table-numbers_of_programs_and_bugs-20231115-row4-gen-no_hole}{97}
\DefMacro{table-numbers_of_programs_and_bugs-20231115-row4-gen-no_gen}{265}
\DefMacro{table-numbers_of_programs_and_bugs-20231115-row4-num_working_templates}{821}
\DefMacro{table-numbers_of_programs_and_bugs-20231115-row4-num_unique_entry_methods_out_of_working_templates}{650}
\DefMacro{table-numbers_of_programs_and_bugs-20231115-row4-num_unique_classes_out_of_working_templates}{137}
\DefMacro{table-numbers_of_programs_and_bugs-20231115-row4-exec-total}{5,914}
\DefMacro{table-numbers_of_programs_and_bugs-20231115-row4-exec-not_compile}{0}
\DefMacro{table-numbers_of_programs_and_bugs-20231115-row4-exec-timeout}{525}
\DefMacro{table-numbers_of_programs_and_bugs-20231115-row4-exec-ok}{5,386}
\DefMacro{table-numbers_of_programs_and_bugs-20231115-row4-exec-skipped_diff}{3}
\DefMacro{table-numbers_of_programs_and_bugs-20231115-row4-exec-skipped_crash}{0}
\DefMacro{table-numbers_of_programs_and_bugs-20231115-row4-exec-final_diff}{0}
\DefMacro{table-numbers_of_programs_and_bugs-20231115-row4-exec-final_crash}{0}
\DefMacro{table-numbers_of_programs_and_bugs-20231115-row5-project}{statistics}
\DefMacro{table-numbers_of_programs_and_bugs-20231115-row5-num_unit_tests}{1,886}
\DefMacro{table-numbers_of_programs_and_bugs-20231115-row5-num_unique_entry_methods}{347}
\DefMacro{table-numbers_of_programs_and_bugs-20231115-row5-num_unique_classes}{25}
\DefMacro{table-numbers_of_programs_and_bugs-20231115-row5-convert-fail}{181}
\DefMacro{table-numbers_of_programs_and_bugs-20231115-row5-convert-timeout}{0}
\DefMacro{table-numbers_of_programs_and_bugs-20231115-row5-convert-not_compile}{0}
\DefMacro{table-numbers_of_programs_and_bugs-20231115-row5-convert-no_original}{0}
\DefMacro{table-numbers_of_programs_and_bugs-20231115-row5-gen-fail}{180}
\DefMacro{table-numbers_of_programs_and_bugs-20231115-row5-gen-timeout}{26}
\DefMacro{table-numbers_of_programs_and_bugs-20231115-row5-gen-no_hole}{0}
\DefMacro{table-numbers_of_programs_and_bugs-20231115-row5-gen-no_gen}{522}
\DefMacro{table-numbers_of_programs_and_bugs-20231115-row5-num_working_templates}{1,183}
\DefMacro{table-numbers_of_programs_and_bugs-20231115-row5-num_unique_entry_methods_out_of_working_templates}{237}
\DefMacro{table-numbers_of_programs_and_bugs-20231115-row5-num_unique_classes_out_of_working_templates}{22}
\DefMacro{table-numbers_of_programs_and_bugs-20231115-row5-exec-total}{9,551}
\DefMacro{table-numbers_of_programs_and_bugs-20231115-row5-exec-not_compile}{0}
\DefMacro{table-numbers_of_programs_and_bugs-20231115-row5-exec-timeout}{166}
\DefMacro{table-numbers_of_programs_and_bugs-20231115-row5-exec-ok}{9,385}
\DefMacro{table-numbers_of_programs_and_bugs-20231115-row5-exec-skipped_diff}{0}
\DefMacro{table-numbers_of_programs_and_bugs-20231115-row5-exec-skipped_crash}{0}
\DefMacro{table-numbers_of_programs_and_bugs-20231115-row5-exec-final_diff}{0}
\DefMacro{table-numbers_of_programs_and_bugs-20231115-row5-exec-final_crash}{0}
\DefMacro{table-numbers_of_programs_and_bugs-20231115-row6-project}{codec}
\DefMacro{table-numbers_of_programs_and_bugs-20231115-row6-num_unit_tests}{1,480}
\DefMacro{table-numbers_of_programs_and_bugs-20231115-row6-num_unique_entry_methods}{379}
\DefMacro{table-numbers_of_programs_and_bugs-20231115-row6-num_unique_classes}{49}
\DefMacro{table-numbers_of_programs_and_bugs-20231115-row6-convert-fail}{19}
\DefMacro{table-numbers_of_programs_and_bugs-20231115-row6-convert-timeout}{0}
\DefMacro{table-numbers_of_programs_and_bugs-20231115-row6-convert-not_compile}{0}
\DefMacro{table-numbers_of_programs_and_bugs-20231115-row6-convert-no_original}{0}
\DefMacro{table-numbers_of_programs_and_bugs-20231115-row6-gen-fail}{76}
\DefMacro{table-numbers_of_programs_and_bugs-20231115-row6-gen-timeout}{43}
\DefMacro{table-numbers_of_programs_and_bugs-20231115-row6-gen-no_hole}{0}
\DefMacro{table-numbers_of_programs_and_bugs-20231115-row6-gen-no_gen}{351}
\DefMacro{table-numbers_of_programs_and_bugs-20231115-row6-num_working_templates}{1,110}
\DefMacro{table-numbers_of_programs_and_bugs-20231115-row6-num_unique_entry_methods_out_of_working_templates}{288}
\DefMacro{table-numbers_of_programs_and_bugs-20231115-row6-num_unique_classes_out_of_working_templates}{39}
\DefMacro{table-numbers_of_programs_and_bugs-20231115-row6-exec-total}{6,195}
\DefMacro{table-numbers_of_programs_and_bugs-20231115-row6-exec-not_compile}{0}
\DefMacro{table-numbers_of_programs_and_bugs-20231115-row6-exec-timeout}{437}
\DefMacro{table-numbers_of_programs_and_bugs-20231115-row6-exec-ok}{5,746}
\DefMacro{table-numbers_of_programs_and_bugs-20231115-row6-exec-skipped_diff}{12}
\DefMacro{table-numbers_of_programs_and_bugs-20231115-row6-exec-skipped_crash}{0}
\DefMacro{table-numbers_of_programs_and_bugs-20231115-row6-exec-final_diff}{0}
\DefMacro{table-numbers_of_programs_and_bugs-20231115-row6-exec-final_crash}{0}
\DefMacro{table-numbers_of_programs_and_bugs-20231115-row7-project}{jfreechart}
\DefMacro{table-numbers_of_programs_and_bugs-20231115-row7-num_unit_tests}{1,337}
\DefMacro{table-numbers_of_programs_and_bugs-20231115-row7-num_unique_entry_methods}{1,152}
\DefMacro{table-numbers_of_programs_and_bugs-20231115-row7-num_unique_classes}{284}
\DefMacro{table-numbers_of_programs_and_bugs-20231115-row7-convert-fail}{3}
\DefMacro{table-numbers_of_programs_and_bugs-20231115-row7-convert-timeout}{53}
\DefMacro{table-numbers_of_programs_and_bugs-20231115-row7-convert-not_compile}{6}
\DefMacro{table-numbers_of_programs_and_bugs-20231115-row7-convert-no_original}{0}
\DefMacro{table-numbers_of_programs_and_bugs-20231115-row7-gen-fail}{249}
\DefMacro{table-numbers_of_programs_and_bugs-20231115-row7-gen-timeout}{63}
\DefMacro{table-numbers_of_programs_and_bugs-20231115-row7-gen-no_hole}{19}
\DefMacro{table-numbers_of_programs_and_bugs-20231115-row7-gen-no_gen}{362}
\DefMacro{table-numbers_of_programs_and_bugs-20231115-row7-num_working_templates}{894}
\DefMacro{table-numbers_of_programs_and_bugs-20231115-row7-num_unique_entry_methods_out_of_working_templates}{772}
\DefMacro{table-numbers_of_programs_and_bugs-20231115-row7-num_unique_classes_out_of_working_templates}{240}
\DefMacro{table-numbers_of_programs_and_bugs-20231115-row7-exec-total}{7,564}
\DefMacro{table-numbers_of_programs_and_bugs-20231115-row7-exec-not_compile}{0}
\DefMacro{table-numbers_of_programs_and_bugs-20231115-row7-exec-timeout}{490}
\DefMacro{table-numbers_of_programs_and_bugs-20231115-row7-exec-ok}{6,949}
\DefMacro{table-numbers_of_programs_and_bugs-20231115-row7-exec-skipped_diff}{123}
\DefMacro{table-numbers_of_programs_and_bugs-20231115-row7-exec-skipped_crash}{1}
\DefMacro{table-numbers_of_programs_and_bugs-20231115-row7-exec-final_diff}{1}
\DefMacro{table-numbers_of_programs_and_bugs-20231115-row7-exec-final_crash}{0}
\DefMacro{table-numbers_of_programs_and_bugs-20231115-row8-project}{zxing}
\DefMacro{table-numbers_of_programs_and_bugs-20231115-row8-num_unit_tests}{1,445}
\DefMacro{table-numbers_of_programs_and_bugs-20231115-row8-num_unique_entry_methods}{341}
\DefMacro{table-numbers_of_programs_and_bugs-20231115-row8-num_unique_classes}{89}
\DefMacro{table-numbers_of_programs_and_bugs-20231115-row8-convert-fail}{69}
\DefMacro{table-numbers_of_programs_and_bugs-20231115-row8-convert-timeout}{0}
\DefMacro{table-numbers_of_programs_and_bugs-20231115-row8-convert-not_compile}{26}
\DefMacro{table-numbers_of_programs_and_bugs-20231115-row8-convert-no_original}{0}
\DefMacro{table-numbers_of_programs_and_bugs-20231115-row8-gen-fail}{47}
\DefMacro{table-numbers_of_programs_and_bugs-20231115-row8-gen-timeout}{88}
\DefMacro{table-numbers_of_programs_and_bugs-20231115-row8-gen-no_hole}{14}
\DefMacro{table-numbers_of_programs_and_bugs-20231115-row8-gen-no_gen}{206}
\DefMacro{table-numbers_of_programs_and_bugs-20231115-row8-num_working_templates}{1,130}
\DefMacro{table-numbers_of_programs_and_bugs-20231115-row8-num_unique_entry_methods_out_of_working_templates}{274}
\DefMacro{table-numbers_of_programs_and_bugs-20231115-row8-num_unique_classes_out_of_working_templates}{73}
\DefMacro{table-numbers_of_programs_and_bugs-20231115-row8-exec-total}{10,174}
\DefMacro{table-numbers_of_programs_and_bugs-20231115-row8-exec-not_compile}{0}
\DefMacro{table-numbers_of_programs_and_bugs-20231115-row8-exec-timeout}{570}
\DefMacro{table-numbers_of_programs_and_bugs-20231115-row8-exec-ok}{9,146}
\DefMacro{table-numbers_of_programs_and_bugs-20231115-row8-exec-skipped_diff}{458}
\DefMacro{table-numbers_of_programs_and_bugs-20231115-row8-exec-skipped_crash}{0}
\DefMacro{table-numbers_of_programs_and_bugs-20231115-row8-exec-final_diff}{0}
\DefMacro{table-numbers_of_programs_and_bugs-20231115-row8-exec-final_crash}{0}
\DefMacro{table-numbers_of_programs_and_bugs-20231115-row9-project}{$\sum$}
\DefMacro{table-numbers_of_programs_and_bugs-20231115-row9-num_unit_tests}{15,664}
\DefMacro{table-numbers_of_programs_and_bugs-20231115-row9-num_unique_entry_methods}{5,414}
\DefMacro{table-numbers_of_programs_and_bugs-20231115-row9-num_unique_classes}{883}
\DefMacro{table-numbers_of_programs_and_bugs-20231115-row9-convert-fail}{598}
\DefMacro{table-numbers_of_programs_and_bugs-20231115-row9-convert-timeout}{53}
\DefMacro{table-numbers_of_programs_and_bugs-20231115-row9-convert-not_compile}{40}
\DefMacro{table-numbers_of_programs_and_bugs-20231115-row9-convert-no_original}{0}
\DefMacro{table-numbers_of_programs_and_bugs-20231115-row9-gen-fail}{926}
\DefMacro{table-numbers_of_programs_and_bugs-20231115-row9-gen-timeout}{435}
\DefMacro{table-numbers_of_programs_and_bugs-20231115-row9-gen-no_hole}{248}
\DefMacro{table-numbers_of_programs_and_bugs-20231115-row9-gen-no_gen}{2,558}
\DefMacro{table-numbers_of_programs_and_bugs-20231115-row9-num_working_templates}{12,167}
\DefMacro{table-numbers_of_programs_and_bugs-20231115-row9-num_unique_entry_methods_out_of_working_templates}{4,053}
\DefMacro{table-numbers_of_programs_and_bugs-20231115-row9-num_unique_classes_out_of_working_templates}{729}
\DefMacro{table-numbers_of_programs_and_bugs-20231115-row9-exec-total}{104,372}
\DefMacro{table-numbers_of_programs_and_bugs-20231115-row9-exec-not_compile}{0}
\DefMacro{table-numbers_of_programs_and_bugs-20231115-row9-exec-timeout}{4,481}
\DefMacro{table-numbers_of_programs_and_bugs-20231115-row9-exec-ok}{99,123}
\DefMacro{table-numbers_of_programs_and_bugs-20231115-row9-exec-skipped_diff}{751}
\DefMacro{table-numbers_of_programs_and_bugs-20231115-row9-exec-skipped_crash}{16}
\DefMacro{table-numbers_of_programs_and_bugs-20231115-row9-exec-final_diff}{1}
\DefMacro{table-numbers_of_programs_and_bugs-20231115-row9-exec-final_crash}{0}

\DefMacro{table-numbers_of_programs_and_bugs-20231108-row0-project}{lang}
\DefMacro{table-numbers_of_programs_and_bugs-20231108-row0-num_unit_tests}{2,060}
\DefMacro{table-numbers_of_programs_and_bugs-20231108-row0-num_unique_entry_methods}{1,201}
\DefMacro{table-numbers_of_programs_and_bugs-20231108-row0-num_unique_classes}{97}
\DefMacro{table-numbers_of_programs_and_bugs-20231108-row0-convert-fail}{129}
\DefMacro{table-numbers_of_programs_and_bugs-20231108-row0-convert-timeout}{0}
\DefMacro{table-numbers_of_programs_and_bugs-20231108-row0-convert-not_compile}{3}
\DefMacro{table-numbers_of_programs_and_bugs-20231108-row0-convert-no_original}{0}
\DefMacro{table-numbers_of_programs_and_bugs-20231108-row0-gen-fail}{38}
\DefMacro{table-numbers_of_programs_and_bugs-20231108-row0-gen-timeout}{70}
\DefMacro{table-numbers_of_programs_and_bugs-20231108-row0-gen-no_hole}{57}
\DefMacro{table-numbers_of_programs_and_bugs-20231108-row0-gen-no_gen}{220}
\DefMacro{table-numbers_of_programs_and_bugs-20231108-row0-num_working_templates}{1,651}
\DefMacro{table-numbers_of_programs_and_bugs-20231108-row0-num_unique_entry_methods_out_of_working_templates}{973}
\DefMacro{table-numbers_of_programs_and_bugs-20231108-row0-num_unique_classes_out_of_working_templates}{80}
\DefMacro{table-numbers_of_programs_and_bugs-20231108-row0-exec-total}{13,971}
\DefMacro{table-numbers_of_programs_and_bugs-20231108-row0-exec-not_compile}{0}
\DefMacro{table-numbers_of_programs_and_bugs-20231108-row0-exec-timeout}{546}
\DefMacro{table-numbers_of_programs_and_bugs-20231108-row0-exec-ok}{13,336}
\DefMacro{table-numbers_of_programs_and_bugs-20231108-row0-exec-skipped_diff}{87}
\DefMacro{table-numbers_of_programs_and_bugs-20231108-row0-exec-skipped_crash}{2}
\DefMacro{table-numbers_of_programs_and_bugs-20231108-row0-exec-final_diff}{0}
\DefMacro{table-numbers_of_programs_and_bugs-20231108-row0-exec-final_crash}{0}
\DefMacro{table-numbers_of_programs_and_bugs-20231108-row1-project}{math}
\DefMacro{table-numbers_of_programs_and_bugs-20231108-row1-num_unit_tests}{1,063}
\DefMacro{table-numbers_of_programs_and_bugs-20231108-row1-num_unique_entry_methods}{590}
\DefMacro{table-numbers_of_programs_and_bugs-20231108-row1-num_unique_classes}{150}
\DefMacro{table-numbers_of_programs_and_bugs-20231108-row1-convert-fail}{14}
\DefMacro{table-numbers_of_programs_and_bugs-20231108-row1-convert-timeout}{0}
\DefMacro{table-numbers_of_programs_and_bugs-20231108-row1-convert-not_compile}{10}
\DefMacro{table-numbers_of_programs_and_bugs-20231108-row1-convert-no_original}{0}
\DefMacro{table-numbers_of_programs_and_bugs-20231108-row1-gen-fail}{262}
\DefMacro{table-numbers_of_programs_and_bugs-20231108-row1-gen-timeout}{27}
\DefMacro{table-numbers_of_programs_and_bugs-20231108-row1-gen-no_hole}{49}
\DefMacro{table-numbers_of_programs_and_bugs-20231108-row1-gen-no_gen}{293}
\DefMacro{table-numbers_of_programs_and_bugs-20231108-row1-num_working_templates}{697}
\DefMacro{table-numbers_of_programs_and_bugs-20231108-row1-num_unique_entry_methods_out_of_working_templates}{418}
\DefMacro{table-numbers_of_programs_and_bugs-20231108-row1-num_unique_classes_out_of_working_templates}{116}
\DefMacro{table-numbers_of_programs_and_bugs-20231108-row1-exec-total}{5,938}
\DefMacro{table-numbers_of_programs_and_bugs-20231108-row1-exec-not_compile}{0}
\DefMacro{table-numbers_of_programs_and_bugs-20231108-row1-exec-timeout}{280}
\DefMacro{table-numbers_of_programs_and_bugs-20231108-row1-exec-ok}{5,544}
\DefMacro{table-numbers_of_programs_and_bugs-20231108-row1-exec-skipped_diff}{114}
\DefMacro{table-numbers_of_programs_and_bugs-20231108-row1-exec-skipped_crash}{0}
\DefMacro{table-numbers_of_programs_and_bugs-20231108-row1-exec-final_diff}{0}
\DefMacro{table-numbers_of_programs_and_bugs-20231108-row1-exec-final_crash}{0}
\DefMacro{table-numbers_of_programs_and_bugs-20231108-row2-project}{text}
\DefMacro{table-numbers_of_programs_and_bugs-20231108-row2-num_unit_tests}{1,696}
\DefMacro{table-numbers_of_programs_and_bugs-20231108-row2-num_unique_entry_methods}{516}
\DefMacro{table-numbers_of_programs_and_bugs-20231108-row2-num_unique_classes}{42}
\DefMacro{table-numbers_of_programs_and_bugs-20231108-row2-convert-fail}{17}
\DefMacro{table-numbers_of_programs_and_bugs-20231108-row2-convert-timeout}{0}
\DefMacro{table-numbers_of_programs_and_bugs-20231108-row2-convert-not_compile}{0}
\DefMacro{table-numbers_of_programs_and_bugs-20231108-row2-convert-no_original}{0}
\DefMacro{table-numbers_of_programs_and_bugs-20231108-row2-gen-fail}{2}
\DefMacro{table-numbers_of_programs_and_bugs-20231108-row2-gen-timeout}{34}
\DefMacro{table-numbers_of_programs_and_bugs-20231108-row2-gen-no_hole}{8}
\DefMacro{table-numbers_of_programs_and_bugs-20231108-row2-gen-no_gen}{396}
\DefMacro{table-numbers_of_programs_and_bugs-20231108-row2-num_working_templates}{1,275}
\DefMacro{table-numbers_of_programs_and_bugs-20231108-row2-num_unique_entry_methods_out_of_working_templates}{443}
\DefMacro{table-numbers_of_programs_and_bugs-20231108-row2-num_unique_classes_out_of_working_templates}{35}
\DefMacro{table-numbers_of_programs_and_bugs-20231108-row2-exec-total}{11,769}
\DefMacro{table-numbers_of_programs_and_bugs-20231108-row2-exec-not_compile}{0}
\DefMacro{table-numbers_of_programs_and_bugs-20231108-row2-exec-timeout}{1,868}
\DefMacro{table-numbers_of_programs_and_bugs-20231108-row2-exec-ok}{9,869}
\DefMacro{table-numbers_of_programs_and_bugs-20231108-row2-exec-skipped_diff}{29}
\DefMacro{table-numbers_of_programs_and_bugs-20231108-row2-exec-skipped_crash}{2}
\DefMacro{table-numbers_of_programs_and_bugs-20231108-row2-exec-final_diff}{0}
\DefMacro{table-numbers_of_programs_and_bugs-20231108-row2-exec-final_crash}{1}
\DefMacro{table-numbers_of_programs_and_bugs-20231108-row3-project}{numbers}
\DefMacro{table-numbers_of_programs_and_bugs-20231108-row3-num_unit_tests}{3,127}
\DefMacro{table-numbers_of_programs_and_bugs-20231108-row3-num_unique_entry_methods}{45}
\DefMacro{table-numbers_of_programs_and_bugs-20231108-row3-num_unique_classes}{4}
\DefMacro{table-numbers_of_programs_and_bugs-20231108-row3-convert-fail}{114}
\DefMacro{table-numbers_of_programs_and_bugs-20231108-row3-convert-timeout}{0}
\DefMacro{table-numbers_of_programs_and_bugs-20231108-row3-convert-not_compile}{0}
\DefMacro{table-numbers_of_programs_and_bugs-20231108-row3-convert-no_original}{0}
\DefMacro{table-numbers_of_programs_and_bugs-20231108-row3-gen-fail}{0}
\DefMacro{table-numbers_of_programs_and_bugs-20231108-row3-gen-timeout}{53}
\DefMacro{table-numbers_of_programs_and_bugs-20231108-row3-gen-no_hole}{0}
\DefMacro{table-numbers_of_programs_and_bugs-20231108-row3-gen-no_gen}{52}
\DefMacro{table-numbers_of_programs_and_bugs-20231108-row3-num_working_templates}{2,961}
\DefMacro{table-numbers_of_programs_and_bugs-20231108-row3-num_unique_entry_methods_out_of_working_templates}{42}
\DefMacro{table-numbers_of_programs_and_bugs-20231108-row3-num_unique_classes_out_of_working_templates}{3}
\DefMacro{table-numbers_of_programs_and_bugs-20231108-row3-exec-total}{29,313}
\DefMacro{table-numbers_of_programs_and_bugs-20231108-row3-exec-not_compile}{0}
\DefMacro{table-numbers_of_programs_and_bugs-20231108-row3-exec-timeout}{62}
\DefMacro{table-numbers_of_programs_and_bugs-20231108-row3-exec-ok}{29,251}
\DefMacro{table-numbers_of_programs_and_bugs-20231108-row3-exec-skipped_diff}{0}
\DefMacro{table-numbers_of_programs_and_bugs-20231108-row3-exec-skipped_crash}{0}
\DefMacro{table-numbers_of_programs_and_bugs-20231108-row3-exec-final_diff}{0}
\DefMacro{table-numbers_of_programs_and_bugs-20231108-row3-exec-final_crash}{0}
\DefMacro{table-numbers_of_programs_and_bugs-20231108-row4-project}{vectorz}
\DefMacro{table-numbers_of_programs_and_bugs-20231108-row4-num_unit_tests}{1,028}
\DefMacro{table-numbers_of_programs_and_bugs-20231108-row4-num_unique_entry_methods}{826}
\DefMacro{table-numbers_of_programs_and_bugs-20231108-row4-num_unique_classes}{159}
\DefMacro{table-numbers_of_programs_and_bugs-20231108-row4-convert-fail}{0}
\DefMacro{table-numbers_of_programs_and_bugs-20231108-row4-convert-timeout}{0}
\DefMacro{table-numbers_of_programs_and_bugs-20231108-row4-convert-not_compile}{0}
\DefMacro{table-numbers_of_programs_and_bugs-20231108-row4-convert-no_original}{0}
\DefMacro{table-numbers_of_programs_and_bugs-20231108-row4-gen-fail}{37}
\DefMacro{table-numbers_of_programs_and_bugs-20231108-row4-gen-timeout}{53}
\DefMacro{table-numbers_of_programs_and_bugs-20231108-row4-gen-no_hole}{95}
\DefMacro{table-numbers_of_programs_and_bugs-20231108-row4-gen-no_gen}{245}
\DefMacro{table-numbers_of_programs_and_bugs-20231108-row4-num_working_templates}{688}
\DefMacro{table-numbers_of_programs_and_bugs-20231108-row4-num_unique_entry_methods_out_of_working_templates}{575}
\DefMacro{table-numbers_of_programs_and_bugs-20231108-row4-num_unique_classes_out_of_working_templates}{145}
\DefMacro{table-numbers_of_programs_and_bugs-20231108-row4-exec-total}{5,174}
\DefMacro{table-numbers_of_programs_and_bugs-20231108-row4-exec-not_compile}{0}
\DefMacro{table-numbers_of_programs_and_bugs-20231108-row4-exec-timeout}{326}
\DefMacro{table-numbers_of_programs_and_bugs-20231108-row4-exec-ok}{4,702}
\DefMacro{table-numbers_of_programs_and_bugs-20231108-row4-exec-skipped_diff}{139}
\DefMacro{table-numbers_of_programs_and_bugs-20231108-row4-exec-skipped_crash}{7}
\DefMacro{table-numbers_of_programs_and_bugs-20231108-row4-exec-final_diff}{0}
\DefMacro{table-numbers_of_programs_and_bugs-20231108-row4-exec-final_crash}{0}
\DefMacro{table-numbers_of_programs_and_bugs-20231108-row5-project}{statistics}
\DefMacro{table-numbers_of_programs_and_bugs-20231108-row5-num_unit_tests}{1,874}
\DefMacro{table-numbers_of_programs_and_bugs-20231108-row5-num_unique_entry_methods}{348}
\DefMacro{table-numbers_of_programs_and_bugs-20231108-row5-num_unique_classes}{25}
\DefMacro{table-numbers_of_programs_and_bugs-20231108-row5-convert-fail}{170}
\DefMacro{table-numbers_of_programs_and_bugs-20231108-row5-convert-timeout}{0}
\DefMacro{table-numbers_of_programs_and_bugs-20231108-row5-convert-not_compile}{0}
\DefMacro{table-numbers_of_programs_and_bugs-20231108-row5-convert-no_original}{0}
\DefMacro{table-numbers_of_programs_and_bugs-20231108-row5-gen-fail}{85}
\DefMacro{table-numbers_of_programs_and_bugs-20231108-row5-gen-timeout}{39}
\DefMacro{table-numbers_of_programs_and_bugs-20231108-row5-gen-no_hole}{0}
\DefMacro{table-numbers_of_programs_and_bugs-20231108-row5-gen-no_gen}{584}
\DefMacro{table-numbers_of_programs_and_bugs-20231108-row5-num_working_templates}{1,120}
\DefMacro{table-numbers_of_programs_and_bugs-20231108-row5-num_unique_entry_methods_out_of_working_templates}{227}
\DefMacro{table-numbers_of_programs_and_bugs-20231108-row5-num_unique_classes_out_of_working_templates}{23}
\DefMacro{table-numbers_of_programs_and_bugs-20231108-row5-exec-total}{9,031}
\DefMacro{table-numbers_of_programs_and_bugs-20231108-row5-exec-not_compile}{0}
\DefMacro{table-numbers_of_programs_and_bugs-20231108-row5-exec-timeout}{196}
\DefMacro{table-numbers_of_programs_and_bugs-20231108-row5-exec-ok}{8,835}
\DefMacro{table-numbers_of_programs_and_bugs-20231108-row5-exec-skipped_diff}{0}
\DefMacro{table-numbers_of_programs_and_bugs-20231108-row5-exec-skipped_crash}{0}
\DefMacro{table-numbers_of_programs_and_bugs-20231108-row5-exec-final_diff}{0}
\DefMacro{table-numbers_of_programs_and_bugs-20231108-row5-exec-final_crash}{0}
\DefMacro{table-numbers_of_programs_and_bugs-20231108-row6-project}{codec}
\DefMacro{table-numbers_of_programs_and_bugs-20231108-row6-num_unit_tests}{1,374}
\DefMacro{table-numbers_of_programs_and_bugs-20231108-row6-num_unique_entry_methods}{371}
\DefMacro{table-numbers_of_programs_and_bugs-20231108-row6-num_unique_classes}{49}
\DefMacro{table-numbers_of_programs_and_bugs-20231108-row6-convert-fail}{19}
\DefMacro{table-numbers_of_programs_and_bugs-20231108-row6-convert-timeout}{0}
\DefMacro{table-numbers_of_programs_and_bugs-20231108-row6-convert-not_compile}{0}
\DefMacro{table-numbers_of_programs_and_bugs-20231108-row6-convert-no_original}{0}
\DefMacro{table-numbers_of_programs_and_bugs-20231108-row6-gen-fail}{76}
\DefMacro{table-numbers_of_programs_and_bugs-20231108-row6-gen-timeout}{33}
\DefMacro{table-numbers_of_programs_and_bugs-20231108-row6-gen-no_hole}{2}
\DefMacro{table-numbers_of_programs_and_bugs-20231108-row6-gen-no_gen}{226}
\DefMacro{table-numbers_of_programs_and_bugs-20231108-row6-num_working_templates}{1,127}
\DefMacro{table-numbers_of_programs_and_bugs-20231108-row6-num_unique_entry_methods_out_of_working_templates}{302}
\DefMacro{table-numbers_of_programs_and_bugs-20231108-row6-num_unique_classes_out_of_working_templates}{36}
\DefMacro{table-numbers_of_programs_and_bugs-20231108-row6-exec-total}{6,258}
\DefMacro{table-numbers_of_programs_and_bugs-20231108-row6-exec-not_compile}{0}
\DefMacro{table-numbers_of_programs_and_bugs-20231108-row6-exec-timeout}{464}
\DefMacro{table-numbers_of_programs_and_bugs-20231108-row6-exec-ok}{5,758}
\DefMacro{table-numbers_of_programs_and_bugs-20231108-row6-exec-skipped_diff}{18}
\DefMacro{table-numbers_of_programs_and_bugs-20231108-row6-exec-skipped_crash}{18}
\DefMacro{table-numbers_of_programs_and_bugs-20231108-row6-exec-final_diff}{0}
\DefMacro{table-numbers_of_programs_and_bugs-20231108-row6-exec-final_crash}{0}
\DefMacro{table-numbers_of_programs_and_bugs-20231108-row7-project}{jfreechart}
\DefMacro{table-numbers_of_programs_and_bugs-20231108-row7-num_unit_tests}{1,287}
\DefMacro{table-numbers_of_programs_and_bugs-20231108-row7-num_unique_entry_methods}{1,123}
\DefMacro{table-numbers_of_programs_and_bugs-20231108-row7-num_unique_classes}{307}
\DefMacro{table-numbers_of_programs_and_bugs-20231108-row7-convert-fail}{11}
\DefMacro{table-numbers_of_programs_and_bugs-20231108-row7-convert-timeout}{50}
\DefMacro{table-numbers_of_programs_and_bugs-20231108-row7-convert-not_compile}{6}
\DefMacro{table-numbers_of_programs_and_bugs-20231108-row7-convert-no_original}{0}
\DefMacro{table-numbers_of_programs_and_bugs-20231108-row7-gen-fail}{195}
\DefMacro{table-numbers_of_programs_and_bugs-20231108-row7-gen-timeout}{57}
\DefMacro{table-numbers_of_programs_and_bugs-20231108-row7-gen-no_hole}{20}
\DefMacro{table-numbers_of_programs_and_bugs-20231108-row7-gen-no_gen}{333}
\DefMacro{table-numbers_of_programs_and_bugs-20231108-row7-num_working_templates}{867}
\DefMacro{table-numbers_of_programs_and_bugs-20231108-row7-num_unique_entry_methods_out_of_working_templates}{761}
\DefMacro{table-numbers_of_programs_and_bugs-20231108-row7-num_unique_classes_out_of_working_templates}{251}
\DefMacro{table-numbers_of_programs_and_bugs-20231108-row7-exec-total}{7,153}
\DefMacro{table-numbers_of_programs_and_bugs-20231108-row7-exec-not_compile}{0}
\DefMacro{table-numbers_of_programs_and_bugs-20231108-row7-exec-timeout}{450}
\DefMacro{table-numbers_of_programs_and_bugs-20231108-row7-exec-ok}{6,687}
\DefMacro{table-numbers_of_programs_and_bugs-20231108-row7-exec-skipped_diff}{16}
\DefMacro{table-numbers_of_programs_and_bugs-20231108-row7-exec-skipped_crash}{0}
\DefMacro{table-numbers_of_programs_and_bugs-20231108-row7-exec-final_diff}{0}
\DefMacro{table-numbers_of_programs_and_bugs-20231108-row7-exec-final_crash}{0}
\DefMacro{table-numbers_of_programs_and_bugs-20231108-row8-project}{zxing}
\DefMacro{table-numbers_of_programs_and_bugs-20231108-row8-num_unit_tests}{1,433}
\DefMacro{table-numbers_of_programs_and_bugs-20231108-row8-num_unique_entry_methods}{347}
\DefMacro{table-numbers_of_programs_and_bugs-20231108-row8-num_unique_classes}{89}
\DefMacro{table-numbers_of_programs_and_bugs-20231108-row8-convert-fail}{53}
\DefMacro{table-numbers_of_programs_and_bugs-20231108-row8-convert-timeout}{0}
\DefMacro{table-numbers_of_programs_and_bugs-20231108-row8-convert-not_compile}{18}
\DefMacro{table-numbers_of_programs_and_bugs-20231108-row8-convert-no_original}{0}
\DefMacro{table-numbers_of_programs_and_bugs-20231108-row8-gen-fail}{71}
\DefMacro{table-numbers_of_programs_and_bugs-20231108-row8-gen-timeout}{55}
\DefMacro{table-numbers_of_programs_and_bugs-20231108-row8-gen-no_hole}{17}
\DefMacro{table-numbers_of_programs_and_bugs-20231108-row8-gen-no_gen}{226}
\DefMacro{table-numbers_of_programs_and_bugs-20231108-row8-num_working_templates}{1,119}
\DefMacro{table-numbers_of_programs_and_bugs-20231108-row8-num_unique_entry_methods_out_of_working_templates}{281}
\DefMacro{table-numbers_of_programs_and_bugs-20231108-row8-num_unique_classes_out_of_working_templates}{75}
\DefMacro{table-numbers_of_programs_and_bugs-20231108-row8-exec-total}{9,722}
\DefMacro{table-numbers_of_programs_and_bugs-20231108-row8-exec-not_compile}{0}
\DefMacro{table-numbers_of_programs_and_bugs-20231108-row8-exec-timeout}{687}
\DefMacro{table-numbers_of_programs_and_bugs-20231108-row8-exec-ok}{8,862}
\DefMacro{table-numbers_of_programs_and_bugs-20231108-row8-exec-skipped_diff}{173}
\DefMacro{table-numbers_of_programs_and_bugs-20231108-row8-exec-skipped_crash}{0}
\DefMacro{table-numbers_of_programs_and_bugs-20231108-row8-exec-final_diff}{0}
\DefMacro{table-numbers_of_programs_and_bugs-20231108-row8-exec-final_crash}{0}
\DefMacro{table-numbers_of_programs_and_bugs-20231108-row9-project}{compress}
\DefMacro{table-numbers_of_programs_and_bugs-20231108-row9-num_unit_tests}{1,063}
\DefMacro{table-numbers_of_programs_and_bugs-20231108-row9-num_unique_entry_methods}{549}
\DefMacro{table-numbers_of_programs_and_bugs-20231108-row9-num_unique_classes}{107}
\DefMacro{table-numbers_of_programs_and_bugs-20231108-row9-convert-fail}{9}
\DefMacro{table-numbers_of_programs_and_bugs-20231108-row9-convert-timeout}{0}
\DefMacro{table-numbers_of_programs_and_bugs-20231108-row9-convert-not_compile}{19}
\DefMacro{table-numbers_of_programs_and_bugs-20231108-row9-convert-no_original}{0}
\DefMacro{table-numbers_of_programs_and_bugs-20231108-row9-gen-fail}{57}
\DefMacro{table-numbers_of_programs_and_bugs-20231108-row9-gen-timeout}{21}
\DefMacro{table-numbers_of_programs_and_bugs-20231108-row9-gen-no_hole}{27}
\DefMacro{table-numbers_of_programs_and_bugs-20231108-row9-gen-no_gen}{120}
\DefMacro{table-numbers_of_programs_and_bugs-20231108-row9-num_working_templates}{888}
\DefMacro{table-numbers_of_programs_and_bugs-20231108-row9-num_unique_entry_methods_out_of_working_templates}{484}
\DefMacro{table-numbers_of_programs_and_bugs-20231108-row9-num_unique_classes_out_of_working_templates}{92}
\DefMacro{table-numbers_of_programs_and_bugs-20231108-row9-exec-total}{7,513}
\DefMacro{table-numbers_of_programs_and_bugs-20231108-row9-exec-not_compile}{0}
\DefMacro{table-numbers_of_programs_and_bugs-20231108-row9-exec-timeout}{211}
\DefMacro{table-numbers_of_programs_and_bugs-20231108-row9-exec-ok}{6,192}
\DefMacro{table-numbers_of_programs_and_bugs-20231108-row9-exec-skipped_diff}{971}
\DefMacro{table-numbers_of_programs_and_bugs-20231108-row9-exec-skipped_crash}{1}
\DefMacro{table-numbers_of_programs_and_bugs-20231108-row9-exec-final_diff}{138}
\DefMacro{table-numbers_of_programs_and_bugs-20231108-row9-exec-final_crash}{0}
\DefMacro{table-numbers_of_programs_and_bugs-20231108-row10-project}{$\sum$}
\DefMacro{table-numbers_of_programs_and_bugs-20231108-row10-num_unit_tests}{16,005}
\DefMacro{table-numbers_of_programs_and_bugs-20231108-row10-num_unique_entry_methods}{5,916}
\DefMacro{table-numbers_of_programs_and_bugs-20231108-row10-num_unique_classes}{1,029}
\DefMacro{table-numbers_of_programs_and_bugs-20231108-row10-convert-fail}{536}
\DefMacro{table-numbers_of_programs_and_bugs-20231108-row10-convert-timeout}{50}
\DefMacro{table-numbers_of_programs_and_bugs-20231108-row10-convert-not_compile}{56}
\DefMacro{table-numbers_of_programs_and_bugs-20231108-row10-convert-no_original}{0}
\DefMacro{table-numbers_of_programs_and_bugs-20231108-row10-gen-fail}{823}
\DefMacro{table-numbers_of_programs_and_bugs-20231108-row10-gen-timeout}{442}
\DefMacro{table-numbers_of_programs_and_bugs-20231108-row10-gen-no_hole}{275}
\DefMacro{table-numbers_of_programs_and_bugs-20231108-row10-gen-no_gen}{2,695}
\DefMacro{table-numbers_of_programs_and_bugs-20231108-row10-num_working_templates}{12,393}
\DefMacro{table-numbers_of_programs_and_bugs-20231108-row10-num_unique_entry_methods_out_of_working_templates}{4,506}
\DefMacro{table-numbers_of_programs_and_bugs-20231108-row10-num_unique_classes_out_of_working_templates}{856}
\DefMacro{table-numbers_of_programs_and_bugs-20231108-row10-exec-total}{105,842}
\DefMacro{table-numbers_of_programs_and_bugs-20231108-row10-exec-not_compile}{0}
\DefMacro{table-numbers_of_programs_and_bugs-20231108-row10-exec-timeout}{5,090}
\DefMacro{table-numbers_of_programs_and_bugs-20231108-row10-exec-ok}{99,036}
\DefMacro{table-numbers_of_programs_and_bugs-20231108-row10-exec-skipped_diff}{1,547}
\DefMacro{table-numbers_of_programs_and_bugs-20231108-row10-exec-skipped_crash}{30}
\DefMacro{table-numbers_of_programs_and_bugs-20231108-row10-exec-final_diff}{138}
\DefMacro{table-numbers_of_programs_and_bugs-20231108-row10-exec-final_crash}{1}

\DefMacro{table-numbers_of_programs_and_bugs-20230505-row0-project}{lang}
\DefMacro{table-numbers_of_programs_and_bugs-20230505-row0-num_unit_tests}{1,930}
\DefMacro{table-numbers_of_programs_and_bugs-20230505-row0-num_unique_entry_methods}{1,147}
\DefMacro{table-numbers_of_programs_and_bugs-20230505-row0-num_unique_classes}{97}
\DefMacro{table-numbers_of_programs_and_bugs-20230505-row0-convert-fail}{96}
\DefMacro{table-numbers_of_programs_and_bugs-20230505-row0-convert-timeout}{0}
\DefMacro{table-numbers_of_programs_and_bugs-20230505-row0-convert-not_compile}{0}
\DefMacro{table-numbers_of_programs_and_bugs-20230505-row0-convert-no_original}{0}
\DefMacro{table-numbers_of_programs_and_bugs-20230505-row0-gen-fail}{40}
\DefMacro{table-numbers_of_programs_and_bugs-20230505-row0-gen-timeout}{109}
\DefMacro{table-numbers_of_programs_and_bugs-20230505-row0-gen-no_hole}{51}
\DefMacro{table-numbers_of_programs_and_bugs-20230505-row0-gen-no_gen}{291}
\DefMacro{table-numbers_of_programs_and_bugs-20230505-row0-num_working_templates}{1,492}
\DefMacro{table-numbers_of_programs_and_bugs-20230505-row0-num_unique_entry_methods_out_of_working_templates}{883}
\DefMacro{table-numbers_of_programs_and_bugs-20230505-row0-num_unique_classes_out_of_working_templates}{78}
\DefMacro{table-numbers_of_programs_and_bugs-20230505-row0-exec-total}{12,514}
\DefMacro{table-numbers_of_programs_and_bugs-20230505-row0-exec-not_compile}{0}
\DefMacro{table-numbers_of_programs_and_bugs-20230505-row0-exec-timeout}{739}
\DefMacro{table-numbers_of_programs_and_bugs-20230505-row0-exec-ok}{11,655}
\DefMacro{table-numbers_of_programs_and_bugs-20230505-row0-exec-skipped_diff}{115}
\DefMacro{table-numbers_of_programs_and_bugs-20230505-row0-exec-skipped_crash}{5}
\DefMacro{table-numbers_of_programs_and_bugs-20230505-row0-exec-final_diff}{0}
\DefMacro{table-numbers_of_programs_and_bugs-20230505-row0-exec-final_crash}{0}
\DefMacro{table-numbers_of_programs_and_bugs-20230505-row1-project}{math}
\DefMacro{table-numbers_of_programs_and_bugs-20230505-row1-num_unit_tests}{600}
\DefMacro{table-numbers_of_programs_and_bugs-20230505-row1-num_unique_entry_methods}{509}
\DefMacro{table-numbers_of_programs_and_bugs-20230505-row1-num_unique_classes}{158}
\DefMacro{table-numbers_of_programs_and_bugs-20230505-row1-convert-fail}{10}
\DefMacro{table-numbers_of_programs_and_bugs-20230505-row1-convert-timeout}{0}
\DefMacro{table-numbers_of_programs_and_bugs-20230505-row1-convert-not_compile}{5}
\DefMacro{table-numbers_of_programs_and_bugs-20230505-row1-convert-no_original}{0}
\DefMacro{table-numbers_of_programs_and_bugs-20230505-row1-gen-fail}{90}
\DefMacro{table-numbers_of_programs_and_bugs-20230505-row1-gen-timeout}{44}
\DefMacro{table-numbers_of_programs_and_bugs-20230505-row1-gen-no_hole}{20}
\DefMacro{table-numbers_of_programs_and_bugs-20230505-row1-gen-no_gen}{159}
\DefMacro{table-numbers_of_programs_and_bugs-20230505-row1-num_working_templates}{406}
\DefMacro{table-numbers_of_programs_and_bugs-20230505-row1-num_unique_entry_methods_out_of_working_templates}{337}
\DefMacro{table-numbers_of_programs_and_bugs-20230505-row1-num_unique_classes_out_of_working_templates}{119}
\DefMacro{table-numbers_of_programs_and_bugs-20230505-row1-exec-total}{3,424}
\DefMacro{table-numbers_of_programs_and_bugs-20230505-row1-exec-not_compile}{0}
\DefMacro{table-numbers_of_programs_and_bugs-20230505-row1-exec-timeout}{205}
\DefMacro{table-numbers_of_programs_and_bugs-20230505-row1-exec-ok}{3,186}
\DefMacro{table-numbers_of_programs_and_bugs-20230505-row1-exec-skipped_diff}{33}
\DefMacro{table-numbers_of_programs_and_bugs-20230505-row1-exec-skipped_crash}{0}
\DefMacro{table-numbers_of_programs_and_bugs-20230505-row1-exec-final_diff}{0}
\DefMacro{table-numbers_of_programs_and_bugs-20230505-row1-exec-final_crash}{0}
\DefMacro{table-numbers_of_programs_and_bugs-20230505-row2-project}{text}
\DefMacro{table-numbers_of_programs_and_bugs-20230505-row2-num_unit_tests}{1,348}
\DefMacro{table-numbers_of_programs_and_bugs-20230505-row2-num_unique_entry_methods}{504}
\DefMacro{table-numbers_of_programs_and_bugs-20230505-row2-num_unique_classes}{41}
\DefMacro{table-numbers_of_programs_and_bugs-20230505-row2-convert-fail}{15}
\DefMacro{table-numbers_of_programs_and_bugs-20230505-row2-convert-timeout}{0}
\DefMacro{table-numbers_of_programs_and_bugs-20230505-row2-convert-not_compile}{0}
\DefMacro{table-numbers_of_programs_and_bugs-20230505-row2-convert-no_original}{0}
\DefMacro{table-numbers_of_programs_and_bugs-20230505-row2-gen-fail}{2}
\DefMacro{table-numbers_of_programs_and_bugs-20230505-row2-gen-timeout}{51}
\DefMacro{table-numbers_of_programs_and_bugs-20230505-row2-gen-no_hole}{40}
\DefMacro{table-numbers_of_programs_and_bugs-20230505-row2-gen-no_gen}{375}
\DefMacro{table-numbers_of_programs_and_bugs-20230505-row2-num_working_templates}{918}
\DefMacro{table-numbers_of_programs_and_bugs-20230505-row2-num_unique_entry_methods_out_of_working_templates}{372}
\DefMacro{table-numbers_of_programs_and_bugs-20230505-row2-num_unique_classes_out_of_working_templates}{34}
\DefMacro{table-numbers_of_programs_and_bugs-20230505-row2-exec-total}{8,445}
\DefMacro{table-numbers_of_programs_and_bugs-20230505-row2-exec-not_compile}{0}
\DefMacro{table-numbers_of_programs_and_bugs-20230505-row2-exec-timeout}{1,712}
\DefMacro{table-numbers_of_programs_and_bugs-20230505-row2-exec-ok}{6,722}
\DefMacro{table-numbers_of_programs_and_bugs-20230505-row2-exec-skipped_diff}{9}
\DefMacro{table-numbers_of_programs_and_bugs-20230505-row2-exec-skipped_crash}{1}
\DefMacro{table-numbers_of_programs_and_bugs-20230505-row2-exec-final_diff}{1}
\DefMacro{table-numbers_of_programs_and_bugs-20230505-row2-exec-final_crash}{0}
\DefMacro{table-numbers_of_programs_and_bugs-20230505-row3-project}{numbers}
\DefMacro{table-numbers_of_programs_and_bugs-20230505-row3-num_unit_tests}{3,079}
\DefMacro{table-numbers_of_programs_and_bugs-20230505-row3-num_unique_entry_methods}{45}
\DefMacro{table-numbers_of_programs_and_bugs-20230505-row3-num_unique_classes}{4}
\DefMacro{table-numbers_of_programs_and_bugs-20230505-row3-convert-fail}{103}
\DefMacro{table-numbers_of_programs_and_bugs-20230505-row3-convert-timeout}{0}
\DefMacro{table-numbers_of_programs_and_bugs-20230505-row3-convert-not_compile}{0}
\DefMacro{table-numbers_of_programs_and_bugs-20230505-row3-convert-no_original}{0}
\DefMacro{table-numbers_of_programs_and_bugs-20230505-row3-gen-fail}{0}
\DefMacro{table-numbers_of_programs_and_bugs-20230505-row3-gen-timeout}{62}
\DefMacro{table-numbers_of_programs_and_bugs-20230505-row3-gen-no_hole}{0}
\DefMacro{table-numbers_of_programs_and_bugs-20230505-row3-gen-no_gen}{84}
\DefMacro{table-numbers_of_programs_and_bugs-20230505-row3-num_working_templates}{2,892}
\DefMacro{table-numbers_of_programs_and_bugs-20230505-row3-num_unique_entry_methods_out_of_working_templates}{42}
\DefMacro{table-numbers_of_programs_and_bugs-20230505-row3-num_unique_classes_out_of_working_templates}{3}
\DefMacro{table-numbers_of_programs_and_bugs-20230505-row3-exec-total}{28,816}
\DefMacro{table-numbers_of_programs_and_bugs-20230505-row3-exec-not_compile}{0}
\DefMacro{table-numbers_of_programs_and_bugs-20230505-row3-exec-timeout}{75}
\DefMacro{table-numbers_of_programs_and_bugs-20230505-row3-exec-ok}{28,741}
\DefMacro{table-numbers_of_programs_and_bugs-20230505-row3-exec-skipped_diff}{0}
\DefMacro{table-numbers_of_programs_and_bugs-20230505-row3-exec-skipped_crash}{0}
\DefMacro{table-numbers_of_programs_and_bugs-20230505-row3-exec-final_diff}{0}
\DefMacro{table-numbers_of_programs_and_bugs-20230505-row3-exec-final_crash}{0}
\DefMacro{table-numbers_of_programs_and_bugs-20230505-row4-project}{vectorz}
\DefMacro{table-numbers_of_programs_and_bugs-20230505-row4-num_unit_tests}{472}
\DefMacro{table-numbers_of_programs_and_bugs-20230505-row4-num_unique_entry_methods}{420}
\DefMacro{table-numbers_of_programs_and_bugs-20230505-row4-num_unique_classes}{108}
\DefMacro{table-numbers_of_programs_and_bugs-20230505-row4-convert-fail}{1}
\DefMacro{table-numbers_of_programs_and_bugs-20230505-row4-convert-timeout}{0}
\DefMacro{table-numbers_of_programs_and_bugs-20230505-row4-convert-not_compile}{0}
\DefMacro{table-numbers_of_programs_and_bugs-20230505-row4-convert-no_original}{0}
\DefMacro{table-numbers_of_programs_and_bugs-20230505-row4-gen-fail}{14}
\DefMacro{table-numbers_of_programs_and_bugs-20230505-row4-gen-timeout}{14}
\DefMacro{table-numbers_of_programs_and_bugs-20230505-row4-gen-no_hole}{51}
\DefMacro{table-numbers_of_programs_and_bugs-20230505-row4-gen-no_gen}{96}
\DefMacro{table-numbers_of_programs_and_bugs-20230505-row4-num_working_templates}{324}
\DefMacro{table-numbers_of_programs_and_bugs-20230505-row4-num_unique_entry_methods_out_of_working_templates}{295}
\DefMacro{table-numbers_of_programs_and_bugs-20230505-row4-num_unique_classes_out_of_working_templates}{97}
\DefMacro{table-numbers_of_programs_and_bugs-20230505-row4-exec-total}{2,296}
\DefMacro{table-numbers_of_programs_and_bugs-20230505-row4-exec-not_compile}{0}
\DefMacro{table-numbers_of_programs_and_bugs-20230505-row4-exec-timeout}{217}
\DefMacro{table-numbers_of_programs_and_bugs-20230505-row4-exec-ok}{2,020}
\DefMacro{table-numbers_of_programs_and_bugs-20230505-row4-exec-skipped_diff}{59}
\DefMacro{table-numbers_of_programs_and_bugs-20230505-row4-exec-skipped_crash}{0}
\DefMacro{table-numbers_of_programs_and_bugs-20230505-row4-exec-final_diff}{0}
\DefMacro{table-numbers_of_programs_and_bugs-20230505-row4-exec-final_crash}{0}
\DefMacro{table-numbers_of_programs_and_bugs-20230505-row5-project}{statistics}
\DefMacro{table-numbers_of_programs_and_bugs-20230505-row5-num_unit_tests}{1,924}
\DefMacro{table-numbers_of_programs_and_bugs-20230505-row5-num_unique_entry_methods}{374}
\DefMacro{table-numbers_of_programs_and_bugs-20230505-row5-num_unique_classes}{27}
\DefMacro{table-numbers_of_programs_and_bugs-20230505-row5-convert-fail}{163}
\DefMacro{table-numbers_of_programs_and_bugs-20230505-row5-convert-timeout}{0}
\DefMacro{table-numbers_of_programs_and_bugs-20230505-row5-convert-not_compile}{0}
\DefMacro{table-numbers_of_programs_and_bugs-20230505-row5-convert-no_original}{0}
\DefMacro{table-numbers_of_programs_and_bugs-20230505-row5-gen-fail}{131}
\DefMacro{table-numbers_of_programs_and_bugs-20230505-row5-gen-timeout}{49}
\DefMacro{table-numbers_of_programs_and_bugs-20230505-row5-gen-no_hole}{0}
\DefMacro{table-numbers_of_programs_and_bugs-20230505-row5-gen-no_gen}{639}
\DefMacro{table-numbers_of_programs_and_bugs-20230505-row5-num_working_templates}{1,122}
\DefMacro{table-numbers_of_programs_and_bugs-20230505-row5-num_unique_entry_methods_out_of_working_templates}{250}
\DefMacro{table-numbers_of_programs_and_bugs-20230505-row5-num_unique_classes_out_of_working_templates}{25}
\DefMacro{table-numbers_of_programs_and_bugs-20230505-row5-exec-total}{8,621}
\DefMacro{table-numbers_of_programs_and_bugs-20230505-row5-exec-not_compile}{0}
\DefMacro{table-numbers_of_programs_and_bugs-20230505-row5-exec-timeout}{258}
\DefMacro{table-numbers_of_programs_and_bugs-20230505-row5-exec-ok}{8,363}
\DefMacro{table-numbers_of_programs_and_bugs-20230505-row5-exec-skipped_diff}{0}
\DefMacro{table-numbers_of_programs_and_bugs-20230505-row5-exec-skipped_crash}{0}
\DefMacro{table-numbers_of_programs_and_bugs-20230505-row5-exec-final_diff}{0}
\DefMacro{table-numbers_of_programs_and_bugs-20230505-row5-exec-final_crash}{0}
\DefMacro{table-numbers_of_programs_and_bugs-20230505-row6-project}{codec}
\DefMacro{table-numbers_of_programs_and_bugs-20230505-row6-num_unit_tests}{1,230}
\DefMacro{table-numbers_of_programs_and_bugs-20230505-row6-num_unique_entry_methods}{375}
\DefMacro{table-numbers_of_programs_and_bugs-20230505-row6-num_unique_classes}{47}
\DefMacro{table-numbers_of_programs_and_bugs-20230505-row6-convert-fail}{15}
\DefMacro{table-numbers_of_programs_and_bugs-20230505-row6-convert-timeout}{0}
\DefMacro{table-numbers_of_programs_and_bugs-20230505-row6-convert-not_compile}{10}
\DefMacro{table-numbers_of_programs_and_bugs-20230505-row6-convert-no_original}{0}
\DefMacro{table-numbers_of_programs_and_bugs-20230505-row6-gen-fail}{76}
\DefMacro{table-numbers_of_programs_and_bugs-20230505-row6-gen-timeout}{38}
\DefMacro{table-numbers_of_programs_and_bugs-20230505-row6-gen-no_hole}{2}
\DefMacro{table-numbers_of_programs_and_bugs-20230505-row6-gen-no_gen}{278}
\DefMacro{table-numbers_of_programs_and_bugs-20230505-row6-num_working_templates}{925}
\DefMacro{table-numbers_of_programs_and_bugs-20230505-row6-num_unique_entry_methods_out_of_working_templates}{298}
\DefMacro{table-numbers_of_programs_and_bugs-20230505-row6-num_unique_classes_out_of_working_templates}{37}
\DefMacro{table-numbers_of_programs_and_bugs-20230505-row6-exec-total}{5,077}
\DefMacro{table-numbers_of_programs_and_bugs-20230505-row6-exec-not_compile}{0}
\DefMacro{table-numbers_of_programs_and_bugs-20230505-row6-exec-timeout}{514}
\DefMacro{table-numbers_of_programs_and_bugs-20230505-row6-exec-ok}{4,562}
\DefMacro{table-numbers_of_programs_and_bugs-20230505-row6-exec-skipped_diff}{1}
\DefMacro{table-numbers_of_programs_and_bugs-20230505-row6-exec-skipped_crash}{0}
\DefMacro{table-numbers_of_programs_and_bugs-20230505-row6-exec-final_diff}{0}
\DefMacro{table-numbers_of_programs_and_bugs-20230505-row6-exec-final_crash}{0}
\DefMacro{table-numbers_of_programs_and_bugs-20230505-row7-project}{jfreechart}
\DefMacro{table-numbers_of_programs_and_bugs-20230505-row7-num_unit_tests}{562}
\DefMacro{table-numbers_of_programs_and_bugs-20230505-row7-num_unique_entry_methods}{505}
\DefMacro{table-numbers_of_programs_and_bugs-20230505-row7-num_unique_classes}{178}
\DefMacro{table-numbers_of_programs_and_bugs-20230505-row7-convert-fail}{5}
\DefMacro{table-numbers_of_programs_and_bugs-20230505-row7-convert-timeout}{29}
\DefMacro{table-numbers_of_programs_and_bugs-20230505-row7-convert-not_compile}{4}
\DefMacro{table-numbers_of_programs_and_bugs-20230505-row7-convert-no_original}{0}
\DefMacro{table-numbers_of_programs_and_bugs-20230505-row7-gen-fail}{76}
\DefMacro{table-numbers_of_programs_and_bugs-20230505-row7-gen-timeout}{32}
\DefMacro{table-numbers_of_programs_and_bugs-20230505-row7-gen-no_hole}{10}
\DefMacro{table-numbers_of_programs_and_bugs-20230505-row7-gen-no_gen}{130}
\DefMacro{table-numbers_of_programs_and_bugs-20230505-row7-num_working_templates}{384}
\DefMacro{table-numbers_of_programs_and_bugs-20230505-row7-num_unique_entry_methods_out_of_working_templates}{347}
\DefMacro{table-numbers_of_programs_and_bugs-20230505-row7-num_unique_classes_out_of_working_templates}{143}
\DefMacro{table-numbers_of_programs_and_bugs-20230505-row7-exec-total}{3,256}
\DefMacro{table-numbers_of_programs_and_bugs-20230505-row7-exec-not_compile}{0}
\DefMacro{table-numbers_of_programs_and_bugs-20230505-row7-exec-timeout}{401}
\DefMacro{table-numbers_of_programs_and_bugs-20230505-row7-exec-ok}{2,841}
\DefMacro{table-numbers_of_programs_and_bugs-20230505-row7-exec-skipped_diff}{11}
\DefMacro{table-numbers_of_programs_and_bugs-20230505-row7-exec-skipped_crash}{2}
\DefMacro{table-numbers_of_programs_and_bugs-20230505-row7-exec-final_diff}{1}
\DefMacro{table-numbers_of_programs_and_bugs-20230505-row7-exec-final_crash}{0}
\DefMacro{table-numbers_of_programs_and_bugs-20230505-row8-project}{compress}
\DefMacro{table-numbers_of_programs_and_bugs-20230505-row8-num_unit_tests}{1,051}
\DefMacro{table-numbers_of_programs_and_bugs-20230505-row8-num_unique_entry_methods}{579}
\DefMacro{table-numbers_of_programs_and_bugs-20230505-row8-num_unique_classes}{110}
\DefMacro{table-numbers_of_programs_and_bugs-20230505-row8-convert-fail}{6}
\DefMacro{table-numbers_of_programs_and_bugs-20230505-row8-convert-timeout}{0}
\DefMacro{table-numbers_of_programs_and_bugs-20230505-row8-convert-not_compile}{15}
\DefMacro{table-numbers_of_programs_and_bugs-20230505-row8-convert-no_original}{0}
\DefMacro{table-numbers_of_programs_and_bugs-20230505-row8-gen-fail}{40}
\DefMacro{table-numbers_of_programs_and_bugs-20230505-row8-gen-timeout}{17}
\DefMacro{table-numbers_of_programs_and_bugs-20230505-row8-gen-no_hole}{16}
\DefMacro{table-numbers_of_programs_and_bugs-20230505-row8-gen-no_gen}{172}
\DefMacro{table-numbers_of_programs_and_bugs-20230505-row8-num_working_templates}{842}
\DefMacro{table-numbers_of_programs_and_bugs-20230505-row8-num_unique_entry_methods_out_of_working_templates}{490}
\DefMacro{table-numbers_of_programs_and_bugs-20230505-row8-num_unique_classes_out_of_working_templates}{96}
\DefMacro{table-numbers_of_programs_and_bugs-20230505-row8-exec-total}{7,217}
\DefMacro{table-numbers_of_programs_and_bugs-20230505-row8-exec-not_compile}{0}
\DefMacro{table-numbers_of_programs_and_bugs-20230505-row8-exec-timeout}{267}
\DefMacro{table-numbers_of_programs_and_bugs-20230505-row8-exec-ok}{6,112}
\DefMacro{table-numbers_of_programs_and_bugs-20230505-row8-exec-skipped_diff}{686}
\DefMacro{table-numbers_of_programs_and_bugs-20230505-row8-exec-skipped_crash}{1}
\DefMacro{table-numbers_of_programs_and_bugs-20230505-row8-exec-final_diff}{151}
\DefMacro{table-numbers_of_programs_and_bugs-20230505-row8-exec-final_crash}{0}
\DefMacro{table-numbers_of_programs_and_bugs-20230505-row9-project}{$\sum$}
\DefMacro{table-numbers_of_programs_and_bugs-20230505-row9-num_unit_tests}{12,196}
\DefMacro{table-numbers_of_programs_and_bugs-20230505-row9-num_unique_entry_methods}{4,458}
\DefMacro{table-numbers_of_programs_and_bugs-20230505-row9-num_unique_classes}{770}
\DefMacro{table-numbers_of_programs_and_bugs-20230505-row9-convert-fail}{414}
\DefMacro{table-numbers_of_programs_and_bugs-20230505-row9-convert-timeout}{29}
\DefMacro{table-numbers_of_programs_and_bugs-20230505-row9-convert-not_compile}{34}
\DefMacro{table-numbers_of_programs_and_bugs-20230505-row9-convert-no_original}{0}
\DefMacro{table-numbers_of_programs_and_bugs-20230505-row9-gen-fail}{469}
\DefMacro{table-numbers_of_programs_and_bugs-20230505-row9-gen-timeout}{416}
\DefMacro{table-numbers_of_programs_and_bugs-20230505-row9-gen-no_hole}{190}
\DefMacro{table-numbers_of_programs_and_bugs-20230505-row9-gen-no_gen}{2,224}
\DefMacro{table-numbers_of_programs_and_bugs-20230505-row9-num_working_templates}{9,305}
\DefMacro{table-numbers_of_programs_and_bugs-20230505-row9-num_unique_entry_methods_out_of_working_templates}{3,314}
\DefMacro{table-numbers_of_programs_and_bugs-20230505-row9-num_unique_classes_out_of_working_templates}{632}
\DefMacro{table-numbers_of_programs_and_bugs-20230505-row9-exec-total}{79,666}
\DefMacro{table-numbers_of_programs_and_bugs-20230505-row9-exec-not_compile}{0}
\DefMacro{table-numbers_of_programs_and_bugs-20230505-row9-exec-timeout}{4,388}
\DefMacro{table-numbers_of_programs_and_bugs-20230505-row9-exec-ok}{74,202}
\DefMacro{table-numbers_of_programs_and_bugs-20230505-row9-exec-skipped_diff}{914}
\DefMacro{table-numbers_of_programs_and_bugs-20230505-row9-exec-skipped_crash}{9}
\DefMacro{table-numbers_of_programs_and_bugs-20230505-row9-exec-final_diff}{153}
\DefMacro{table-numbers_of_programs_and_bugs-20230505-row9-exec-final_crash}{0}

\DefMacro{table-numbers_of_programs_and_bugs-caption-common}{Numbers of programs and bugs. U.E.M. means ``Unique Entry Methods''; U.C. means ``Unique Classes''; Tmpl. means ``\Template programs''; TO means ``Timeout''; N.C. means ``Not Compile''; N.O. means ``No Original program''; N.H. means ``No generated programs due to no (reachable) Hole''; N.G. means ``No Generated programs out of other reasons''; W.T. means ``Working \Templates''; U.C.W.T. means ``Unique Classes out of Working \Templates''; U.E.M.W.T. means ``Unique Entry Methods out of Working \Templates''; Skp. means ``Skipped''}
\DefMacro{table-numbers_of_programs_and_bugs-20230904-caption}{\UseMacro{table-numbers_of_programs_and_bugs-caption-common}.\label{tab:numbers-of-programs-and-bugs-20230904} results-20230904}
\DefMacro{table-numbers_of_programs_and_bugs-20230423-caption}{\UseMacro{table-numbers_of_programs_and_bugs-caption-common}.\label{tab:numbers-of-programs-and-bugs-20230423} results-20230423}
\DefMacro{table-numbers_of_programs_and_bugs-20230409_20230411-caption}{\UseMacro{table-numbers_of_programs_and_bugs-caption-common}.\label{tab:numbers-of-programs-and-bugs-20230409_20230411} results-20230409 \& results-20230411}

\DefMacro{table-numbers_of_programs_and_bugs-20231009-caption}{\UseMacro{table-numbers_of_programs_and_bugs-caption-common}.\label{tab:numbers-of-programs-and-bugs-20231009} results-20231009}
\DefMacro{table-numbers_of_programs_and_bugs-20230910-caption}{\UseMacro{table-numbers_of_programs_and_bugs-caption-common}.\label{tab:numbers-of-programs-and-bugs-20230910} results-20230910}
\DefMacro{table-numbers_of_programs_and_bugs-20230412_20230415-caption}{\UseMacro{table-numbers_of_programs_and_bugs-caption-common}.\label{tab:numbers-of-programs-and-bugs-20230412_20230415} results-20230412 \& results-20230415}

\DefMacro{table-numbers_of_programs_and_bugs-20230419-caption}{\UseMacro{table-numbers_of_programs_and_bugs-caption-common}.\label{tab:numbers-of-programs-and-bugs-20230419} results-20230419}
\DefMacro{table-numbers_of_programs_and_bugs-20230418-caption}{\UseMacro{table-numbers_of_programs_and_bugs-caption-common}.\label{tab:numbers-of-programs-and-bugs-20230418} results-20230418}
\DefMacro{table-numbers_of_programs_and_bugs-20230417-caption}{\UseMacro{table-numbers_of_programs_and_bugs-caption-common}.\label{tab:numbers-of-programs-and-bugs-20230417} results-20230417}

\DefMacro{table-numbers_of_programs_and_bugs-20231018-caption}{\UseMacro{table-numbers_of_programs_and_bugs-caption-common}.\label{tab:numbers-of-programs-and-bugs-20231018} results-20231018}
\DefMacro{table-numbers_of_programs_and_bugs-20231013-caption}{\UseMacro{table-numbers_of_programs_and_bugs-caption-common}.\label{tab:numbers-of-programs-and-bugs-20231013} results-20231013}
\DefMacro{table-numbers_of_programs_and_bugs-20230421-caption}{\UseMacro{table-numbers_of_programs_and_bugs-caption-common}.\label{tab:numbers-of-programs-and-bugs-20230421} results-20230421}

\DefMacro{table-numbers_of_programs_and_bugs-20231002-caption}{\UseMacro{table-numbers_of_programs_and_bugs-caption-common}.\label{tab:numbers-of-programs-and-bugs-20231002} results-20231002}
\DefMacro{table-numbers_of_programs_and_bugs-20230930-caption}{\UseMacro{table-numbers_of_programs_and_bugs-caption-common}.\label{tab:numbers-of-programs-and-bugs-20230930} results-20230930}
\DefMacro{table-numbers_of_programs_and_bugs-20230928_2-caption}{\UseMacro{table-numbers_of_programs_and_bugs-caption-common}.\label{tab:numbers-of-programs-and-bugs-20230928_2} results-20230928\_2}
\DefMacro{table-numbers_of_programs_and_bugs-20230928-caption}{\UseMacro{table-numbers_of_programs_and_bugs-caption-common}.\label{tab:numbers-of-programs-and-bugs-20230928} results-20230928}
\DefMacro{table-numbers_of_programs_and_bugs-20230924-caption}{\UseMacro{table-numbers_of_programs_and_bugs-caption-common}.\label{tab:numbers-of-programs-and-bugs-20230924} results-20230924}

\DefMacro{table-numbers_of_programs_and_bugs-20231115-caption}{\UseMacro{table-numbers_of_programs_and_bugs-caption-common}.\label{tab:numbers-of-programs-and-bugs-20231115} results-20231115}
\DefMacro{table-numbers_of_programs_and_bugs-20231108-caption}{\UseMacro{table-numbers_of_programs_and_bugs-caption-common}.\label{tab:numbers-of-programs-and-bugs-20231108} results-20231108}
\DefMacro{table-numbers_of_programs_and_bugs-20230505-caption}{\UseMacro{table-numbers_of_programs_and_bugs-caption-common}.\label{tab:numbers-of-programs-and-bugs-20230505} results-20230505}

\DefMacro{table-numbers_of_programs_and_bugs-col-project}{Project}
\DefMacro{table-numbers_of_programs_and_bugs-col-num_unit_tests}{\makecell{\# Unit\\Tests}}
\DefMacro{table-numbers_of_programs_and_bugs-col-num_unique_entry_methods}{\# U.E.M.}
\DefMacro{table-numbers_of_programs_and_bugs-col-num_unique_classes}{\# U.C.}
\DefMacro{table-numbers_of_programs_and_bugs-col-convert}{Convert (\# Tmpl.)}
\DefMacro{table-numbers_of_programs_and_bugs-col-gen}{Generation (\# Tmpl.)}
\DefMacro{table-numbers_of_programs_and_bugs-col-exec}{Execution (\# Generated Programs)}
\DefMacro{table-numbers_of_programs_and_bugs-col-convert-no_original}{N.O.}
\DefMacro{table-numbers_of_programs_and_bugs-col-convert-fail}{Fail}
\DefMacro{table-numbers_of_programs_and_bugs-col-convert-timeout}{TO}
\DefMacro{table-numbers_of_programs_and_bugs-col-convert-not_compile}{N.C.}
\DefMacro{table-numbers_of_programs_and_bugs-col-gen-fail}{Fail}
\DefMacro{table-numbers_of_programs_and_bugs-col-gen-timeout}{TO}
\DefMacro{table-numbers_of_programs_and_bugs-col-gen-no_hole}{N.H.}
\DefMacro{table-numbers_of_programs_and_bugs-col-gen-no_gen}{N.G.}
\DefMacro{table-numbers_of_programs_and_bugs-col-num_working_templates}{\# W.T.}
\DefMacro{table-numbers_of_programs_and_bugs-col-num_unique_entry_methods_out_of_working_templates}{\makecell{\# U.E.\\M.W.T.}}
\DefMacro{table-numbers_of_programs_and_bugs-col-num_unique_classes_out_of_working_templates}{\makecell{\# U.C.\\W.T.}}
\DefMacro{table-numbers_of_programs_and_bugs-col-exec-total}{Total}
\DefMacro{table-numbers_of_programs_and_bugs-col-exec-not_compile}{N.C.}
\DefMacro{table-numbers_of_programs_and_bugs-col-exec-timeout}{TO}
\DefMacro{table-numbers_of_programs_and_bugs-col-exec-ok}{OK}
\DefMacro{table-numbers_of_programs_and_bugs-col-exec-skipped_diff}{\makecell{Skp.\\Diff}}
\DefMacro{table-numbers_of_programs_and_bugs-col-exec-skipped_crash}{\makecell{Skp.\\Crash}}
\DefMacro{table-numbers_of_programs_and_bugs-col-exec-final_diff}{\makecell{Final\\Diff}}
\DefMacro{table-numbers_of_programs_and_bugs-col-exec-final_crash}{\makecell{Final\\Crash}}

\DefMacro{table-hole_features-row0-project}{codec}
\DefMacro{table-hole_features-row0-template-total}{11,098}
\DefMacro{table-hole_features-row0-template-ARRAY}{2,476}
\DefMacro{table-hole_features-row0-template-COND}{2,719}
\DefMacro{table-hole_features-row0-template-LOOP}{4,102}
\DefMacro{table-hole_features-row0-template-OBJARG}{10,414}
\DefMacro{table-hole_features-row0-generated-total}{43,147}
\DefMacro{table-hole_features-row0-generated-ARRAY}{19,760}
\DefMacro{table-hole_features-row0-generated-COND}{23,244}
\DefMacro{table-hole_features-row0-generated-LOOP}{23,474}
\DefMacro{table-hole_features-row0-generated-OBJARG}{39,529}
\DefMacro{table-hole_features-row1-project}{compress}
\DefMacro{table-hole_features-row1-template-total}{8,269}
\DefMacro{table-hole_features-row1-template-ARRAY}{1,036}
\DefMacro{table-hole_features-row1-template-COND}{4,133}
\DefMacro{table-hole_features-row1-template-LOOP}{1,818}
\DefMacro{table-hole_features-row1-template-OBJARG}{7,794}
\DefMacro{table-hole_features-row1-generated-total}{62,431}
\DefMacro{table-hole_features-row1-generated-ARRAY}{7,585}
\DefMacro{table-hole_features-row1-generated-COND}{35,774}
\DefMacro{table-hole_features-row1-generated-LOOP}{15,713}
\DefMacro{table-hole_features-row1-generated-OBJARG}{58,578}
\DefMacro{table-hole_features-row2-project}{jfreechart}
\DefMacro{table-hole_features-row2-template-total}{11,041}
\DefMacro{table-hole_features-row2-template-ARRAY}{215}
\DefMacro{table-hole_features-row2-template-COND}{2,073}
\DefMacro{table-hole_features-row2-template-LOOP}{554}
\DefMacro{table-hole_features-row2-template-OBJARG}{10,656}
\DefMacro{table-hole_features-row2-generated-total}{61,883}
\DefMacro{table-hole_features-row2-generated-ARRAY}{1,894}
\DefMacro{table-hole_features-row2-generated-COND}{13,294}
\DefMacro{table-hole_features-row2-generated-LOOP}{4,117}
\DefMacro{table-hole_features-row2-generated-OBJARG}{59,587}
\DefMacro{table-hole_features-row3-project}{lang}
\DefMacro{table-hole_features-row3-template-total}{18,602}
\DefMacro{table-hole_features-row3-template-ARRAY}{3,491}
\DefMacro{table-hole_features-row3-template-COND}{9,156}
\DefMacro{table-hole_features-row3-template-LOOP}{6,050}
\DefMacro{table-hole_features-row3-template-OBJARG}{16,136}
\DefMacro{table-hole_features-row3-generated-total}{117,061}
\DefMacro{table-hole_features-row3-generated-ARRAY}{28,995}
\DefMacro{table-hole_features-row3-generated-COND}{78,002}
\DefMacro{table-hole_features-row3-generated-LOOP}{53,117}
\DefMacro{table-hole_features-row3-generated-OBJARG}{100,349}
\DefMacro{table-hole_features-row4-project}{math}
\DefMacro{table-hole_features-row4-template-total}{20,692}
\DefMacro{table-hole_features-row4-template-ARRAY}{3,964}
\DefMacro{table-hole_features-row4-template-COND}{10,116}
\DefMacro{table-hole_features-row4-template-LOOP}{5,311}
\DefMacro{table-hole_features-row4-template-OBJARG}{18,097}
\DefMacro{table-hole_features-row4-generated-total}{153,912}
\DefMacro{table-hole_features-row4-generated-ARRAY}{33,044}
\DefMacro{table-hole_features-row4-generated-COND}{84,054}
\DefMacro{table-hole_features-row4-generated-LOOP}{43,810}
\DefMacro{table-hole_features-row4-generated-OBJARG}{139,857}
\DefMacro{table-hole_features-row5-project}{numbers}
\DefMacro{table-hole_features-row5-template-total}{18,021}
\DefMacro{table-hole_features-row5-template-ARRAY}{826}
\DefMacro{table-hole_features-row5-template-COND}{6,473}
\DefMacro{table-hole_features-row5-template-LOOP}{2,449}
\DefMacro{table-hole_features-row5-template-OBJARG}{1,971}
\DefMacro{table-hole_features-row5-generated-total}{98,218}
\DefMacro{table-hole_features-row5-generated-ARRAY}{8,260}
\DefMacro{table-hole_features-row5-generated-COND}{63,496}
\DefMacro{table-hole_features-row5-generated-LOOP}{15,599}
\DefMacro{table-hole_features-row5-generated-OBJARG}{9,322}
\DefMacro{table-hole_features-row6-project}{statistics}
\DefMacro{table-hole_features-row6-template-total}{10,010}
\DefMacro{table-hole_features-row6-template-ARRAY}{11}
\DefMacro{table-hole_features-row6-template-COND}{4,396}
\DefMacro{table-hole_features-row6-template-LOOP}{159}
\DefMacro{table-hole_features-row6-template-OBJARG}{9,188}
\DefMacro{table-hole_features-row6-generated-total}{44,181}
\DefMacro{table-hole_features-row6-generated-ARRAY}{37}
\DefMacro{table-hole_features-row6-generated-COND}{34,634}
\DefMacro{table-hole_features-row6-generated-LOOP}{810}
\DefMacro{table-hole_features-row6-generated-OBJARG}{38,622}
\DefMacro{table-hole_features-row7-project}{text}
\DefMacro{table-hole_features-row7-template-total}{11,532}
\DefMacro{table-hole_features-row7-template-ARRAY}{2,423}
\DefMacro{table-hole_features-row7-template-COND}{5,105}
\DefMacro{table-hole_features-row7-template-LOOP}{3,506}
\DefMacro{table-hole_features-row7-template-OBJARG}{10,752}
\DefMacro{table-hole_features-row7-generated-total}{66,395}
\DefMacro{table-hole_features-row7-generated-ARRAY}{18,975}
\DefMacro{table-hole_features-row7-generated-COND}{46,965}
\DefMacro{table-hole_features-row7-generated-LOOP}{31,068}
\DefMacro{table-hole_features-row7-generated-OBJARG}{61,473}
\DefMacro{table-hole_features-row8-project}{vectorz}
\DefMacro{table-hole_features-row8-template-total}{23,005}
\DefMacro{table-hole_features-row8-template-ARRAY}{6,031}
\DefMacro{table-hole_features-row8-template-COND}{11,277}
\DefMacro{table-hole_features-row8-template-LOOP}{6,904}
\DefMacro{table-hole_features-row8-template-OBJARG}{21,718}
\DefMacro{table-hole_features-row8-generated-total}{175,814}
\DefMacro{table-hole_features-row8-generated-ARRAY}{49,594}
\DefMacro{table-hole_features-row8-generated-COND}{94,726}
\DefMacro{table-hole_features-row8-generated-LOOP}{58,759}
\DefMacro{table-hole_features-row8-generated-OBJARG}{165,803}
\DefMacro{table-hole_features-row9-project}{zxing}
\DefMacro{table-hole_features-row9-template-total}{10,925}
\DefMacro{table-hole_features-row9-template-ARRAY}{1,660}
\DefMacro{table-hole_features-row9-template-COND}{3,767}
\DefMacro{table-hole_features-row9-template-LOOP}{2,375}
\DefMacro{table-hole_features-row9-template-OBJARG}{10,145}
\DefMacro{table-hole_features-row9-generated-total}{63,136}
\DefMacro{table-hole_features-row9-generated-ARRAY}{13,440}
\DefMacro{table-hole_features-row9-generated-COND}{30,993}
\DefMacro{table-hole_features-row9-generated-LOOP}{19,417}
\DefMacro{table-hole_features-row9-generated-OBJARG}{57,703}
\DefMacro{table-hole_features-row10-project}{$\sum$}
\DefMacro{table-hole_features-row10-template-total}{143,195}
\DefMacro{table-hole_features-row10-template-ARRAY}{22,133}
\DefMacro{table-hole_features-row10-template-COND}{59,215}
\DefMacro{table-hole_features-row10-template-LOOP}{33,228}
\DefMacro{table-hole_features-row10-template-OBJARG}{116,871}
\DefMacro{table-hole_features-row10-generated-total}{886,178}
\DefMacro{table-hole_features-row10-generated-ARRAY}{181,584}
\DefMacro{table-hole_features-row10-generated-COND}{505,182}
\DefMacro{table-hole_features-row10-generated-LOOP}{265,884}
\DefMacro{table-hole_features-row10-generated-OBJARG}{730,823}

\DefMacro{table-hole_features-caption}{Number of templates and programs with different language features. Cond. is Conditional Statements. R.A. is Reference Arguments.\label{tab:hole-features}\vspace{-8pt}}
\DefMacro{table-hole_features-col-project}{Project}
\DefMacro{table-hole_features-col-template}{\# \Template}
\DefMacro{table-hole_features-col-generated}{\# \GeneratedPrograms}
\DefMacro{table-hole_features-col-template-total}{Total}
\DefMacro{table-hole_features-col-template-ARRAY}{Arrays}
\DefMacro{table-hole_features-col-template-COND}{Cond.}
\DefMacro{table-hole_features-col-template-LOOP}{Loops}
\DefMacro{table-hole_features-col-template-OBJARG}{R.A.}
\DefMacro{table-hole_features-col-generated-total}{Total}
\DefMacro{table-hole_features-col-generated-ARRAY}{Arrays}
\DefMacro{table-hole_features-col-generated-COND}{Cond.}
\DefMacro{table-hole_features-col-generated-LOOP}{Loops}
\DefMacro{table-hole_features-col-generated-OBJARG}{R.A.}

\DefMacro{table-bug_features-row0-project}{codec}
\DefMacro{table-bug_features-row0-bugCases-total}{5}
\DefMacro{table-bug_features-row0-bugCases-ARRAY}{1}
\DefMacro{table-bug_features-row0-bugCases-COND}{3}
\DefMacro{table-bug_features-row0-bugCases-LOOP}{1}
\DefMacro{table-bug_features-row0-bugCases-OBJARG}{5}
\DefMacro{table-bug_features-row0-bugs-total}{3}
\DefMacro{table-bug_features-row0-bugs-ARRAY}{1}
\DefMacro{table-bug_features-row0-bugs-COND}{2}
\DefMacro{table-bug_features-row0-bugs-LOOP}{1}
\DefMacro{table-bug_features-row0-bugs-OBJARG}{3}
\DefMacro{table-bug_features-row1-project}{compress}
\DefMacro{table-bug_features-row1-bugCases-total}{6}
\DefMacro{table-bug_features-row1-bugCases-ARRAY}{1}
\DefMacro{table-bug_features-row1-bugCases-COND}{5}
\DefMacro{table-bug_features-row1-bugCases-LOOP}{1}
\DefMacro{table-bug_features-row1-bugCases-OBJARG}{6}
\DefMacro{table-bug_features-row1-bugs-total}{2}
\DefMacro{table-bug_features-row1-bugs-ARRAY}{1}
\DefMacro{table-bug_features-row1-bugs-COND}{1}
\DefMacro{table-bug_features-row1-bugs-LOOP}{1}
\DefMacro{table-bug_features-row1-bugs-OBJARG}{2}
\DefMacro{table-bug_features-row2-project}{jfreechart}
\DefMacro{table-bug_features-row2-bugCases-total}{0}
\DefMacro{table-bug_features-row2-bugCases-ARRAY}{0}
\DefMacro{table-bug_features-row2-bugCases-COND}{0}
\DefMacro{table-bug_features-row2-bugCases-LOOP}{0}
\DefMacro{table-bug_features-row2-bugCases-OBJARG}{0}
\DefMacro{table-bug_features-row2-bugs-total}{0}
\DefMacro{table-bug_features-row2-bugs-ARRAY}{0}
\DefMacro{table-bug_features-row2-bugs-COND}{0}
\DefMacro{table-bug_features-row2-bugs-LOOP}{0}
\DefMacro{table-bug_features-row2-bugs-OBJARG}{0}
\DefMacro{table-bug_features-row3-project}{lang}
\DefMacro{table-bug_features-row3-bugCases-total}{24}
\DefMacro{table-bug_features-row3-bugCases-ARRAY}{6}
\DefMacro{table-bug_features-row3-bugCases-COND}{18}
\DefMacro{table-bug_features-row3-bugCases-LOOP}{14}
\DefMacro{table-bug_features-row3-bugCases-OBJARG}{23}
\DefMacro{table-bug_features-row3-bugs-total}{2}
\DefMacro{table-bug_features-row3-bugs-ARRAY}{1}
\DefMacro{table-bug_features-row3-bugs-COND}{2}
\DefMacro{table-bug_features-row3-bugs-LOOP}{2}
\DefMacro{table-bug_features-row3-bugs-OBJARG}{2}
\DefMacro{table-bug_features-row4-project}{math}
\DefMacro{table-bug_features-row4-bugCases-total}{21}
\DefMacro{table-bug_features-row4-bugCases-ARRAY}{11}
\DefMacro{table-bug_features-row4-bugCases-COND}{19}
\DefMacro{table-bug_features-row4-bugCases-LOOP}{10}
\DefMacro{table-bug_features-row4-bugCases-OBJARG}{13}
\DefMacro{table-bug_features-row4-bugs-total}{6}
\DefMacro{table-bug_features-row4-bugs-ARRAY}{4}
\DefMacro{table-bug_features-row4-bugs-COND}{6}
\DefMacro{table-bug_features-row4-bugs-LOOP}{4}
\DefMacro{table-bug_features-row4-bugs-OBJARG}{4}
\DefMacro{table-bug_features-row5-project}{numbers}
\DefMacro{table-bug_features-row5-bugCases-total}{0}
\DefMacro{table-bug_features-row5-bugCases-ARRAY}{0}
\DefMacro{table-bug_features-row5-bugCases-COND}{0}
\DefMacro{table-bug_features-row5-bugCases-LOOP}{0}
\DefMacro{table-bug_features-row5-bugCases-OBJARG}{0}
\DefMacro{table-bug_features-row5-bugs-total}{0}
\DefMacro{table-bug_features-row5-bugs-ARRAY}{0}
\DefMacro{table-bug_features-row5-bugs-COND}{0}
\DefMacro{table-bug_features-row5-bugs-LOOP}{0}
\DefMacro{table-bug_features-row5-bugs-OBJARG}{0}
\DefMacro{table-bug_features-row6-project}{statistics}
\DefMacro{table-bug_features-row6-bugCases-total}{1}
\DefMacro{table-bug_features-row6-bugCases-ARRAY}{0}
\DefMacro{table-bug_features-row6-bugCases-COND}{0}
\DefMacro{table-bug_features-row6-bugCases-LOOP}{0}
\DefMacro{table-bug_features-row6-bugCases-OBJARG}{0}
\DefMacro{table-bug_features-row6-bugs-total}{1}
\DefMacro{table-bug_features-row6-bugs-ARRAY}{0}
\DefMacro{table-bug_features-row6-bugs-COND}{0}
\DefMacro{table-bug_features-row6-bugs-LOOP}{0}
\DefMacro{table-bug_features-row6-bugs-OBJARG}{0}
\DefMacro{table-bug_features-row7-project}{text}
\DefMacro{table-bug_features-row7-bugCases-total}{2}
\DefMacro{table-bug_features-row7-bugCases-ARRAY}{1}
\DefMacro{table-bug_features-row7-bugCases-COND}{2}
\DefMacro{table-bug_features-row7-bugCases-LOOP}{2}
\DefMacro{table-bug_features-row7-bugCases-OBJARG}{2}
\DefMacro{table-bug_features-row7-bugs-total}{2}
\DefMacro{table-bug_features-row7-bugs-ARRAY}{1}
\DefMacro{table-bug_features-row7-bugs-COND}{2}
\DefMacro{table-bug_features-row7-bugs-LOOP}{2}
\DefMacro{table-bug_features-row7-bugs-OBJARG}{2}
\DefMacro{table-bug_features-row8-project}{vectorz}
\DefMacro{table-bug_features-row8-bugCases-total}{107}
\DefMacro{table-bug_features-row8-bugCases-ARRAY}{42}
\DefMacro{table-bug_features-row8-bugCases-COND}{95}
\DefMacro{table-bug_features-row8-bugCases-LOOP}{91}
\DefMacro{table-bug_features-row8-bugCases-OBJARG}{93}
\DefMacro{table-bug_features-row8-bugs-total}{5}
\DefMacro{table-bug_features-row8-bugs-ARRAY}{3}
\DefMacro{table-bug_features-row8-bugs-COND}{5}
\DefMacro{table-bug_features-row8-bugs-LOOP}{5}
\DefMacro{table-bug_features-row8-bugs-OBJARG}{5}
\DefMacro{table-bug_features-row9-project}{zxing}
\DefMacro{table-bug_features-row9-bugCases-total}{8}
\DefMacro{table-bug_features-row9-bugCases-ARRAY}{8}
\DefMacro{table-bug_features-row9-bugCases-COND}{2}
\DefMacro{table-bug_features-row9-bugCases-LOOP}{8}
\DefMacro{table-bug_features-row9-bugCases-OBJARG}{0}
\DefMacro{table-bug_features-row9-bugs-total}{1}
\DefMacro{table-bug_features-row9-bugs-ARRAY}{1}
\DefMacro{table-bug_features-row9-bugs-COND}{1}
\DefMacro{table-bug_features-row9-bugs-LOOP}{1}
\DefMacro{table-bug_features-row9-bugs-OBJARG}{0}
\DefMacro{table-bug_features-row10-project}{$\sum$}
\DefMacro{table-bug_features-row10-bugCases-total}{174}
\DefMacro{table-bug_features-row10-bugCases-ARRAY}{70}
\DefMacro{table-bug_features-row10-bugCases-COND}{144}
\DefMacro{table-bug_features-row10-bugCases-LOOP}{127}
\DefMacro{table-bug_features-row10-bugCases-OBJARG}{142}
\DefMacro{table-bug_features-row10-bugs-total}{10*}
\DefMacro{table-bug_features-row10-bugs-ARRAY}{7*}
\DefMacro{table-bug_features-row10-bugs-COND}{8*}
\DefMacro{table-bug_features-row10-bugs-LOOP}{8*}
\DefMacro{table-bug_features-row10-bugs-OBJARG}{9*}

\DefMacro{table-bug_features-caption}{Impact of Java language features on bugs. Cond. is Conditional Statements. R.A. is Reference Arguments. *Bugs may repeat across projects, and we show the number of unique bugs across all projects. \label{tab:bug-features}\vspace{-8pt}}
\DefMacro{table-bug_features-col-project}{Project}
\DefMacro{table-bug_features-col-bugCases}{\# Failures due to Bugs}
\DefMacro{table-bug_features-col-bugs}{\# Bugs (Unique)}
\DefMacro{table-bug_features-col-bugCases-total}{Total}
\DefMacro{table-bug_features-col-bugCases-ARRAY}{Arrays}
\DefMacro{table-bug_features-col-bugCases-COND}{Cond.}
\DefMacro{table-bug_features-col-bugCases-LOOP}{Loops}
\DefMacro{table-bug_features-col-bugCases-OBJARG}{R.A.}
\DefMacro{table-bug_features-col-bugs-total}{Total}
\DefMacro{table-bug_features-col-bugs-ARRAY}{Arrays}
\DefMacro{table-bug_features-col-bugs-COND}{Cond.}
\DefMacro{table-bug_features-col-bugs-LOOP}{Loops}
\DefMacro{table-bug_features-col-bugs-OBJARG}{R.A.}

\DefMacro{table-hole_types_comparison-row0-Kind}{\#Templates}
\DefMacro{table-hole_types_comparison-row0-Id}{13,090}
\DefMacro{table-hole_types_comparison-row0-Val}{12,139}
\DefMacro{table-hole_types_comparison-row0-ArithmeticAndShift}{13,606}
\DefMacro{table-hole_types_comparison-row0-RelationAndLogic}{13,682}
\DefMacro{table-hole_types_comparison-row0-All}{12,061}
\DefMacro{table-hole_types_comparison-row1-Kind}{\#Programs}
\DefMacro{table-hole_types_comparison-row1-Id}{105,759}
\DefMacro{table-hole_types_comparison-row1-Val}{87,230}
\DefMacro{table-hole_types_comparison-row1-ArithmeticAndShift}{59,917}
\DefMacro{table-hole_types_comparison-row1-RelationAndLogic}{64,906}
\DefMacro{table-hole_types_comparison-row1-All}{102,670}
\DefMacro{table-hole_types_comparison-row2-Kind}{\#Failures}
\DefMacro{table-hole_types_comparison-row2-Id}{106}
\DefMacro{table-hole_types_comparison-row2-Val}{29}
\DefMacro{table-hole_types_comparison-row2-ArithmeticAndShift}{39}
\DefMacro{table-hole_types_comparison-row2-RelationAndLogic}{70}
\DefMacro{table-hole_types_comparison-row2-All}{131}
\DefMacro{table-hole_types_comparison-row3-Kind}{\#Bugs}
\DefMacro{table-hole_types_comparison-row3-Id}{3}
\DefMacro{table-hole_types_comparison-row3-Val}{1}
\DefMacro{table-hole_types_comparison-row3-ArithmeticAndShift}{3}
\DefMacro{table-hole_types_comparison-row3-RelationAndLogic}{4}
\DefMacro{table-hole_types_comparison-row3-All}{4}

\DefMacro{table-hole_types_comparison-caption}{Results when using a specific set of types of \holes.\label{tab:hole-types-comparison}\vspace{-8pt}}
\DefMacro{table-hole_types_comparison-col-Kind}{}
\DefMacro{table-hole_types_comparison-col-Id}{\scriptsize \texttt{<\aType{}>Id()}}
\DefMacro{table-hole_types_comparison-col-Val}{\scriptsize\texttt{<\aType{}>Val()}}
\DefMacro{table-hole_types_comparison-col-ArithmeticAndShift}{\makecell{\scriptsize \texttt{arithmetic()} \\ \scriptsize \& \texttt{shift()}}}
\DefMacro{table-hole_types_comparison-col-RelationAndLogic}{\makecell{\scriptsize \texttt{relation()} \\ \scriptsize \& \texttt{logic()}}}
\DefMacro{table-hole_types_comparison-col-All}{All}

\DefMacro{table-numbers_of_holes-row0-project}{vectorz}
\DefMacro{table-numbers_of_holes-row0-primitive_id}{1,585,341}
\DefMacro{table-numbers_of_holes-row0-primitive_val}{498,470}
\DefMacro{table-numbers_of_holes-row0-array}{187,333}
\DefMacro{table-numbers_of_holes-row0-arithmetic}{468,965}
\DefMacro{table-numbers_of_holes-row0-shift}{3,115}
\DefMacro{table-numbers_of_holes-row0-relation}{207,971}
\DefMacro{table-numbers_of_holes-row0-logic}{7,960}
\DefMacro{table-numbers_of_holes-row0-loop}{112,319}
\DefMacro{table-numbers_of_holes-row0-total}{2,959,155}
\DefMacro{table-numbers_of_holes-row1-project}{math}
\DefMacro{table-numbers_of_holes-row1-primitive_id}{969,150}
\DefMacro{table-numbers_of_holes-row1-primitive_val}{391,390}
\DefMacro{table-numbers_of_holes-row1-array}{83,093}
\DefMacro{table-numbers_of_holes-row1-arithmetic}{275,034}
\DefMacro{table-numbers_of_holes-row1-shift}{6,392}
\DefMacro{table-numbers_of_holes-row1-relation}{114,672}
\DefMacro{table-numbers_of_holes-row1-logic}{10,547}
\DefMacro{table-numbers_of_holes-row1-loop}{67,306}
\DefMacro{table-numbers_of_holes-row1-total}{1,850,278}
\DefMacro{table-numbers_of_holes-row2-project}{lang}
\DefMacro{table-numbers_of_holes-row2-primitive_id}{984,373}
\DefMacro{table-numbers_of_holes-row2-primitive_val}{464,883}
\DefMacro{table-numbers_of_holes-row2-array}{73,260}
\DefMacro{table-numbers_of_holes-row2-arithmetic}{93,359}
\DefMacro{table-numbers_of_holes-row2-shift}{1,318}
\DefMacro{table-numbers_of_holes-row2-relation}{190,413}
\DefMacro{table-numbers_of_holes-row2-logic}{15,680}
\DefMacro{table-numbers_of_holes-row2-loop}{77,790}
\DefMacro{table-numbers_of_holes-row2-total}{1,823,286}
\DefMacro{table-numbers_of_holes-row3-project}{text}
\DefMacro{table-numbers_of_holes-row3-primitive_id}{246,909}
\DefMacro{table-numbers_of_holes-row3-primitive_val}{86,419}
\DefMacro{table-numbers_of_holes-row3-array}{8,648}
\DefMacro{table-numbers_of_holes-row3-arithmetic}{34,378}
\DefMacro{table-numbers_of_holes-row3-shift}{0}
\DefMacro{table-numbers_of_holes-row3-relation}{48,976}
\DefMacro{table-numbers_of_holes-row3-logic}{3,462}
\DefMacro{table-numbers_of_holes-row3-loop}{11,943}
\DefMacro{table-numbers_of_holes-row3-total}{428,792}
\DefMacro{table-numbers_of_holes-row4-project}{compress}
\DefMacro{table-numbers_of_holes-row4-primitive_id}{178,754}
\DefMacro{table-numbers_of_holes-row4-primitive_val}{89,706}
\DefMacro{table-numbers_of_holes-row4-array}{9,296}
\DefMacro{table-numbers_of_holes-row4-arithmetic}{17,012}
\DefMacro{table-numbers_of_holes-row4-shift}{2,668}
\DefMacro{table-numbers_of_holes-row4-relation}{25,710}
\DefMacro{table-numbers_of_holes-row4-logic}{3,325}
\DefMacro{table-numbers_of_holes-row4-loop}{10,163}
\DefMacro{table-numbers_of_holes-row4-total}{326,471}
\DefMacro{table-numbers_of_holes-row5-project}{zxing}
\DefMacro{table-numbers_of_holes-row5-primitive_id}{93,637}
\DefMacro{table-numbers_of_holes-row5-primitive_val}{74,610}
\DefMacro{table-numbers_of_holes-row5-array}{10,368}
\DefMacro{table-numbers_of_holes-row5-arithmetic}{23,204}
\DefMacro{table-numbers_of_holes-row5-shift}{1,282}
\DefMacro{table-numbers_of_holes-row5-relation}{17,740}
\DefMacro{table-numbers_of_holes-row5-logic}{3,364}
\DefMacro{table-numbers_of_holes-row5-loop}{6,937}
\DefMacro{table-numbers_of_holes-row5-total}{224,205}
\DefMacro{table-numbers_of_holes-row6-project}{codec}
\DefMacro{table-numbers_of_holes-row6-primitive_id}{35,414}
\DefMacro{table-numbers_of_holes-row6-primitive_val}{36,528}
\DefMacro{table-numbers_of_holes-row6-array}{5,420}
\DefMacro{table-numbers_of_holes-row6-arithmetic}{8,103}
\DefMacro{table-numbers_of_holes-row6-shift}{1,597}
\DefMacro{table-numbers_of_holes-row6-relation}{3,831}
\DefMacro{table-numbers_of_holes-row6-logic}{373}
\DefMacro{table-numbers_of_holes-row6-loop}{1,335}
\DefMacro{table-numbers_of_holes-row6-total}{91,266}
\DefMacro{table-numbers_of_holes-row7-project}{statistics}
\DefMacro{table-numbers_of_holes-row7-primitive_id}{31,753}
\DefMacro{table-numbers_of_holes-row7-primitive_val}{7,271}
\DefMacro{table-numbers_of_holes-row7-array}{141}
\DefMacro{table-numbers_of_holes-row7-arithmetic}{7,825}
\DefMacro{table-numbers_of_holes-row7-shift}{0}
\DefMacro{table-numbers_of_holes-row7-relation}{5,110}
\DefMacro{table-numbers_of_holes-row7-logic}{507}
\DefMacro{table-numbers_of_holes-row7-loop}{235}
\DefMacro{table-numbers_of_holes-row7-total}{52,607}
\DefMacro{table-numbers_of_holes-row8-project}{jfreechart}
\DefMacro{table-numbers_of_holes-row8-primitive_id}{6,253}
\DefMacro{table-numbers_of_holes-row8-primitive_val}{2,920}
\DefMacro{table-numbers_of_holes-row8-array}{287}
\DefMacro{table-numbers_of_holes-row8-arithmetic}{1,190}
\DefMacro{table-numbers_of_holes-row8-shift}{12}
\DefMacro{table-numbers_of_holes-row8-relation}{539}
\DefMacro{table-numbers_of_holes-row8-logic}{78}
\DefMacro{table-numbers_of_holes-row8-loop}{182}
\DefMacro{table-numbers_of_holes-row8-total}{11,279}
\DefMacro{table-numbers_of_holes-row9-project}{numbers}
\DefMacro{table-numbers_of_holes-row9-primitive_id}{6,441}
\DefMacro{table-numbers_of_holes-row9-primitive_val}{2,082}
\DefMacro{table-numbers_of_holes-row9-array}{22}
\DefMacro{table-numbers_of_holes-row9-arithmetic}{891}
\DefMacro{table-numbers_of_holes-row9-shift}{140}
\DefMacro{table-numbers_of_holes-row9-relation}{1,065}
\DefMacro{table-numbers_of_holes-row9-logic}{154}
\DefMacro{table-numbers_of_holes-row9-loop}{123}
\DefMacro{table-numbers_of_holes-row9-total}{10,795}
\DefMacro{table-numbers_of_holes-row10-project}{$\sum$}
\DefMacro{table-numbers_of_holes-row10-primitive_id}{4,138,025}
\DefMacro{table-numbers_of_holes-row10-primitive_val}{1,654,279}
\DefMacro{table-numbers_of_holes-row10-array}{377,868}
\DefMacro{table-numbers_of_holes-row10-arithmetic}{929,961}
\DefMacro{table-numbers_of_holes-row10-shift}{16,524}
\DefMacro{table-numbers_of_holes-row10-relation}{616,027}
\DefMacro{table-numbers_of_holes-row10-logic}{45,450}
\DefMacro{table-numbers_of_holes-row10-loop}{288,333}
\DefMacro{table-numbers_of_holes-row10-total}{7,778,134}

\DefMacro{table-numbers_of_holes-caption}{Project information and number of holes per hole type. PrimitiveId contains all the \CodeIn{<\aType{}>Id} \holes, and PrimitiveVal contains all the \CodeIn{<\aType{}>Val} \holes, where \aType is one of the primitive types in Java. \# Loops is the number of loop limiters. \label{tab:numbers-of-holes}\vspace{-8pt}}
\DefMacro{table-numbers_of_holes-col-project}{Project}
\DefMacro{table-numbers_of_holes-col-holes}{\# \Holes}
\DefMacro{table-numbers_of_holes-col-primitive_id}{PrimitiveId}
\DefMacro{table-numbers_of_holes-col-primitive_val}{PrimitiveVal}
\DefMacro{table-numbers_of_holes-col-array}{Array}
\DefMacro{table-numbers_of_holes-col-arithmetic}{Arithmetic}
\DefMacro{table-numbers_of_holes-col-shift}{Shift}
\DefMacro{table-numbers_of_holes-col-relation}{Relation}
\DefMacro{table-numbers_of_holes-col-logic}{Logic}
\DefMacro{table-numbers_of_holes-col-loop}{\# Loops}
\DefMacro{table-numbers_of_holes-col-total}{$\sum$}

\DefMacro{coverage-increase-javatailor-c1-function}{1.6}
\DefMacro{javatailor-coverage-c1-function}{77.3}
\DefMacro{coverage-increase-javatailor-c1-line}{3.3}
\DefMacro{javatailor-coverage-c1-line}{74.4}
\DefMacro{coverage-increase-javatailor-c2-function}{3.1}
\DefMacro{javatailor-coverage-c2-function}{71.5}
\DefMacro{coverage-increase-javatailor-c2-line}{4.0}
\DefMacro{javatailor-coverage-c2-line}{68.5}
\DefMacro{coverage-increase-javatailor-hotspot-function}{-0.9}
\DefMacro{javatailor-coverage-hotspot-function}{40.3}
\DefMacro{coverage-increase-javatailor-hotspot-line}{0.4}
\DefMacro{javatailor-coverage-hotspot-line}{47.6}

\DefMacro{table-tool_comparison-row0-tool}{\JITfuzz}
\DefMacro{table-tool_comparison-row0-programs}{45,352}
\DefMacro{table-tool_comparison-row0-failures}{2,115}
\DefMacro{table-tool_comparison-row0-bugs}{*1}
\DefMacro{table-tool_comparison-row0-coverage-c1-function}{70.8}
\DefMacro{table-tool_comparison-row0-coverage-c1-line}{69.7}
\DefMacro{table-tool_comparison-row0-coverage-c2-function}{67.2}
\DefMacro{table-tool_comparison-row0-coverage-c2-line}{64.3}
\DefMacro{table-tool_comparison-row0-coverage-hotspot-function}{36.6}
\DefMacro{table-tool_comparison-row0-coverage-hotspot-line}{44.6}
\DefMacro{table-tool_comparison-row1-tool}{\ToolTestBased}
\DefMacro{table-tool_comparison-row1-programs}{96,626}
\DefMacro{table-tool_comparison-row1-failures}{97}
\DefMacro{table-tool_comparison-row1-bugs}{0}
\DefMacro{table-tool_comparison-row1-coverage-c1-function}{78.9}
\DefMacro{table-tool_comparison-row1-coverage-c1-line}{77.7}
\DefMacro{table-tool_comparison-row1-coverage-c2-function}{74.6}
\DefMacro{table-tool_comparison-row1-coverage-c2-line}{72.5}
\DefMacro{table-tool_comparison-row1-coverage-hotspot-function}{39.4}
\DefMacro{table-tool_comparison-row1-coverage-hotspot-line}{48.0}
\DefMacro{coverage-increase-c1-function}{8.1}
\DefMacro{coverage-increase-c1-line}{8.0}
\DefMacro{coverage-increase-c2-function}{7.4}
\DefMacro{coverage-increase-c2-line}{8.2}
\DefMacro{coverage-increase-hotspot-function}{2.9}
\DefMacro{coverage-increase-hotspot-line}{3.4}

\DefMacro{table-tool_comparison-caption}{Comparison of \JITfuzz and \Tool. {*}The bug was already found before using \JITfuzz so we did not report it again.\label{tab:tool-comparison}\vspace{-8pt}}
\DefMacro{table-tool_comparison-col-tool}{}
\DefMacro{table-tool_comparison-col-programs}{\# Programs}
\DefMacro{table-tool_comparison-col-failures}{\# Fail.}
\DefMacro{table-tool_comparison-col-bugs}{\# Bugs}
\DefMacro{table-tool_comparison-col-coverage}{Coverage (\%)}
\DefMacro{table-tool_comparison-col-coverage-c1}{C1}
\DefMacro{table-tool_comparison-col-coverage-c1-function}{Func.}
\DefMacro{table-tool_comparison-col-coverage-c1-line}{Line}
\DefMacro{table-tool_comparison-col-coverage-c2}{C2}
\DefMacro{table-tool_comparison-col-coverage-c2-function}{Func.}
\DefMacro{table-tool_comparison-col-coverage-c2-line}{Line}
\DefMacro{table-tool_comparison-col-coverage-hotspot}{\HotSpot}
\DefMacro{table-tool_comparison-col-coverage-hotspot-function}{Func.}
\DefMacro{table-tool_comparison-col-coverage-hotspot-line}{Line}

\DefMacro{table-variant_comparison-row0-tool}{\ToolTestBased}
\DefMacro{table-variant_comparison-row0-templates}{10,714}
\DefMacro{table-variant_comparison-row0-programs}{90,916}
\DefMacro{table-variant_comparison-row0-failures}{66}
\DefMacro{table-variant_comparison-row0-bugs}{3.3}
\DefMacro{table-variant_comparison-row0-coverage-c1-line}{83.8}
\DefMacro{table-variant_comparison-row0-coverage-c2-line}{79.0}
\DefMacro{table-variant_comparison-row0-coverage-hotspot-line}{49.5}
\DefMacro{table-variant_comparison-row1-tool}{\ToolOnlyRandoop}
\DefMacro{table-variant_comparison-row1-templates}{14,854}
\DefMacro{table-variant_comparison-row1-programs}{14,854}
\DefMacro{table-variant_comparison-row1-failures}{24}
\DefMacro{table-variant_comparison-row1-bugs}{0.7}
\DefMacro{table-variant_comparison-row1-coverage-c1-line}{83.6}
\DefMacro{table-variant_comparison-row1-coverage-c2-line}{78.4}
\DefMacro{table-variant_comparison-row1-coverage-hotspot-line}{47.9}
\DefMacro{table-variant_comparison-row2-tool}{\ToolPoolBased}
\DefMacro{table-variant_comparison-row2-templates}{11,921}
\DefMacro{table-variant_comparison-row2-programs}{99,644}
\DefMacro{table-variant_comparison-row2-failures}{129}
\DefMacro{table-variant_comparison-row2-bugs}{5.3}
\DefMacro{table-variant_comparison-row2-coverage-c1-line}{84.4}
\DefMacro{table-variant_comparison-row2-coverage-c2-line}{79.6}
\DefMacro{table-variant_comparison-row2-coverage-hotspot-line}{50.0}
\DefMacro{table-variant_comparison-row3-tool}{\ToolWoRandoop}
\DefMacro{table-variant_comparison-row3-templates}{10,185}
\DefMacro{table-variant_comparison-row3-programs}{89,977}
\DefMacro{table-variant_comparison-row3-failures}{56}
\DefMacro{table-variant_comparison-row3-bugs}{4.0}
\DefMacro{table-variant_comparison-row3-coverage-c1-line}{84.0}
\DefMacro{table-variant_comparison-row3-coverage-c2-line}{78.7}
\DefMacro{table-variant_comparison-row3-coverage-hotspot-line}{49.2}

\DefMacro{table-variant_comparison-caption}{Comparison of \Tool variants. \# Tmpl. is the number of templates, \# Gen. is the number of generated programs, \# Fail. is the number of failures. All the numbers are averages.\label{tab:variant-comparison}\vspace{-8pt}}
\DefMacro{table-variant_comparison-col-variant}{}
\DefMacro{table-variant_comparison-col-templates}{\# Tmpl.}
\DefMacro{table-variant_comparison-col-programs}{\# Gen.}
\DefMacro{table-variant_comparison-col-failures}{\# Fail.}
\DefMacro{table-variant_comparison-col-bugs}{\makecell{\# Avg.\\Bugs}}
\DefMacro{table-variant_comparison-col-coverage}{Coverage (\%)}
\DefMacro{table-variant_comparison-col-coverage-c1}{C1}
\DefMacro{table-variant_comparison-col-coverage-c1-function}{Func.}
\DefMacro{table-variant_comparison-col-coverage-c1-line}{Line}
\DefMacro{table-variant_comparison-col-coverage-c2}{C2}
\DefMacro{table-variant_comparison-col-coverage-c2-function}{Func.}
\DefMacro{table-variant_comparison-col-coverage-c2-line}{Line}
\DefMacro{table-variant_comparison-col-coverage-hotspot}{\HotSpot}
\DefMacro{table-variant_comparison-col-coverage-hotspot-function}{Func.}
\DefMacro{table-variant_comparison-col-coverage-hotspot-line}{Line}

\DefMacro{table-javatailor_results-row0-Kind}{Number}
\DefMacro{table-javatailor_results-row0-NoDiff}{5}
\DefMacro{table-javatailor_results-row0-NonDeterministic}{18}
\DefMacro{table-javatailor_results-row0-DiffText}{39}
\DefMacro{table-javatailor_results-row0-VerifyError}{2}
\DefMacro{table-javatailor_results-row0-DiffException}{24}
\DefMacro{table-javatailor_results-row0-NoException}{7}
\DefMacro{table-javatailor_results-row0-Misc}{7}
\DefMacro{table-javatailor_results-total}{102}
\DefMacro{table-javatailor_results-numKinds}{7}

\DefMacro{table-javatailor_results-caption}{Number of cases of each group reported from \JavaTailor.\label{tab:javatailor-results}\vspace{-8pt}}
\DefMacro{table-javatailor_results-col-Kind}{Group}
\DefMacro{table-javatailor_results-col-NoDiff}{NoRep.}
\DefMacro{table-javatailor_results-col-NonDeterministic}{NonDet.}
\DefMacro{table-javatailor_results-col-DiffText}{DiffText}
\DefMacro{table-javatailor_results-col-VerifyError}{VerifyError}
\DefMacro{table-javatailor_results-col-DiffException}{DiffException}
\DefMacro{table-javatailor_results-col-NoException}{NoException}
\DefMacro{table-javatailor_results-col-Misc}{Misc.}

\DefMacro{table-bugs-row0-jvm}{GraalVM}
\DefMacro{table-bugs-row0-id}{\href{https://github.com/oracle/graal/issues/6403}{GR-45498}}
\DefMacro{table-bugs-row0-type}{Diff}
\DefMacro{table-bugs-row0-affected_versions}{17, 20}
\DefMacro{table-bugs-row0-status}{Fixed}
\DefMacro{table-bugs-row0-cve}{-}
\DefMacro{table-bugs-row0-duplicate}{-}
\DefMacro{table-bugs-row1-jvm}{HotSpot}
\DefMacro{table-bugs-row1-id}{\href{https://bugs.openjdk.java.net/browse/JDK-8301663}{JDK-8301663}}
\DefMacro{table-bugs-row1-type}{Diff}
\DefMacro{table-bugs-row1-affected_versions}{18, 19, 19.0.2}
\DefMacro{table-bugs-row1-status}{Fixed}
\DefMacro{table-bugs-row1-cve}{-}
\DefMacro{table-bugs-row1-duplicate}{\href{https://bugs.openjdk.java.net/browse/JDK-8288064}{JDK-8288064}}
\DefMacro{table-bugs-row2-id}{\href{https://bugs.openjdk.java.net/browse/JDK-8303946}{JDK-8303946}}
\DefMacro{table-bugs-row2-type}{Diff}
\DefMacro{table-bugs-row2-affected_versions}{8, 11, 17, 19, 20, 21}
\DefMacro{table-bugs-row2-status}{Confirmed}
\DefMacro{table-bugs-row2-cve}{-}
\DefMacro{table-bugs-row2-duplicate}{-}
\DefMacro{table-bugs-row3-id}{\href{https://bugs.openjdk.java.net/browse/JDK-8304336}{JDK-8304336}}
\DefMacro{table-bugs-row3-type}{Diff}
\DefMacro{table-bugs-row3-affected_versions}{17, 19, 20, 21}
\DefMacro{table-bugs-row3-status}{Fixed}
\DefMacro{table-bugs-row3-cve}{\href{https://www.cve.org/CVERecord?id=CVE-2023-22044}{CVE-2023-22044}}
\DefMacro{table-bugs-row3-duplicate}{-}
\DefMacro{table-bugs-row4-id}{\href{https://bugs.openjdk.java.net/browse/JDK-8305946}{JDK-8305946}}
\DefMacro{table-bugs-row4-type}{Crash}
\DefMacro{table-bugs-row4-affected_versions}{17, 19, 20, 21}
\DefMacro{table-bugs-row4-status}{Fixed}
\DefMacro{table-bugs-row4-cve}{\href{https://www.cve.org/CVERecord?id=CVE-2023-22045}{CVE-2023-22045}}
\DefMacro{table-bugs-row4-duplicate}{-}
\DefMacro{table-bugs-row5-id}{\href{https://bugs.openjdk.java.net/browse/JDK-8325216}{JDK-8325216}}
\DefMacro{table-bugs-row5-type}{Crash}
\DefMacro{table-bugs-row5-affected_versions}{17, 18, 19, 20, 21}
\DefMacro{table-bugs-row5-status}{Fixed}
\DefMacro{table-bugs-row5-cve}{-}
\DefMacro{table-bugs-row5-duplicate}{\href{https://bugs.openjdk.java.net/browse/JDK-8319793}{JDK-8319793}}
\DefMacro{table-bugs-row6-jvm}{OpenJ9}
\DefMacro{table-bugs-row6-id}{\href{https://github.com/eclipse-openj9/openj9/issues/17066}{17066}}
\DefMacro{table-bugs-row6-type}{Crash}
\DefMacro{table-bugs-row6-affected_versions}{8, 11, 17, 18}
\DefMacro{table-bugs-row6-status}{Fixed}
\DefMacro{table-bugs-row6-cve}{-}
\DefMacro{table-bugs-row6-duplicate}{-}
\DefMacro{table-bugs-row7-id}{\href{https://github.com/eclipse-openj9/openj9/issues/17129}{17129}}
\DefMacro{table-bugs-row7-type}{Diff}
\DefMacro{table-bugs-row7-affected_versions}{8, 11, 17, 18}
\DefMacro{table-bugs-row7-status}{Fixed}
\DefMacro{table-bugs-row7-cve}{-}
\DefMacro{table-bugs-row7-duplicate}{-}
\DefMacro{table-bugs-row8-id}{\href{https://github.com/eclipse-openj9/openj9/issues/17139}{17139}}
\DefMacro{table-bugs-row8-type}{Diff}
\DefMacro{table-bugs-row8-affected_versions}{8, 11, 17, 18}
\DefMacro{table-bugs-row8-status}{Fixed}
\DefMacro{table-bugs-row8-cve}{-}
\DefMacro{table-bugs-row8-duplicate}{-}
\DefMacro{table-bugs-row9-id}{\href{https://github.com/eclipse-openj9/openj9/issues/17171}{17171}}
\DefMacro{table-bugs-row9-type}{Crash}
\DefMacro{table-bugs-row9-affected_versions}{11, 17, 18}
\DefMacro{table-bugs-row9-status}{Fixed}
\DefMacro{table-bugs-row9-cve}{-}
\DefMacro{table-bugs-row9-duplicate}{-}
\DefMacro{table-bugs-row10-id}{\href{https://github.com/eclipse-openj9/openj9/issues/17212}{17212}}
\DefMacro{table-bugs-row10-type}{Crash}
\DefMacro{table-bugs-row10-affected_versions}{8, 11, 17, 18}
\DefMacro{table-bugs-row10-status}{Fixed}
\DefMacro{table-bugs-row10-cve}{-}
\DefMacro{table-bugs-row10-duplicate}{\href{https://github.com/eclipse-openj9/openj9/issues/15363}{15363}}
\DefMacro{table-bugs-row11-id}{\href{https://github.com/eclipse-openj9/openj9/issues/17249}{17249}}
\DefMacro{table-bugs-row11-type}{Diff}
\DefMacro{table-bugs-row11-affected_versions}{8, 11, 17, 18}
\DefMacro{table-bugs-row11-status}{Fixed}
\DefMacro{table-bugs-row11-cve}{-}
\DefMacro{table-bugs-row11-duplicate}{-}
\DefMacro{table-bugs-row12-id}{\href{https://github.com/eclipse-openj9/openj9/issues/17250}{17250}}
\DefMacro{table-bugs-row12-type}{Diff}
\DefMacro{table-bugs-row12-affected_versions}{17, 18}
\DefMacro{table-bugs-row12-status}{Fixed}
\DefMacro{table-bugs-row12-cve}{-}
\DefMacro{table-bugs-row12-duplicate}{-}
\DefMacro{table-bugs-row13-id}{\href{https://github.com/eclipse-openj9/openj9/issues/18802}{18802}}
\DefMacro{table-bugs-row13-type}{Crash}
\DefMacro{table-bugs-row13-affected_versions}{8, 11, 17, 21}
\DefMacro{table-bugs-row13-status}{Fixed}
\DefMacro{table-bugs-row13-cve}{-}
\DefMacro{table-bugs-row13-duplicate}{\href{https://github.com/eclipse-openj9/openj9/issues/17045}{17045}}
\DefMacro{table-bugs-row14-id}{\href{https://github.com/eclipse-openj9/openj9/issues/18803}{18803}}
\DefMacro{table-bugs-row14-type}{Crash}
\DefMacro{table-bugs-row14-affected_versions}{11, 17, 21}
\DefMacro{table-bugs-row14-status}{Fixed}
\DefMacro{table-bugs-row14-cve}{-}
\DefMacro{table-bugs-row14-duplicate}{-}
\DefMacro{total-num-hotspot-bugs}{five}
\DefMacro{total-num-openj9-bugs}{nine}
\DefMacro{total-num-graalvm-bugs}{one}
\DefMacro{total-num-bugs}{15}
\DefMacro{total-num-confirmed-bugs}{15}
\DefMacro{total-num-new-bugs}{11}
\DefMacro{Total-num-new-bugs}{11}
\DefMacro{total-num-duplicate-bugs}{four}
\DefMacro{total-num-cves}{two}

\DefMacro{table-bugs-caption}{Detected bugs in \HotSpot, \OpenJNine and \GraalVM using \Tool; \NumOfPreviouslyUnknownBugs bugs were previously unknown.\label{tab:bugs}\vspace{-8pt}}
\DefMacro{table-bugs-col-id}{Bug ID}
\DefMacro{table-bugs-col-type}{Type}
\DefMacro{table-bugs-col-priority}{Priority}
\DefMacro{table-bugs-col-component}{Component}
\DefMacro{table-bugs-col-affected_versions}{JDK Versions}
\DefMacro{table-bugs-col-jvm}{JVM}
\DefMacro{table-bugs-col-status}{Status}
\DefMacro{table-bugs-col-cve}{CVE}
\DefMacro{table-bugs-col-duplicate}{Duplicates}

\DefMacro{figure-example-caption}{An example (a)~\template extracted from the program in Figure~\ref{fig:example:original}, and (b)~a concrete program generated from the \template by filling in the \holes, which crashed OpenJ9 \JIT compiler.\label{fig:example}\vspace{-10pt}}
\DefMacro{figure-example-original-caption}{An existing program from the \commonstext project~\cite{commonstextStrBuilderSource} used as a source for \template extraction.\label{fig:example:original}\vspace{-10pt}}
\DefMacro{figure-example-template-caption}{\label{fig:example:template}}
\DefMacro{figure-example-generated-caption}{\label{fig:example:generated}}

\DefMacro{figure-overview-caption}{The overview of \Tool. Dotted-dashed lines: \ApproachtestBased approach; dashed lines: \ApproachpoolBased approach. \label{fig:framework:overview}\vspace{-15pt}}

\DefMacro{figure-overlap-caption}{The overlap of bugs detected by \Tool variants. \ToolOnlyRandoop: no templates/generated programs; \ToolTestBased: \Tool with \ApproachTestBased approach; \ToolPoolBased: \Tool with \ApproachPoolBased approach; \ToolWoRandoop: enhanced \JAttack.\label{fig:overlap}}

\DefMacro{figure-evaluation-bugone}{\BugOne.\label{fig:evaluation:bugone}}
\DefMacro{figure-evaluation-bugtwo}{\BugTwo.\label{fig:evaluation:bugtwo}}
\DefMacro{figure-evaluation-bugthree}{\BugThree.\label{fig:evaluation:bugthree}}
\DefMacro{figure-evaluation-bugfour}{\BugFour.\label{fig:evaluation:bugfour}}
\DefMacro{figure-evaluation-detectedbugs}{Examples of minimized programs of detected bugs.\label{fig:evaluation:detectedbugs}\vspace{-8pt}}

\DefMacro{algo-extraction-fontsize}{footnotesize}
\DefMacro{algo-extraction-caption}{\Template extraction algorithm.\label{algo:extraction}\vspace{-10pt}}

\DefMacro{algo-pruning-fontsize}{footnotesize}
\DefMacro{algo-pruning-caption}{Pruning algorithm.\label{algo:pruning}}

\newcommand{\RQComponent}{RQ1\xspace}
\newcommand{\RQComparison}{RQ2\xspace}
\newcommand{\RQDetectedBugs}{RQ3\xspace}

\newcommand{\LeJitTotalNumOfTemplates}{\UseMacro{table-hole_features-row10-template-total}\xspace}
\newcommand{\LeJitTotalNumOfGenerated}{\UseMacro{table-hole_features-row10-generated-total}\xspace}

\newcommand{\TAveNumOfTemplates}{\UseMacro{table-variant_comparison-row0-templates}\xspace}
\newcommand{\TAveNumOfPrograms}{\UseMacro{table-variant_comparison-row0-programs}\xspace}
\newcommand{\TAveNumOfFailures}{\UseMacro{table-variant_comparison-row0-failures}\xspace}
\newcommand{\TAveNumOfBugs}{\UseMacro{table-variant_comparison-row0-bugs}\xspace}
\newcommand{\ORAveNumOfBugs}{\UseMacro{table-variant_comparison-row1-bugs}\xspace}
\newcommand{\PAveNumOfFailures}{\UseMacro{table-variant_comparison-row2-failures}\xspace}
\newcommand{\PAveNumOfBugs}{\UseMacro{table-variant_comparison-row2-bugs}\xspace}

\newcommand{\LineCoverageIncreaseJavaTailorCOne}{\UseMacro{coverage-increase-javatailor-c1-line}\%\xspace}
\newcommand{\LineCoverageIncreaseJavaTailorCTwo}{\UseMacro{coverage-increase-javatailor-c2-line}\%\xspace}
\newcommand{\LineCoverageIncreaseJavaTailorHotSpot}{\UseMacro{coverage-increase-javatailor-hotspot-line}\%\xspace}

\newcommand{\JITfuzzNumOfFailures}{\UseMacro{table-tool_comparison-row0-failures}\xspace}
\newcommand{\JITfuzzNumOfPrograms}{\UseMacro{table-tool_comparison-row0-programs}\xspace}
\newcommand{\ToolNumOfFailures}{\UseMacro{table-tool_comparison-row1-failures}\xspace}
\newcommand{\ToolNumOfPrograms}{\UseMacro{table-tool_comparison-row1-programs}\xspace}
\newcommand{\LineCoverageIncreaseCOne}{\UseMacro{coverage-increase-c1-line}\%\xspace}
\newcommand{\LineCoverageIncreaseCTwo}{\UseMacro{coverage-increase-c2-line}\%\xspace}
\newcommand{\LineCoverageIncreaseHotSpot}{\UseMacro{coverage-increase-hotspot-line}\%\xspace}

\newcommand{\JavaTailorNumOfFailures}{\UseMacro{table-javatailor_results-total}\xspace}
\newcommand{\JavaTailorNumOfKinds}{\UseMacro{table-javatailor_results-numKinds}\xspace}

\newcommand{\TotalNumOfCVEs}{\UseMacro{total-num-cves}\xspace}
\newcommand{\TotalNumOfBugs}{\UseMacro{total-num-bugs}\xspace}
\newcommand{\TotalNumOfBugsInHotSpot}{\UseMacro{total-num-hotspot-bugs}\xspace}
\newcommand{\TotalNumOfBugsInOpenJNine}{\UseMacro{total-num-openj9-bugs}\xspace}
\newcommand{\TotalNumOfBugsInGraalVM}{\UseMacro{total-num-graalvm-bugs}\xspace}
\newcommand{\NumOfConfirmedBugs}{\UseMacro{total-num-confirmed-bugs}\xspace}
\newcommand{\NumOfPreviouslyKnownBugs}{\UseMacro{total-num-duplicate-bugs}\xspace}
\newcommand{\NumOfPreviouslyUnknownBugs}{\UseMacro{total-num-new-bugs}\xspace}
\newcommand{\CapitalizedNumOfPreviouslyUnknownBugs}{\UseMacro{Total-num-new-bugs}\xspace}
\newcommand{\NumOfBugsFoundInPreliminaryExperiments}{two\xspace}
\newcommand{\NumOfBugsFoundInEvalOfTemplateTypes}{three\xspace}

\newcommand{\LeJitPoolBasedRunDuration}{six days\xspace}
\newcommand{\JITfuzzSingleProgramTimeout}{50 seconds\xspace}
\newcommand{\ToolComparisonEndToEndTimeout}{six days\xspace}

\newcommand{\LeJitFilterOutRate}{96\%}
\newcommand{\JAttackFilterOutRate}{60\%}

\newcommand{\NumOfProjectsMigratedToJUnitFour}{five\xspace}
\newcommand{\TotalNumOfProjectsJITfuzz}{nine\xspace}
\newcommand{\TotalNumOfProjectsUsable}{62\xspace}
\newcommand{\NumOfProjectsUsed}{ten\xspace}
\newcommand{\NumProjectsSearch}{1,000\xspace}
\newcommand{\NumStars}{20\xspace}
\newcommand{\NumProjectsAtFist}{1,793\xspace}
\newcommand{\NumProjectsFiltered}{161\xspace}

\newcommand{\NumGenPerTemplate}{ten\xspace}
\newcommand{\RandoopOutputLimit}{5,000\xspace}
\newcommand{\RandoopTimeLimit}{30 minutes\xspace}
\newcommand{\NumItrs}{100,000\xspace}
\newcommand{\GenTimeout}{three-minute\xspace}
\newcommand{\ExecTimeout}{one-minute\xspace}
\newcommand{\NumRepeatedRuns}{three\xspace}

\newcommand{\JITfuzzVersion}{3dc8f91\xspace}
\newcommand{\JavaTailorVersion}{bf9421f\xspace}

\newcommand{\commonscodec}{codec\xspace}
\newcommand{\commonsmath}{math\xspace}
\newcommand{\commonstext}{text\xspace}

\newcommand{\DataURL}{\url{https://figshare.com/s/04c63cc25ee5c38c2491}}
\newcommand{\LejitURL}{\url{https://github.com/EngineeringSoftware/lejit}}

\begin{document}

\title{\Title}

\author{Zhiqiang Zang}
\orcid{0000-0001-7285-7677}
\affiliation{%
\institution{The University of Texas at Austin}
\city{Austin}
\country{USA}
}
\email{zhiqiang.zang@utexas.edu}

\author{Fu-Yao Yu}
\orcid{0009-0007-6666-8715}
\affiliation{%
\institution{The University of Texas at Austin}
\city{Austin}
\country{USA}
}
\email{fu.yao.yu@utexas.edu}

\author{Aditya Thimmaiah}
\orcid{0009-0002-1917-7386}
\affiliation{%
\institution{The University of Texas at Austin}
\city{Austin}
\country{USA}
}
\email{auditt@utexas.edu}

\author{August Shi}
\orcid{0000-0001-8239-3124}
\affiliation{%
\institution{The University of Texas at Austin}
\city{Austin}
\country{USA}
}
\email{august@utexas.edu}

\author{Milos Gligoric}
\orcid{0000-0002-5894-7649}
\affiliation{%
\institution{The University of Texas at Austin}
\city{Austin}
\country{USA}
}
\email{gligoric@utexas.edu}

\begin{abstract}
We present \Tool, a template-based framework for testing Java
just-in-time (\JIT) compilers.
Like recent template-based frameworks, \Tool executes a template---a
program with holes to be filled---to generate concrete programs given
as inputs to Java \JIT compilers.
\Tool automatically generates template programs from existing Java
code by converting expressions to holes, as well as generating
necessary glue code (i.e., code that generates instances of
non-primitive types) to make generated templates executable.
We have successfully used \Tool to test a range of popular Java \JIT
compilers, revealing \TotalNumOfBugsInHotSpot bugs in \HotSpot,
\TotalNumOfBugsInOpenJNine bugs in \OpenJNine, and
\TotalNumOfBugsInGraalVM bug in \GraalVM. All of these bugs have been
confirmed by Oracle and IBM developers, and
\NumOfPreviouslyUnknownBugs of these bugs were previously unknown,
including \TotalNumOfCVEs CVEs (Common Vulnerabilities and Exposures).
Our comparison with several existing approaches shows that \Tool is
complementary to them
and is a powerful technique for ensuring Java \JIT compiler
correctness.
\end{abstract}



\maketitle

\section{Introduction}
\label{sec:intro}

Compilers are the cornerstone of software development, and their
correctness is of utmost importance.
For years, the compiler testing community has primarily focused on
static compilers, such as GCC, LLVM, and javac. Tools like
\Csmith~\cite{YangPLDI11CSmith} and
Hephaestus~\cite{Chaliasos22FindingTypingCompilerBugs} have been shown
effective in discovering bugs in static
compilers~\cite{GCCBugListFoundByCsmith, LLVMBugListFoundByCsmith}.
However, these tools are ineffective in discovering bugs in
just-in-time (\JIT) compilers. \JIT compilers, or \JIT for short,
dynamically (i.e., at runtime) rewrite parts of programs to optimize
program execution based on profiling data. Testing such compilers
requires carefully crafted inputs that trigger \JIT compilation and
provide challenging code snippets for the optimizing compilers.

Recently, there has been substantial work on testing \JIT compilers,
which can be organized in three main categories: grammar-based,
mutation-based, and template-based techniques. The first
group~\cite{SirerETAL00UsingProductionGrammarsInSoftwareTesting,
YoshikawaETAL03RandomGeneratorForJIT, JavaFuzzerAzulGitHubPage}
generates test inputs based on language grammars.
The second group~\cite{ChenPLDI16DifferentialTestingOfJVM,
ChenICSE19DeepDifferentialTestingOfJVMImplementations,
Vikram21BonsaiFuzzing, Zhao22JavaTailor} commonly starts with a set
of seed programs that are evolved with a predefined set of mutation
operators.
The third group~\cite{ChingKatz93TestingAPLCompiler,
ZhangPLDI17SkeletalProgramEnumeration, zang22jattack} combines
manual testing (writing templates) with fuzzing, i.e., filling \holes
in \templates based on a set of manually predefined choices that can
be different for each \hole.

These three groups are complementary and each group provides some
benefits over the others. Consider \JAttack~\cite{zang22jattack}, a
\template-based technique that recently discovered two CVEs (Common
Vulnerabilities and Exposures) in Oracle's \JIT. Each \emph{\template
program} is a valid Java program with \holes, and each \hole is
written in an embedded domain-specific language that specifies the set
of expressions that can potentially fill the hole. \JAttack generates
programs by fuzzing \holes during the execution of a given \template.
The advantage of this technique is that developers have full control
of the space that should be tested and the way programs should be
modified.  At the same time, \JAttack requires substantial developers'
engagement, as both \template program design and hole values are
written manually.

In this paper, we present a novel framework, dubbed \Tool, for
automatically generating \template programs from existing code.
\Tool generates templates by rewriting existing expressions to holes,
as well as generating necessary glue code (e.g., code that creates
instances of non-primitive types on which methods can be invoked) to
make those templates executable.
Execution of generated templates, which randomly fills the holes,
creates concrete programs that are used as inputs for Java \JIT
testing.
As a result, \Tool sits in between mutation-based techniques and
\template-based techniques. Unlike existing \template-based
techniques, templates are automatically extracted from any existing
code. Unlike mutation-based techniques, each \hole has its own set of
values and can be filled dynamically (rather than statically), and
multiple \holes are filled simultaneously during the execution of the
template. Subsequently, \Tool is in a way similar to higher order
mutation~\cite{Jia08HigherOrderMutationTesting}, but holes are filled
dynamically through executing \templates.

\Tool is designed to enable generation of a program \template from any
existing method. One of the key challenges was to enable templates for
methods that accept instances of complex types as arguments, including
an instance on which an instance method is to be invoked.
Our key insight in this direction is to capture instances of various
types during testing of methods from which \templates are to be
extracted; tests used can be either existing manually-written tests or
automatically-generated tests.

Unlike several existing tools for testing Java runtime
environments~\cite{JavaBook}, \Tool generates source code rather than
bytecode.
Some advantages of focusing on source code rather than bytecode
include: 1)~eliminating the need to worry about invalid classfiles, as
those obtained from source files always pass the early check of the class format performed by bytecode
verifiers, allowing for ``deeper'' testing; 2)~simplifying every step
during the bug reporting phase:
a bug reported as a source code snippet instead of bytecode is easier
to understand, minimize, and report, and it also facilitates
confirmation, fixing, and integrating in test suites by compiler
developers;
3)~decreasing the likelihood of revealed bugs
resulting in false positives, since these programs
result in valid bytecode generated via a Java compiler as opposed to
random sequences of bytecode instructions.

We used \Tool to test several \JIT compilers: Oracle \HotSpot, IBM
\OpenJNine, and Oracle \GraalVM. We used differential
testing~\cite{Mckeeman98DifferentialTesting} to detect crash and inconsistency
between \JIT compilers. We extracted \templates from \NumOfProjectsUsed
open-source Java projects available on GitHub, although our technique
can extract \templates from any other code. Our runs discovered
\TotalNumOfBugsInHotSpot bugs in \HotSpot, \TotalNumOfBugsInOpenJNine
bugs in \OpenJNine, and \TotalNumOfBugsInGraalVM bug in \GraalVM.
\CapitalizedNumOfPreviouslyUnknownBugs out of the \TotalNumOfBugs bugs
were previously unknown, including \TotalNumOfCVEs CVEs. All bugs have
been confirmed by compiler developers.

We further compared \Tool with \JITfuzz~\cite{wu23jitfuzz} and
\JavaTailor~\cite{Zhao22JavaTailor}, the state-of-the-art testing
tools for Java \JIT and \JVMs, respectively.
Our experiments show that \Tool increased code coverage of C1 compiler
by \LineCoverageIncreaseCOne and C2 compiler by
\LineCoverageIncreaseCTwo~\cite{JavaHotSpotVMWhitePaper} compared to
\JITfuzz, and increased by \LineCoverageIncreaseJavaTailorCOne and
\LineCoverageIncreaseJavaTailorCTwo compared to \JavaTailor, when
testing OpenJDK \HotSpot.
Additionally, using \JITfuzz and \JavaTailor we have not discovered
any of the bugs found by \Tool.

\vspace{5pt}
\noindent
The main contributions of this paper include:

\begin{itemize}[topsep=3pt,itemsep=3pt,partopsep=0ex,parsep=0ex,leftmargin=*]
\item \textbf{Framework}. We designed and implemented a framework
for extracting \templates from existing code by converting
expressions into \holes and capturing instances of complex types
during test execution.  Captured instances enable execution of
\templates that produce concrete programs used as inputs for
compiler testing.
\item \textbf{Implementation}. We have implemented \Tool for Java
and built it around a recent publicly available framework
(\JAttack). We have also developed several variants of \Tool to
help us understand the benefits of \templates and captured
instances used for arguments.
\item \textbf{Evaluation}. We have performed extensive evaluation of
\Tool. We have extracted \LeJitTotalNumOfTemplates \templates from
\NumOfProjectsUsed open-source Java projects on GitHub. We then
used \JAttack to generate \LeJitTotalNumOfGenerated concrete
programs. We have used the generated programs to test three
compilers -- Oracle \HotSpot, IBM \OpenJNine, and Oracle \GraalVM.
Additionally, we compare \Tool with \JITfuzz and \JavaTailor the
state-of-the-art tools for testing Java runtime environments.
\item
\textbf{Analysis}. We performed an in-depth analysis of \templates and
\generatedprograms to understand how the presence of various Java
language features, e.g., arrays, conditional statements, loops,
etc., affect \Tool's bug detection capabilities. We also studied the
impact of various types of \templates on the result, and we find
types of \holes that play an important role in bug detection.
\item
\textbf{Results}. Our results show the effectiveness of \Tool. We have
discovered \TotalNumOfBugs bugs, including
\TotalNumOfBugsInHotSpot bugs in \HotSpot,
\TotalNumOfBugsInOpenJNine bugs in \OpenJNine, and
\TotalNumOfBugsInGraalVM bug in \GraalVM.
\CapitalizedNumOfPreviouslyUnknownBugs of the bugs are previously
unknown, including \TotalNumOfCVEs CVEs. All bugs have been
confirmed by compiler developers.
Our results also show that \Tool is complementary to the
state-of-the-art techniques, which did not discover any of the
bugs found by \Tool.

\end{itemize}

\noindent
Our tool is publicly available at \LejitURL.

\section{Example}
\label{sec:example}

We demonstrate the capabilities of \Tool, using an example program
that involves a bug we detected in the \OpenJNine \JIT compiler.
Figure~\ref{fig:example:original} shows a snippet of the example
program.

\begin{figure}[!t]
\begin{lstlisting}[language=example]
package org.apache.commons.text;...
public class StrBuilder implements ... { ...
  static final int CAPACITY = 32;
  char[] buffer; private int size;
  private String newLine, nullText;
  public StrBuilder(final String str) {
      if (str(*@\annotation{1}{annot:hole:1:original}@*) == null) { buffer = new char[CAPACITY(*@\annotation{2}{annot:hole:2:original}@*)]; } ... }
  public StrBuilder trim() {
    if (size == 0(*@\annotation{3}{annot:hole:3:original}@*)) { return this; }
    int len = size(*@\annotation{4}{annot:hole:4:original}@*);
    final char[] buf = buffer(*@\annotation{5}{annot:hole:5:original}@*);
    int pos = 0(*@\annotation{6}{annot:hole:6:original}@*);
    while (pos < len && buf[pos] <= ' '(*@\annotation{7}{annot:hole:7:original}@*)) { pos++; }
    while (pos < len && buf[len - 1] <= ' '(*@\annotation{8}{annot:hole:8:original}@*)) { len--; } ...
    return this; } }
\end{lstlisting}
\vspace{-10pt}
\caption{\UseMacro{figure-example-original-caption}}
\end{figure}
\begin{figure}[!t]
\begin{subfigure}[b]{0.50\linewidth}
\begin{lstlisting}[language=example-half-width]
package org.apache.commons.text;...
import jattack.annotation.*;
import static jattack.Boom.*;
public class StrBuilder implements ... { ...
  static final int CAPACITY = 32;
  char[] buffer; private int size;
  private String newLine, nullText;
  public StrBuilder(final String str) {
    if (refId(String.class).eval()(*@\annotation{1}{annot:hole:1:template}@*) == null) {
      buffer = new char[intId().eval()(*@\annotation{2}{annot:hole:2:template}@*)]; } ... }

  @Entry
  public StrBuilder trim() {
    if (relation(intId(), intVal()).eval()(*@\annotation{3}{annot:hole:3:template}@*)) {
      return this; }
    int len = intId().eval()(*@\annotation{4}{annot:hole:4:template}@*);
    final char[] buf = refId(char[].class).eval()(*@\annotation{5}{annot:hole:5:template}@*);
    int pos = intVal().eval()(*@\annotation{6}{annot:hole:6:template}@*);
    int _lim1 = 0;
    while (logic(
              relation(intId(), intId()),
              relation(charArrAcc(
                          refId(char[].class),
                          intId()),
                        charVal()))
            .eval()(*@\annotation{7}{annot:hole:7:template}@*) && _lim1++ < 1000) {(*@\label{line:hole:7:template}@*)
      pos++; }
    int _lim2 = 0;
    while (logic(
              relation(intId(), intId()),
              relation(charArrAcc(
                          refId(char[].class),
                          arithmetic(intId(),
                                      intVal())),
                        charVal()))
            .eval()(*@\annotation{8}{annot:hole:8:template}@*) && _lim2++ < 1000) {
      len--; } ...
    return this; }

  @Argument(0)
  public static StrBuilder _arg0() {
    StrBuilder sb1 = new StrBuilder("date");(*@\label{line:codeseq:start}@*)
    sb1.append((Object) 10.0);
    sb1.appendSeparator("d");
    Object[] arr = new Object[] { 1.0 };
    return sb1.append("resourceBundle", arr); } }(*@\label{line:codeseq:end}@*)
\end{lstlisting}
\vspace{-5pt}
\caption{\UseMacro{figure-example-template-caption}}
\end{subfigure}
\begin{subfigure}[b]{0.47\linewidth}
\begin{lstlisting}[language=example-half-width]
package org.apache.commons.text;...
import jattack.annotation.*;
import jattack.csutil.Helper;
import jattack.csutil.checksum.WrappedChecksum;
import jattack.exception.UnfilledHoleException;
import static jattack.Boom.*;
public class StrBuilder implements ... { ...
  static final int CAPACITY = 32;
  char[] buffer; private int size;
  private String newLine, nullText;
  public StrBuilder(final String str) {
    if (nullText(*@\annotation{1}{annot:hole:1:generated}@*) == null) {
      buffer = new char[CAPACITY(*@\annotation{2}{annot:hole:2:generated}@*)]; } ... }

  public StrBuilder trim() {
    if (size <= -1838784853(*@\annotation{3}{annot:hole:3:generated}@*)) { return this; }
    int len = CAPACITY(*@\annotation{4}{annot:hole:4:generated}@*);
    final char[] buf = buffer(*@\annotation{5}{annot:hole:5:generated}@*);
    int pos = 809931165(*@\annotation{6}{annot:hole:6:generated}@*);
    int _lim1 = 0;
    while ((len > _lim1 && buf[size] != 'Z')(*@\annotation{7}{annot:hole:7:generated}@*)
             && _lim1++ < 1000) { pos++; }
    int _lim2 = 0;
    while ((pos <= _lim2
               || buf[size - 1312433786] > '0')(*@\annotation{8}{annot:hole:8:generated}@*)
             && _lim2++ < 1000) { len--; } ...
    return this; }

  public static StrBuilder _arg0() {
    StrBuilder sb1 = new StrBuilder("date");
    sb1.append((Object) 10.0);
    sb1.appendSeparator("d");
    Object[] arr = new Object[] { 1.0 };
    return sb1.append("resourceBundle", arr); }

  public static void main(String[] args) {(*@\label{line:main}@*)
    WrappedChecksum cs = new WrappedChecksum();
    StrBuilder rcvr = _arg0();(*@\label{line:invoke-argument-method}@*)
    cs.update(rcvr);
    for (int i = 0; i < 100_000; ++i) {(*@\label{line:for-loop-start}@*)
      try {
        cs.update(rcvr.trim());
      } catch (UnfilledHoleException e) {
        throw e;
      } catch(Throwable e) {
        cs.update(e.getClass().getName()); } }(*@\label{line:for-loop-end}@*)
    cs.update(StrBuilder.class);
    Helper.write(cs.getValue()); } }
\end{lstlisting}
\vspace{-5pt}
\caption{\UseMacro{figure-example-generated-caption}}
\end{subfigure}
\vspace{-5pt}
\caption{\UseMacro{figure-example-caption}}
\end{figure}

\Tool extracts a \template from this program by replacing expressions
with \holes, as shown in Figure~\ref{fig:example:template}. A
\emph{\hole} is a placeholder to be filled with concrete expressions
during program generation. Each \hole is expressed as an API call,
which defines the type and range of values that can be used to fill
the hole, e.g., the first \hole \CodeIn{refId(String.class)}
(line~\ref{annot:hole:1:template}) in the \template represents any
available variable with type \CodeIn{String} at this execution
point~\cite{zang22jattack}.

There are eight \holes displayed in the \template, two in the
constructor and six in the method \CodeIn{trim}. Each \hole, labeled
with a circled number, is converted from the expression in the
original program with the same number. For example, the first \hole
\CodeIn{refId(String.class)} is converted from the local variable
\CodeIn{str} (line~\ref{annot:hole:1:original}) in
Figure~\ref{fig:example:original}. The next \hole \CodeIn{intId()}
(line~\ref{annot:hole:2:template}) in
Figure~\ref{fig:example:template}, which represents any available
\CodeIn{int} variable, is converted from the \CodeIn{int} field
\CodeIn{CAPACITY} (line~\ref{annot:hole:2:original}) in
Figure~\ref{fig:example:original}. The third \hole
(line~\ref{annot:hole:3:template}) in
Figure~\ref{fig:example:template}, represents a relational expression
that connects an integer variable and an integer literal (between
\CodeIn{Integer.MIN\_VALUE} and \CodeIn{Integer.MAX\_VALUE}) using a
relational operator (\CodeIn{<}, \CodeIn{<=}, \CodeIn{>}, \CodeIn{>=},
\CodeIn{==}, \CodeIn{!=}). This \hole is converted from the
\CodeIn{if} condition \CodeIn{size == 0}
(line~\ref{annot:hole:3:original}) in
Figure~\ref{fig:example:original}.  Similarly, the next three \holes:
an integer variable hole (line~\ref{annot:hole:4:template}), a char
array variable hole (line~\ref{annot:hole:5:template}), and an integer
literal hole (line~\ref{annot:hole:6:template}) in
Figure~\ref{fig:example:template}, are converted from \CodeIn{size}
(line~\ref{annot:hole:4:original}), \CodeIn{buffer}
(line~\ref{annot:hole:5:original}), and \CodeIn{0}
(line~\ref{annot:hole:6:original}) in
Figure~\ref{fig:example:original}, respectively. The last two \holes
are converted from the \CodeIn{while} condition \CodeIn{pos < len \&\&
buf[pos] <= \textquotesingle\ \textquotesingle} and \CodeIn{pos <
len \&\& buf[len - 1] <= \textquotesingle\ \textquotesingle},
respectively. The hole 7 (line~\ref{annot:hole:7:template}) in
Figure~\ref{fig:example:template} represents a logical relational
expression that connects two relational expressions using a logical
operator (\CodeIn{\&\&}, \CodeIn{||}). The first relational expression
connects two integer variables using one of the relational operators.
The second relational expression connects a char array access
expression and an integer variable. The char array access expression
selects an available variable of type \CodeIn{char[]} as the array and
utilizes an integer variable as the index value to retrieve the
corresponding element from the array. The last hole
(line~\ref{annot:hole:8:template}) in
Figure~\ref{fig:example:template} represents a similar
expression but it uses an arithmetic expression as the index of the
char array access expression. The arithmetic expression applies one of
the arithmetic operators (\CodeIn{+}, \CodeIn{-}, \CodeIn{*},
\CodeIn{/}, \CodeIn{\%}) on an integer variable and an integer literal
(i.e., constant).

A \template must have an \entrymethod that is the start of the
execution, annotated with \CodeIn{@Entry} as shown in the \template
(method \CodeIn{trim}). One of the key challenges is
to obtain an
argument for the method \CodeIn{trim}, i.e., an instance on which
the method is to be invoked, so that the \template can
be executed.
In our example, since the \entrymethod
\CodeIn{trim} is an instance method, the only required input is an
instance of the \template class \CodeIn{StrBuilder} that declares the
method. To provide inputs to the \entrymethod, \Tool inserts a public
static \emph{\argumentmethod}, annotated with \CodeIn{@Argument}
(method \CodeIn{\_arg0}). Thus, the argument method \CodeIn{\_arg0}
instantiates a \CodeIn{StrBuilder} and returns the instance after
invoking a sequence of methods
(line~\ref{line:codeseq:start}--\ref{line:codeseq:end}) in
Figure~\ref{fig:example:template}.
Our key insight in this direction is to capture the sequence of
methods during testing of the \entrymethod \CodeIn{trim}; tests used
can be either existing manually-written tests or automatically-generated
tests.  The sequence of methods to return an instance
of class \CodeIn{StrBuilder}
(line~\ref{line:codeseq:start}--\ref{line:codeseq:end}) in
Figure~\ref{fig:example:template} is obtained from a generated test.

Following \JAttack, \Tool generates programs by executing the
\template from the \entrymethod defined in the \template. When \Tool
reaches an unfilled \hole the first time, it randomly picks a valid
expression within the bounded search space defined by the \hole.
Once \Tool has filled all reachable \holes, it outputs a
\generatedprogram. Figure~\ref{fig:example:generated} shows an example
\generatedprogram from the \template in
Figure~\ref{fig:example:template}. In the figure, the \hole API and
the concrete expression generated to fill it share the same circled
number, indicating a match between them. The \generatedprogram can be
executed directly, as \Tool also generates a \CodeIn{main} method
(line~\ref{line:main}) in Figure~\ref{fig:example:generated}, that
invokes the \entrymethod using an instantiation of the \template class
returned from \CodeIn{\_arg0} (line~\ref{line:invoke-argument-method})
in Figure~\ref{fig:example:generated}. The \CodeIn{main} method
repeatedly invokes the \entrymethod in a \CodeIn{for} loop
(line~\ref{line:for-loop-start}--\ref{line:for-loop-end}) in
Figure~\ref{fig:example:generated}. The large number of iterations is
necessary to trigger \JIT optimizations in Java since the \JIT
compiler triggers and starts to optimize code only when a method
becomes ``hot'', i.e., frequently executed. To encode the program
behavior during execution, the \CodeIn{main} method hashes and saves
the argument values, return values, or any thrown exceptions from each
iteration, and the final class state (i.e., static fields) of the
\template. These hashes are used to generate a checksum, which is the
final output of the execution.

To perform differential testing~\cite{Mckeeman98DifferentialTesting},
\Tool executes every \generatedprogram using \JIT compilers in various
JVM implementations and compare their outputs.
The program in Figure~\ref{fig:example:generated} gave the same output
using \HotSpot and \GraalVM, but it crashed the \OpenJNine JIT
compiler.
The IBM developers confirmed that the crash is due to a bug in the
\OpenJNine \JIT compiler on handling array index out-of-bounds.

\section{\Tool Framework}
\label{sec:framework}

The \Tool framework has five key phases: (a)~\emph{\Phasecollection},
(b)~\emph{\Phaseextraction}, (c)~\emph{\Phasegeneration},
(d)~\emph{\Phasetesting}, and (e)~\emph{\Phasepruning}, as illustrated
in Figure~\ref{fig:framework:overview}.
First (Section~\ref{sec:collection}), \Tool collects a list of methods
from the given
code and obtains tests that can be used to create
meaningful inputs to these methods.
Then (Section~\ref{sec:extraction}), treating each method in the list
as an \entrymethod, \Tool selects the input that can be used to invoke
the method, and extracts a \template from the Java class that defines
the method.
Next (Section~\ref{sec:generation}),
\Tool executes each \template with the selected inputs to the
\entrymethod to generate concrete Java programs.
After that (Section~\ref{sec:testing}), \Tool executes the
\generatedprograms through the \entrymethod with the same selected
inputs, using different Java \JIT compilers for differential
testing~\cite{Mckeeman98DifferentialTesting}.
Finally (Section~\ref{sec:pruning}), \Tool prunes the detected crash
and/or consistency as to minimize false positives, and then reports
detected bugs.

\begin{figure}[!t]
\begin{scriptsize}
\centering

\newlength{\nodedistance}
\nodedistance=0.72cm

\tikzstyle{process} = [rectangle,
rounded corners,
minimum width=0.8cm,
minimum height=0.5cm,
text centered,
text width=0.9cm,
draw=black,
fill=white]
\tikzstyle{process_short} = [rectangle,
rounded corners,
minimum width=0.8cm,
minimum height=0.5cm,
text centered,
text width=0.9cm,
draw=black,
fill=white]
\tikzstyle{process_text} = [rectangle, draw=none, fill=none, text centered]
\tikzstyle{process_text_bold} = [process_text, font={\bfseries}, text height = 0.2cm]

\tikzstyle{arrow} = [semithick,->,>=stealth]
\tikzstyle{dashed_arrow} = [semithick,->,>=stealth, dashed]
\tikzstyle{dotted_arrow} = [semithick,->,>=stealth, dash pattern={on 7pt off 2pt on 1pt off 3pt}]
\tikzstyle{line} = [semithick]
\tikzstyle{point} = [circle, inner sep=0pt, minimum size=0pt, fill=white] %

\colorlet{backcolor1}{red!18}
\colorlet{backcolor2}{yellow!15}
\colorlet{backcolor3}{blue!10}
\colorlet{backcolor4}{green!10}
\colorlet{backcolor5}{orange!15}
\colorlet{backcolor1-light}{red!10}
\colorlet{backcolor1-dark}{red!27}

\begin{tikzpicture}[node distance=\nodedistance and \nodedistance]

\node (gen1) [process, text width=1.0cm] {Extended \JAttack};
\node (extract1) [process, text width=1.2cm, above left=0.01\nodedistance and 1.8\nodedistance of gen1] {Expressions $\rightarrow$ \Holes};
\node (extract2) [process, text width=1.2cm, below left=0.01\nodedistance and 1.8\nodedistance of gen1] {Arguments};

\node (collection1) [process, left=1.4\nodedistance of extract2] {Test Finder};
\node (proj_in) [process_text, above left=0.08\nodedistance and 1.2\nodedistance of collection1] {Project};

\node (test1) [process_short, right=2\nodedistance of gen1] {\HotSpot};
\node (test1b) [process_short, above=0.2\nodedistance of test1] {\OpenJNine};
\node (test1c) [process_short, below=0.2\nodedistance of test1] {\GraalVM};
\node (test2) [process, right=1.3\nodedistance of test1, minimum width=1.7\nodedistance] {Reference \JVM};
\node (report) [process_text, right=1\nodedistance of test2] {Bugs};

\path let \p1 = (extract1.west) in node [point] (extract1_westtop) at (\x1, \y1 + 0.15\nodedistance) {};
\path let \p1 = (extract1.west) in node [point] (extract1_westbottom) at (\x1, \y1 - 0.15\nodedistance) {};
\path let \p1 = (collection1.north) in node [point] (collection1_northleft) at (\x1, \y1) {};

\coordinate (projSplit) at ([xshift=0.4\nodedistance] proj_in.east);
\draw [line] (proj_in) -- (projSplit);
\draw [arrow] (projSplit) |- (collection1);
\draw [dashed_arrow] (projSplit) |- (extract1_westtop) node[above right=1.5\nodedistance and -1.0\nodedistance of collection1] {All Methods};

\draw [dotted_arrow] (collection1_northleft) |- (extract1_westbottom) node[above right=0.65\nodedistance and -0.7\nodedistance of collection1] {Unit Tests};

\begin{scope}[transform canvas={yshift=0.15\nodedistance}]
\draw [dotted_arrow] (collection1) -- (extract2) node [midway,above] {Input $L$};
\end{scope}
\begin{scope}[transform canvas={yshift=-0.15\nodedistance}]
\draw [dashed_arrow] (collection1) -- (extract2) node [midway,below] {Pool $I$};
\end{scope}

\coordinate (genSplit) at ([xshift=1.6\nodedistance]gen1.east);
\draw [line] (gen1) -- (genSplit) node[midway,above=0.3\nodedistance] {Generated} node[midway,above=-0.1\nodedistance] {Programs};
\draw [arrow] (genSplit) |- (test1b);
\draw [arrow] (genSplit) |- (test1);
\draw [arrow] (genSplit) |- (test1c);

\coordinate (testJoin) at ([xshift=0.2\nodedistance]test1.east);
\draw [arrow] (testJoin) -- (test2);
\draw [line] (test1) -| (testJoin);
\draw [line] (test1b) -| (testJoin);
\draw [line] (test1c) -| (testJoin);

\draw [arrow] (test1) -- (test2) node[midway,above=0.3\nodedistance] {Diff /} node[midway,above=-0.01\nodedistance] {Crash};
\draw [arrow] (test2) -- (report);

\coordinate (genJoin) at ([xshift=-1.6\nodedistance]gen1.west);
\draw [arrow] (genJoin) -- (gen1) node[midway,above] {Templates};
\draw [line] (extract1) -| (genJoin);
\draw [line] (extract2) -| (genJoin);

\path let \p1 = (proj_in.south east), \p2 = (collection1.south west), \p3 = ($(\p1)!0.5!(\p2)$) in node [point] (anchor1) at (\x3, \y1 + 2.5\nodedistance) {};
\path let \p1 = (collection1.south east), \p2 = (extract2.south west), \p3 = ($(\p1)!0.5!(\p2)$) in node [point] (anchor2) at (\x3, \y1 - 0.5\nodedistance) {};

\path let \p1 = (anchor1), \p2 = (anchor2) in node [point] (anchor3) at (\x2, \y1) {};
\path let \p1 = (anchor2), \p2 = (extract2.north east), \p3 = (gen1.south west), \p4 = ($(\p2)!0.5!(\p3)$) in node [point] (anchor4) at (\x4, \y1) {};

\path let \p1 = (anchor3), \p2 = (anchor4) in node [point] (anchor5) at (\x2, \y1) {};
\path let \p1 = (anchor4), \p2 = (gen1.north east), \p3 = (test1.south west), \p4 = ($(\p2)!0.5!(\p3)$) in node [point] (anchor6) at (\x4, \y1) {};

\path let \p1 = (anchor5), \p2 = (anchor6) in node [point] (anchor7) at (\x2, \y1) {};
\path let \p1 = (anchor6), \p2 = (test1.north east), \p3 = (test2.south west), \p4 = ($(\p2)!0.5!(\p3)$) in node [point] (anchor8) at (\x4, \y1) {};

\path let \p1 = (anchor7), \p2 = (anchor8) in node [point] (anchor9) at (\x2, \y1) {};
\path let \p1 = (anchor8), \p2 = (test2.north east), \p3 = (report.south west), \p4 = ($(\p2)!0.5!(\p3)$) in node [point] (anchor10) at (\x4, \y1) {};

\path let \p1 = (anchor1), \p2 = (anchor2), \p3 = ($(\p1)!0.5!(\p2)$) in node [point] (anchor-c1a) at (\x1, \y3 + 1.2\nodedistance) {};
\path let \p1 = (anchor1), \p2 = (anchor2), \p3 = ($(\p1)!0.5!(\p2)$) in node [point] (anchor-c1b) at (\x2, \y3 - 0.2\nodedistance) {};
\path let \p1 = (anchor1), \p2 = (anchor2), \p3 = ($(\p1)!0.5!(\p2)$) in node [point] (anchor-c2a) at (\x1, \y3 - 0.2\nodedistance) {};
\path let \p1 = (anchor1), \p2 = (anchor2), \p3 = ($(\p1)!0.5!(\p2)$) in node [point] (anchor-c2b) at (\x2, \y2 + 0.2\nodedistance) {};
\path let \p1 = (anchor-c2a), \p2 = (anchor-c2b) in node [point] (anchor-c2c) at (\x1, \y2) {};

\begin{scope}[on background layer]
\fill [fill=backcolor2,rounded corners] (anchor3)
rectangle (anchor4); \fill [fill=backcolor3,rounded corners]
(anchor5) rectangle (anchor6); \fill [fill=backcolor4,rounded
corners] (anchor7) rectangle (anchor8);
\fill[fill=backcolor5,rounded corners] (anchor9) rectangle
(anchor10);

\fill [fill=backcolor1-light,rounded corners] (anchor1) rectangle
(anchor2);
\fill [fill=backcolor1,rounded corners] (anchor-c1a) rectangle
(anchor-c1b);
\fill [fill=backcolor1-dark,rounded corners] (anchor-c2a) rectangle
(anchor2);

\end{scope}

\node (extract_text) [process_text_bold] at ($(anchor3)!0.5!(anchor4)$) [above=1.3\nodedistance] {Extraction};
\node (collection_text) [process_text_bold] at ($(anchor1)!0.5!(anchor2)$) [above=1.3\nodedistance] {Collection};
\node (generation_text) [process_text_bold] at ($(anchor5)!0.5!(anchor6)$) [above=1.3\nodedistance] {Generation};
\node (testing_text) [process_text_bold] at ($(anchor7)!0.5!(anchor8)$) [above=1.3\nodedistance] {Testing};
\node (pruning_text) [process_text_bold] at ($(anchor9)!0.5!(anchor10)$) [above=1.3\nodedistance] {Pruning};

\node (static_text) at (anchor-c1a) [below right= -0.2\nodedistance and -0.2\nodedistance] {\underline{Static}};
\node (dynamic_text) at (anchor-c2a) [below right= -0.2\nodedistance and -0.2\nodedistance] {\underline{Dynamic}};

\end{tikzpicture}

\end{scriptsize}
\caption{\UseMacro{figure-overview-caption}}
\end{figure}

\subsection{\PhaseCollection}
\label{sec:collection}

In the collection phase, \Tool collects a list of methods from
the given code to be used as \template \entrymethods.  \Tool then
obtains tests for each of the methods, which will be used to obtain
objects that can be used as arguments to the entry method.

We developed two approaches to collect a list of \entrymethods and to
obtain code sequences that create arguments for \entrymethods.

\MyPara{\ApproachTestBased}
We use automated test generation to generate a large number of unit
tests for all the classes in the given code.
We utilize \emph{the last method call} in the unit test as the
\entrymethod.  As such, we can save the code sequence leading up to
the method call as a way to construct arguments for that method. This
approach ideally results in the same number of \entrymethods as the
number of unit tests generated,
and each \entrymethod is associated with the saved code sequence as
the input to the method.

\MyPara{\ApproachPoolBased}
Instead of using generated unit tests as in the previous approach,
we save all prefixes of generated tests in
the \ApproachpoolBased approach; each prefix creates
an object that we add to an object pool.
This object pool stores all code sequences produced during a test generation run,
where each code sequence ultimately returns an instance of a class
defined within the project (the object returned by the final method
call in the sequence). We organize the pool using a mapping that
associates each class with all the code sequences that can instantiate
that class. This approach parses all the Java classes in the given
project and obtains all the methods from these classes, and then uses
all of the methods as \entrymethods.

Note that the \ApproachpoolBased approach creates a superset of
objects created by the \ApproachtestBased approach, but the entry
methods are different (as described above).  We compare these two
approaches in our evaluation to discover if they are complementary and
if each leads to valuable inputs during compiler testing.

\subsection{\PhaseExtraction}
\label{sec:extraction}

For every \entrymethod in the list provided by the \Phasecollection
phase, \Tool creates a \template from the class that declares the
method. Figure~\ref{algo:extraction} shows the overall algorithm for
\Tool to extract a \template from a given \entrymethod.
The input to the function \CodeIn{Extract} is the entry method
$\aEntry$, and either the collected inputs to $\aEntry$, if using the
\ApproachtestBased approach, or the pool of inputs for all methods, if
using the \ApproachpoolBased approach, is represented by $L$ or $I$,
respectively. The output is the extracted \template \aTemplate.

\begin{figure}[t]
\begin{\UseMacro{algo-extraction-fontsize}}
\begin{subfigure}[b]{0.59\linewidth}

\begin{algorithmic}[1]
\Input $M$: the entry method
\Input $L$: the collected input to the entry method (\ApproachtestBased only)
\Input $I$: the pool of inputs to all methods (\ApproachpoolBased only)
\Output the extracted \template
\Function{Extract}{$M$, $L$, $I$}
\State $C$ $\gets$ \Call{GetClassDeclaring}{$M$} \label{line:algo:extraction:get-class}
\State \aTemplate $\gets$ \Call{Clone}{$C$} \label{line:algo:extraction:clone-class}
\ForAll{$e$ in \Call{GetAllExprs}{\aTemplate}} \label{line:algo:extraction:convert:start}
\State $e^\prime$ $\gets$ \Call{Convert}{$e$, $0$} \label{line:algo:extraction:call-convert}
\State \Call{Replace}{$e$, $e^\prime$} \label{line:algo:extraction:replace}
\EndFor \label{line:algo:extraction:convert:end}
\ForAll{$l$ in \Call{GetAllLoops}{\aTemplate}} \label{line:algo:extraction:loop-limiter:start}
\State \Call{InsertLoopLimiter}{$l$}
\EndFor \label{line:algo:extraction:loop-limiter:end}
\If{$L$}
\State $m$ $\gets$ \Call{CreateArgumentsMethod}{$L$} \label{line:algo:extraction:create-argument:test:start}
\State \Call{InsertMethod}{$m$, \aTemplate} \label{line:algo:extraction:create-argument:test:end}
\EndIf
\If{$I$}
\ForAll{$p$ in \Call{GetAllParams}{$M$}} \Comment{including the receiver} \label{line:algo:extraction:create-argument:pool:start}
\State \aType $\gets$ \Call{ResolveType}{$p$}
\If{\Call{IsPrimitive}{\aType}}
\State $i$ $\gets$ ``\CodeInAlgo{<\aType{}>Val()}'' \label{line:algo:extraction:create-argument:pool:primitive}
\ElsIf{$I$.contains(\aType)}
\State $i$ $\gets$ \Call{RandomInputOfTypeFromPool}{\aType, $I$} \label{line:algo:extraction:create-argument:pool:pick-from-pool}
\Else
\State $i$ $\gets$ ``\CodeInAlgo{null}'' \label{line:algo:extraction:create-argument:pool:null}
\EndIf
\State $m$ $\gets$ \Call{CreateArgumentMethod}{$p$, $i$}
\State \Call{InsertMethod}{$m$, \aTemplate}
\EndFor \label{line:algo:extraction:create-argument:pool:end}
\EndIf
\State \Return \aTemplate \label{line:algo:extraction:return}
\EndFunction
\rememberlines
\end{algorithmic}

\end{subfigure}
\begin{subfigure}[b]{0.4\linewidth}

\begin{algorithmic}[1]
\resumenumbering
\Input \aExp: the original expression
\Input $\overline{d}$: the depth of \aExp
\Output the \hole API
\Function{Convert}{\aExp, $\overline{d}$}
\State \aType $\gets$ \Call{ResolveType}{\aExp}
\Switch{\Call{GetCategoryOfExpr}{\aExp}}
\Case{Identifier}
\State $h$ $\gets$ ``\CodeInAlgo{<\aType{}>Id()}''
\EndCase
\Case{Literal}
\State $h$ $\gets$ ``\CodeInAlgo{<\aType{}>Val()}''
\EndCase
\Case{Relation}
\State $\overline{l}$ $\gets$ \Call{Convert}{\aExp.left, $d + 1$} \label{line:algo:extraction:relation:left}
\State $\overline{r}$ $\gets$ \Call{Convert}{\aExp.right, $d + 1$} \label{line:algo:extraction:relation:right}
\State $h$ $\gets$ ``\CodeInAlgo{relation(<$\overline{l}$>, <$\overline{r}$>)}'' \label{line:algo:extraction:relation:combine}
\EndCase
\State \dots
\EndSwitch
\If{$\overline{d}$ = 0} \label{line:algo:extraction:convert:return:start}
\State \Return ``\CodeInAlgo{<$h$>.eval()}''
\Else
\State \Return ``\CodeInAlgo{<$h$>}''
\EndIf \label{line:algo:extraction:convert:return:end}
\EndFunction
\end{algorithmic}

\end{subfigure}
\end{\UseMacro{algo-extraction-fontsize}}
\vspace{-10pt}
\caption{\UseMacro{algo-extraction-caption}}
\end{figure}

The function \CodeIn{Extract} starts by finding the original class
$\aClass$ that declares the \entrymethod in the given Java code
(line~\ref{line:algo:extraction:get-class}) and initializes \template
\aTemplate as a clone of $\aClass$
(line~\ref{line:algo:extraction:clone-class}). Next, for the class,
\Tool recursively converts every expression in every method (obtained
from \CodeIn{GetAllExprs($\aTemplate$)}) into a hole. Next,
\CodeIn{Extract} replaces each expression $\aExp$ in \aTemplate with a
\hole \API call (i.e., Java method that represents a hole) by calling
the function \CodeIn{Convert}
(line~\ref{line:algo:extraction:call-convert}) and then replacing the
expression into the \hole in place
(line~\ref{line:algo:extraction:replace}).
Rather than convert each expression into a hole, \Tool can also
selectively create holes for some types of \holes. Although we
empirically evaluate impact of various types of holes, we assume in
this algorithm that we convert each expression into a hole w.l.o.g.

The function \CodeIn{Convert} takes an expression $\aExp$ and its
depth $\overline{d}$ as input and returns a \hole API. It resolves the
type of $\aExp$ as $\aType$, then converts $\aExp$ into a \hole API by
recursively replacing each sub-expression of $\aExp$ with the proper
\hole API call. If $\aExp$ is an identifier, it is converted into a
\CodeIn{<$\aType$>Id} \hole API call.

\MyPara{Example}
The variable \CodeIn{str} of \CodeIn{String} type in the class from
Figure~\ref{fig:example:original} (line~\ref{annot:hole:1:original})
is converted into \CodeIn{refId(String.class)} in the \template from
Figure~\ref{fig:example:template} (line~\ref{annot:hole:1:template}).
Another \CodeIn{int} variable \CodeIn{size}
(line~\ref{annot:hole:4:original}) is converted into \CodeIn{intId()}
in the same \template (line~\ref{annot:hole:4:template}).

\vspace{3pt}
\noindent
If $\aExp$ is a literal, it is converted into a \CodeIn{<$\aType$>Val}
\hole API call.

\MyPara{Example}
The integer number \CodeIn{0} in the class from
Figure~\ref{fig:example:original} (line~\ref{annot:hole:6:original})
is converted into \CodeIn{intVal()} in the \template from
Figure~\ref{fig:example:template} (line~\ref{annot:hole:6:template}).

\vspace{3pt}
\noindent
If $\aExp$ is not a terminal, its sub-expressions are recursively
converted into \hole \API calls. For a relational expression $\aExp$,
the left and right sub-expressions are converted into the \hole \API
call $\overline{l}$ (line~\ref{line:algo:extraction:relation:left} in
Figure~\ref{algo:extraction}) and $\overline{r}$
(line~\ref{line:algo:extraction:relation:right}), respectively, before
$\overline{l}$ and $\overline{r}$ are combined using the
\CodeIn{relation} \hole API
(line~\ref{line:algo:extraction:relation:combine}).
The operator of the relational expression is ignored because a \hole
API uses all available operators by default if no operator argument is
provided.

\MyPara{Example}
Consider the relational expression \CodeIn{size == 0} in the class
from Figure~\ref{fig:example:original}
(line~\ref{annot:hole:3:original}). The left sub-expression
\CodeIn{size} is converted into \CodeIn{intId()}, and the right
sub-expression is converted into \CodeIn{intVal()}. Then the two
results are combined as \CodeIn{relation(intId(), intVal())} in the
\template from Figure~\ref{fig:example:template}
(line~\ref{annot:hole:3:template}).

\vspace{3pt}
\noindent
Other expressions that are non-terminals, e.g., arithmetic, logical,
array access, etc., are converted in a similar way as a relational
expression; we do not list all of them in the algorithm.
Once the given expression $\aExp$ is converted into a \hole API call,
\CodeIn{Convert} checks if the current depth is $0$. If so, it appends
the \CodeIn{eval()} call to the \hole API call and returns the
resulting call as the output of the function
(line~\ref{line:algo:extraction:convert:return:start}--\ref{line:algo:extraction:convert:return:end}).
Note that \CodeIn{eval()} first triggers the \hole API call, which
returns an expression that fills the \hole.  Then \CodeIn{eval()} is
called on the returned expression
that evaluates to the type that the \hole represents (e.g.,
\CodeIn{int}).
Thus, there is only one \CodeIn{eval()} for the outermost \hole API
call.

A \hole as a loop condition might introduce an infinite loop in the
\template class \aTemplate if the \hole is filled in with some
expression that is always evaluated to \CodeIn{true} at the
\Phasegeneration phase. Therefore, \CodeIn{Extract} inserts a loop
limiter to restrict the maximum iterations that one loop can be
executed
(line~\ref{line:algo:extraction:loop-limiter:start}--\ref{line:algo:extraction:loop-limiter:end}
in Figure~\ref{algo:extraction}).

\MyPara{Example} Consider hole 7 in the \template from
Figure~\ref{fig:example:template} (line~\ref{line:hole:7:template}),
which is a loop condition.
To prevent the infinite loop that may occur due to filling in random
values, a loop limiter \CodeIn{\_lim1++ < 1000} is appended to the
\CodeIn{logic} \hole to restrict the maximum iterations to one
thousand times.

\vspace{3pt}
Once the \holes are created, \CodeIn{Extract} then creates and inserts
argument methods into \aTemplate, according to the selected approach
in the \Phasecollection phase.

\MyPara{\ApproachTestBased} A public static \CodeIn{@Arguments} method
is added to the \template class \aTemplate
(line~\ref{line:algo:extraction:create-argument:test:start}--\ref{line:algo:extraction:create-argument:test:end}
in Figure~\ref{algo:extraction}). The method returns an array of all
the inputs to the \entrymethod in the order of method parameters.
Consider the \entrymethod with signature:
\vspace{-2pt}
\begin{lstlisting}[language=jattack-display]
public Quaternion multiply(final double alpha)
\end{lstlisting}
\vspace{-2pt}
in the class \CodeIn{org.apache.commons.math4.complex.Quaternion} from
open-source project \commonsmath~\cite{commonsmathQuaternionSource}.
The following \CodeIn{@Arguments} method is generated:
\begin{lstlisting}[language=jattack-display]
@Arguments
public static Object[] _args() throws Throwable {
  Quaternion quaternion = new Quaternion(
      35.0, (double) 0, 57.29577951308232, -1.0);
  return new Object[] { quaternion, (double) 17 };
}
\end{lstlisting}
where \CodeIn{quaternion} and \CodeIn{17} are the inputs collected in
the \Phasecollection phase, i.e., extracted from a generated unit test
with the \entrymethod \CodeIn{multiply} as the last method
call.

\MyPara{\ApproachPoolBased}
A public static \CodeIn{@Argument} method is created to provide an
input for each parameter (including the receiver) of the \entrymethod
according to the type \aType of the parameter
(line~\ref{line:algo:extraction:create-argument:pool:start}--\ref{line:algo:extraction:create-argument:pool:end}
in Figure~\ref{algo:extraction}).
If a primitive input is required, then a \CodeIn{<\aType{}>Val} \hole
API will be used
(line~\ref{line:algo:extraction:create-argument:pool:primitive});
otherwise a reference input with the required type is randomly picked
from the object pool provided from the \Phasecollection phase
(line~\ref{line:algo:extraction:create-argument:pool:pick-from-pool}).
If the pool does not contain the type, \CodeIn{null} is used
(line~\ref{line:algo:extraction:create-argument:pool:null}).
Consider the \entrymethod \CodeIn{trim} in the \template from
Figure~\ref{fig:example:template}, since it is an instance method
without any parameter, only a single \CodeIn{@Argument} method
\CodeIn{\_arg0}
(line~\ref{line:codeseq:start}--\ref{line:codeseq:end}) is created.
The method \CodeIn{\_arg0} uses a randomly picked code sequence from
the object pool and returns an instance of \CodeIn{StrBuilder} that
can be used to invoke the instance \entrymethod \CodeIn{trim}.

Finally, \CodeIn{Extract} returns the \template \aTemplate
(line~\ref{line:algo:extraction:return} in
Figure~\ref{algo:extraction}). \Tool repeats the procedure to extract
a \template for every \entrymethod provided in the list from the
\Phasecollection phase.

\subsection{\PhaseGeneration}
\label{sec:generation}

In the \Phasegeneration phase, \Tool obtains concrete programs from
every \template extracted from the previous phase.  \Tool generates
programs through an execution-based model. Given a \template
\aTemplate,
the initial global state is captured first.  Then, the \entrymethod of
the \template is repeatedly executed, stopping when all \holes are
filled or the maximum iterations \(N\) has been reached.
Next, the technique outputs a \generatedprogram by filling every \hole
with corresponding concrete code.  This process repeats \(M\) times to
generate \(M\) programs, with the \template state reset after each
program generation.
\Tool builds on \JAttack to support the \Phasegeneration phase.

\MyPara{Extending \template support}
\Tool enhances \JAttack in several aspects.
(1)~\JAttack allows only a static method as the \entrymethod. On the
other hand, \Tool introduces the receiver object for the \entrymethod,
which allows an instance method as the \entrymethod via passing the
receiver's value from \CodeIn{@Argument} or \CodeIn{@Arguments}
methods.
(2)~\JAttack does not support non-primitive static fields in
\templates, as it resets a static field by saving and recovering the
value. To resolve this, \Tool resets states of \template classes by
re-invoking static initializers (\CodeIn{clinit})~\cite{bell14vmvm},
thus allowing non-primitive static fields to be reset in \templates.
(3)~\JAttack crashes due to \CodeIn{UnfilledHoleException} immediately when
encountering \holes in static initializers, while \Tool adds extra
logic to handle those exceptions when loading (including
re-initializing) \template classes.
(4)~Certain \holes in constructors are not supported well by \JAttack.
For a \CodeIn{<\aType{}>Id} \hole inside \CodeIn{super()} or
\CodeIn{this()} calls from a constructor, \JAttack can fill the hole
with a field accessed from \CodeIn{uninitializedThis}, which fails
bytecode verification. \Tool overcomes the limitation by tracking at
which execution point in a constructor (when all
\CodeIn{INVOKESPECIAL} and \CodeIn{NEW} bytecode instructions are
paired up) \CodeIn{uninitializedThis} gets initialized and can be
used.
(5)~\Tool introduces a number of new \hole APIs for type casting and
improves the checksum utility of \JAttack to avoid hash collisions
when hashing an object graph.

\MyPara{Improving generation procedure}
\Tool makes two changes to the original generation procedure of
\JAttack.
(1)~One of the advantages of \JAttack's execution-based generation
over static generation is that it knows exactly what gets executed in
a \generatedprogram and such information can help generate better
programs. However, \JAttack does not leverage the information in its
implementation.
It simply outputs any \generatedprogram as long as the program
compiles. Instead, \Tool skips certain \generatedprograms that are
less likely to trigger \JIT optimizations. For instance, \Tool will
skip a \generatedprogram if the execution stops even before entering
the \entrymethod due to an exception thrown from argument methods.
(2)~\JAttack renames the class with a unique suffix in every
\generatedprogram, e.g., Gen1, Gen2, etc. However, such renaming
breaks circular dependencies between the \template class and other
classes in the same project, which makes many \generatedprograms not
compilable. \Tool disables renaming and keeps the original class name
of the \template for all \generatedprograms. When executing a
\generatedprogram in the testing phase, \Tool ensures that the
compiled classfile of the \generatedprogram appears in the classpath
before all the other classes of the original Java source, such that
the \generatedprogram, rather than the original class with the same
name in the project, will be used.

\subsection{\PhaseTesting}
\label{sec:testing}

For differential testing, \Tool executes each \generatedprogram with
various implementations and levels of \JIT, i.e., different
\emph{\JITconfigs}.
We define a \JITconfig as a tuple (vendor name, compiler name, version
number, JVM options),
for example: (Oracle, \HotSpot, 20, \CodeIn{-XX:TieredStopAtLevel=1})
and (IBM, \OpenJNine, 17.0.6, \CodeIn{-Xjit:optlevel=hot}).
Each \generatedprogram is executed repeatedly with a large number of
iterations (to trigger \JIT compilation) and outputs a checksum value
in the end. This checksum value is calculated by hashing the arguments
provided to the \entrymethod, the output of the return value from the
\entrymethod in each iteration, and the final state (i.e., static fields) of the entire
class~\cite{zang22jattack}. Then, \Tool compares the checksum values
from different \JITconfigs and reports a failure if it observes any
difference. Additionally, \Tool reports a failure if the program
crashes on some \JITconfigs.

\subsection{\PhasePruning}
\label{sec:pruning}

Not every failure indicates a real issue with \JIT. During our
experiments, we find that most of the failures reported due to
observed inconsistent checksum values across \JITconfigs were caused by
either (1)~non-deterministic features of the \generatedprogram
itself, e.g., random numbers, current timestamps, hashcode, etc.,
(2)~the inconsistency between \JITconfigs themselves, e.g., system
property \CodeIn{java.vm.name} and \CodeIn{java.vm.info}, which
contain the \JVM's version information and Java options used, or
(3)~discrepancies between JVM implementations from different vendors,
such as \HotSpot and \OpenJNine, which disagree on the maximum array
size. To alleviate this problem, \JAttack reruns the failing program
twice using the interpreter mode (\CodeIn{-Xint}) of a single
\JITconfig, while keeping the rest of the \JITconfig intact,
and reports the failure only when the two reruns using interpreter
mode give the exact same outputs.
However, this solution can only filter out false positive \JIT bugs
caused by (1) but not (2) or (3). \Tool improves the filtering by a)
using various \JVMs (e.g., \HotSpot and \OpenJNine) and b) increasing
the number of reruns of the failing program from one to three
times. If any of the reruns using interpreter mode still shows
inconsistent checksum values across various \JVMs, \Tool considers the
failure to be not related to \JIT and thus ignore it as a false
positive.

When a \generatedprogram crashes while being executed using a
particular \JITconfig, \JAttack always labels it as a bug. However,
not all crashes are caused by issues with the JVM and therefore not
all are worth reporting to developers. Some crashes are
\CodeIn{UnfilledHoleException}, which occur due to unfilled \holes in
the program, which are left as API method calls during the
\Phasegeneration phase but are reached during execution in the
\Phasetesting phase (due to non-determinism). Such cases may be caused
by incorrect \JIT compilation that leads to a mismatch in program
behavior between the \Phasegeneration and \Phasetesting phases, which
we want to report as a bug. However, many of these cases result from
the aforementioned three reasons that cause inconsistent checksum
values between \JITconfigs. For example, non-deterministic features
such as current timestamps may have inconsistent values between the
\Phasegeneration phase and \Phasetesting phase. This inconsistency can
cause disagreement in code paths taken between the \Phasegeneration
phase and the \Phasetesting phase, e.g., when evaluating \CodeIn{if}
conditions on timestamps, which can result in unfilled \holes that
were not reached during \Phasegeneration but were reached during
\Phasetesting.
To address this issue, \Tool reruns the \generatedprogram with various
\JVMs using interpreter mode if the program reports a crash due to
\CodeIn{UnfilledHoleException}. If the program does not crash during
the rerun, then \Tool reports a bug. However, if the program still
crashes during the rerun, then \Tool considers the crash as a false
positive and skips reporting the failure.

While the pruning approach is simple, with manual inspection on a
number of cases, we found it sufficiently useful. We also compared our
pruning with original \JAttack's pruning. Our pruning filtered around
\LeJitFilterOutRate, while \JAttack filters out around
\JAttackFilterOutRate, out of total failures.

\subsection{Implementation}
\label{sec:implementation}

We implement collection of entry methods, extraction, and pruning as
standalone tools.  We extend Randoop~\cite{Pacheco07Randoop} to obtain
objects used as arguments for non-primitive types.  Finally, we extend
\JAttack~\cite{zang22jattack} to support generation and testing
phases.

\section{Evaluation}
\label{sec:eval}

We assess the value of \Tool by answering the following research
questions:

\begin{enumerate}[label={\textbf{RQ\arabic*}:}, leftmargin=*]
\item What are the contributions of the major components of \Tool?
\item How effective is \Tool compared with the state-of-the-art techniques?
\item What is the impact of \holes in various Java language features
on \Tool's bug detection?
\item What is the impact of different types of \templates on \Tool's
bug detection?
\item What critical bugs does \Tool detect in Java \JIT compilers?
\end{enumerate}

\noindent
We first describe the experiment setup (Section~\ref{sec:eval:setup})
and then answer each of the research questions
(sections~\ref{sec:components}-\ref{sec:bugs}).

\subsection{Setup}
\label{sec:eval:setup}

\MyPara{Collection}
We use open-source projects as the main input to \Tool for extracting
\templates.  An alternative was to generate Java programs using one of
the techniques for testing traditional Java
compilers~\cite{GligoricICSE10UDITA, DanielFSE07ASTGen}, but
open-source projects cover a much broader range of Java features.
We search GitHub~\cite{GitHub} for \NumProjectsSearch Java open-source
projects with the most stars, and we also include projects with at
least \NumStars stars that belong to several popular organizations,
e.g., Apache, Google, etc. In total, we collected \NumProjectsAtFist
projects. We further filter by keeping the projects that (1)~use the
Maven~\cite{Maven} build system; (2)~have a license that permits our
use; and (3)~have tests. After this step, there were
\NumProjectsFiltered projects. Then, we attempt to build each project
from its source and create a fat jar~\cite{FatJarDefinition} for each
project.
We filter out any projects that cannot be packaged this way. Lastly,
we exclude some projects that are not compatible with \Tool's
toolchain, e.g., ASM~\cite{asmwebpage},
JavaParser~\cite{javaparserwebpage}, and Randoop~\cite{Pacheco07Randoop}.
Eventually, there
are \TotalNumOfProjectsUsable projects for use.

We run \Tool once using the \ApproachpoolBased approach with all the
\TotalNumOfProjectsUsable projects but stop \Tool early, before the
\Phasegeneration phase, in order to extract \templates and collect
\holes. We next select the top \NumOfProjectsUsed projects with the
most \holes and loop limiters (used to avoid introducing infinite
loops; see Section~\ref{sec:extraction}) in the extracted \templates.
Table~\ref{tab:numbers-of-holes} shows the \NumOfProjectsUsed
open-source Java projects and associated numbers of \holes and loop
limiters; we show the number of holes for each \emph{hole type}.

\begin{table}[!t]
\setlength\tabcolsep{3.0pt}
\centering
\caption{\UseMacro{table-numbers_of_holes-caption}}
\begin{footnotesize}

\begin{tabular}{lrrrrrrrrr}
\toprule
\multirow{2}{*}{\textbf{\UseMacro{table-numbers_of_holes-col-project}}} & \multicolumn{7}{c}{\textbf{\UseMacro{table-numbers_of_holes-col-holes}}} & \multirow{2}{*}{\textbf{\UseMacro{table-numbers_of_holes-col-loop}}} & \multirow{2}{*}{\textbf{\UseMacro{table-numbers_of_holes-col-total}}}\\
\cmidrule(lr){2-8}
& \textbf{\UseMacro{table-numbers_of_holes-col-primitive_id}} & \textbf{\UseMacro{table-numbers_of_holes-col-primitive_val}} & \textbf{\UseMacro{table-numbers_of_holes-col-array}} & \textbf{\UseMacro{table-numbers_of_holes-col-arithmetic}} & \textbf{\UseMacro{table-numbers_of_holes-col-shift}} & \textbf{\UseMacro{table-numbers_of_holes-col-relation}} & \textbf{\UseMacro{table-numbers_of_holes-col-logic}} &  & \\
\midrule
\UseMacro{table-numbers_of_holes-row0-project} & \UseMacro{table-numbers_of_holes-row0-primitive_id} & \UseMacro{table-numbers_of_holes-row0-primitive_val} & \UseMacro{table-numbers_of_holes-row0-array} & \UseMacro{table-numbers_of_holes-row0-arithmetic} & \UseMacro{table-numbers_of_holes-row0-shift} & \UseMacro{table-numbers_of_holes-row0-relation} & \UseMacro{table-numbers_of_holes-row0-logic} & \UseMacro{table-numbers_of_holes-row0-loop} & \UseMacro{table-numbers_of_holes-row0-total}\\
\UseMacro{table-numbers_of_holes-row1-project} & \UseMacro{table-numbers_of_holes-row1-primitive_id} & \UseMacro{table-numbers_of_holes-row1-primitive_val} & \UseMacro{table-numbers_of_holes-row1-array} & \UseMacro{table-numbers_of_holes-row1-arithmetic} & \UseMacro{table-numbers_of_holes-row1-shift} & \UseMacro{table-numbers_of_holes-row1-relation} & \UseMacro{table-numbers_of_holes-row1-logic} & \UseMacro{table-numbers_of_holes-row1-loop} & \UseMacro{table-numbers_of_holes-row1-total}\\
\UseMacro{table-numbers_of_holes-row2-project} & \UseMacro{table-numbers_of_holes-row2-primitive_id} & \UseMacro{table-numbers_of_holes-row2-primitive_val} & \UseMacro{table-numbers_of_holes-row2-array} & \UseMacro{table-numbers_of_holes-row2-arithmetic} & \UseMacro{table-numbers_of_holes-row2-shift} & \UseMacro{table-numbers_of_holes-row2-relation} & \UseMacro{table-numbers_of_holes-row2-logic} & \UseMacro{table-numbers_of_holes-row2-loop} & \UseMacro{table-numbers_of_holes-row2-total}\\
\UseMacro{table-numbers_of_holes-row3-project} & \UseMacro{table-numbers_of_holes-row3-primitive_id} & \UseMacro{table-numbers_of_holes-row3-primitive_val} & \UseMacro{table-numbers_of_holes-row3-array} & \UseMacro{table-numbers_of_holes-row3-arithmetic} & \UseMacro{table-numbers_of_holes-row3-shift} & \UseMacro{table-numbers_of_holes-row3-relation} & \UseMacro{table-numbers_of_holes-row3-logic} & \UseMacro{table-numbers_of_holes-row3-loop} & \UseMacro{table-numbers_of_holes-row3-total}\\
\UseMacro{table-numbers_of_holes-row4-project} & \UseMacro{table-numbers_of_holes-row4-primitive_id} & \UseMacro{table-numbers_of_holes-row4-primitive_val} & \UseMacro{table-numbers_of_holes-row4-array} & \UseMacro{table-numbers_of_holes-row4-arithmetic} & \UseMacro{table-numbers_of_holes-row4-shift} & \UseMacro{table-numbers_of_holes-row4-relation} & \UseMacro{table-numbers_of_holes-row4-logic} & \UseMacro{table-numbers_of_holes-row4-loop} & \UseMacro{table-numbers_of_holes-row4-total}\\
\UseMacro{table-numbers_of_holes-row5-project} & \UseMacro{table-numbers_of_holes-row5-primitive_id} & \UseMacro{table-numbers_of_holes-row5-primitive_val} & \UseMacro{table-numbers_of_holes-row5-array} & \UseMacro{table-numbers_of_holes-row5-arithmetic} & \UseMacro{table-numbers_of_holes-row5-shift} & \UseMacro{table-numbers_of_holes-row5-relation} & \UseMacro{table-numbers_of_holes-row5-logic} & \UseMacro{table-numbers_of_holes-row5-loop} & \UseMacro{table-numbers_of_holes-row5-total}\\
\UseMacro{table-numbers_of_holes-row6-project} & \UseMacro{table-numbers_of_holes-row6-primitive_id} & \UseMacro{table-numbers_of_holes-row6-primitive_val} & \UseMacro{table-numbers_of_holes-row6-array} & \UseMacro{table-numbers_of_holes-row6-arithmetic} & \UseMacro{table-numbers_of_holes-row6-shift} & \UseMacro{table-numbers_of_holes-row6-relation} & \UseMacro{table-numbers_of_holes-row6-logic} & \UseMacro{table-numbers_of_holes-row6-loop} & \UseMacro{table-numbers_of_holes-row6-total}\\
\UseMacro{table-numbers_of_holes-row7-project} & \UseMacro{table-numbers_of_holes-row7-primitive_id} & \UseMacro{table-numbers_of_holes-row7-primitive_val} & \UseMacro{table-numbers_of_holes-row7-array} & \UseMacro{table-numbers_of_holes-row7-arithmetic} & \UseMacro{table-numbers_of_holes-row7-shift} & \UseMacro{table-numbers_of_holes-row7-relation} & \UseMacro{table-numbers_of_holes-row7-logic} & \UseMacro{table-numbers_of_holes-row7-loop} & \UseMacro{table-numbers_of_holes-row7-total}\\
\UseMacro{table-numbers_of_holes-row8-project} & \UseMacro{table-numbers_of_holes-row8-primitive_id} & \UseMacro{table-numbers_of_holes-row8-primitive_val} & \UseMacro{table-numbers_of_holes-row8-array} & \UseMacro{table-numbers_of_holes-row8-arithmetic} & \UseMacro{table-numbers_of_holes-row8-shift} & \UseMacro{table-numbers_of_holes-row8-relation} & \UseMacro{table-numbers_of_holes-row8-logic} & \UseMacro{table-numbers_of_holes-row8-loop} & \UseMacro{table-numbers_of_holes-row8-total}\\
\UseMacro{table-numbers_of_holes-row9-project} & \UseMacro{table-numbers_of_holes-row9-primitive_id} & \UseMacro{table-numbers_of_holes-row9-primitive_val} & \UseMacro{table-numbers_of_holes-row9-array} & \UseMacro{table-numbers_of_holes-row9-arithmetic} & \UseMacro{table-numbers_of_holes-row9-shift} & \UseMacro{table-numbers_of_holes-row9-relation} & \UseMacro{table-numbers_of_holes-row9-logic} & \UseMacro{table-numbers_of_holes-row9-loop} & \UseMacro{table-numbers_of_holes-row9-total}\\
\midrule
\UseMacro{table-numbers_of_holes-row10-project} & \UseMacro{table-numbers_of_holes-row10-primitive_id} & \UseMacro{table-numbers_of_holes-row10-primitive_val} & \UseMacro{table-numbers_of_holes-row10-array} & \UseMacro{table-numbers_of_holes-row10-arithmetic} & \UseMacro{table-numbers_of_holes-row10-shift} & \UseMacro{table-numbers_of_holes-row10-relation} & \UseMacro{table-numbers_of_holes-row10-logic} & \UseMacro{table-numbers_of_holes-row10-loop} & \UseMacro{table-numbers_of_holes-row10-total}\\
\bottomrule
\end{tabular}

\end{footnotesize}
\end{table}

In the \ApproachtestBased approach, we configure the test generation
to obtain \RandoopOutputLimit unit tests for each project or for
\RandoopTimeLimit, whichever comes earlier. In the \ApproachpoolBased
approach, we always run test generation for \RandoopTimeLimit. Lastly,
we use Eclipse Temurin 11.0.18 (Adoptium OpenJDK build) in the
\Phasecollection phase, including building open-source projects, test
generation, and running \Tool itself. We select this lower version of
Java in order to maximize compatibility with open-source projects and
\Tool's toolchain such as Randoop, i.e., being able to compile most
projects into fat jars and to run Randoop with the projects, with a
Java version.

\MyPara{Generation}
We generate \NumGenPerTemplate programs from each \template, with a
\GenTimeout timeout. We also set a \ExecTimeout timeout in the
\Phasetesting phase for executing each \generatedprogram. (We change
the value to \JITfuzzSingleProgramTimeout when later comparing against
\JITfuzz for fair comparison.)
We use Oracle JDK 17.0.6 to execute \templates in the \Phasegeneration
phase. We select a different JDK version for additional differential
testing between the \Phasegeneration and \Phasetesting phases.

\MyPara{Testing}
We test a wide range of JDKs with different vendors and versions
during our experiments. When evaluating \Tool alone including its
variants, we use Oracle JDK 20 (\HotSpot default and level 1), IBM
Semeru 17.0.6.0 (\OpenJNine default and hot level), and GraalVM
Enterprise Edition 22.3.1 (\GraalVM default) for differential testing.
When comparing \Tool with \JITfuzz and \JavaTailor, we also include a
custom build of OpenJDK jdk-17.0.6+10 (\HotSpot default and level 1)
(see Section~\ref{sec:comparison}).
We collect code coverage over the JVM code using the custom build of
OpenJDK. We separately re-run \generatedprograms to collect code
coverage when evaluating \Tool alone including its variants. We
collect code coverage on the fly when comparing \Tool with \JITfuzz
and \JavaTailor. \JITfuzz uses coverage, so we use the same setup for
all the tools.

\MyPara{Pruning} We use reference \JITconfigs to rerun three times
every failing program, i.e., a \generatedprogram that either has
different outputs or has crashed \JVM in the \Phasetesting phase.
We report such a failing program as a bug if
the rerun using reference \JITconfigs does not show any difference or
crash (see Section \ref{sec:pruning}).
We use \HotSpot with \CodeIn{-XX:TieredStopAtLevel=0} and \OpenJNine
with \CodeIn{-Xnojit} as reference \JITconfigs when pruning failures.

\MyPara{Machine}
We run all experiments on a 64-bit Ubuntu 18.04.1 desktop with an
Intel(R) Core(TM) i7-8700 CPU @3.20GHz and 64GB RAM.

\subsection{Contribution of Major Components}
\label{sec:components}

We evaluate \Tool using both \ApproachtestBased and \ApproachpoolBased
approaches (Section~\ref{sec:framework}), named \ToolTestBased and
\ToolPoolBased, respectively.
Additionally, we define two baselines (\ToolOnlyRandoop and
~\ToolWoRandoop) to help us understand the benefit of using \templates
and creating instances for entry methods.

\begin{table}[!t]
\centering
\caption{\UseMacro{table-variant_comparison-caption}}
\begin{footnotesize}

\begin{tabular}{lrrrrrrr}
\toprule
\multirow{2}{*}{\textbf{\UseMacro{table-variant_comparison-col-variant}}} & \multirow{2}{*}{\textbf{\UseMacro{table-variant_comparison-col-templates}}} & \multirow{2}{*}{\textbf{\UseMacro{table-variant_comparison-col-programs}}} & \multirow{2}{*}{\textbf{\UseMacro{table-variant_comparison-col-failures}}} & \multirow{2}{*}{\textbf{\UseMacro{table-variant_comparison-col-bugs}}} & \multicolumn{3}{c}{\textbf{\UseMacro{table-variant_comparison-col-coverage}}}\\
\cmidrule(lr){6-8}
&  &  &  &  & \textbf{\UseMacro{table-variant_comparison-col-coverage-c1}} & \textbf{\UseMacro{table-variant_comparison-col-coverage-c2}} & \textbf{\UseMacro{table-variant_comparison-col-coverage-hotspot}}\\
\midrule
\UseMacro{table-variant_comparison-row0-tool} & \UseMacro{table-variant_comparison-row0-templates} & \UseMacro{table-variant_comparison-row0-programs} & \UseMacro{table-variant_comparison-row0-failures} & \UseMacro{table-variant_comparison-row0-bugs} & \UseMacro{table-variant_comparison-row0-coverage-c1-line} & \UseMacro{table-variant_comparison-row0-coverage-c2-line} & \UseMacro{table-variant_comparison-row0-coverage-hotspot-line}\\
\UseMacro{table-variant_comparison-row1-tool} & \UseMacro{table-variant_comparison-row1-templates} & \UseMacro{table-variant_comparison-row1-programs} & \UseMacro{table-variant_comparison-row1-failures} & \UseMacro{table-variant_comparison-row1-bugs} & \UseMacro{table-variant_comparison-row1-coverage-c1-line} & \UseMacro{table-variant_comparison-row1-coverage-c2-line} & \UseMacro{table-variant_comparison-row1-coverage-hotspot-line}\\
\cmidrule{1-8}
\UseMacro{table-variant_comparison-row2-tool} & \UseMacro{table-variant_comparison-row2-templates} & \UseMacro{table-variant_comparison-row2-programs} & \UseMacro{table-variant_comparison-row2-failures} & \UseMacro{table-variant_comparison-row2-bugs} & \UseMacro{table-variant_comparison-row2-coverage-c1-line} & \UseMacro{table-variant_comparison-row2-coverage-c2-line} & \UseMacro{table-variant_comparison-row2-coverage-hotspot-line}\\
\UseMacro{table-variant_comparison-row3-tool} & \UseMacro{table-variant_comparison-row3-templates} & \UseMacro{table-variant_comparison-row3-programs} & \UseMacro{table-variant_comparison-row3-failures} & \UseMacro{table-variant_comparison-row3-bugs} & \UseMacro{table-variant_comparison-row3-coverage-c1-line} & \UseMacro{table-variant_comparison-row3-coverage-c2-line} & \UseMacro{table-variant_comparison-row3-coverage-hotspot-line}\\
\bottomrule
\end{tabular}

\end{footnotesize}
\vspace{-5pt}
\end{table}

The variant \ToolOnlyRandoop follows the same \Phasecollection phase
as \ToolTestBased that collects the last method call as the
\entrymethod from every unit test generated.
However, \ToolOnlyRandoop does not extract any \templates, and thus not
generate any programs from \templates.  Instead, it uses existing code
and directly goes to the \Phasetesting phase and executes the
\entrymethod a large number of times, using just the arguments in the
test.
This baseline (indirectly) shows the power of automatically generated
tests, obtained on a randomly selected set of projects,
for discovering Java \JIT compiler bugs.

We design the variant \ToolWoRandoop,
which extracts a \template for every method in a given project. The
only difference (compared to \ToolPoolBased) is that \ToolWoRandoop
does not collect the object pool,
when extracting \templates, but it rather searches for public
constructors, or static methods without parameters or with only
primitive parameters, or \CodeIn{null} to construct reference
arguments.
\ToolWoRandoop is a superset of the original \template extraction
technique presented in \JAttack~\cite{zang22jattack}; \ToolWoRandoop
supports more types of \holes and more entry methods than \JAttack.
\ToolWoRandoop shares the same \Phasegeneration and \Phasetesting
phases with \ToolPoolBased.

\begin{wrapfigure}{l}{0.33\linewidth}
\vspace{-5pt}
\centering
\includegraphics[width=\linewidth]{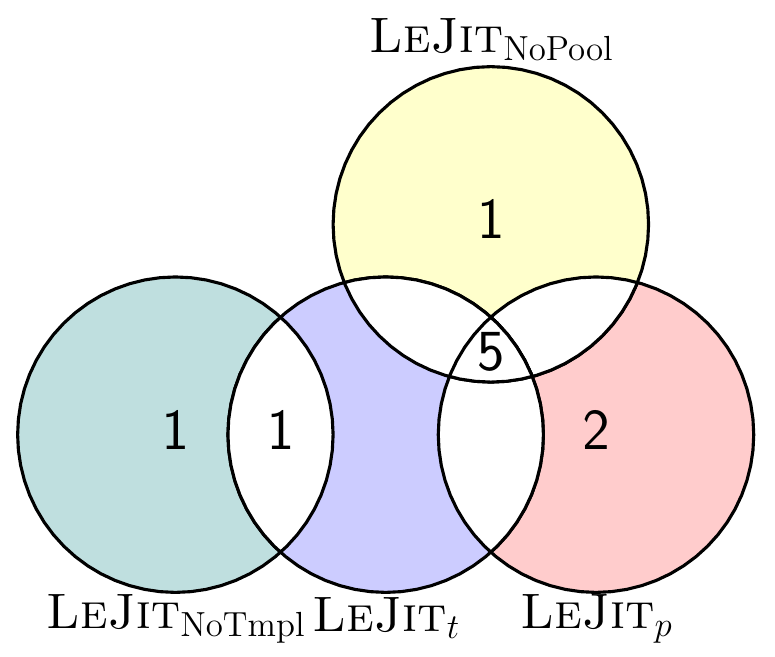}
\caption{\UseMacro{figure-overlap-caption}}
\vspace{-5pt}
\end{wrapfigure}

Table~\ref{tab:variant-comparison} compares the numbers of
\generatedprograms, failures, and unique bugs reported.
Note that all the numbers in the table represent averages from
\NumRepeatedRuns runs. The various \Tool variants exhibit differences
in their running times. Specifically, the slowest variant
(\ToolPoolBased) required around \LeJitPoolBasedRunDuration for a
single run, on average.
The first two rows compare \ToolTestBased and \ToolOnlyRandoop.
\ToolTestBased executes much more programs than \ToolOnlyRandoop,
since \ToolOnlyRandoop does not extract \templates to generate
programs. \ToolTestBased also slightly increases code coverage in
\COne and \CTwo (separate optimizing compilers within \HotSpot),
as well as in the entire \HotSpot. However, it was interesting to observe
that \ToolOnlyRandoop can even find an average of
\UseMacro{table-variant_comparison-row1-bugs} bugs per run.
As seen from the second two rows, \ToolPoolBased and \ToolWoRandoop
execute a comparable number of programs and both find a few bugs.
\ToolPoolBased achieves both higher coverage and higher number of bugs
on average.

Figure~\ref{fig:overlap} shows all the bugs we found from the variants
and the overlap of different variants. We do not include
\NumOfBugsFoundInPreliminaryExperiments bugs found in our preliminary
experiments and \NumOfBugsFoundInEvalOfTemplateTypes bugs found during
experimenting with various template types
(Section~\ref{sec:eval:template-types}).
We can see that both automated generation of instances and template
extraction contribute to detecting the bugs.
\ToolTestBased and \ToolPoolBased together miss two bugs that
\ToolOnlyRandoop and \ToolWoRandoop find. Interestingly, \ToolOnlyRandoop
without any \template or \holes finds two bugs, which shows the
effectiveness of traditional automated test generation even for domain
that is not originally targeted.

\subsection{Comparison with State-of-the-art}
\label{sec:comparison}

We compare \Tool with \JITfuzz~\cite{wu23jitfuzz} (version
\CodeIn{\JITfuzzVersion}), a state-of-the-art technique for automated
Java \JIT testing. Additionally, we compare \Tool with
\JavaTailor~\cite{Zhao22JavaTailor} (version
\CodeIn{\JavaTailorVersion}), a history-driven test program synthesis
for testing \JVMs.
Although \JavaTailor does not target \JIT per se, it is worth learning
about the relation and potential overlap between \Tool and
\JavaTailor.

\MyPara{\JITfuzz vs. \Tool}
To compare against \JITfuzz, for a given project, we need to provide
\JITfuzz an \emph{initial class} as the seed, as well as a test class
as a starting point to execute the mutated programs from those seeds.
Following the same methodology as in the previous
work~\cite{wu23jitfuzz}, we first identify ten classes in the given
project with the highest cyclomatic complexity, and we pick the
initial class randomly from these ten classes. Since the original work
did not mention how the test class should be selected, we randomly
pick a test class that imports and instantiates the initial class.
To ensure a fair comparison, we run \JITfuzz with the same
\NumOfProjectsUsed projects (Section~\ref{sec:eval:setup}). During our
preliminary experiment, we found that \JITfuzz does not support tests
using JUnit 5~\cite{JUnit5}, so we manually migrated the picked test
classes to JUnit 4~\cite{JUnit4} in \NumOfProjectsMigratedToJUnitFour
projects (other projects already used JUnit 4).

We use \ToolTestBased, i.e., with the \ApproachtestBased approach, in
order to better control end-to-end running time by specifying the
number of generated tests.
(The \ApproachpoolBased approach uses all the available methods in the
projects, which makes it hard to estimate the time needed.)
We run both tools for the same length of time, around
\ToolComparisonEndToEndTimeout, which is longer than used in the \JITfuzz
evaluation~\cite{wu23jitfuzz} and in recommended
practice~\cite{Klees18EvaluatingFuzzTesting}, while also matching the
end-to-end running time of \ToolTestBased.
We use the same timeout, \JITfuzzSingleProgramTimeout, which is the
default setting of \JITfuzz, for executing each single
\generatedprogram.
\JITfuzz requires custom debug builds of OpenJDK with AFL++ toolchain
to work, because it needs to collect runtime coverage of \JIT source
code~\cite{wu23jitfuzz}. Thus, we build OpenJDK jdk-17.0.6+10 from
source~\cite{OpenJKDRepo} with \CodeIn{--enable-debug} and
\CodeIn{--enable-native-coverage} and use the debug build as \JIT
under test. Note that \Tool works on both debug and release builds. We
use a debug build for fair comparison (we already ran \Tool on
multiple released binaries in Section~\ref{sec:components}).
Also, \JITfuzz does not use differential testing but detects only
crashes, so we do not use \OpenJNine and \GraalVM for \Tool for a fair
comparison; instead we use only default level and level 1 of \HotSpot
from the custom debug build of OpenJDK for differential testing
required by \Tool.
We collect code coverage
of \COne (\CodeIn{src/hotspot/share/c1/*}), \CTwo
(\CodeIn{src/hotspot/share/opto/*}), and the entire \HotSpot
(\CodeIn{src/hotspot/*}).

\begin{table}[!t]
\centering
\caption{\UseMacro{table-tool_comparison-caption}}
\begin{footnotesize}

\begin{tabular}{lrrrrrrrrr}
\toprule
\multirow{3}{*}{\textbf{\UseMacro{table-tool_comparison-col-tool}}} & \multirow{3}{*}{\textbf{\UseMacro{table-tool_comparison-col-programs}}} & \multirow{3}{*}{\textbf{\UseMacro{table-tool_comparison-col-failures}}} & \multirow{3}{*}{\textbf{\UseMacro{table-tool_comparison-col-bugs}}} & \multicolumn{6}{c}{\textbf{\UseMacro{table-tool_comparison-col-coverage}}}\\
\cmidrule(lr){5-10}
&  &  &  & \multicolumn{2}{c}{\textbf{\UseMacro{table-tool_comparison-col-coverage-c1}}} & \multicolumn{2}{c}{\textbf{\UseMacro{table-tool_comparison-col-coverage-c2}}} & \multicolumn{2}{c}{\textbf{\UseMacro{table-tool_comparison-col-coverage-hotspot}}}\\
\cmidrule(lr){5-6}\cmidrule(lr){7-8}\cmidrule(lr){9-10}
&  &  &  & \textbf{\UseMacro{table-tool_comparison-col-coverage-c1-function}} & \textbf{\UseMacro{table-tool_comparison-col-coverage-c1-line}} & \textbf{\UseMacro{table-tool_comparison-col-coverage-c2-function}} & \textbf{\UseMacro{table-tool_comparison-col-coverage-c2-line}} & \textbf{\UseMacro{table-tool_comparison-col-coverage-hotspot-function}} & \textbf{\UseMacro{table-tool_comparison-col-coverage-hotspot-line}}\\
\midrule
\UseMacro{table-tool_comparison-row0-tool} & \UseMacro{table-tool_comparison-row0-programs} & \UseMacro{table-tool_comparison-row0-failures} & \UseMacro{table-tool_comparison-row0-bugs} & \UseMacro{table-tool_comparison-row0-coverage-c1-function} & \UseMacro{table-tool_comparison-row0-coverage-c1-line} & \UseMacro{table-tool_comparison-row0-coverage-c2-function} & \UseMacro{table-tool_comparison-row0-coverage-c2-line} & \UseMacro{table-tool_comparison-row0-coverage-hotspot-function} & \UseMacro{table-tool_comparison-row0-coverage-hotspot-line}\\
\UseMacro{table-tool_comparison-row1-tool} & \UseMacro{table-tool_comparison-row1-programs} & \UseMacro{table-tool_comparison-row1-failures} & \UseMacro{table-tool_comparison-row1-bugs} & \UseMacro{table-tool_comparison-row1-coverage-c1-function} & \UseMacro{table-tool_comparison-row1-coverage-c1-line} & \UseMacro{table-tool_comparison-row1-coverage-c2-function} & \UseMacro{table-tool_comparison-row1-coverage-c2-line} & \UseMacro{table-tool_comparison-row1-coverage-hotspot-function} & \UseMacro{table-tool_comparison-row1-coverage-hotspot-line}\\
\bottomrule
\end{tabular}

\end{footnotesize}
\vspace{-5pt}
\end{table}

Table~\ref{tab:tool-comparison} compares the results from both tools.
Note that all the numbers in the table represent averages from
\NumRepeatedRuns runs. \JITfuzz reports \JITfuzzNumOfFailures failures
out of \JITfuzzNumOfPrograms programs that have been generated and
executed.
On the other hand, \Tool executes \ToolNumOfPrograms programs and
reports \ToolNumOfFailures failures.
We then analyze and inspect the failures from both tools. Both tools
do not detect new bugs.
All the \JITfuzzNumOfFailures failures reported by \JITfuzz are
assertion failures (which are checked on debug builds only). We group
the assertion failures by stack traces and error lines in source code
within \HotSpot, and there are only two unique failures. Both are
duplicates of an existing bug JDK-8280126~\cite{BugJDK8280126} on
optimizing irreducible loops.
We do not find any bug from \Tool's failures. We believe the reason
for this finding is that we collect code coverage on the fly for the
debug build, which impacts the way JIT optimizes generated programs.
\Tool detects a number of \HotSpot bugs in other experiments we
perform using non-debug builds (Section~\ref{sec:components}).
\Tool increases line coverage of \COne by \LineCoverageIncreaseCOne, \CTwo
by \LineCoverageIncreaseCTwo, and \HotSpot by
\LineCoverageIncreaseHotSpot compared to \JITfuzz.

\MyPara{\JavaTailor vs. \Tool}
\JavaTailor~\cite{Zhao22JavaTailor} performs history-driven test
program synthesis to test \JVM implementations. More precisely,
\JavaTailor uses previously reported bugs as seeds to synthesize
diverse test programs by combining ingredients from historical
bug-revealing programs. \JavaTailor was shown efficient for testing
\JVM implementations and here we explore if it can also discover
\JIT-related bugs.

We ran \JavaTailor three times in the default configuration until
completion ($\sim$8h each run). We inspected the three runs in detail
and concluded that findings are similar across runs, thus no further
runs were warranted. We used two versions of Java, as \JavaTailor also
performs differential testing: IBM Semeru 17.0.6.0 (\OpenJNine default
level) and a custom build of OpenJDK jdk-17.0.6+10 (\HotSpot default
level) like in the previous section.
We pick these two versions because \HotSpot and \OpenJNine were used
by \JavaTailor's authors in their evaluation, and we use the custom
build of OpenJDK because we need to collect code coverage of \JIT
compilers and compare with \Tool.

As a result of each run, \JavaTailor outputs a diff log. We could not
find any existing scripts for processing the diff logs, so we wrote
our own to help us classify failures and perform inspection.
Additionally, we wrote scripts to help us try to reproduce each of the
reported failures.

\begin{table}[!t]
\centering
\caption{\UseMacro{table-javatailor_results-caption}}
\begin{footnotesize}

\begin{tabular}{lrrrrrrr}
\toprule
\textbf{\UseMacro{table-javatailor_results-col-Kind}} & \textbf{\UseMacro{table-javatailor_results-col-NoDiff}} & \textbf{\UseMacro{table-javatailor_results-col-NonDeterministic}} & \textbf{\UseMacro{table-javatailor_results-col-DiffText}} & \textbf{\UseMacro{table-javatailor_results-col-VerifyError}} & \textbf{\UseMacro{table-javatailor_results-col-DiffException}} & \textbf{\UseMacro{table-javatailor_results-col-NoException}} & \textbf{\UseMacro{table-javatailor_results-col-Misc}}\\
\midrule
\UseMacro{table-javatailor_results-row0-Kind} & \UseMacro{table-javatailor_results-row0-NoDiff} & \UseMacro{table-javatailor_results-row0-NonDeterministic} & \UseMacro{table-javatailor_results-row0-DiffText} & \UseMacro{table-javatailor_results-row0-VerifyError} & \UseMacro{table-javatailor_results-row0-DiffException} & \UseMacro{table-javatailor_results-row0-NoException} & \UseMacro{table-javatailor_results-row0-Misc}\\
\bottomrule
\end{tabular}

\end{footnotesize}
\vspace{-5pt}
\end{table}

\JavaTailor reported
\JavaTailorNumOfFailures differences in the diff log (and each diff
corresponds to one class file that is executed with two JVMs and
produces different results). We semi-automatically classified the
reported cases into \JavaTailorNumOfKinds groups.
Table~\ref{tab:javatailor-results} shows number of cases of each group.
\UseMacro{table-javatailor_results-col-NoDiff} includes cases that
show no differences when we tried to reproduce the difference.
\UseMacro{table-javatailor_results-col-NonDeterministic} includes
cases that non-deterministically pass or fail (e.g., due to elapsed
time being in the output) and are not revealing any bug.
\UseMacro{table-javatailor_results-col-DiffText} includes cases that
are only reported with different text across \JVMs, but the reported
issue is actually the same.
\UseMacro{table-javatailor_results-col-VerifyError} includes cases
when bytecode verification failed in both \JVMs, but the messages were
different. \UseMacro{table-javatailor_results-col-DiffException}
includes cases when exceptions are printed in a different order across
\JVMs. \UseMacro{table-javatailor_results-col-NoException} includes
cases when only one of the \JVMs throws an exception, but our further
inspection showed that these cases were caused by flags that have
different default values across \JVMs.
\UseMacro{table-javatailor_results-col-Misc} includes single instance
failures that do not fit into any other group we defined; we found one
bug in this group, but the same bug was previously
reported~\cite{BugOpenJ913242}.

Regarding code coverage, \Tool increases code coverage of \COne by
\LineCoverageIncreaseJavaTailorCOne, \CTwo by
\LineCoverageIncreaseJavaTailorCTwo, and the entire \HotSpot by
\LineCoverageIncreaseJavaTailorHotSpot, compared to \JavaTailor.

In conclusion, \JavaTailor can discover \JVM bugs, but none were
related to \JIT.  We also found that the default reporting has many
false positives.
We find \Tool and \JavaTailor complementary, and each could
potentially benefit from the other; we leave the combination of the
two for future work.

\subsection{Impact of \Holes in Various Language Features on \Tool's Bug Detection}

In order to understand
how \holes in different language features contribute to bug detection
of \Tool, we perform in-depth analysis on the features within
extracted \templates and \generatedprograms.

We analyze three language constructs, i.e., arrays, conditional
statements, and loops, and one other language feature, i.e., reference
arguments.
If a filled \hole is inside a language construct (e.g., a \hole is
inside a loop), then we say the \generatedprogram that contains the
filled \hole \emph{has} the language construct, and we also say the
associated \template from which the \generatedprogram is generated
\emph{has} the language construct.
Similarly, we also measure how many \templates and \generatedprograms
use an \entrymethod that needs an argument of non-primitive
(reference) type, which means the arguments are obtained by generated
tests. We say such \templates and \generatedprograms have reference
arguments.
Table~\ref{tab:hole-features} shows the numbers of \templates and
\generatedprograms that use the four language features.  We can see
that a substantial number of \templates and programs need
non-primitive arguments.

Similarly, we say a bug \emph{has} a language feature if any
\generatedprogram that exposes the bug (i.e., failure due to bug) has
the language feature. Note that we do not claim that the presence of a
feature implies that the bug is related to the feature or the feature
is the root cause of the bug. Table~\ref{tab:bug-features} shows the
numbers of failures due to bugs and unique bugs that use various
language features.
In conclusion, \Tool well explores the four language features and
\holes in each of these features contribute to the unique bugs
discovered.

\begin{table}[!t]
\centering
\caption{\UseMacro{table-hole_features-caption}}
\begin{footnotesize}

\begin{tabular}{lrrrrrrrrrr}
\toprule
\multirow{2}{*}{\textbf{\UseMacro{table-hole_features-col-project}}} & \multicolumn{5}{c}{\textbf{\UseMacro{table-hole_features-col-template}}} & \multicolumn{5}{c}{\textbf{\UseMacro{table-hole_features-col-generated}}}\\
\cmidrule(lr){2-6}
\cmidrule(lr){7-11}
& \textbf{\UseMacro{table-hole_features-col-template-total}} & \textbf{\UseMacro{table-hole_features-col-template-ARRAY}} & \textbf{\UseMacro{table-hole_features-col-template-COND}} & \textbf{\UseMacro{table-hole_features-col-template-LOOP}} & \textbf{\UseMacro{table-hole_features-col-template-OBJARG}} & \textbf{\UseMacro{table-hole_features-col-generated-total}} & \textbf{\UseMacro{table-hole_features-col-generated-ARRAY}} & \textbf{\UseMacro{table-hole_features-col-generated-COND}} & \textbf{\UseMacro{table-hole_features-col-generated-LOOP}} & \textbf{\UseMacro{table-hole_features-col-generated-OBJARG}}\\
\midrule
\UseMacro{table-hole_features-row0-project} & \UseMacro{table-hole_features-row0-template-total} & \UseMacro{table-hole_features-row0-template-ARRAY} & \UseMacro{table-hole_features-row0-template-COND} & \UseMacro{table-hole_features-row0-template-LOOP} & \UseMacro{table-hole_features-row0-template-OBJARG} & \UseMacro{table-hole_features-row0-generated-total} & \UseMacro{table-hole_features-row0-generated-ARRAY} & \UseMacro{table-hole_features-row0-generated-COND} & \UseMacro{table-hole_features-row0-generated-LOOP} & \UseMacro{table-hole_features-row0-generated-OBJARG}\\
\UseMacro{table-hole_features-row1-project} & \UseMacro{table-hole_features-row1-template-total} & \UseMacro{table-hole_features-row1-template-ARRAY} & \UseMacro{table-hole_features-row1-template-COND} & \UseMacro{table-hole_features-row1-template-LOOP} & \UseMacro{table-hole_features-row1-template-OBJARG} & \UseMacro{table-hole_features-row1-generated-total} & \UseMacro{table-hole_features-row1-generated-ARRAY} & \UseMacro{table-hole_features-row1-generated-COND} & \UseMacro{table-hole_features-row1-generated-LOOP} & \UseMacro{table-hole_features-row1-generated-OBJARG}\\
\UseMacro{table-hole_features-row2-project} & \UseMacro{table-hole_features-row2-template-total} & \UseMacro{table-hole_features-row2-template-ARRAY} & \UseMacro{table-hole_features-row2-template-COND} & \UseMacro{table-hole_features-row2-template-LOOP} & \UseMacro{table-hole_features-row2-template-OBJARG} & \UseMacro{table-hole_features-row2-generated-total} & \UseMacro{table-hole_features-row2-generated-ARRAY} & \UseMacro{table-hole_features-row2-generated-COND} & \UseMacro{table-hole_features-row2-generated-LOOP} & \UseMacro{table-hole_features-row2-generated-OBJARG}\\
\UseMacro{table-hole_features-row3-project} & \UseMacro{table-hole_features-row3-template-total} & \UseMacro{table-hole_features-row3-template-ARRAY} & \UseMacro{table-hole_features-row3-template-COND} & \UseMacro{table-hole_features-row3-template-LOOP} & \UseMacro{table-hole_features-row3-template-OBJARG} & \UseMacro{table-hole_features-row3-generated-total} & \UseMacro{table-hole_features-row3-generated-ARRAY} & \UseMacro{table-hole_features-row3-generated-COND} & \UseMacro{table-hole_features-row3-generated-LOOP} & \UseMacro{table-hole_features-row3-generated-OBJARG}\\
\UseMacro{table-hole_features-row4-project} & \UseMacro{table-hole_features-row4-template-total} & \UseMacro{table-hole_features-row4-template-ARRAY} & \UseMacro{table-hole_features-row4-template-COND} & \UseMacro{table-hole_features-row4-template-LOOP} & \UseMacro{table-hole_features-row4-template-OBJARG} & \UseMacro{table-hole_features-row4-generated-total} & \UseMacro{table-hole_features-row4-generated-ARRAY} & \UseMacro{table-hole_features-row4-generated-COND} & \UseMacro{table-hole_features-row4-generated-LOOP} & \UseMacro{table-hole_features-row4-generated-OBJARG}\\
\UseMacro{table-hole_features-row5-project} & \UseMacro{table-hole_features-row5-template-total} & \UseMacro{table-hole_features-row5-template-ARRAY} & \UseMacro{table-hole_features-row5-template-COND} & \UseMacro{table-hole_features-row5-template-LOOP} & \UseMacro{table-hole_features-row5-template-OBJARG} & \UseMacro{table-hole_features-row5-generated-total} & \UseMacro{table-hole_features-row5-generated-ARRAY} & \UseMacro{table-hole_features-row5-generated-COND} & \UseMacro{table-hole_features-row5-generated-LOOP} & \UseMacro{table-hole_features-row5-generated-OBJARG}\\
\UseMacro{table-hole_features-row6-project} & \UseMacro{table-hole_features-row6-template-total} & \UseMacro{table-hole_features-row6-template-ARRAY} & \UseMacro{table-hole_features-row6-template-COND} & \UseMacro{table-hole_features-row6-template-LOOP} & \UseMacro{table-hole_features-row6-template-OBJARG} & \UseMacro{table-hole_features-row6-generated-total} & \UseMacro{table-hole_features-row6-generated-ARRAY} & \UseMacro{table-hole_features-row6-generated-COND} & \UseMacro{table-hole_features-row6-generated-LOOP} & \UseMacro{table-hole_features-row6-generated-OBJARG}\\
\UseMacro{table-hole_features-row7-project} & \UseMacro{table-hole_features-row7-template-total} & \UseMacro{table-hole_features-row7-template-ARRAY} & \UseMacro{table-hole_features-row7-template-COND} & \UseMacro{table-hole_features-row7-template-LOOP} & \UseMacro{table-hole_features-row7-template-OBJARG} & \UseMacro{table-hole_features-row7-generated-total} & \UseMacro{table-hole_features-row7-generated-ARRAY} & \UseMacro{table-hole_features-row7-generated-COND} & \UseMacro{table-hole_features-row7-generated-LOOP} & \UseMacro{table-hole_features-row7-generated-OBJARG}\\
\UseMacro{table-hole_features-row8-project} & \UseMacro{table-hole_features-row8-template-total} & \UseMacro{table-hole_features-row8-template-ARRAY} & \UseMacro{table-hole_features-row8-template-COND} & \UseMacro{table-hole_features-row8-template-LOOP} & \UseMacro{table-hole_features-row8-template-OBJARG} & \UseMacro{table-hole_features-row8-generated-total} & \UseMacro{table-hole_features-row8-generated-ARRAY} & \UseMacro{table-hole_features-row8-generated-COND} & \UseMacro{table-hole_features-row8-generated-LOOP} & \UseMacro{table-hole_features-row8-generated-OBJARG}\\
\UseMacro{table-hole_features-row9-project} & \UseMacro{table-hole_features-row9-template-total} & \UseMacro{table-hole_features-row9-template-ARRAY} & \UseMacro{table-hole_features-row9-template-COND} & \UseMacro{table-hole_features-row9-template-LOOP} & \UseMacro{table-hole_features-row9-template-OBJARG} & \UseMacro{table-hole_features-row9-generated-total} & \UseMacro{table-hole_features-row9-generated-ARRAY} & \UseMacro{table-hole_features-row9-generated-COND} & \UseMacro{table-hole_features-row9-generated-LOOP} & \UseMacro{table-hole_features-row9-generated-OBJARG}\\
\midrule
\UseMacro{table-hole_features-row10-project} & \UseMacro{table-hole_features-row10-template-total} & \UseMacro{table-hole_features-row10-template-ARRAY} & \UseMacro{table-hole_features-row10-template-COND} & \UseMacro{table-hole_features-row10-template-LOOP} & \UseMacro{table-hole_features-row10-template-OBJARG} & \UseMacro{table-hole_features-row10-generated-total} & \UseMacro{table-hole_features-row10-generated-ARRAY} & \UseMacro{table-hole_features-row10-generated-COND} & \UseMacro{table-hole_features-row10-generated-LOOP} & \UseMacro{table-hole_features-row10-generated-OBJARG}\\
\bottomrule
\end{tabular}

\end{footnotesize}
\vspace{-5pt}
\end{table}

\begin{table}[!t]
\centering
\caption{\UseMacro{table-bug_features-caption}}
\begin{footnotesize}

\begin{tabular}{lrrrrrrrrrr}
\toprule
\multirow{2}{*}{\textbf{\UseMacro{table-bug_features-col-project}}} & \multicolumn{5}{c}{\textbf{\UseMacro{table-bug_features-col-bugCases}}} & \multicolumn{5}{c}{\textbf{\UseMacro{table-bug_features-col-bugs}}}\\
\cmidrule(lr){2-6}
\cmidrule(lr){7-11}
& \textbf{\UseMacro{table-bug_features-col-bugCases-total}} & \textbf{\UseMacro{table-bug_features-col-bugCases-ARRAY}} & \textbf{\UseMacro{table-bug_features-col-bugCases-COND}} & \textbf{\UseMacro{table-bug_features-col-bugCases-LOOP}} & \textbf{\UseMacro{table-bug_features-col-bugCases-OBJARG}} & \textbf{\UseMacro{table-bug_features-col-bugs-total}} & \textbf{\UseMacro{table-bug_features-col-bugs-ARRAY}} & \textbf{\UseMacro{table-bug_features-col-bugs-COND}} & \textbf{\UseMacro{table-bug_features-col-bugs-LOOP}} & \textbf{\UseMacro{table-bug_features-col-bugs-OBJARG}}\\
\midrule
\UseMacro{table-bug_features-row0-project} & \UseMacro{table-bug_features-row0-bugCases-total} & \UseMacro{table-bug_features-row0-bugCases-ARRAY} & \UseMacro{table-bug_features-row0-bugCases-COND} & \UseMacro{table-bug_features-row0-bugCases-LOOP} & \UseMacro{table-bug_features-row0-bugCases-OBJARG} & \UseMacro{table-bug_features-row0-bugs-total} & \UseMacro{table-bug_features-row0-bugs-ARRAY} & \UseMacro{table-bug_features-row0-bugs-COND} & \UseMacro{table-bug_features-row0-bugs-LOOP} & \UseMacro{table-bug_features-row0-bugs-OBJARG}\\
\UseMacro{table-bug_features-row1-project} & \UseMacro{table-bug_features-row1-bugCases-total} & \UseMacro{table-bug_features-row1-bugCases-ARRAY} & \UseMacro{table-bug_features-row1-bugCases-COND} & \UseMacro{table-bug_features-row1-bugCases-LOOP} & \UseMacro{table-bug_features-row1-bugCases-OBJARG} & \UseMacro{table-bug_features-row1-bugs-total} & \UseMacro{table-bug_features-row1-bugs-ARRAY} & \UseMacro{table-bug_features-row1-bugs-COND} & \UseMacro{table-bug_features-row1-bugs-LOOP} & \UseMacro{table-bug_features-row1-bugs-OBJARG}\\
\UseMacro{table-bug_features-row2-project} & \UseMacro{table-bug_features-row2-bugCases-total} & \UseMacro{table-bug_features-row2-bugCases-ARRAY} & \UseMacro{table-bug_features-row2-bugCases-COND} & \UseMacro{table-bug_features-row2-bugCases-LOOP} & \UseMacro{table-bug_features-row2-bugCases-OBJARG} & \UseMacro{table-bug_features-row2-bugs-total} & \UseMacro{table-bug_features-row2-bugs-ARRAY} & \UseMacro{table-bug_features-row2-bugs-COND} & \UseMacro{table-bug_features-row2-bugs-LOOP} & \UseMacro{table-bug_features-row2-bugs-OBJARG}\\
\UseMacro{table-bug_features-row3-project} & \UseMacro{table-bug_features-row3-bugCases-total} & \UseMacro{table-bug_features-row3-bugCases-ARRAY} & \UseMacro{table-bug_features-row3-bugCases-COND} & \UseMacro{table-bug_features-row3-bugCases-LOOP} & \UseMacro{table-bug_features-row3-bugCases-OBJARG} & \UseMacro{table-bug_features-row3-bugs-total} & \UseMacro{table-bug_features-row3-bugs-ARRAY} & \UseMacro{table-bug_features-row3-bugs-COND} & \UseMacro{table-bug_features-row3-bugs-LOOP} & \UseMacro{table-bug_features-row3-bugs-OBJARG}\\
\UseMacro{table-bug_features-row4-project} & \UseMacro{table-bug_features-row4-bugCases-total} & \UseMacro{table-bug_features-row4-bugCases-ARRAY} & \UseMacro{table-bug_features-row4-bugCases-COND} & \UseMacro{table-bug_features-row4-bugCases-LOOP} & \UseMacro{table-bug_features-row4-bugCases-OBJARG} & \UseMacro{table-bug_features-row4-bugs-total} & \UseMacro{table-bug_features-row4-bugs-ARRAY} & \UseMacro{table-bug_features-row4-bugs-COND} & \UseMacro{table-bug_features-row4-bugs-LOOP} & \UseMacro{table-bug_features-row4-bugs-OBJARG}\\
\UseMacro{table-bug_features-row5-project} & \UseMacro{table-bug_features-row5-bugCases-total} & \UseMacro{table-bug_features-row5-bugCases-ARRAY} & \UseMacro{table-bug_features-row5-bugCases-COND} & \UseMacro{table-bug_features-row5-bugCases-LOOP} & \UseMacro{table-bug_features-row5-bugCases-OBJARG} & \UseMacro{table-bug_features-row5-bugs-total} & \UseMacro{table-bug_features-row5-bugs-ARRAY} & \UseMacro{table-bug_features-row5-bugs-COND} & \UseMacro{table-bug_features-row5-bugs-LOOP} & \UseMacro{table-bug_features-row5-bugs-OBJARG}\\
\UseMacro{table-bug_features-row6-project} & \UseMacro{table-bug_features-row6-bugCases-total} & \UseMacro{table-bug_features-row6-bugCases-ARRAY} & \UseMacro{table-bug_features-row6-bugCases-COND} & \UseMacro{table-bug_features-row6-bugCases-LOOP} & \UseMacro{table-bug_features-row6-bugCases-OBJARG} & \UseMacro{table-bug_features-row6-bugs-total} & \UseMacro{table-bug_features-row6-bugs-ARRAY} & \UseMacro{table-bug_features-row6-bugs-COND} & \UseMacro{table-bug_features-row6-bugs-LOOP} & \UseMacro{table-bug_features-row6-bugs-OBJARG}\\
\UseMacro{table-bug_features-row7-project} & \UseMacro{table-bug_features-row7-bugCases-total} & \UseMacro{table-bug_features-row7-bugCases-ARRAY} & \UseMacro{table-bug_features-row7-bugCases-COND} & \UseMacro{table-bug_features-row7-bugCases-LOOP} & \UseMacro{table-bug_features-row7-bugCases-OBJARG} & \UseMacro{table-bug_features-row7-bugs-total} & \UseMacro{table-bug_features-row7-bugs-ARRAY} & \UseMacro{table-bug_features-row7-bugs-COND} & \UseMacro{table-bug_features-row7-bugs-LOOP} & \UseMacro{table-bug_features-row7-bugs-OBJARG}\\
\UseMacro{table-bug_features-row8-project} & \UseMacro{table-bug_features-row8-bugCases-total} & \UseMacro{table-bug_features-row8-bugCases-ARRAY} & \UseMacro{table-bug_features-row8-bugCases-COND} & \UseMacro{table-bug_features-row8-bugCases-LOOP} & \UseMacro{table-bug_features-row8-bugCases-OBJARG} & \UseMacro{table-bug_features-row8-bugs-total} & \UseMacro{table-bug_features-row8-bugs-ARRAY} & \UseMacro{table-bug_features-row8-bugs-COND} & \UseMacro{table-bug_features-row8-bugs-LOOP} & \UseMacro{table-bug_features-row8-bugs-OBJARG}\\
\UseMacro{table-bug_features-row9-project} & \UseMacro{table-bug_features-row9-bugCases-total} & \UseMacro{table-bug_features-row9-bugCases-ARRAY} & \UseMacro{table-bug_features-row9-bugCases-COND} & \UseMacro{table-bug_features-row9-bugCases-LOOP} & \UseMacro{table-bug_features-row9-bugCases-OBJARG} & \UseMacro{table-bug_features-row9-bugs-total} & \UseMacro{table-bug_features-row9-bugs-ARRAY} & \UseMacro{table-bug_features-row9-bugs-COND} & \UseMacro{table-bug_features-row9-bugs-LOOP} & \UseMacro{table-bug_features-row9-bugs-OBJARG}\\
\midrule
\UseMacro{table-bug_features-row10-project} & \UseMacro{table-bug_features-row10-bugCases-total} & \UseMacro{table-bug_features-row10-bugCases-ARRAY} & \UseMacro{table-bug_features-row10-bugCases-COND} & \UseMacro{table-bug_features-row10-bugCases-LOOP} & \UseMacro{table-bug_features-row10-bugCases-OBJARG} & \UseMacro{table-bug_features-row10-bugs-total} & \UseMacro{table-bug_features-row10-bugs-ARRAY} & \UseMacro{table-bug_features-row10-bugs-COND} & \UseMacro{table-bug_features-row10-bugs-LOOP} & \UseMacro{table-bug_features-row10-bugs-OBJARG}\\
\bottomrule
\end{tabular}

\end{footnotesize}
\vspace{-5pt}
\end{table}

\subsection{Impact of \Template Types on \Tool's Bug Detection}
\label{sec:eval:template-types}

\begin{table}[!t]
\centering
\caption{\UseMacro{table-hole_types_comparison-caption}}
\begin{footnotesize}

\begin{tabular}{lrrrrr}
\toprule
\textbf{\UseMacro{table-hole_types_comparison-col-Kind}} & \textbf{\UseMacro{table-hole_types_comparison-col-Id}} & \textbf{\UseMacro{table-hole_types_comparison-col-Val}} & \textbf{\UseMacro{table-hole_types_comparison-col-ArithmeticAndShift}} & \textbf{\UseMacro{table-hole_types_comparison-col-RelationAndLogic}} & \textbf{\UseMacro{table-hole_types_comparison-col-All}}\\
\midrule
\UseMacro{table-hole_types_comparison-row0-Kind} & \UseMacro{table-hole_types_comparison-row0-Id} & \UseMacro{table-hole_types_comparison-row0-Val} & \UseMacro{table-hole_types_comparison-row0-ArithmeticAndShift} & \UseMacro{table-hole_types_comparison-row0-RelationAndLogic} & \UseMacro{table-hole_types_comparison-row0-All}\\
\UseMacro{table-hole_types_comparison-row1-Kind} & \UseMacro{table-hole_types_comparison-row1-Id} & \UseMacro{table-hole_types_comparison-row1-Val} & \UseMacro{table-hole_types_comparison-row1-ArithmeticAndShift} & \UseMacro{table-hole_types_comparison-row1-RelationAndLogic} & \UseMacro{table-hole_types_comparison-row1-All}\\
\UseMacro{table-hole_types_comparison-row2-Kind} & \UseMacro{table-hole_types_comparison-row2-Id} & \UseMacro{table-hole_types_comparison-row2-Val} & \UseMacro{table-hole_types_comparison-row2-ArithmeticAndShift} & \UseMacro{table-hole_types_comparison-row2-RelationAndLogic} & \UseMacro{table-hole_types_comparison-row2-All}\\
\UseMacro{table-hole_types_comparison-row3-Kind} & \UseMacro{table-hole_types_comparison-row3-Id} & \UseMacro{table-hole_types_comparison-row3-Val} & \UseMacro{table-hole_types_comparison-row3-ArithmeticAndShift} & \UseMacro{table-hole_types_comparison-row3-RelationAndLogic} & \UseMacro{table-hole_types_comparison-row3-All}\\
\bottomrule
\end{tabular}

\end{footnotesize}
\vspace{-5pt}
\end{table}

To explore how different types of \templates affect \Tool's bug
detection, we extract different sets of \templates from the same
\NumOfProjectsUsed projects (Section~\ref{sec:eval:setup}). We
modified the \template extraction algorithm
(Figure~\ref{algo:extraction}) so that each extracted set of
\templates contains a single set of specific types of \holes out of
(1)~\CodeIn{<\aType{}>Id}, (2)~\CodeIn{<\aType{}>Val},
(3)~\CodeIn{arithmetic()} and \CodeIn{shift()},
(4)~\CodeIn{<\aType{}>relation} and \CodeIn{<\aType{}>logic}. We also
extract a set of \templates with all types of \holes, which is the
default setting. Other than the \Phaseextraction phase, we use the
same methodology and configuration as described in
Section~\ref{sec:eval:setup} to run \Tool with the five sets of
\templates.
Table~\ref{tab:hole-types-comparison} shows the numbers of \templates,
\generatedprograms, reported failures and bugs from all five sets of
\templates. The set of \templates with all types of \holes reports the
most number of bugs. Out of the other four sets of \templates with
only a single set of \holes, the \CodeIn{relation()} and
\CodeIn{logic()} \holes reports the most number of bugs, but even the
simplest set of holes, i.e., constant replacement
(\CodeIn{<\aType{}>Val}), plays an important role.
Furthermore, we discovered \NumOfBugsFoundInEvalOfTemplateTypes
additional \JIT bugs using these various sets of templates.

\subsection{Detected Bugs}
\label{sec:bugs}

Table~\ref{tab:bugs} lists the bugs that \Tool detected.
So far, we have discovered and reported \TotalNumOfBugs
bugs, \NumOfPreviouslyUnknownBugs of which are previously unknown,
including \TotalNumOfCVEs CVEs.
We show (in Figure~\ref{fig:evaluation:detectedbugs}) and describe
four bugs
that encompass a variety of \JIT issues.

\begin{table}[!t]
\centering
\caption{\UseMacro{table-bugs-caption}}
\begin{footnotesize}

\begin{tabular}{lrrrrrr}
\toprule
\textbf{\UseMacro{table-bugs-col-jvm}} & \textbf{\UseMacro{table-bugs-col-id}} & \textbf{\UseMacro{table-bugs-col-type}} & \textbf{\UseMacro{table-bugs-col-affected_versions}} & \textbf{\UseMacro{table-bugs-col-status}} & \textbf{\UseMacro{table-bugs-col-cve}} & \textbf{\UseMacro{table-bugs-col-duplicate}}\\
\midrule
\multirow{1}{*}{\UseMacro{table-bugs-row0-jvm}} & \UseMacro{table-bugs-row0-id} & \UseMacro{table-bugs-row0-type} & \UseMacro{table-bugs-row0-affected_versions} & \UseMacro{table-bugs-row0-status} & \UseMacro{table-bugs-row0-cve} & \UseMacro{table-bugs-row0-duplicate}\\
\cmidrule{1-7}
\multirow{5}{*}{\UseMacro{table-bugs-row1-jvm}} & \UseMacro{table-bugs-row1-id} & \UseMacro{table-bugs-row1-type} & \UseMacro{table-bugs-row1-affected_versions} & \UseMacro{table-bugs-row1-status} & \UseMacro{table-bugs-row1-cve} & \UseMacro{table-bugs-row1-duplicate}\\
& \UseMacro{table-bugs-row2-id} & \UseMacro{table-bugs-row2-type} & \UseMacro{table-bugs-row2-affected_versions} & \UseMacro{table-bugs-row2-status} & \UseMacro{table-bugs-row2-cve} & \UseMacro{table-bugs-row2-duplicate}\\
& \UseMacro{table-bugs-row3-id} & \UseMacro{table-bugs-row3-type} & \UseMacro{table-bugs-row3-affected_versions} & \UseMacro{table-bugs-row3-status} & \UseMacro{table-bugs-row3-cve} & \UseMacro{table-bugs-row3-duplicate}\\
& \UseMacro{table-bugs-row4-id} & \UseMacro{table-bugs-row4-type} & \UseMacro{table-bugs-row4-affected_versions} & \UseMacro{table-bugs-row4-status} & \UseMacro{table-bugs-row4-cve} & \UseMacro{table-bugs-row4-duplicate}\\
& \UseMacro{table-bugs-row5-id} & \UseMacro{table-bugs-row5-type} & \UseMacro{table-bugs-row5-affected_versions} & \UseMacro{table-bugs-row5-status} & \UseMacro{table-bugs-row5-cve} & \UseMacro{table-bugs-row5-duplicate}\\
\cmidrule{1-7}
\multirow{9}{*}{\UseMacro{table-bugs-row6-jvm}} & \UseMacro{table-bugs-row6-id} & \UseMacro{table-bugs-row6-type} & \UseMacro{table-bugs-row6-affected_versions} & \UseMacro{table-bugs-row6-status} & \UseMacro{table-bugs-row6-cve} & \UseMacro{table-bugs-row6-duplicate}\\
& \UseMacro{table-bugs-row7-id} & \UseMacro{table-bugs-row7-type} & \UseMacro{table-bugs-row7-affected_versions} & \UseMacro{table-bugs-row7-status} & \UseMacro{table-bugs-row7-cve} & \UseMacro{table-bugs-row7-duplicate}\\
& \UseMacro{table-bugs-row8-id} & \UseMacro{table-bugs-row8-type} & \UseMacro{table-bugs-row8-affected_versions} & \UseMacro{table-bugs-row8-status} & \UseMacro{table-bugs-row8-cve} & \UseMacro{table-bugs-row8-duplicate}\\
& \UseMacro{table-bugs-row9-id} & \UseMacro{table-bugs-row9-type} & \UseMacro{table-bugs-row9-affected_versions} & \UseMacro{table-bugs-row9-status} & \UseMacro{table-bugs-row9-cve} & \UseMacro{table-bugs-row9-duplicate}\\
& \UseMacro{table-bugs-row10-id} & \UseMacro{table-bugs-row10-type} & \UseMacro{table-bugs-row10-affected_versions} & \UseMacro{table-bugs-row10-status} & \UseMacro{table-bugs-row10-cve} & \UseMacro{table-bugs-row10-duplicate}\\
& \UseMacro{table-bugs-row11-id} & \UseMacro{table-bugs-row11-type} & \UseMacro{table-bugs-row11-affected_versions} & \UseMacro{table-bugs-row11-status} & \UseMacro{table-bugs-row11-cve} & \UseMacro{table-bugs-row11-duplicate}\\
& \UseMacro{table-bugs-row12-id} & \UseMacro{table-bugs-row12-type} & \UseMacro{table-bugs-row12-affected_versions} & \UseMacro{table-bugs-row12-status} & \UseMacro{table-bugs-row12-cve} & \UseMacro{table-bugs-row12-duplicate}\\
& \UseMacro{table-bugs-row13-id} & \UseMacro{table-bugs-row13-type} & \UseMacro{table-bugs-row13-affected_versions} & \UseMacro{table-bugs-row13-status} & \UseMacro{table-bugs-row13-cve} & \UseMacro{table-bugs-row13-duplicate}\\
& \UseMacro{table-bugs-row14-id} & \UseMacro{table-bugs-row14-type} & \UseMacro{table-bugs-row14-affected_versions} & \UseMacro{table-bugs-row14-status} & \UseMacro{table-bugs-row14-cve} & \UseMacro{table-bugs-row14-duplicate}\\
\bottomrule
\end{tabular}

\end{footnotesize}
\vspace{-5pt}
\end{table}

\MyPara{\BugOne}
A mis-compilation occurred when the \OpenJNine \JIT performed a
modular operation with parameter passing
(Figure~\ref{fig:evaluation:bugone}). We discovered the bug using a
\template created from \commonsmath
\cite{commonsmathQuaternionSource}.
The issue lies in an incorrect reuse of a register whose value
changes after a floating-point remainder operation.

\MyPara{\BugTwo}
From a \template extracted from
\commonsmath~\cite{commonsmathAdamNordsieckTransformerSource}, we
discovered a \HotSpot \JIT mis-compilation bug where the range check
for array accesses was incorrectly eliminated, which missed throwing
exceptions and produced incorrect results
(Figure~\ref{fig:evaluation:bugtwo}).
Upon reporting the bug, Oracle developers promptly confirmed the issue.
They classified the bug as a CVE and rolled out the fix in the next
Critical Patch Update.

\begin{figure}
\begin{subfigure}[b]{0.49\linewidth}
\begin{lstlisting}[language=bug]
public class C {
  double q0, q1, q2, q3;
  C(double a0, double a1, double a2, double a3) {
    q0 = a3; q1 = a1; q2 = 0; q3 = 0; }
  static double m(double d) {
    C c = new C(0, 1.0, 0, d % d);
    return c.q1; }
  public static void main(String[] args) {
    double sum = 0;
    for (int i = 0; i < 100_000; ++i) {
      // m(1.0) expected to be 1.0 returns 0.0
      sum += m(1.0); }
    // expected 100000.0
    System.out.println(sum); } }
\end{lstlisting}
\vspace{-5pt}
\caption{\UseMacro{figure-evaluation-bugone}}
\end{subfigure}
\begin{subfigure}[b]{0.49\linewidth}
\begin{lstlisting}[language=bug]
public class C {
  static void m(int n) {
    int[] a = new int[n];
    for (int i = 0; i < 1; i++) {
      int x = a[i % -1]; } }
  public static void main(String[] args) {
    int count = 0;
    for (int i = 0; i < 1000; ++i) {
      try { m(0);
      } catch (ArrayIndexOutOfBoundsException e) {
        count += 1; } }
    System.out.println(count); } } // expect 1000
\end{lstlisting}
\vspace{-5pt}
\caption{\UseMacro{figure-evaluation-bugtwo}}
\end{subfigure}
\begin{subfigure}[b]{0.49\linewidth}
\begin{lstlisting}[language=bug]
public class C {
  static int m(int len) {
    int[] arr = new int[8];
    for (int i = 10000000, j = 0;
         (boolean) (i >= 1) && j < 100; i--, j++) {
      // should not enter inner loop.
      for (int k = 0; len < arr.length; ++k) {
        int x = 1 / 0; }
    } return 0; }
  public static void main(String[] args) {
    int sum = 0;
    for (int i = 0; i < 100_000; ++i) {
      try { m(13);
      } catch (ArithmeticException e) { sum += 1; }
    } System.out.println(sum); } } // expected 0
\end{lstlisting}
\vspace{-5pt}
\caption{\UseMacro{figure-evaluation-bugthree}}
\end{subfigure}
\begin{subfigure}[b]{0.49\linewidth}
\begin{lstlisting}[language=bug]
static import java.nio.charset.StandardCharsets;
public class C {
  static int m(String s) {
    byte[] arr = s.getBytes(ISO_8859_1);
    return arr[2]; }
  public static void main(String[] args) {
    long sum = 0;
    for (int i = 0; i < 10_000_000; ++i) {
      sum += m("\u8020\000\000\020"); }
    System.out.println(sum); } } // expected 0
\end{lstlisting}
\vspace{-5pt}
\caption{\UseMacro{figure-evaluation-bugfour}}
\end{subfigure}
\vspace{-5pt}
\caption{\UseMacro{figure-evaluation-detectedbugs}}
\end{figure}

\MyPara{\BugThree}
Execution of \OpenJNine \JIT-compiled code faced a situation where a
loop condition was incorrectly evaluated as true, enabling the loop
body to run (Figure~\ref{fig:evaluation:bugthree}). However, the loop
body should never execute, and this correct behavior was observed in
non-\JIT executions. \Tool flagged this issue as a bug using a
\template from \commonscodec~\cite{commonscodecBinaryCodecSource}.
IBM developers confirmed the bug within a day.

\MyPara{\BugFour}
An incorrect output occurred when using \GraalVM \JIT compilation with
the String \CodeIn{getBytes} method
(Figure~\ref{fig:evaluation:bugfour}).
The \generatedprogram by \Tool emerged from a \template based off code
from \commonscodec~\cite{commonscodecStringUtilsSource}.
The developers confirmed the bug within one day.

Bugs detected with \Tool are presented in an easily digestible manner.
\Generatedprograms are easy to minimize and understand, because \Tool
extracts \templates from real-world Java programs and the minimum
example programs we submitted are Java source code. Developers were
able to quickly understand our reports and reproduce or further
minimize source code as needed. Many reports were confirmed by the
first 48 hours. In contrast, bytecode files generated by some tools
require substantial effort to understand by compiler
developers~\cite{CommentOnByteCodeHardToDebug}.

\section{Limitations and Future Work}
\label{sec:discussion}

When we compared \Tool with
\JITfuzz, we used only the \ApproachtestBased approach, and we were
unable to successfully run another tool:
\JOpFuzzer~\cite{jia23jopfuzzer}.
Also, when we evaluated \Tool variants, we did not attempt to match
end-to-end duration of \ToolOnlyRandoop, or \ToolWoRandoop,
and end-to-end duration of \ToolTestBased, or \ToolPoolBased.
The randomness could impact the experiment results, so we run each
experiment \NumRepeatedRuns times, as we already described.

Our current implementation of the \Tool \ApproachtestBased approach
collects arguments to \templates by extending one tool for test
generation. The same methodology can also work for manually-written
tests or other tools~\cite{Fraser11EvoSuite}, which we plan to explore
in the future.
Also, it will be easy to migrate \Tool to test other software systems
that also take Java programs as input, such as refactoring tools.

\Tool has enhanced \JAttack to support non-primitive static fields in
\templates by re-initializing the \template class, but \Tool is
limited on re-initializing other classes in dependencies.
While \JAttack's exception handling has been enhanced to handle
exceptions thrown from class loading and initializing, \Tool does not
handle errors directly thrown from JVM in the \Phasegeneration and
\Phasetesting phase, e.g., \CodeIn{StackOverflowError},
\CodeIn{OutOfMemoryError}, etc.
These shortcomings sometimes left unfilled \holes that were supposed
to be filled, leading to false positives in the testing phase.
We will further explore re-initializing all classes in dependencies
and better handling of \JVM errors.

\MyPara{Ethical considerations}
To avoid ``spamming'' open-source community, we submit a bug report
only when we can reproduce the bug on the latest release of the
affected JDKs.
We also tried our best to detect duplicates and minimize the programs
that reproduce the bugs.

\section{Related Work}
\label{sec:related}

We describe closely related work on compiler
testing~\cite{Chen20SurveyofCompilerTesting} using grammar-based,
mutation-based and template-based approaches, and test input generation
in general.

\MyPara{Grammar-based}
These tools use the grammar production rules of the language to
generate test programs. \Csmith~\cite{YangPLDI11CSmith},
Orange~\cite{Nakamura16EMIForCCompilers} and
YARPGen~\cite{Livinskii20YARPGen} are grammar-based tools for
generating C/C++ test programs.
Lava~\cite{SirerETAL00UsingProductionGrammarsInSoftwareTesting} uses
grammar production rules to generate Java bytecode test programs for
testing the JVM. Java* Fuzzer~\cite{JavaFuzzerAzulGitHubPage} is a
grammar-based tool that generates random Java programs for testing
\JIT compilers. Yoshikawa et
al.~\cite{YoshikawaETAL03RandomGeneratorForJIT} proposed an approach
for generating test programs following grammar for testing Java
JIT compilers. Unlike all these techniques, \Tool generates Java
programs by utilizing the structures of real-world Java programs and
explores possibilities of expressions by inserting and refilling
\holes.

\MyPara{Mutation-based}
A tool that implements a mutation-based approach generates new test
programs by mutating existing programs, with several techniques and
tools specifically for Java~\cite{ChenPLDI16DifferentialTestingOfJVM,
ChenICSE19DeepDifferentialTestingOfJVMImplementations,
Vikram21BonsaiFuzzing, Chaliasos22FindingTypingCompilerBugs,
Zhao22JavaTailor, Gao22VECT}.
Classfuzz~\cite{ChenPLDI16DifferentialTestingOfJVM} and
Classming~\cite{ChenICSE19DeepDifferentialTestingOfJVMImplementations}
are coverage-guided fuzzers that mutate seeds by modifying their
syntactic structures and
control/data flows.
VECT~\cite{Gao22VECT} improves \JavaTailor~\cite{Zhao22JavaTailor} via
grouping by code vectorization code ingredients, which are collected
from history bug-revealing programs in OpenJDK tests, to insert into
seed programs.
\JOpFuzzer~\cite{jia23jopfuzzer} uses \JIT optimization options as a
new dimension for test input in conjunction with profile data of JIT
compilers for guiding the fuzzing process. It relies on Java* Fuzzer
to generate seed programs.
\textsc{ComFuzz}~\cite{Ye23ComFuzz} is a recent compiler fuzzing
framework that leverages existing test cases that have generated
compiler bugs and deep learning to learn language features that are
more likely to generate such bugs.
SJFuzz~\cite{Wu23SJFuzz} is a JVM fuzzer that aims to address the
lack of guidance in fuzzing due to the absence of well-designed seeds
and mutator scheduling. It aims to automate the process of JVM
differential testing via mutating class files using control flow
mutators to identify inter-JVM discrepancies.
Artemis~\cite{Li23Artemis} explores various \JIT compilation traces as
differential testing, via diverse combinations of
optimized/de-optimized method invocations.
Although \Tool is similar in nature to concepts in mutation
testing~\cite{DeMillo78MutationTesting}, via extracting \templates and
then using them to generate concrete programs, \Tool has several
differences. (1)~Mutation testing uses a predefined set of mutation
operators. However, each \hole in a \template has its own set of
values (and the set can be dynamically determined; see the next
point). (2)~\Tool fills \holes dynamically (rather than statically),
which brings unique advantages: (a) uses meaningful variables to fill
\holes; (b) allows \hole construction with runtime information, e.g.
length of array; (c) allows \Tool to establish which \holes are in
dead code and, in the \Phasegeneration phase, focus on exploring the
reachable \holes. (3)~\Tool can fill multiple \holes simultaneously,
which is more similar to higher-order
mutation~\cite{Jia08HigherOrderMutationTesting}, during the execution
of the \template.  (4)~\Tool introduces an approach for obtaining
objects necessary for running templates.

\MyPara{Template-based}
Ching and Katz~\cite{ChingKatz93TestingAPLCompiler} use templates
constructed from existing code bases to generate concrete programs for
testing the APL-to-C compiler COMPC.
Zhang et al.~\cite{ZhangPLDI17SkeletalProgramEnumeration} statically
mutate \CodeIn{int} variable occurrences in existing programs for
testing C/C++ compilers.
The original \JAttack work presented an automated \template extraction
technique~\cite{zang22jattack}. \Tool's \ApproachpoolBased extraction
is similar to this previous approach, but contains several key
differences. (1)~\JAttack converts expressions with limited types,
i.e., \CodeIn{boolean}, \CodeIn{int}, \CodeIn{double}, and skips
expressions in constructors, while \Tool converts all expressions to
all \holes supported; (2)~\JAttack converts only static methods while
\Tool can convert any method into a \template; (3)~\JAttack searches
in all classes of the project for a public constructor or static
method, with no parameter or only primitive parameters, to create
reference argument for the \template, and uses \CodeIn{null} if it
cannot find such a constructor or static method. In contrast, \Tool
uses test generation to generate method sequences that create desired
objects.

\MyPara{Test input generation}
JUnit tests can be automatically generated using
\Randoop~\cite{Pacheco07Randoop} and EvoSuite~\cite{Fraser11EvoSuite}
for a given set of classes under test. In theory, the tests generated
from such tools can be directly used to test Java \JIT, and we
implemented such a prototype using \Randoop and evaluated its
bug-discovery effectiveness within our evaluation of \Tool (see
Section~\ref{sec:components}).
Popularized by QuickCheck~\cite{Claessen00QuickCheck}, another
approach is to allow developers to write generators from which valid
test inputs can be obtained.
Several approaches~\cite{BoyapatiISSTA02Korat, DanielFSE07ASTGen,
GligoricICSE10UDITA, CelikETAL17intKorat}
use the generators to exhaustively enumerate all possible paths
through the generators up to a given bound.
Theoretically, these tools can be used to test Java \JIT as well, but
they would require developers to manually write generators for Java
programs that include various language features; writing those
generators manually can be a tedious process.
In contrast, \Tool allows generating programs by concretizing
\templates extracted from open-source projects without developer
intervention.

\section{Conclusion}
\label{sec:conclusion}

We presented a framework, \Tool, which enables fully automated
end-to-end template-based testing of \JIT compilers.  \Tool can create
a template from any Java method, and it automatically inserts holes
and generates necessary arguments for the template.  To obtain
instances of complex types needed for extracted templates, \Tool uses
novel techniques built on automated test generation.
We have extensively evaluated \Tool by generating \TAveNumOfPrograms
programs and discovered \TotalNumOfBugs bugs in three popular and
widely used compilers.
Our findings show the power of automating template extraction via
\Tool and the power of scaling experiments without humans in the loop,
as well as complementary power compared to the state-of-the-art tools
for \JIT and \JVM testing techniques.
We believe that \Tool should become an integral part of continuous
testing for any Java \JIT compiler.

\begin{acks}
We thank Nader Al Awar, Yu Liu, Pengyu Nie, Jiyang Zhang, and the
anonymous reviewers for their comments and feedback.
This work is partially supported by a Google Faculty Research Award, a
grant from the Army Research Office, and the US National Science
Foundation under Grant Nos.~CCF-2107291, CCF-2217696, CCF-2313027.
\end{acks}

\bibliography{bib}

\end{document}